\documentclass[11pt]{book}

\usepackage[latin1]{inputenc} %accents dans les fichiers
\usepackage[T1]{fontenc} %accents dans le dvi
\usepackage{fancyhdr}
\usepackage{graphicx,multicol}

\usepackage{amsfonts}
\usepackage{amssymb}
\usepackage{amsmath}
\usepackage{comment}
\usepackage{mathrsfs}
\usepackage{appendix}

\renewcommand{\baselinestretch}{1.2}

\newcommand{\beq}{\begin{equation}}
\newcommand{\eeq}{\end{equation}}
\newcommand{\bea}{\begin{eqnarray}}
\newcommand{\eea}{\end{eqnarray}}
\newcommand{\ba}{\begin{array}}
\newcommand{\ea}{\end{array}}
\newcommand{\bi}{\begin{itemize}}
\newcommand{\ei}{\end{itemize}}
\newcommand{\bc}{\begin{center}}
\newcommand{\ec}{\end{center}}

\newcommand{\ie}{{\it{i.e.}\quad}}
\newcommand{\eg}{{\it{e.g. }}}
\newcommand{\tbf}[1]{{\mathbf{#1}}}
\renewcommand{\rm}[1]{{\mathrm{#1}}}

\newcommand{\eps}{\varepsilon}
\newcommand{\e}{\varepsilon}
\renewcommand{\epsilon}{\varepsilon}
\newcommand{\al}{\alpha}
\newcommand{\be}{\beta}
\newcommand{\la}{\lambda}
\newcommand{\lam}{\lambda}
\newcommand{\D}{\Delta}
\newcommand{\Ra}{\rightarrow}

\newcommand{\LRa}{\leftrightarrow}
\newcommand{\IM}[1]{\rm{Im}\left\lbrace#1\right\rbrace }

\newcommand{\eV}{\,{\rm {eV}}}
\newcommand{\MeV}{\,{\rm {MeV}}}
\newcommand{\GeV}{\,{\rm{GeV}}}
\newcommand{\TeV}{\,{\rm {TeV}}}
\newcommand{\sla}[1]{\slash{} \hspace{-0.2cm}#1}
\newcommand{\ol}{\overline}

\newcommand{\diag}{{\rm{\diag}}}
\def \unit{\leavevmode\hbox{\small1\kern-3.6pt\normalsize1}}

\newcommand{\aell}[1]{a_{\ell_{#1}}}
\newcommand{\aelld}[1]{a_{\ell_{#1}}^{\dagger}}

\newcommand{\atau}{a_{\tau^{c}}}
\newcommand{\ataud}{a_{\tau^{c}}^{\dagger}}
\newcommand{\abtau}{a_{\bar{\tau^{c}}}}
\newcommand{\abtaud}{a_{\bar{\tau^{c}}}^{\dagger}}
\newcommand{\aN}{a_{N}}
\newcommand{\aNd}{a_{N}^{\dagger}}
\newcommand{\aphi}{a_{\phi}}
\newcommand{\aphid}{a_{\phi}^{\dagger}}
\newcommand{\fell}[1]{f_{\ell}^{#1}}
\newcommand{\fbell}[1]{f_{\bar{\ell}}^{#1}}
\newcommand{\tk}[1]{\tilde{\kappa}_{#1}}

\renewcommand{\thefigure}{(\arabic{chapter}.\arabic{figure})}

\newcommand{\captionfonts}{\small}

\makeatletter  % Allow the use of @ in command names
\long\def\@makecaption#1#2{%
  \vskip\abovecaptionskip
  \sbox\@tempboxa{{\captionfonts #1: #2}}%
  \ifdim \wd\@tempboxa >\hsize
    {\captionfonts #1: #2\par}
  \else
    \hbox to\hsize{\hfil\box\@tempboxa\hfil}%
  \fi
  \vskip\belowcaptionskip}
\makeatother   % Cancel the effect of \makeatletter

\def\unit{\relax{\rm 1\kern-.26em I}}

\fancypagestyle{plain}{
	\fancyhead{}
	
}

\begin{document}
\begin{titlepage}

\includegraphics[trim=1mm 1mm 1mm 1mm, clip,scale=0.3]{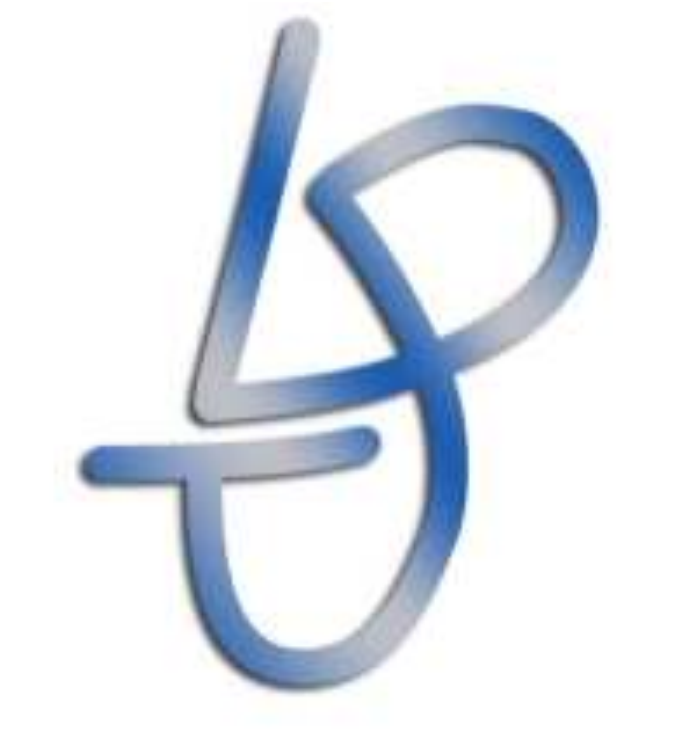}
\hfill
\includegraphics[trim=1mm 1mm 1mm 1mm, clip, scale=0.1]{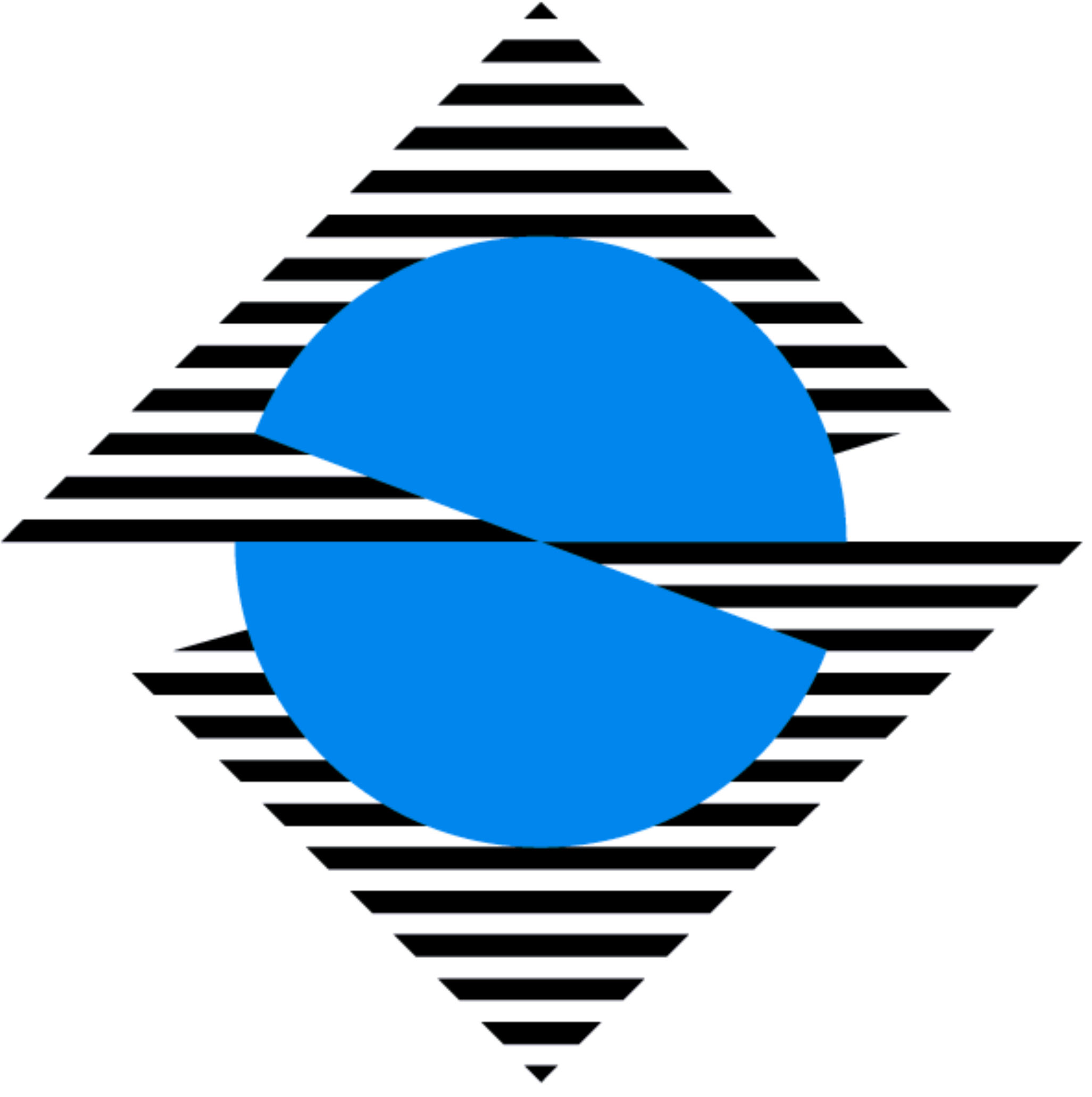}
\vspace*{2cm}
\begin{center}
\fbox{{\bf UNIVERSIT\'E PARIS-SUD XI}}

\vskip1cm

{\bf TH\`ESE}
\vskip1cm

Sp\'ecialit\'e: {\bf  PHYSIQUE TH\'EORIQUE}\\
\vskip0.3cm
Pr\'esentée\\
 pour obtenir le grade de
\vskip0.75cm
\large {\bf Docteur de l'Université Paris XI}

\vskip0.75cm

par

\vskip0.5cm

{\sc \bf François-Xavier Josse-Michaux}

\vskip1cm

Sujet: 

\vskip0.2cm
\Large{\bf Recent developments in thermal leptogenesis:\\
the role of flavours in various seesaw realisations.}

\end{center}
\vskip0.75cm
\begin{flushleft}

Soutenue le mercredi 25 juin 2008 devant la commission d'examen:\\
\vskip0.75cm
$\begin{array}{lll}
\mbox{MM}. & \mbox{Asmaa Abada}, & \mbox{directrice de th\`ese},\\
   & \mbox{Sacha Davidson}, & \\
   & \mbox{Ulrich Ellwanger}, & \mbox{Président du jury},\\ 
&  \mbox{Belen Gavela}, & \mbox{rapporteur},\\
  & \mbox{Thomas Hambye}, & \\  
  & \mbox{Michael Pl\"umacher} &\mbox{rapporteur}.
   \end{array}$

\end{flushleft}
\end{titlepage}

\thispagestyle{empty}
\vspace*{2.5cm}
~\\
% \newpage
% \cleardoublepage
\thispagestyle{empty}
\thispagestyle{empty}
\vspace*{2.5cm}
~\\
\newpage
\thispagestyle{empty}

\includegraphics[trim=1mm 1mm 1mm 1mm, clip, scale=0.3]{logo-LPT.pdf}
\hfill
\includegraphics[trim=1mm 1mm 1mm 1mm, clip, scale=0.1]{logoP11.pdf}
\vspace*{2cm}
\begin{center}
\fbox{{\bf UNIVERSIT\'E PARIS-SUD XI}}

\vskip1cm

{\bf TH\`ESE}
\vskip1cm

Sp\'ecialit\'e: {\bf  PHYSIQUE TH\'EORIQUE}\\
\vskip0.3cm
Pr\'esentée\\
 pour obtenir le grade de
\vskip0.75cm
\large {\bf Docteur de l'Université Paris XI}

\vskip0.75cm

par

\vskip0.5cm

{\sc \bf François-Xavier Josse-Michaux}

\vskip1cm

Sujet: 

\vskip0.2cm

\Large{\bf Recent developments in thermal leptogenesis:\\
the role of flavours in various seesaw realisations.}

\end{center}
\vskip0.75cm
\begin{flushleft}

Soutenue le mercredi 25  juin 2008 devant la commission d'examen:\\
\vskip0.75cm
$\begin{array}{lll}
\mbox{MM}. & \mbox{Asmaa Abada}, & \mbox{directrice de th\`ese},\\
   & \mbox{Sacha Davidson}, & \\
   & \mbox{Ulrich Ellwanger}, & \mbox{Président du jury},\\ 
&  \mbox{Belen Gavela}, & \mbox{rapporteur},\\
  & \mbox{Thomas Hambye}, & \\  
  & \mbox{Michael Pl\"umacher} &\mbox{rapporteur}.
   \end{array}$

\end{flushleft}

\newpage
\thispagestyle{empty}
\vspace*{2.5cm}
~\\
% \cleardoublepage
\thispagestyle{empty}
\thispagestyle{empty}
%\vspace*{2.5cm}
~\\

\chapter*{Remerciements}
\addcontentsline{toc}{chapter}{Remerciements}
Ce petit manuscrit pour clore ces trois années de thèse, ce petit chapitre pour remercier ceux grâce à qui ce travail à pu aboutir.\\
Avant d'arriver sur Paris, je ne savais pas encore ce que je voulais faire, si même je voulais devenir physicien. Alors merci à mes prof. de Paris 7, en particulier à Jean-Pierre Gazeau, pour ces cours de maths vivifiants du lundi matin, ainsi que pour le stage de License; merci à Jim Bartlett, à Pascal David et à Luc Valentin. Ces personnes m'ont donné envie de faire de la Physique, d'enseigner la Physique.\\
Si un jour cette possibilité m'est offerte, je le dois à ma directrice de thèse, Asmaa Abada. Merci Asmaa pour tes conseils et ton soutien, pour ta compréhension, merci pour ta gentillesse. Merci de m'avoir poussé à m'affirmer un peu plus. Merci de m'avoir enrôlé dans tes fructueux-savoureux projets, et de m'avoir permis de rencontrer de talentueux physiciens.\\
Merci donc à mes collaborateurs, en particulier à Sacha Davidson pour nos nombreux échanges, et qui de plus a accepté de faire partie de mon jury de thèse. Merci à Alessandro Ibarra, Marta Losada et à Antonio Riotto, en particulier de m'avoir accueilli au CERN pendant quelques jours. Merci à Stéphane Lavignac et à Pierre Hosteins-gageons que ce papier sorte un jour!\footnote{Le papier est sorti! Il pèse 950 Ko et mesure 32 pages. Souhaitons-lui un bel avenir!}\\
Je tiens à remercier Ulrich Ellwanger pour avoir accepté de présider mon jury de thèse, ainsi que les rapporteurs Belen Gavela -notamment pour ses remarques concernant ce manuscrit- et Michael Plümacher pour avoir accepté cette tache. Merci enfin à Thomas Hambye d'avoir accepté de faire partie de mon jury de thèse. J'ai hâte ou j'espère pouvoir travailler un jour avec vous.\\
Merci à Ana Margarida Teixeira, pour ses très nombreux coups de main, et en particulier, merci infiniment pour le travail titanesque que tu as accomplis en corrigeant ce manuscrit, qui te doit beaucoup.\\
Salutations enfin aux camarades-collègues/amis, la jeunesse du labo. Vu la croissance du nombre de thésards au LPT, bientôt le block entier n'y suffira pas: ce sera tant mieux! Salut donc aux troisième années alexeï, benoît, yacine\footnote{Salutations spéciales aux collègues fumeurs.}, notamment pour les discussion politiques, salut aux deuxième année chloé, mathieu, pour les rires partagés.\\
Salut à mes collègue de promo 2008 du LPT: mathieu, razvan$^{1}$ et dérogation à l'ordre alphabétique pour nicolas : salut mon ami, et merci, tu étais là quand j'en avais besoin.\\
Salut aux "jeunes" maintenant, même si on devrait pas, car c'est bien connu les jeunes y respectent rien : salut aux "-1", benjamin-ph$^{1}$ et à benjamin-th$^{1}$, salut à florian, et bien joué pour avoir lancé le sinje, dont il faut dire je suis pas peu fier du nom (devrais-je?), et salut à jean.\\
Salut aux "-2": andreas$^{1}$, charles, charles-christophe, mounir, s-florian.\\
Salutations également aux postdocs que j'ai cotoyé: paramita, ritesh, irene, nils.\\
C'est sympa de causer avec vous.\\ 
Salutations aussi à mes anciens camarades du magistère, devenus amis:
justine, marine, julien, fabio \& steph, xav \& marie, miguel et marie-anne, maria et à ceux qu'on voit plus aussi: salut pascale et lihua! C'est vrai que ces années de magistère étaient chouettes, c'était bien de trouver des gens sympas avec qui causer et faire la fête. Maintenant qu'on va tous bouger à droite à gauche, j'espère qu'on continuera à ce voir!\\
Salut à mes potes de Rennes pierre, romain, malo, françois (grolo), erwann, tugdual...En particulier salut à mon colloc pierre, avec qui j'aurai vécu pendant la majeure partie de cette thèse, qui a du supporter mes tasses de cafés moisies, les mégots de cigarettes et la vaisselle s'empilant dans l'évier, sans trop broncher. Pierre, romain, merci les mecs de m'avoir soutenu dans les coups durs, et d'avoir été là quand il le fallait - ne serait-ce que pour faire la fête.\\
Si ces trois années ont été riches de joies, elles n'auront pas été avares de peines ni de douleurs. Aussi grandes furent-elles, tu as toujours été près de moi : merci, Constance, mon petit bout de femme. Sans toi, je ne sais pas si j'en serais là.

\newpage
\thispagestyle{empty}
\vspace*{2.5cm}
~\\

\begin{flushleft}
\hspace{8cm}`` Ayez une connaissance exacte\\
\hspace{8cm}et de d\'etail de tout ce qui vous environne;\\
\hspace{8cm}sachez o\`u il y a une f\^oret, un petit bois,\\
\hspace{8cm}une rivi\`ere, un ruisseau,\\
\hspace{8cm}un terrain aride et pierreux\\
\hspace{8cm}un lieu mar\'ecageux et malsain,\\
\hspace{8cm}une montagne, une colline,\\
\hspace{8cm}une petite \'el\'evation,\\
\hspace{8cm}un vallon, un pr\'ecipice,\\
\hspace{8cm}un d\'efil\'e, un champ ouvert,\\
\hspace{8cm}enfin tout ce qui peut servir ou nuire\\
\hspace{8cm}aux troupe que vous commandez.\\
\hspace{8cm}S'il arrive que vous soyez hors d'\'etat\\
\hspace{8cm}de pouvoir \^etre instruit par vous-meme\\
\hspace{8cm}de l'avantage ou du desavantage du terrain,\\
\hspace{8cm}ayez des guides locaux sur lesquels\\
\hspace{8cm}vous puissiez compter s\^urement."\\
\vspace*{1.5cm}
\hspace{8cm}\textbf{Sun-Tzu}, \textit{l'Art de la guerre}, art. 7.
\end{flushleft}
\newpage
\thispagestyle{empty}
\vspace*{2.5cm}
~\\

\newpage
\thispagestyle{empty}
\vspace*{2.5cm}
~\\

\begin{flushleft}
\hspace{10cm}\textit{à mon frère et à mes soeurs,}\\
\hspace{10cm}\textit{pierre, julie et charlotte.}
\end{flushleft}

\newpage
\thispagestyle{empty}
\vspace*{2.5cm}
~\\

\tableofcontents
%%%%%%%%%%%%%%%%%%%%%%%%%%%%%%%%%%%%%%%%%%%%%%%%%%%%%%%%%%%%%%%%%%%%%%%%%%%%%%%%%%%%%%%%%%%%%%%%%%%%%%%%%%%%%%%%%%%%%%%%%%%%%%%%%%%%%%%%%%%%%%%%%%%%%%%%%%%%%%%%%%
\renewcommand{\thechapter}{}
\renewcommand{\chaptername}{}
\chapter[Introduction]{Introduction}
%\addcontentsline{toc}{chapter}{Introduction}
The Standard Model of particle physics (SM) is a renormalisable quantum field theory, based on the $SU(3)\times SU(2)\times U(1)$ gauge group. This model is very successful, since it can explain \textit{almost} all observations. However, a few points  are unexplained. For example, among many other questions, we do not understand why there are three (and only three) generations of fermions, why the fermions exhibit such a mass pattern, why there is such a flavour structure among quarks or leptons. We do not know why the electrical charge is quantified, what is the origin of $CP$ violation and the anomally cancellation in the SM looks rather miraculous. If the Higgs mechanism indeed provides masses to the fermions, the stability of the Higgs mass under radiative corrections is a problem. There are also a few observations that possibly provide evidence for physics beyond the SM. Among the latter, is the origin of neutrino masses. About $3/4$ of our Universe is constituted of Dark Energy, the exact nature of it we do not grasp. Only $1/4$ of our Universe is made of matter, but only $15\%$ of this matter is known, the remaining part, the Dark Matter, being yet unknown.  Moreover, we do not know why is there matter at all, if the Universe was initially matter-antimatter symmetric.\\
In this thesis, we focus on two of these observations, namely the matter-antimatter asymmetry of the Universe and the fact that neutrinos are massive.\\
%While the former observation, properly saying, is not problem of the SM of particle, but rather a problem of the SM of cosmology, the matter asymmetry cannot be accommodated within the SM of particles, as we will see in chapter 1.\\
The dominance of matter over antimatter is an obvious and happy constatation that we do every day. Observations tell us however that matter dominates, but the domination is really small. This tiny number, the baryon asymmetry of the Universe, is unexplained in the SM. In chapter 1 we will discuss the most important observations, and present some of the possible explanations.\\
The other failure of the SM we are interested in stems from the observation of a deficit in solar and atmospheric neutrino fluxes. As we will explain in chapter 2,  these deficits imply that neutrinos are massive particles, and this really constitutes a first experimental evidence of physics beyond the SM.\\ %Actually, the neutrino have been "build" massless, as its mass scale is very smaller than the other particle-so small that it still have not been observed.\\
Among the different mechanisms advocated to explain neutrino masses, we consider in more detail the class of seesaw models, where some heavy fields which couple to light neutrinos are added to the SM particle content.\\
The seesaw model constitutes so far the preferred solution to the non-zero neutrino masses, mainly for two reasons. First, these heavy fields naturally emerge in larger gauge groups which attempt to unify the SM interactions. The second reason is that these fields provide an explanation for the observed matter-antimatter asymmetry, through the leptogenesis mechanism.\\
Chapters 3, 4, 5 and 6 are devoted to the study of leptogenesis, in which the heavy fields introduced can generate during their decay an excess of leptons over anti-leptons. The lepton asymmetry is then partly reprocessed into a baryon asymmetry. Chapter 3 can be viewed as an introduction to the work I have achieved during my thesis, which is discussed in chapters 4, 5 and 6.\\
In chapter 3, we discuss in detail leptogenesis, in particular the constraints under which does this mechanism accounts for the observed baryon asymmetry. Chapter 3 relies on the old wisdom based on the one-flavour approximation. This picture, even if yielding good estimates of the lepton asymmetry, is nevertheless incorrect.\\
Indeed, as explained in chapter 4, since the interactions involving charged leptons are usually in-equilibrium for the temperatures relevant in leptogenesis, lepton flavours have to be considered. We show in this chapter under which conditions the flavours must be included, and we show what are the good objects to study in this framework.\\
In chapter 5, we study the effects of lepton flavours in leptogenesis in the type I seesaw, where fermion singlets are added to the SM particle content. We show in this chapter how flavours affect leptogenesis. This chapter constitutes the main results obtained during my thesis, which are important results in all leptogenesis models, since the inclusion of flavours profoundly modifies the constraints derived in the one-flavour approximation.\\
Finally, in chapter 6 we investigate the effects of flavours in the type II seesaw, in which scalar triplets are added to the SM. The framework of this study is a supersymmetric grand-unified (SUSY-GUT) model, in which scalar triplets as well as fermion singlets naturally appear. In this chapter we study the influence of lepton flavours, as well as the effect of including second lightest fermion singlet in leptogenesis. We further correct the GUT relation between charged lepton and down-type quark masses. These inclusions and corrections are found to be fruitfull, since they render  this model of SUSY-GUT leptogenesis a viable mechanism to account for the observed baryon asymmetry.  
\newpage
%%%%%%%%%%%%%%%%%%%%%%%%%%%%%%%%%%%%%%%%%%%%%%%%%%%%%%%%%%%%%%%%%%%%%%%%%%%%%%%%%%%%%%%%%%%%%%%%%%%%%%%%%%%%%%%%%%%%%%%%%%%%%%%%%%%%%%%%%%%%%%%%%%%%%%%%%%%%%%%%%

\setcounter{chapter}{0}
\renewcommand{\thechapter}{\arabic{chapter}}
\renewcommand{\chaptername}{Chapter}
\chapter{Baryon Asymmetry of the Universe}
It is an every-day observation that matter dominates over anti-matter. On Earth, matter is present everywhere around us, whereas antimatter is only significantly produced in accelerators. %with $n_{\bar{b}}/n_{b}\simeq 10^{-4}$. 
The Solar system or our galaxie as well, are clearly matter-antimatter asymmetric. One could think that the Universe contains as much matter as anti-matter, each kind living apart in distinct regions. However, if it was so, one would expect annihilations $X+\bar{X}\Ra 2\gamma$ to occur on matter-antimatter borders, and therefore high-energy photons would be produced with a clear signature. As we have not observed such annihilation, we must conclude that our Universe is globally matter-antimatter asymmetric.\\
More precisely, two distinct cosmological observations, based on two very different physics, give us similar probes: in a volume which contains $10^{10}$ photons, the number of particles exceeds that of antiparticles by 6:
\bea
\eta_{B}=\frac{n_{b}-n_{\bar{b}}}{n_{\gamma}}\simeq 6 \times 10^{-10} \ .
\eea
This quantity $\eta_{B}$ is defined as the Baryon Asymmetry of the Universe (BAU)
\section{Evidence for a baryon asymmetry}
Let us briefly discuss the experimental evidences of the BAU.
\subsection{Big-Bang Nucleosynthesis}
\begin{figure}[htb]
\begin{center}
\includegraphics[height=9cm]{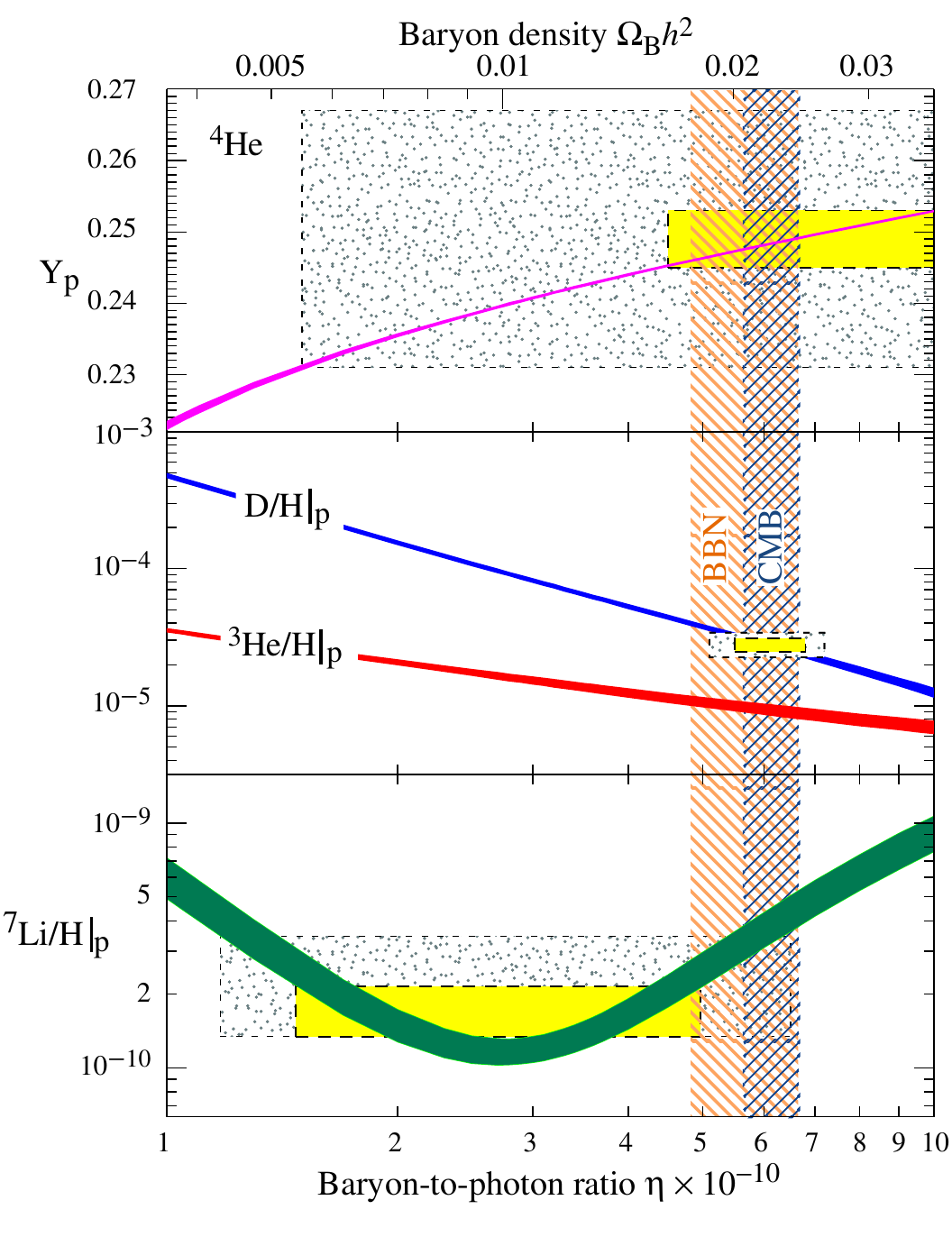} 
\caption{Abundance of light elements. Figure taken from~\cite{0601514}.}
\label{BBN}
\end{center}
\end{figure}

%\textbf{\large{revoir discussion deuterium/helium\\et donnees d'apres steigman de base + 0803.3465}}\\
Big-Bang Nucleosynthesis (BBN)~\cite{KolbTurner,BBN} is a remarkably trustful window on the early Universe, since it is based on well-known Standard Model physics. It predicts the abundance of light elements $D,\hbox{$^{3}He$},\hbox{$^{4}He$},\hbox{$^{7}He$}$ as a function of a small set of parameters, among which the matter-antimatter asymmetry $\eta_{B}$.\\
In the early Universe, at $T\simeq 20 \MeV$, baryon-antibaryon pairs annihilate, leaving any baryon excess frozen-out and constant in a comoving volume. This amount of baryons consists of protons and neutrons, which are in thermal equilibrium together with $e^{\pm}$ and neutrinos. For $T\gtrsim 1 \MeV$, these species undergo weak interactions:
\bea
 n+\nu_{e}\LRa p+e^{-}\,,\quad p+\ol{\nu_{e}}\LRa n+e^{+}\,, \quad n\LRa p+e^{-}+\ol{\nu_{e}}, \nonumber
\eea
which maintain the neutron-to-proton ratio at its equilibrium value:
\bea
\label{NPeq}
\frac{n_{n}^{eq}}{n_{p}^{eq}}\simeq e^{-\frac{m_{n}-m_{p}}{T}}\simeq e^{-\frac{1.293 \MeV}{T}} \ . 
\eea
At high temperatures, protons and neutrons are most likely free particles, as light nuclei are disfavoured, due to the high energetic gamma rays of the thermal bath.\\
Since the rate of weak interactions, $\Gamma_{n\LRa p}\simeq G_{F}^{2}\,T^{5}$, decreases faster than the Hubble expansion rate, $H(T)\simeq 1.66\,\sqrt{g_{*}\,G_{N}}\,T^{2}$, the electroweak interactions will freeze as the temperature of the Universe cools down. This occurs at $T_{fr}\simeq 0.8 \MeV$ and at this temperature, according to eq.(\ref{NPeq}), the neutron-to-proton ratio is about $n_{n}/n_{p}\simeq 1/6$.\\
During the cooling process, light nuclei become energetically favoured, as the temperature drops below their different binding energies. $\hbox{$^{4}He$}$, the more stable of them, is likely to be formed. However,  direct production, $2p+2n \Ra \hbox{$^{4}He$} +\gamma$, is highly inefficient due to the small cross-section of the electroweak processes, and also due to the relative low density of species. Therefore $\hbox{$^{4}He$}$ is more likely produced through a series of chain reactions, which are depicted in fig.\ref{BBNreaction}:
\begin{figure}[htb]
\begin{center}
\includegraphics[height=6cm]{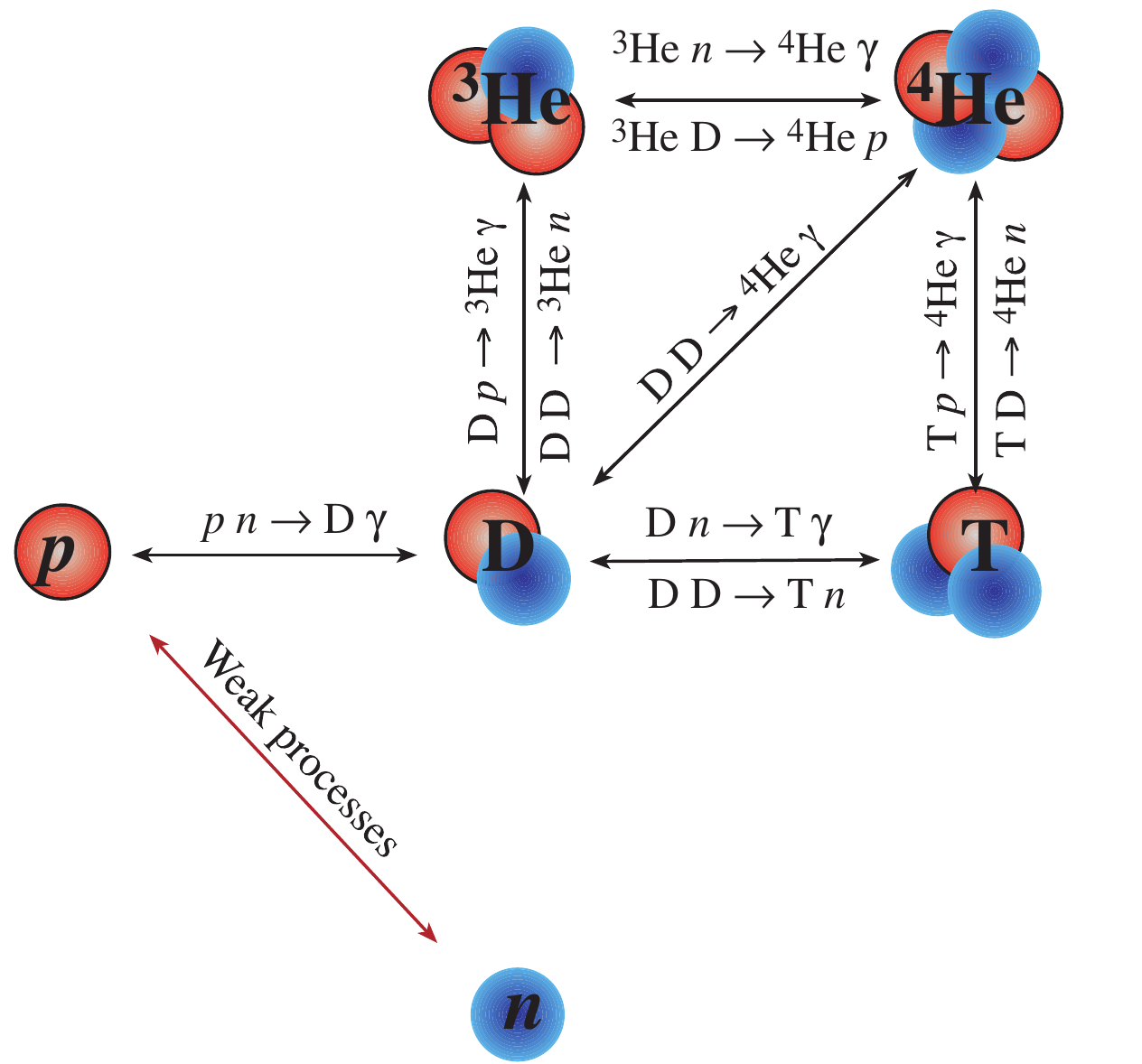} 
\caption{Schematic representation of the nucleosynthesis chain reaction. Picture taken from~\cite{StrumiaBaryoviaLepto}.}
\label{BBNreaction}
\end{center}
\end{figure}\\
%\bea
%p+n &\Ra & D+\gamma , \nonumber \\
%D+p &\Ra & \hbox{$^{3}He$}+\gamma , \nonumber \\
%D+n &\Ra & T+\gamma , \nonumber \\
%\hbox{$^{3}He$}+n &\Ra & \hbox{$^{4}He$}+\gamma , \nonumber\\
%T+p&\Ra & \hbox{$^{4}He$}+\gamma . \nonumber 
%\eea
The deuterium production is a prior to the chain reaction, through the process
\bea
p+n &\Ra & D+\gamma \, . \nonumber 
\eea
However, given its small binding energy, $E_{B}\simeq 2.2 \MeV$, as long as the gamma rays of the thermal bath are energetic enough to photo-dissociate $D$ via  $\gamma + D\Ra p+n$, no nuclei can be efficiently produced. This is the so-called "deuterium bottleneck".\\
The chain reaction begins when the deuterium to baryon ratio is $\sim 1$:
\bea
\frac{n_{D}}{n_{b}}\simeq \eta_{B}\left(\frac{T_{nuc}}{m_{p}}\right)^{3/2} e^{\frac{E_{B}}{T_{nuc}}}\simeq 1 \ . \nonumber
\eea
This happens for $T\simeq 0.06 \MeV$. Once deuterium production starts, the chain reaction rapidly occurs, until $\hbox{$^{4}He$}$ is synthethised. All neutrons are to be converted in $\hbox{$^{4}He$}$, which mass fraction is then roughly found to be:
\bea
Y_{p}=\frac{2 n_{n}}{n_{n}+n_{p}}\simeq 1/4 . \nonumber
\eea
Precise computations of this  mass fraction show a tiny dependance on the baryon asymmetry~\cite{BBN2}. A fit of $Y_{p}$, accurate over the value $\eta_{B}\simeq 6\times 10^{-10}$, gives~\cite{BBN3}:
\bea
Y_{p}\simeq 0.2485\pm 0.0006+0.0016(\eta_{10}-6)\, , 
\eea
where $\eta_{10}=10^{10} \eta_{B}$, is the normalised baryon-to-photon ratio.\\ 
In fact, the $\hbox{$^{4}He$}$ mass fraction is not a good indicator of the baryon-to-photon ratio. Rather, it strongly depends on the expansion rate, and so it is a good indicator of the time at which nucleosynthesis occured. Indeed, since in a first approximation all neutrons are captured in this nuclei, the later this capture happens, the longer time neutrons have to decay, and hence a lower mass fraction is obtained.\\
The prefered "baryometer" turns out to be deuterium, since its abundance directly results from the competition between production and destruction processes. The primordial $D$ to $H$ ratio is accurately fitted around its central value by~\cite{BBN3}:
\bea
y_{D}=2.64\left(1\pm 0.03\right)\left(\frac{6}{\eta_{10}}\right)^{1.6} \, ,
\eea
where $y_{D}=10^{5}(D/H)$. This ratio clearly denotes a stronger dependence on the baryon asymmetry.\\
Moreover, the post-BBN evolution of the deuterium is easy to track, the latter being only destroyed in stars. Therefore, observations of $D$ in "old" regions of the Universe yield the value of its primordial abundance:
\bea
y_{D}=2.68^{+0.27}_{-0.25} \, .
\eea
Using this observation, the baryon-to-photon ratio is found to be
\bea
\eta_{10}=6.0\pm 0.4\, ,
\eea
which is in excellent agreement with the value predicted by the cosmic microwave background observation, as we will see in the next section.\\
Comparatively, the $\hbox{$^{4}He$}$ primordial mass fraction is inferred to be
\bea
Y_{P}=0.240\pm 0.006 \, ,
\eea
which, together with the $D$ abundance, provides the value 
\bea
\eta_{10}=5.7\pm 0.4\, .
\eea
Similarly to deuterium and defining the number abundance of various nuclei relative to the Hydrogen abundance, $y_{3}=10^{5}(\hbox{$^{3}He$}/H)$ and $y_{Li}=10^{10}(\hbox{$^{7}Li$}/H)$, the mass fraction of $\hbox{$^{3}He$}$  and $\hbox{$^{7}Li$}$ are found to be~\cite{BBN2}:
\bea
y_{3} &\simeq &3.1(1\pm 0.01)\,\eta_{10}^{-0.6} ,\nonumber \\
y_{Li} &\simeq &\frac{\eta_{10}^{2}}{8.5} \, . \nonumber
\eea
From these fits, and the corresponding primordial abundances inferred by observation, $\eta_{10}$ can be deduced. However, the determinations of the primordial abundance for $\hbox{$^{3}He$}$  and $\hbox{$^{7}Li$}$ are more complex and model dependent, hence the inferred $\eta_{10}$ should be regarded with less confidence.\\
Fig.\ref{BBN} shows how the analytical fits for the various mass fractions, together with the primordial abundances inferred from observation, can result in a determination of the baryon asymmetry.
\\
In conclusion, BBN tells us that the baryon-to-photon ratio is non-zero, but small: 
\bea
\eta_{B} = 5.7\pm 0.4 \times 10^{-10} \,.
\eea
\subsection{Cosmic Microwave Background}
\begin{figure}[h!]
\begin{center}
\includegraphics[height=6cm]{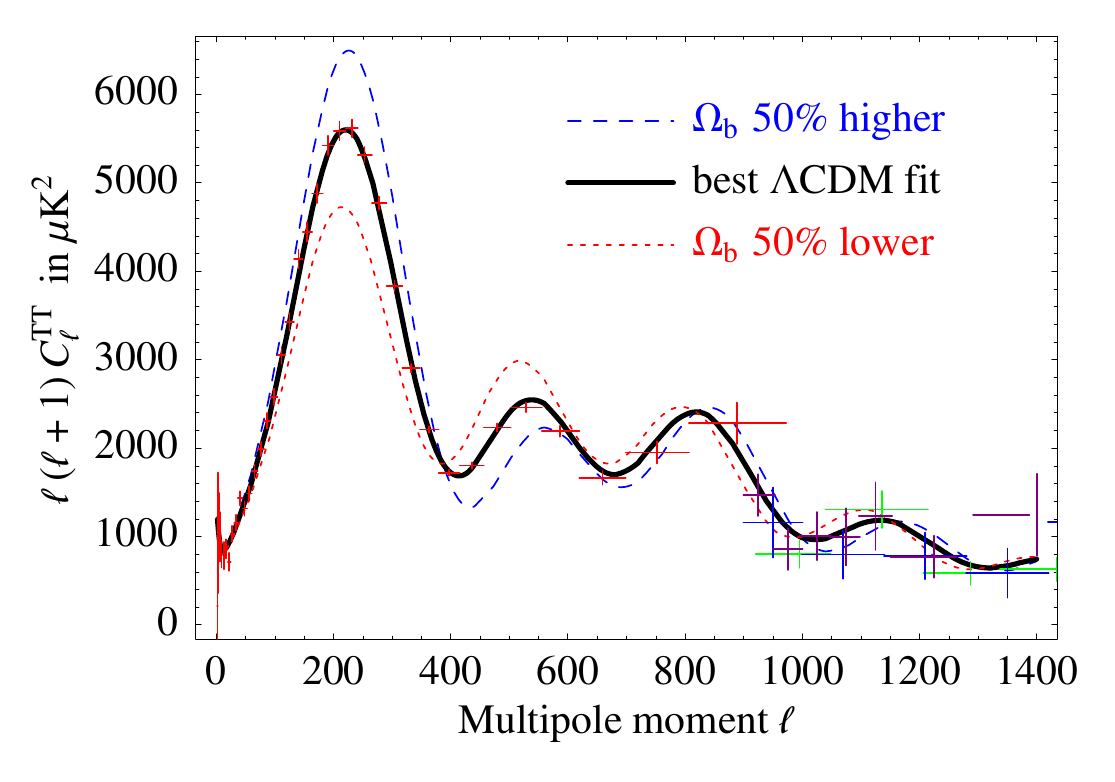} 
\caption{Temperature angular power spectrum. Figure taken from~\cite{StrumiaBaryoviaLepto}.}
\label{CMBdeux}
\end{center}
\end{figure}

The observation of the CMB from the WMAP satellite~\cite{WMAPexp} gives a very precise determination of the baryon-to-photon ratio, which is consistent with the BBN value given above. However, while BBN started around $T\simeq 0.08\MeV$, $t\simeq 3$ min and was almost complete  at $t\simeq 20$ min, the CMB photons come from matter-radiation decoupling, which occured at recombination, $T\sim 1 \eV$ or equivalently $t\simeq 4\,\times 10^{5}$ years~\cite{CMB}.\\
As the temperature drops below $T\simeq 13.6 \eV$, photons are no longer energetic enough to photo-dissociate Hydrogen, hence the reaction
\bea
e^{-}+p\LRa H+\gamma 
\eea
deviates from equilibrium, resulting in the absorption of free electrons by protons and ionized atoms. The photons, which were in equilibrium with electrons, via scatterings and annihilations processes:
\bea
e{-}+e^{+} \LRa \gamma+\gamma \,,\quad e^{-}+\gamma \LRa e^{-}+\gamma \, ,
\eea
then decouple and become free streaming particles. The observation of these photons provides us a snapshot of the Universe at that time.\\
Before decoupling, the photons and the baryons are strongly coupled, and form a fluid which is liable to interactions with opposite effects. On the one hand, in a gravitational well, gravity pushes this fluid down the well, whereas on the other hand the radiation pressure tends to push it out. The fluid thus  goes through accoustic oscillations, which last until the baryons and the photons decouple.\\
After their last scattering, photons freely propagate, and the surface of this last scattering appears to us as a sphere of homogeneous temperature $T_{0}\simeq 2.273 K$ corresponding to a decoupling temperature $T_{decoupling}\simeq 0.26 \eV$. However, the temperature distribution is not fully homogeneous, and the study of the anisotropies provides an accurate determination of a set of cosmological parameters. More specifically, the angular distributions of rescaled anisotropies
\bea
\frac{\Delta T(\theta,\phi)}{T_{mean}}=\frac{T(\theta,\phi)-T_{mean}}{T_{mean}}
\eea
are decomposed on spherical harmonics
\bea
\frac{\Delta T(\theta,\phi)}{T_{mean}}=\sum_{\ell=1}^{\infty}\sum_{m=-\ell}^{\ell}\,a_{\ell\,m}\,Y_{m}^{\ell}(\theta,\phi) \, ,
\eea
so that the study of the anisotropies is translated into the study of the angular power spectrum $C_{\ell}=\left\langle \vert a_{\ell,m} \vert^{2} \right\rangle$. The cosmological parameters are extracted after a fit of the power spectrum, over a set of \textit{priors}, which are theoretical assumptions~\cite{WMAP5a}. By modifying the values of the differents parameters, the relative height and space between the different peaks vary. For example, fig.\ref{CMBdeux} show the different power spectra obtained by varying  $\eta_{B}$ around its central value, increasing or lowering the baryon asymmetry up to $50\%$.
\\
We clearly see how the first accoustic peak, located at $\ell\simeq 200$, is influenced by the value of $\eta_{B}$. This first peak is precisely the one which provides the constraint on the baryon asymmetry.\\
Using WMAP 5 years only, the baryon asymmetry is found to be~\cite{WMAP5a}
\bea
\eta_{B}=6.225\pm 0.17 \times 10^{-10}\,.
\eea

%The WMAP five years data power spectrum is shown in fig. \ref{CMBWMAP}.
%\begin{figure}[htb]
%\begin{center}
%\includegraphics[height=5cm]{CMBWMAP.pdf} 
%\caption{\footnotesize{Temperature angular power spectrum. d'apres WMAP collabration $http://lambda.gsfc.nasa.gov/product/map/dr3/pub_papers/fiveyear/parameters/wmap_5yr_params.pdf$}}
%\label{CMBWMAP}
%\end{center}
%\end{figure}
The combination of different observations leads~\cite{BBN3} to the value
\bea
\eta_{B}=6.11^{+0.26}_{-0.27}\,\times 10^{-10}\, ,
\eea
at $95\%$ confidence level. This is the value we will use throughout this thesis.
\section{Generating the Baryon number of the Universe}
The observed baryon asymmetry calls for explanation. One could argue that this asymmetry was an initial condition of the Universe, or that is has been produced in the very early stages of its evolution. Assuming that initially the Universe was asymmetric goes against a naturalness principle, and is not satisfactory. Furthermore, an inflationary period is believed to have occurred in the Standard Model of Cosmology, (see, \eg~\cite{Liddle:1999mq}), during which the scale factor $a(t)$ of the Universe is increased by a factor $\sim 10^{30}$. Consequently, any asymmetry existing prior to inflation would have been extremely diluted, by a factor $\sim 10^{-30}$. 
%Therefore, either the Universe was extremely matter-asymmetric initially, or one has to invoke a mechanism dramatically efficient in producing the asymmetry. This solution is clearly not satisfactory.\\
Therefore, one has to suppose that the Universe was initially matter-symmetric, and find a way to produce $\eta_{B}\neq 0$ after the inflation period. We know that the annihilation of baryons and antibaryons is not fully efficient, and so one can wonder whether such an asymmetry could be produced at the freezing of annihilations. Following Appendix B~\footnote{Cf. also  chapter 5 and 6 of \cite{KolbTurner}.}, we have that for $1 \MeV \lesssim T \lesssim 1 \GeV$, baryons and antibaryons are in thermal equilibrium, with 
\bea
\frac{n_{b}^{eq}}{n_{\gamma}}\simeq\frac{n_{\ol{b}}^{eq}}{n_{\gamma}}\simeq \left(\frac{m_{p}}{T}\right)^2 K_{2}(\frac{m_{p}}{T})\simeq \left(\frac{m_{p}}{T}\right)^{3/2}\,e^{-m_{p}/T} .
\eea 
Baryons and anti-baryons annihilate and their density decreases until the reactions freeze-out. Since the annihilation rate is $\Gamma_{a}\simeq n_{b}^{eq} \, \left\langle \sigma \vert v\vert \right\rangle$, with a thermally averages rate $\left\langle \sigma \vert v\vert \right\rangle \sim 1/m_{\pi}^{2}$, the decoupling occurs when $\Gamma_{annihilation} \simeq H(T)$, at a temperature $T_{D}\simeq 20 \MeV$. At this temperature, the baryon-to-photon ratio is
\bea
\frac{n_{b}^{eq}}{n_{\gamma}}\simeq\frac{n_{\bar{b}}^{eq}}{n_{\gamma}}\simeq 10^{-18} \ll 10^{-10} \, ,
\eea
which is far below the observed matter-antimatter number density:
the Standard Model of Cosmology cannot explain the baryon asymmetry. Therefore, a model is needed, in which this number  is generated after inflation and before the freeze-out of baryon-antibaryon annihilations.
\subsection*{Sakharov's conditions}
It has been shown that three conditions, known as Sakharov's conditions~\cite{Sakharov}, are necessary to produce a baryon asymmetry in the Universe, starting from a symmetric initial state:
\begin{itemize}
\item baryon number must be violated,
\item $C$ (Charge conjugation) and $CP$ (charge$\times$parity) must not be conserved,
\item the system needs to undergo an out-of-equilibrium period.
\end{itemize}
The first condition is somewhat obvious, if starting from $B=0$ one aims at generating $B\neq0$.\\
The second condition requires both $C$ and $CP$ to be violated. Indeed, for a given process $X\LRa i$ in which the particle $i$ bears a non-zero baryon number $N_{b,i}$ and $X$ is any particle, if $C$ were conserved, one would have $\mathcal{M}(X\Ra i)=\mathcal{M}(\bar{X}\Ra \bar{i})$ and  therefore the global baryon number variation $\Delta B$ would be approximately given by the difference between baryon number creation $\Delta b$ and antibaryon number creation $\Delta \bar{b}$:
\bea
\Delta B=\Delta b-\Delta \bar{b}\propto N_{b,i}\sum_{{X}}\vert \mathcal{M}(X\Ra i)\vert^2 + N_{b,\bar{i}}\sum_{{X}}\vert \mathcal{M}(\bar{X}\Ra \bar{i})\vert^2 ,
\eea
As $N_{b,i}=-N_{b,\bar{i}}$, one trivially observes that  no baryon number can be generated.\\
If $CP$ is conserved, then  $\mathcal{M}(X\Ra i)=\mathcal{M}(i\Ra X)=\mathcal{M}(\bar{X}\Ra \bar{i})$, the last equality coming from $CPT$ invariance. Again, it turns out that no excess of baryons over antibaryons can be generated.\\
Finally, a departure from thermal equilibrium is needed. Indeed, looking at the Boltzmann equation for the number densities, neglecting the expansion term, one has
\bea
\frac{d n_{i}}{dt}=\sum_{{X}}\int d\Pi_{X}d\Pi_{i}\left(n_{X}(1\pm n_{i})\,\Gamma(X\Ra i)-n_{i}\,(1\pm n_{X})\,\Gamma(i\Ra b)\right),
\eea
where $(1\pm n)$ are Pauli blocking or stimulating factors, for fermions $(+)$ or bosons $(-)$.
If the different species $i$ and $X$ were in thermal equilibrium all along their evolution, we would have
\bea
n_{X}(1\pm n_{i})=n_{i}\,(1\pm n_{X}) ,
\eea
which, combined with the unitary condition:
\bea
\sum_{{X}}\vert \mathcal{M}(X\Ra a) \vert = \sum_{{X}}\vert \mathcal{M}(a\Ra X) \vert
\eea
would ensure that $d n_{i}/dt=0$.\\
Another way to see how these three conditions arise is to consider how the baryon number behaves under $C$, $P$ and $T$ transformations. Being odd under $C$, we have 
\bea
\left\langle B \right\rangle=\left\langle C\,C^{-1} B \right\rangle=\left\langle C^{-1} B C \right\rangle=-\left\langle  B \right\rangle ,
\eea
and so $C$ conservation implies $\left\langle  B \right\rangle=0$.\\
Similarly, $B$ is odd under $CP$ so that the same argument holds.\\
Finally, in thermal equilibrium the system is described by a density matrix $\rho(t)=e^{-\be H}$, so that one has
\bea
\left\langle  B(t) \right\rangle =tr(e^{-\be H} B(t))=tr(e^{-\be H}\,e^{-iHt}B(0)\,e^{iHt})=\left\langle  B(0) \right\rangle .
\eea
Given that initially $B(0)=0$, no baryon asymmetry can be generated if the system is in thermal equilibrium.\\
These three conditions are the basic requirements that a model should fulfil in order to produce a non-vanishing baryon asymmetry.
\section{Models of Baryogenesis}
Sakharov's conditions can be satisfied in many different ways.\\
The first attempts to explain the BAU were based on Grand-Unified Theories (GUT)~\cite{GUTbaryo}. Indeed, $C$ and $CP$ violation are easily fulfilled, at least qualitatively if the Yukawa couplings are complex and the departure from thermal equilibrium is found to be naturally satisfied during the freeze-out of a heavy particle. Moreover, $B$ violation, which was not known to occur in the SM, is a common by-product of any GUT theory.
\subsection{GUT baryogenesis}
The early realisations of GUT baryogenesis were based on the $SU(5)$ gauge group, in which a superheavy field, whose mass is related to the various breaking scales (usually $10^{12}~\GeV\lesssim~M_{X}~\lesssim M_{GUT}~\simeq~10^{16}~\GeV$), decays when the temperature drops below its mass. This implies a Boltzmann suppression of inverse-reactions, and thus out-of-equilibrium decays. Under the condition that the $X$ couplings to its decay-products are complex, these decays can produce a non-zero Baryon number, which was supposed to remain constant during the subsequent cooling of the Universe.\\
However, as already stated by Dimopoulos and Suskind in~\cite{GUTbaryo}, "a quantum mechanical source of baryon-number violation (...) is possible (...) to seriously alter the results of this [paper]".\\
Indeed, in 1976, 't Hooft~\cite{thooft} showed that non-perturbative effects, which he called instantons,  and which emerge from the SM gauge structure, are a possible source of $B$ violation. This violation occurs due to the tunneling over an energy-barrier of the instantons. However, this instanton-effect was calculated in zero-temperature field theory, and the probability for such a $B$ violation to occur was shown to be exponentially suppressed.\\
Taking temperature into account, Kuzmin, Rubakov and Shapovnikov showed in 1985~\cite{KRS} that while the tunnelling through the energy barrier was exponentially suppressed, a non-static gauge field configuration, lying on the top of the energy barrier, allows $B$ violating processes to quantitatively occur. More precisely, they showed that this field configurations, called sphalerons, are in equilibrium for $10^{2} \GeV \lesssim T \lesssim 10^{12} \GeV$. These processes violate $B+L$, the excess of which they rapidly wash-out, whilst conserving the orthogonal $B-L$. We refer the reader to appendix A, where sphalerons are discussed in more detail.\\
The GUT early attempts were $B-L$ conserving, and given that any $B+L$ number created before sphalerons come into equilibrium is completely erased, such GUT-baryogenesis were ruled-out. It is customary now to effectively add another condition to Sakharov's three, which could be formulated as "Prevent $B$ number creation from being erased by sphalerons". There are many ways to do so. For example, GUT models in which a $B-L$ number is created are not affected by sphalerons erasure, as we will see in chapter 6.
\subsection{Electroweak baryogenesis}
Another class of models consists in the electroweak-baryogenesis mechanism~\cite{baryo}. While GUT-baryogenesis was based on the out-of-equilibrium decays of a superheavy field, electroweak-baryogenesis relies on the possibility of having a strong departure from thermal equilibrium during the electroweak (EW) phase transition, when the Higgs field(s) acquires a non-zero vaccum expectation value (vev). The mechanism can be roughly described as follows. Assuming the electroweak phase-transition to be of first order, then, during the transition, two degenerate vacua will coexist, one corresponding to the broken phase, with $\left\langle \phi \right\rangle \neq 0$, while the other corresponds to the symmetric phase $\left\langle \phi \right\rangle = 0$. As  the temperature drops down, the regions of broken phase grow until they fill all the space, a phenomena called  "bubble nucleation".\\
In the SM, the only source of $CP$ violation resides in the $\delta_{KM}$ phase of the $CKM$\footnote{Cabibbo-Kobayashi-Maskawa.} mixing matrix for quarks, \ie originates from the non-degeneracy of up and down quark type masses. As fermion mass generation is closely related to the Higgs mechanism, it is natural to think that $CP$ will be violated during the EW phase-transition, when going from the unbroken phase $T \geq T_{EW}$ into a bubble of true vacuum $T\leq T_{EW}$.\\
Since sphalerons are in equilibrium for $T \geq T_{EW}$, but highly suppressed for lower temperatures, the creation of a net $B$ number in the unbroken phase, close to the bubble frontier, can result in a non-zero $B$ number in the broken phase if the phase transition is fast enough compared to the characteristic time of the sphaleron erasure.\\
Actually, it has been shown that electroweak-baryogenesis does not work in the SM.\\
The first reason lies in not having enough sources of $CP$ violation to account for the observed baryon asymmetry~\cite{FailsSMbaryo}. The second reason comes from the fact that the strong first-order phase-transition needed for this mechanism to occur requires the Higgs mass to be small, $m_{H}\lesssim 40 \GeV$ in the SM. Given the LEP result, $m_{H}^{SM} \gtrsim 114 \GeV$~\cite{LEP}, the electroweak-scale SM baryogenesis is clearly ruled out.\\
Supersymmetric extensions of the SM could solve this problem~\cite{SUSYbaryo}, by increasing the number of $CP$ violating sources, and providing a sufficiently first-order phase-transition. For the minimal supersymmetric (MSSM) extensions to work, some fine-tuning is required~\cite{MSSMbaryo}. Electroweak-baryogenesis in singlet extensions of the MSSM is found to be viable~\cite{NMSSMbaryo}.\\
We do not consider this class of mechanism in this thesis.
\subsection{Affleck-Dine baryogenesis}
This class of models relies on flat directions of the supersymmetric scalar potentials~\cite{AffleckDine}. During inflation, fermion condensates can form along flat directions of the superpotential and can develop large vevs. After inflation these condensates coherently oscillate around their minima, and baryon as well as lepton number can be stored in these oscillations.\\
Albeit being an interesting possibility, we will also not study Affleck-Dine baryogenesis here.
\subsection{Baryogenesis via leptogenesis}
There is another class of models for baryogenesis, which has received an increasing amount of attention since the original paper of Fukugita and Yanagida~\cite{Fukugita}: the leptogenesis mechanism.\\
$SU(2)$ sphalerons violate $B+L$ via the effective interaction term (cf. appendix A)
\bea
\mathcal{O}=\prod_{i} q_{L}^{i}\,q_{L}^{i}\,q_{L}^{i}\,\ell_{L}^{i}\, .
\eea
On the other hand, the $B-L$ direction is conserved by these sphalerons. Hence the proposal of leptogenesis~\cite{lepto},\cite{leptorept}: a lepton number could be generated from the out-of-equilibrium decays of a heavy field which couples to leptons. Then this lepton number is converted, at least partly, into a baryon number.\\
This mechanism has the interesting feature of relating the observed baryon asymmetry of the Universe to the yet unexplained fact that the neutrinos are massive particles.\\
Indeed, in seesaw models, as we will see in the following chapter, the smallness of the light neutrino masses is usually explained through the addition of a heavy field, whose mass scale is typically of order of $10^{10}\GeV$. This field, which is assumed to be a Majorana spinor, can decay into leptons and anti-leptons, and can furthermore distinguish between them, granted that their couplings are complex.\\
If the seesaw indeed provides mass to the light neutrinos, leptogenesis occurs, at least qualitatively.\\
We will see in the next chapters that under a few assumptions this mechanism can naturally accommodate the observed baryon asymmetry.
%There is another class of model that explain the observed BAU: leptogenesis scenarii. We saw that sphalerons processes violate and erase any $B+L$ asymmetry, but conserve $B-L$ one. Hence the proposal of Fukugita and Yanadiga: instead of producing a baryon asymmetry possibly ereased by sphalerons, one could either produce a lepton asymmetry by out-of-equilibrium decays of a superheavy particle, that will be at least partly convert by sphalerons interactions. This class of model has the interesting feature that parameters governing lepton asymmetry production are somehow linked to light neutrino masses and mixings. 

%There also exist non-seesaw leptogenesis scenario.\\
%Dirac leptogenesis (aka neutrinogenesis): besides right-handed neutrinos, additionnal particles are introduced, together with additionnal symmetry $U(1)_{N}$ which avoid standard Dirac mass for neutrinos. Lepton asymmetry is produced by out-of-equilibrium decays of the extra fields, in a similar manner than in seesaw leptogenesis. Dirac leptogenesis predicts non-observation of neutrinoless double beta decay, hence could be tested in the future.\\
%Soft leptogenesis: this scenario is based on supersymmetric extension of the SM. Decays of sneutrinos will produce lepton asymmetry, thanks to $CP$ violation arising from soft supersymmetry breaking terms.\\
%In this manuscript we will develop the seesaw-leptogenesis scenario in details, particularly the thermal scenario where right-handed neutrinos are produced by scatterings of the thermal bath.
\newpage
\chapter{Massive neutrinos}
Neutrinos have a very long history. Their existance was first postulated in 1930 by Pauli~\cite{Pauli} to rescue the principle of energy-momentum conservation in $\beta$ decays. Thus, neutrinos are even \textit{older} than the Standard Model. In 1962, muon neutrinos were observed~\cite{MuonNeutrino}. Finaly, tau neutrinos were observed in 2000 by the experiment DONUT at Fermilab~\cite{TauNeutrino}, and so the tau neutrino became the last observed particle of the SM. Despite this long history, very little is known about neutrinos when compared to other species: neutrinos are elusive particles.\\
Only left-handed neutrinos have been observed. This is because if a right-handed $\nu_{R}$ would exist, it would be a singlet under the SM gauge group.%, since the colour and electric charges of $\nu$'s are null and right-handed species are singlets under $SU(2)$.
Furthermore, given the experimental possibilities in the 70's, when the SM was being built, (almost) no evidence of neutrino masses were observed.\\
Therefore in the SM only the left-handed neutrino component was included.
% $\nu_{L}\equiv (1,2,-1)$ under $SU(3)\times SU(2)_{L}\times U(1)_{Y} (\equiv G_{321})$. They are weak isospin $+1/2$ of $SU(2)$ Lepton doublet, and therefore interact with $Z,W^{\pm}$ gauge bosons, via neutral and charged currents. Given the LEP result on the $Z$ width, we know there are three active neutrinos: $\nu_{e}$, $\nu_{\mu}$ and $\nu_{\tau}$. The lack of a right-handed component forbid us to write a conventionnal Dirac-type mass term $m (\bar{\psi_{L}} \psi_{R}+\bar{\psi_{R}} \psi_{L})$.\\
However, this construction faced, from 60's onward, two major problems:  the solar and atmospheric neutrino anomalies.
\section{Evidence for neutrino oscillations}
The measurements of neutrino fluxes coming from different sources revealed a discrepency between the expected and the observed signals. This anomaly was observed both in the solar sector~\cite{Solar} and in the flux of muon neutrinos coming from Earth's atmosphere~\cite{Atmosphere}, and constitutes one of the first evidences of physics beyond the Standard Model. Here we just briefly discuss this question. We refer the reader to the review of Strumia and Vissani~\cite{neutrinoreview} for a quasi-exhaustive presentation of physics where neutrinos are involved. 
\subsection*{Solar neutrinos}
Historically, this was the first neutrino anomaly. Electron neutrinos are produced in the core of the Sun, according to the fusion reaction
\bea
4\,p+2\,e^{-}\Ra \hbox{$^{4}He$}+2\,\nu_{e} \, ,
\eea
and given the small neutrino cross section, they escape from the core. A flux of $\nu_{e}$ is then measured on Earth. The first evidence came from the Homestake experiment~\cite{Solar}, which uses Chlorine through the reaction in a water tank
\bea
\nu_{e}+\hbox{$^{37}Cl$} \Ra \hbox{$^{35}Ar$}+e^{-} \ .
\eea
Counting the number of $\hbox{$^{35}Ar$}$, it was possible to determine the flux of solar $\nu_{e}$. This flux was measured to be only $1/3$ of the expected one, hence the solar neutrino anomaly. At that time, it was not clear whether new physics was needed, or if solar models had to be revised. Later on, the SNO experiment~\cite{Solar} was built being sensitive to all neutrino flavours, and also able to discriminate $\nu_{e}$ from $\nu_{\mu,\tau}$. The total neutrino flux was measured to be in agreement with solar model predictions, while the deficit in $\nu_{e}$ was confirmed: whereas only  $\nu_{e}$ are expected, all neutrinos flavours were observed. Being confident on the fact that only the electron flavour is created in the Sun, the disappearance of $\nu_{e}$ and the appearance of muon and tau flavours remained to be explained.
\subsection*{Atmospheric neutrinos}
Another evidence comes from the detection of neutrinos produced in the upper atmosphere. High energetic cosmic rays enter the atmosphere, and interact with nuclei, thus producing pions. These pions decay into muons, according to
\bea
\pi^{+}\Ra \mu^{+}+\nu_{\mu}\, , \nonumber \\
\pi^{-}\Ra \mu^{-}+\bar{\nu_{\mu}} \ . \nonumber
\eea
The high energetic muons subsequently decay
\bea
\mu^{-}\Ra e^{-}+\nu_{\mu}+\bar{\nu_{e}} \, ,\nonumber \\
\mu^{+}\Ra e^{+}+\bar{\nu_{\mu}}+\nu_{e} \ , \nonumber
\eea
leading to about a rate for $\nu_{\mu}$ which is roughly twice that of $\nu_{e}$. Atmospheric neutrino experiments  measure the value 
\bea
R=\frac{N(\nu_{\mu})^{obs}/N(\nu_{e})^{obs}}{N(\nu_{\mu})^{MC}/N(\nu_{e})^{MC}}\, ,
\eea
which compares the observed ratio to the expected one. While $R=1$ is expected, $R\simeq 0.65$ is observed~\cite{Atmosphere}. This points towards an appearance of $\nu_{e}$ and/or a disappearance of $\nu_{\mu}$. The SuperKamiokande experiment~\cite{Atmosphere} settled this question: the deficit from athmospheric neutrinos results from $\nu_{\mu}$ disappearance.

\subsection*{Interpretation}
Different solutions were advocated to explain these flavour conversions, such as neutrino oscillations~\cite{Osci}, neutrino decays~(cf ref. 56 of \cite{neutrinoreview}),  or decoherence of propagating $\nu$~(ref. 44 of \cite{neutrinoreview}): the SK experiment showed that no-oscillation solutions were ruled-out at $4\sigma$~\cite{Osci2}.\\
The oscillatory solution has as immediate consequence that neutrinos are massive particles.
\section{Neutrino mixing in the $3$ flavour scheme.}
A neutrino involved in EW interaction is not a mass eigenstate. Labelling the mass basis eigenstates by $\nu_{i}$, $i=1,2,3$ and the flavour basis eigenstate by $\nu_{\al}$, $\al=e,\mu,\tau$, they are related through a unitary transformation:
\bea
\vert \nu_{\al} \rangle=\sum_{i=1}^{3} U_{\al i}^{*}\vert \nu_{i} \rangle \ ,
\eea
The  matrix $U$, called the PMNS\footnote{Pontecorvo, Maki, Nakagawa and Sakata.}~\cite{PMNS} mixing matrix, is a $3\times3$ unitary matrix  which depends on $3$ real mixing angles and on $6$ $CP$-odd phases. Some of these phases can be rotated away by a redefinition of the fields. After this, for a general $n\times n$ unitary matrix, if neutrinos are pure Dirac spinors  $n(n-1)/2$ mixing angles and $n(n-1)/2-(n-1)$ phases remain, while  $n(n-1)/2$ mixing angles and $n(n-1)/2$ phases if neutrinos are Majorana spinors. It is customary to express $U$ as a product of three rotation matrices:  
\beq
U=\left(\begin{array}{ccc}
1 & 0 & 0 \\ 
0 & c_{23} & s_{23} \\ 
0 & -s_{23} & c_{23}
\end{array}\right).\left(\begin{array}{ccc}
c_{13} & 0 & s_{13}\,e^{-i \delta} \\ 
0 & 1 & 0 \\ 
-s_{13}\,e^{i \delta} & 0 & c_{13}
\end{array}\right).\left(\begin{array}{ccc}
c_{12} & s_{12} & 0 \\ 
-s_{12} & c_{12} & 0 \\ 
0 & 0 & 1
\end{array}\right).\rm{diag}\left(e^{i\phi_{1}},e^{i\phi_{2}},1\right) \, ,
\eeq
with $c_{ij}=\cos{(\theta_{ij})}$ and $s_{ij}=\sin{(\theta_{ij})}$. $\theta_{ij}$ are real mixing angles and $\delta$ the "Dirac" $CP$ phase. If neutrinos are Majorana fields, there are two supplementary phases $\phi_{1},\phi_{2}$, the "Majorana" $CP$ phases, which are absent if neutrinos are Dirac fields.\\
Given this, the probability for a neutrino that propagates with a mean energy $E$ to oscillate from a flavour $\al$ to a flavour $\be$, after having travelled a distance $L$,  is given by
\bea
P(\nu_{\al}\Ra\nu_{\be})&=&\Big\vert \sum_{j} U^{*}_{\al\,j}U_{\be\,j}\,e^{-i\frac{m_{j}^{2}\,L}{2 E_{j}}} \Big\vert^{2} \nonumber \\
=\delta_{\al\be}&-&4\sum_{i>j}\rm{Re}\left(U^{*}_{\al\,i}\,U_{\al\,j}\,U_{\be\,i}\,U^{*}_{\be\,j}\right)\times \rm{sin}^{2}\left(\frac{\Delta m_{i\,j}^{2}\,L}{4\,E} \right) \nonumber \\
&+&2\sum_{i>j}\rm{Im}\left(U^{*}_{\al\,i}\,U_{\al\,j}\,U_{\be\,i}\,U^{*}_{\be\,j}\right)\times \rm{sin}{\left(\frac{\Delta m_{i\,j}^{2}\,L}{4\,E} \right)} \, ,
\eea
where $\Delta m_{i\,j}^{2}=m_{j}^{2}-m_{i}^{2}$.\\
The determination of the mixing parameters results from the measurement of neutrino fluxes coming from different sources and at different distances from the emission source. Together with the natural sources, man-made sources such as reactors~\cite{Reactornu} or neutrino beams~\cite{Beamnu}, are complementary in determining $U$.\\
In the atmospheric sector, the disappearance of $\nu_{\mu}$ is explained by an oscillation $\nu_{\mu}\LRa \nu_{\tau}$, whose detection allows to constrain $\theta_{23}$ and $\D m_{23}^{2}$, therefore called atmospheric mixing angle and (difference of squared) mass(es). In the solar sector, the disappearance of $\nu_{e}$ is explained by $\nu_{e}\LRa \nu_{\mu}$ oscillations, the observation of the latter constraining the solar sector, $\theta_{12}$ and $\D m_{12}^{2}$. Finally the last sector, the so-called Chooz mixing angle $\theta_{13}$ and squared mass difference $\D m_{13}^{2}$, intervenes both in solar and atmospheric oscillations and is very constrained by the CHOOZ reactor experiment~\cite{Reactornu}.\\

\begin{figure}[h!]
\hspace{-1cm}\includegraphics[height=6cm]{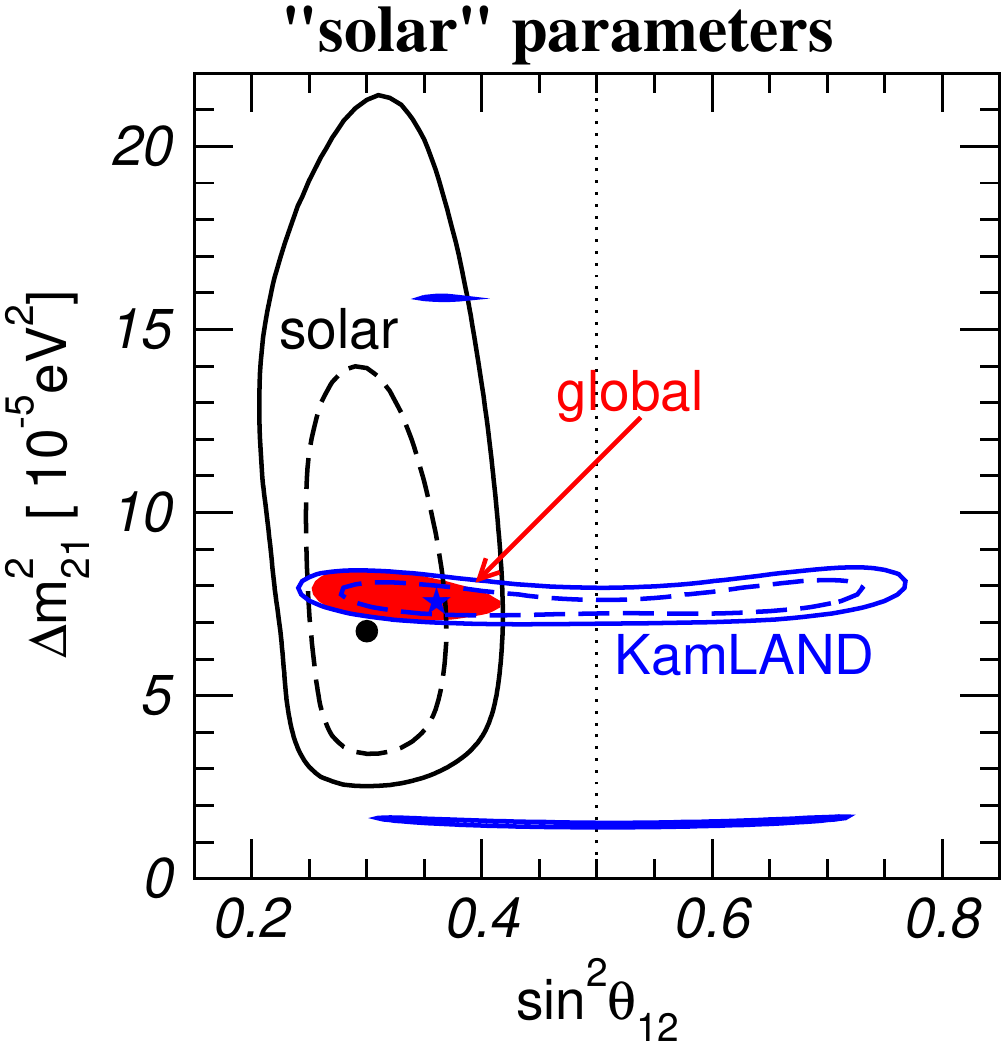}\quad \includegraphics[height=6cm]{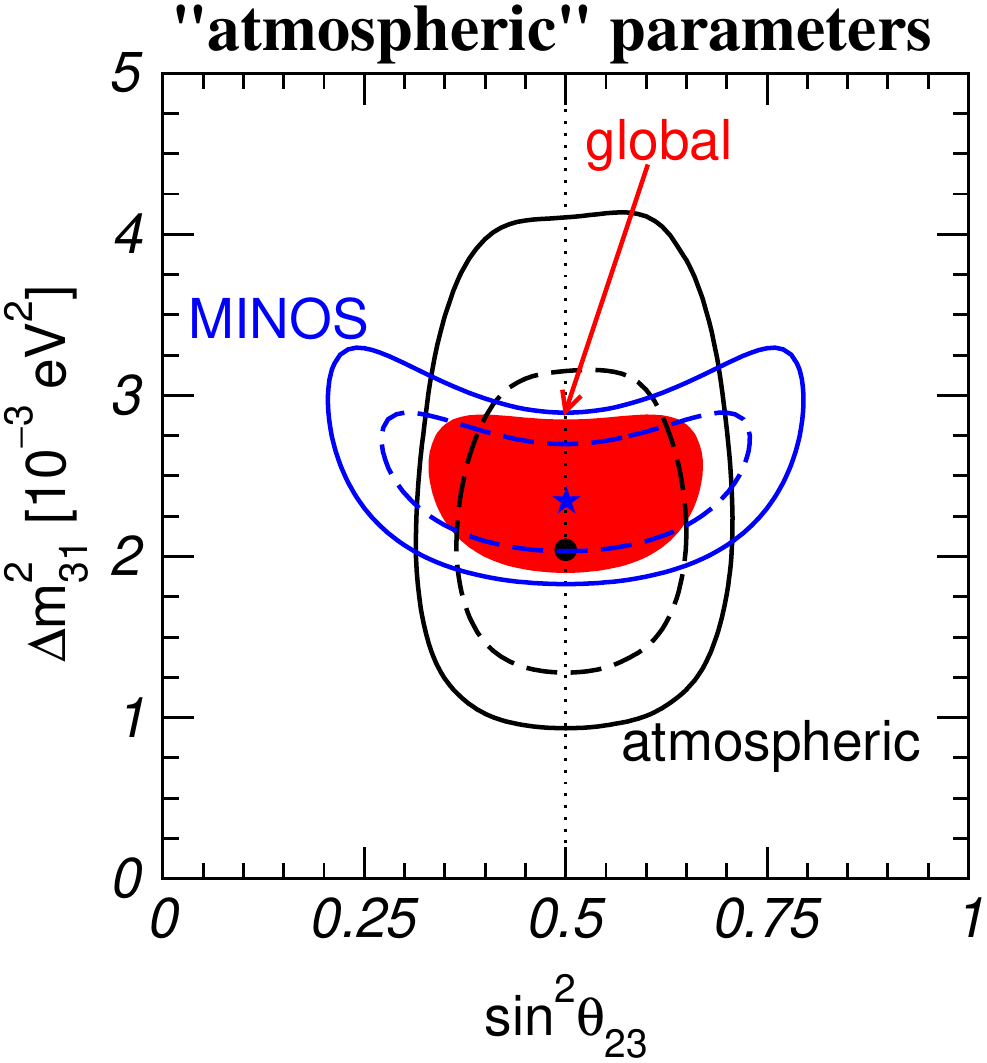}\quad \includegraphics[height=6cm]{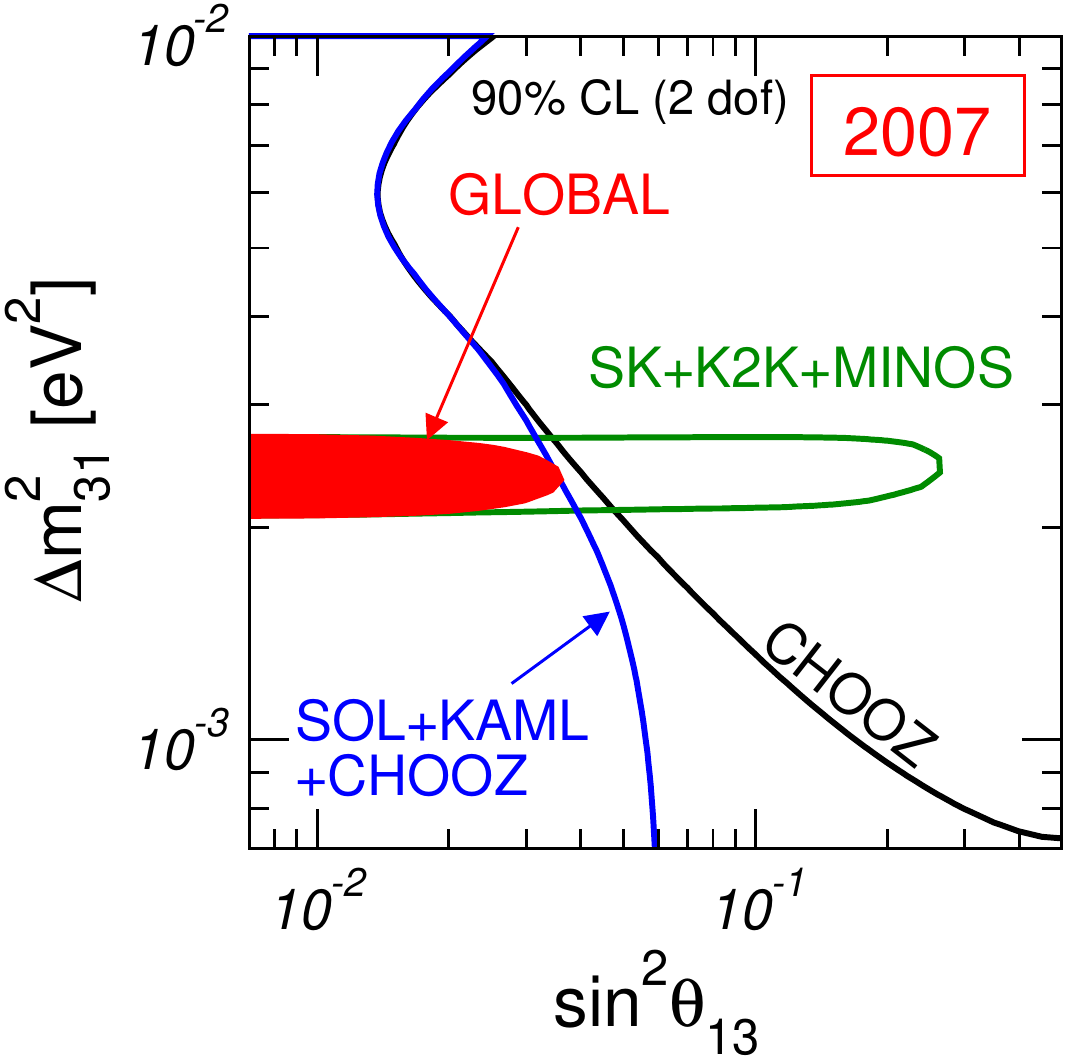} 
  \caption{Values of squared masses differences and mixings angles determine by global fit of neutrino oscillation data. Figures taken from~\cite{Schwetz}.}
\label{OsciMix}
\end{figure}

%In the atmospheric sector, the disappearance of $\nu_{\mu}$ is explained by an oscillation of the mass eigenstate $\nu_{2}\LRa \nu_{3}$, which observation allows to constrain $\theta_{23}$ and $\D m_{23}^{2}$ that are therefore called atmospheric mixing angle and (difference of squared) mass(es). In the solar sector, the disappearance of $\nu_{e}$ is explained by and $\nu_{1}\LRa \nu_{2}$ oscillation, the observation of it constraining the solar sector $\theta_{12}$ and $\D m_{12}^{2}$. Finally the last sector, which intervenes both in solar and atmospheric oscillations, is yet constrained by the CHOOZ reactor~\cite{Reactornu}.\\
We display in fig.\ref{OsciMix} the status of neutrino mixing in the relevant $(\D m_{ij}^{2}-\theta_{ij})$ plane for the different sectors. It shows the complementarity of natural and man-made sources of neutrinos, leading to the determination of the several parameters.
\\
The global fit to these oscillation parameters, taken from \cite{Schwetz}, is given in the following table:
\begin{table}[htb] \centering
    \begin{tabular}{|c|c|c|}        
		\hline	
        Parameter & 
        Best fit & 
       3$\sigma$
        \\     
		\hline
        $\Delta m^2_{12}/10^{-5}\eV$
        & 7.6 &  7.1--8.3 \\
        $|\Delta m^2_{13}|/10^{-3}\eV$
        & 2.4 &  2.0--2.8 \\
        $\sin^2\theta_{12}$
        & 0.32 &  0.26--0.40\\
        $\sin^2\theta_{23}$
        & 0.50 &  0.34--0.67\\
        $\sin^2\theta_{13}$
        & 0.007 &   $\leq$ 0.050 \\
        \hline
    \end{tabular}
%\caption{ \label{tab:summary} 
 % Best-fit values and 3$\sigma$ levels of
  % neutrino oscillation parameters from global data.}
\end{table}\\
The atmospheric mixing angle is observed to be maximal, while the solar angle is large but non-maximal, at a  very high level of confidence. The last mixing angle $\theta_{13}$ is still only upper-constrained. The precise value of this mixing angle is very important, since it is necessary for it to have a non-zero value if $CP$ violation coming from the Dirac $CP$ phase is to be experimentally observed.
%we see that in the PMNS matrix, the $CP$ violating phase $\delta$ appears only times $\rm{sin}(\theta_{13})$. The latter being 0, no $CP$ violation in neutrino oscillation could be measured.\\
The observation of a non-zero value for $\theta_{13}$ could also allow to determine the neutrino mass ordering. As of today the sign of $\Delta m^2_{13}\simeq \Delta m^2_{23}$ remains unknown, leaving two possibilities for the neutrino spectrum:    either the mass ordering is normal hierarchic, with $m_{1}\leq m_{2} \leq m_{3}$, or it can be inverse    hierarchic with  $m_{3}\leq m_{1}\leq m_{2}$. Considering the oscillation probability, we see that only the third   term 
\bea
\propto 2\sum_{i>j}\rm{Im}\left(U^{*}_{\al\,i}\,U_{\al\,j}\,U_{\be\,i}\,U^{*}_{\be\,j}\right)\times \rm{sin}{\left(\frac{\Delta m_{i\,j}^{2}\,L}{4\,E} \right)}
\eea
can lift this degeneracy. Since the Jarlskog invariant~\cite{Jarlskog},
\bea
\vert U^{*}_{\al\,i}\,U_{\al\,j}\,U_{\be\,i}\,U^{*}_{\be\,j}\vert \equiv J= s_{12}\,c_{23}\,s_{23}\,c_{13}^{3}\,s_{13}\,\rm{sin}(\delta)\, ,
\eea
depends on $\rm{sin}(\theta_{13})\rm{sin}(\delta)$, we see that a non-zero value of $\theta_{13}$ is crucial.\\
The oscillation experiments provided the first proof that neutrinos are massive. Nevertheless, these oscillations neither depend on the neutrino mass scale nor on the nature of neutrinos, that is, on having Dirac or Majorana neutrinos. In the last case, two additionnal $CP$ violating phases $\phi_{1,2}$ exist, which do not enter in the oscillation probability.\\% Moreover, we implicitly used here the fact that the LEP result on the $Z$-width constrains the number of light interacting neutrinos to be $N_{\nu}=2.984\pm0.008$. However\\
Hence, non-oscillation experiments are required in order to complete the picture.

%\begin{figure}[htb]
%\begin{center}
%  \includegraphics[width=13cm,height=6cm]{./figt/SpectreNeutrino.pdf}
%    \caption{Schematic view of neutrino mixing}
%\label{Spectrum}
%\end{center}
%\end{figure}
\section{Non-oscillation experiments}
%Let us briefly discuss these important issues.
\subsubsection*{Cosmology}
It may not be straightforward how cosmological observations could constrain neutrino parameters. Nevertheless, robust constraints are derived, when different observations are combined~\cite{CosmoNu}.
%Since neutrino are massive, they carry a fraction of the total mass density, $\Omega_{m}$, which is precisely measured by WMAP, for instance. This mass fraction reads:
%\bea
%\Omega_{\nu}=\frac{\Sigma m_{\nu}}{94\,h^{2}\eV}\ , 
%\eea
%and is small. 
Were neutrinos  massive enough, they would have contribute to distort the photon power spectrum during recombination. Observations of the CMB alone~\cite{WMAP5b} give an upper-bound on $m_{\nu}$:
\bea
\Sigma m_{\nu}\lesssim 1.3 \eV\,(95\% \rm{C.L.})\, .
\eea
The effect of neutrinos is not seen on the CMB, but rather on large scale structure formation. Being massive, neutrinos carry a fraction of the total mass density, $\Omega_{m}$, which has been precisely measured by WMAP, for instance. However this mass fraction is small. Furthermore, neutrinos have an important free-streaming length $\ell_{F}$, since they are only charged under $SU(2)$. Therefore, on scales that are smaller than $\ell_{F}$, neutrinos are unable to cluster, that is, to participate in the formation of large-scale structures, the clustering on these scales being somewhat delayed. As the temperature of the Universe lowers, neutrinos become non-relativistic and their free-streaming length is reduced. Hence neutrinos begin to cluster, as ordinary matter. Therefore, the heavier the neutrinos are, the earlier do they cluster, allowing large-scale structure formation at higher temperature (earlier times).\\
The constraint on neutrino mass from WMAP 5 year analysis, including many different observations, reads~\cite{WMAP5b}
\bea
\Sigma m_{\nu}\lesssim 0.61\eV\quad(95\% \rm{C.L.}) \, .
\eea
%In the era of precision cosmology, it could be somewhat amusing if the neutrino mass was determined by cosmological observations.
\subsubsection*{Beta decay}
The beta decay experiment constitutes the best direct test of neutrino masses. The effect of neutrino masses is to distort the end-point of the beta-decay energy spectrum of the beta decay. Indeed, in the decay
\bea
d\Ra u+e^{-}+\ol{\nu_{e}}\, ,
\eea
the energy of the electron is $E_{e}=Q-E_{\nu}$, which is maximal for $E_{e}=Q-m_{\nu}$. Here $Q$ stands for the energy released in the $\beta$ decay, \eg for tritium beta decay, $\hbox{$^{3}H$}\Ra \hbox{$^{3}He$}+e+\ol{\nu_{e}}$, $Q=m_{\tiny{ {\hbox{$^{3}H$}}}}-m_{\tiny{{\hbox{$^{3}He$}}}}\simeq 18.6$ keV.\\
Around the end-point, the energy spectrum of the electron is $\propto \sqrt{(Q-E_{e})^{2}-m_{\nu_{e}}^{2}}$, and so $m_{\nu_{e}}\neq 0$ will imply a deviation from the line $Q-E_{e}$. So far, the best constraint comes from MAINZ~\cite{Mainz} and TROITSK~\cite{Troitsk} experiments:
\bea
m_{\nu_{e}}\lesssim 2.2 \eV \, .
\eea
The future beta decay experiment, Katrin~\cite{Katrin}, which is schedules to begin data taking on 2010, is expected to reach a sensivity of $0.2\eV$ to neutrino masses.
\subsubsection*{Neutrinoless double-beta decay}
\begin{figure}[h!]
\begin{center}
\includegraphics[trim =20mm 21cm 120mm 25mm, clip,scale=1]{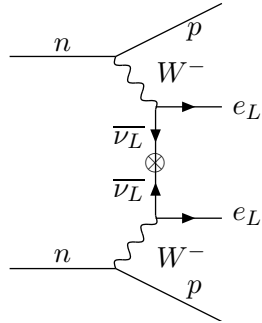} 
\caption{Feynman diagram for neutrinoless double-beta decay.}
\end{center}
\label{0NUBB}
\end{figure}

The experiments of neutrinoless double-beta decay~\cite{ONUBB} are very important for neutrino physics, since the observation of such a process would imply that neutrinos are Majorana particles. As can be seen in fig.\ref{0NUBB}, $0\nu\,\be\be$ consists in the reaction
\bea
(A,Z)\Ra (A,Z+2)+2\,e^{-}\, ,
\eea
which violates lepton number by two units.
\\

The decay rate for this process, 
\bea
\Gamma_{0_{\nu}\,\be\,\be}\propto \vert \mathcal{M} \vert^{2}\,\vert m_{ee} \vert ^{2} \, ,
\eea
depends on $m_{ee}$ which is the $ee$ entry of the neutrino mass matrix
\bea
m_{ee}&=&\sum_{i} U_{ei}^{2}\,m_{i}\nonumber \\&=&\rm{cos}^{2}(\theta_{13})\left(m_{1}\,e^{2\rm{i}\,\phi_{1}}\,\rm{cos}^{2}(\theta_{12})+m_{2}\,e^{2\rm{i}\,\phi_{2}}\,\rm{sin}^{2}(\theta_{12})\right)+m_{3}\,\rm{sin}^{2}(\theta_{13}) \, .
\eea
We see that these processes also depends on the Majorana $CP$ violating phases $\phi_{1,2}$. However, there is an uncertainty of order $\mathcal{O}(3)$ on the nuclear matrix elements, which further depend on the nuclei used~\cite{neutrinoreview}.\\
The status of $0\nu\,\be\be$ is controversial, since the observation of a positive signal has been claimed by members of the Heidelberg-Moscow experiment~\cite{Heidelberg-Moscow}, corresponding to $\vert m_{ee}\vert \lesssim 0.1-0.9 \eV$, but this signal has not been confirmed by, for example NEMO or CUORICINO~\cite{ONUBB}, which obtained the bounds $\vert m_{ee}\vert \lesssim 0.7-1.2 \eV$ and $\vert m_{ee}\vert \lesssim 0.2-0.9 \eV$, respectively. Hence the importance of future $0\nu\,\be\be$ experiments.
\section{Neutrino mass models}
The observation of neutrino oscillations lead us to the conclusion that neutrinos are massive particles, which necessarily implies to somehow extend the Standard Model. In fact, these observations provided the first evidence of physics beyond the SM.\\
It has been noticed by Weinberg~\cite{Weinberg} that, given the quantum numbers of lepton and Higgs fields, the Lagrangian can contain the following non-renormalisable dimension 5 operator
\bea
\label{Wein}
\mathcal{L}^{d=5}=\frac{1}{2}\,f_{\al\,\be}\left(\ol{\ell_{L\,\al}^{c}}\,\tilde{\phi}^{*}\right)\left(\tilde{\phi}^{\dagger}\,\ell_{L\,\be}\right)\, ,
\eea
which violates lepton number by two units. Actually, this operator is the only dimension 5 operator built out of SM fields, and invariant under the SM gauge group.\\
After the Higgs develops a vev, this term gives a Majorana mass to the light neutrinos $m_{\nu}\simeq f_{\al\be}\,v^{2}$, the smallness of $m_{\nu}$ coming from the smallness of $f$. In Weinberg's paper, the coefficient $f_{\al\be}$ was expected to be $\propto M^{-1}\lesssim 10^{-12}/m_{W}$, hence producing a neutrino mass $m_{\nu}\simeq 10^{-5}-10^{-1}\eV$.\\ 
This operator may result from an underlying theory, where light neutrinos couple to some  field, stabilised at some high scale $M$, the integration of which would give rise to the Weinberg operator eq.(\ref{Wein}). Thus, the appealing feature that light neutrino masses could be a window to (very) high energy physics.\\
There is other possibilities to generate neutrino mass. For example, a natural solution is to introduce right-handed neutrinos, which couple to the left ones via an additional Yukawa coupling $\la$, the smallness of $m_{\nu}$ being explained by the smallness of $\la$. This coupling should then be $\lesssim 10^{-12}$.\\
Many other possibilities and models can account for neutrino masses. Given the SM field content, let us see what are the possible extensions.\\
For that purpose we consider the following fields, whose quantum numbers under $SU(3)\times SU(2)_{L}\times U(1)_{Y}$ are given below:
\begin{center}
\begin{tabular}{|c|c|c|}
\hline field & label & $G_{321}$ \\ 
\hline lepton doublet & $\ell_{L}$ & $(\tbf{1},\tbf{2},\tbf{-1/2})$ \\ 
\hline charged lepton singlet & $e_{R}$ & $(\tbf{1},\tbf{1},\tbf{-1})$ \\ 
\hline neutral lepton singlet & $\nu_{R}$ & $(\tbf{1},\tbf{1},0)$ \\ 
\hline Higgs doublet & $\phi_{d}$ & $(\tbf{1},\tbf{2},-1/2)$ \\ 
\hline Higgs doublet$^{*}$ & $\phi_{u}$ & $(\tbf{1},\tbf{2},1/2)$ \\ 
\hline 
\end{tabular} 
\end{center}
In the above notation, we label the SM Higgs field $\phi_{d}$, while $\phi_{u} \equiv \tilde{\phi_{d}}=\rm{i}\,\sigma_{2}\,\phi_{d}^{*}$, $\sigma_{2}$ being the Pauli matrix.\\
Given the fields considered, it is straightforward to determine the possible extensions of the SM which would give masses to the neutrinos either at tree level or due to radiative corrections:
\begin{center}
\begin{tabular}{|c|c|c|c|}
\hline
\multicolumn{4}{|c|}{Extension of the Higgs sector}\\

\hline  Product & $G_{321}$ & field & mass term \\
%\hline $\ol{\ell_{L}} \times \e_{R}$ & $(\tbf{1},\tbf{2},\tbf{1/2})$ & scalar $\phi_{d}$ & charged lepton mass \\
\hline $\ell_{L}\otimes \ell_{L}$ & $(\tbf{1},\tbf{1},-1)$ & singlet scalar $\eta^{+}$ & Majorana (LH) mass \\
\hline $\ell_{L} \otimes \ell_{L}$ & $(\tbf{1},\tbf{3},-1)$ & triplet scalar $\D$ & Majorana (LH)  mass \\
\hline $e_{R} \otimes e_{R}$ & $(\tbf{1},\tbf{1},-2)$ & singlet scalar $k^{++}$ & Majorana (RH)  mass \\
%\hline 
%\multicolumn{4}{c}{}\\
\hline
\multicolumn{4}{|c|}{Extension of the fermionic sector}\\
\hline $\ell_{L}\otimes\phi_{u}$ & $(\tbf{1},\tbf{1},\tbf{0})$ & singlet $\nu_{R}$ & neutrino Dirac type mass \\
\hline $\ell_{L}\otimes\phi_{u}$ & $(\tbf{1},\tbf{3},\tbf{0})$ & triplet $\Sigma$ & Majorana (LH) neutrino mass \\
\hline
\end{tabular} 
\end{center}
In the above table, RH and LH denote right and left handedness, respectively.
In addition to the mass terms above, the right-handed neutrinos can also have a Majorana mass term $\ol{\nu_{R}}_{i}\,\nu_{R\,j} M_{ij}$, appearing as a bare mass term in the SM potential.
% or from the coupling of the RHns with singlet neutral scalar field.\\
%Every model providing a mass to neutrinos has to explain their smallness; different mechanisms are argued.\\
As stated above, the minimal extension of the SM consists in including right-handed neutrinos (RHn) which couple to left-handed ones via $\la_{ij}\,\ell_{L}\,\phi_{u}\,\ol{\nu_{R}}+h.c.$. When $\phi$ acquires a vev, this term provides a conventional Dirac type mass for neutrinos $m_{\nu}\left(\nu_{L}\,\ol{\nu_{R}}+\ol{\nu_{L}}\,\nu_{R}\right)$. For neutrino masses of $m_{\nu}\lesssim 0.5\eV$, the neutrino Yukawa coupling should be $\la \lesssim 10^{-12} \sim 10^{-6} h_{e}$, where $h_{e}$ is the Yukawa coupling of the electron. Such a model is perfectly viable; after all $h_{e}\sim 10^{-6}\,h_{t}$. This model, called $\nu$MSM, has been studied for example in~\cite{nuMSM}, and has several cosmological implications.\\
Nevertheless, one could argue (even if this cannot be seen has a proof on itself) that since the inclusion of right-handed neutrinos induces a violation of the lepton number, it is quite natural to link it to some high energy scale at which $B-L$ is broken. Another argument is that such small couplings, even if possible, are non-natural. We adopt this point of view and do not consider this scenario.\\
In the Zee~\cite{Zee} or Zee-Babu~\cite{Babu} models, which are based on the inclusion of the fields $\eta^{+}$ and $k^{++}$, neutrino masses emerge from radiative effects at either one or two loops, and the suppression of neutrino masses comes from these loop factors. We do not consider here these models, but examine the class of seesaw models explained hereafter, in which neutrino masses are generated at tree level and which up to now provides the preferred explanation for neutrino masses.
\section{Seesaw models}
Common to all these models, is having a suppression of the neutrino mass which  indirectly originates from the introduction of a new mass scale, typically far higher than the electroweak scale.
\subsection{Fermion singlet: type I seesaw}
This mechanism was originally the first of the seesaw class~\cite{Seesaw}, and originates from grand unification theory. Indeed, in $S0(10)$ models, the minimal representation which is complex and anomaly free, is a $\tbf{16}$ spinorial representation. This representation contains all SM fermions fields of one generation plus one extra field, singlet under the SM gauge group, which can be identified as a right-handed neutrino.\\
In these GUT models, typically the RHn is not a gauge singlet. It couples to some Higgs representation, which acquires its vev at a high scale, of order $10^{12}-10^{16}\GeV$. Integrating out this heavy field at the electroweak scale gives the effective term eq.(\ref{Wein}).\\
Such a GUT embedding is nevertheless not necessary at all, since the Majorana mass term can be put by hand as a bare mass term in the Lagrangian, the right-handed neutrino being singlet under $G_{321}$. In such a case, the neutrino Yukawa sector reads
\bea
\mathcal{L}_{\nu}=-\la_{\al\,i}^{\dagger}\,\nu_{R}^{i}\,\ol{\ell}_{\al}\,\phi_{u}^{*}-\frac{1}{2}\,M_{ij}\ol{\nu_{R}^{c}}^{i}\nu_{R}^{j}+h.c. \,,
\eea
where $\psi^{c}=C\,\ol{\psi}^{T}$ is the charge conjugated of $\psi$.\\
After the SM Higgs acquires its vev,$\left\langle \phi_{u}\right\rangle =v\simeq 174 \GeV$, the first term provides a Dirac mass for the neutrinos 
\bea
\mathcal{L}_{\nu}\supset -m_{D}^{\al\,i}\,\ol{\nu_{L}}^{i}\,\nu_{R}^{j}\,,
\eea
with 
\bea
m_{D}=\la\,v/\sqrt{2}\,.
\eea
Using the identity 
\bea
\ol{\nu_{L}}\,\nu_{R}=\frac{1}{2}\left(\ol{\nu_{L}}\,\nu_{R}+\ol{\nu_{R}^{c}}\,\nu_{L}^{c}\right) \nonumber
\eea
the above mass term can be written
\bea
\mathcal{L}_{\nu}=-\frac{1}{2}\,\left(\ol{\nu_{L}},\ol{\nu_{R}^{c}}\right)\left(
\begin{array}{cc}
0 & m_{D} \\
m_{D}^{T} & M
\end{array} \right)\,\left(\begin{array}{c} \nu_{L}^{c}  \\ \nu_{R} \end{array}\right)\, . \nonumber 
\eea
Assuming $M\gg m_{D}$, the diagonalisation of the matrix above gives us two eigenstates per generation, which are Majorana spinors. At leading order, one state is given by 
\bea
N= \nu_{R}+\nu_{R}^{c} \nonumber
\eea
with mass $M$. For now we will call this state \textit{the} right-handed neutrino. The other state is at leading order
\bea
\nu= \nu_{L}+\nu_{L}^{c} \, ,\nonumber 
\eea
mostly composed of the left-handed neutrino, as implied from low energy observations. The mass of this light neutrino is given by the seesaw relation~\cite{Seesaw}: 
\bea
m_{\nu}^{ij}=v^{2}\la_{ik}\,M^{-1}_{k}\,\la_{j k}\, ,
\eea
in the basis where $M$ is diagonal.\\
The light neutrino mass matrix is diagonalised by the PMNS matrix $U$ introduced earlier:
\bea
D_{m}=U^{T}\,m_{\nu}\,U \, ,
\eea
where $D_{m}=\rm{diag}(m_{1},m_{2},m_{3})$.\\
In addition to the nine low-energy parameters, enlarging the fermionic sector by three right-handed neutrinos provides nine extra parameters: three masses + three mixing angles + three $CP$ violating phases. There are different ways of  parametrising  these unknown parameters, namely the bottom-up approach, where one fixes the low-energy sector to reconstruct the high-energy one, or the converse top-down approach in which one fixes the high-energy sector. These two parametrisations are useful depending on whether one considers either an effective or a full theory, respectively. In this thesis we use the intermediate Casas-Ibarra~\cite{CasasIbarra} parametrisation, where one fixes both low and high energy parameters to infer the neutrino Yukawa coupling $\la$. This useful parametrisation derives from the fact that since $\la.\la^{\dagger}$ is an hermitian matrix, it can be diagonalised by an orthogonal matrix $R$, satisfying $R.R^{T}=1$. The matrix $R$ is parametrised in terms of three complex angles. Here we use the following parametrisation
\bea
\label{Rmatrix}
R=\left[\begin{array}{ccc}
c_{3}c_{2} & c_{3}s_{2} & s_{3} \\
-c_{1}s_{2}-s_{1}s_{3}c_{2} & c_{1}c_{2}-s_{1}s_{3}s_{2} & s_{1}c_{3} \\
s_{1}s_{2}-c_{1}s_{3}c_{2} & -s_{1}c_{2}-c_{1}s_{3}s_{2} & c_{1}c_{3} 
\end{array}\right] \, ,
\eea
where, as for the PMNS matrix, $c(s)_{ij}=\rm{cos(sin)}(z_{ij})$, the $z_{ij}$s being complex angles. One can therefore reconstruct $\la$, according to
\bea
\la_{1 \al}=\left(\sqrt{M}.R.\sqrt{m}.U^{\dagger}\right)_{1\al} \ ,
\eea
where $M=\rm{diag}(M_{1},M_{2},M_{3})$ is the heavy neutrino mass matrix. Hence, in this parametrisation, among the $18$ parameters, $15$ are free: in the low energy sector, once the solar and atmospheric $\D m^{2}$ and mixing angles are fixed, the neutrino mass scale -$\rm{min}(m_{i})$- and the mass ordering -$\rm{sign}(\D m^{2}_{13})$- remain free, as well as the mixing angle $\theta_{13}$ and the three $CP$ phases $\delta-\phi_{1,2}$. In the high energy sector, the $R$ matrix and the 3 right-handed neutrinos masses are unconstrained, hence there are 9 free parameters~\footnote{Actually there are 3 more degrees of freedom which are the $CP$ parities of the RHns, and one should use $R^{\prime}=R.{\rm{diag}}(\pm 1,\pm 1,\pm 1)$, with $R$ given by eq.(\ref{Rmatrix}). Here we consider the case $R^{\prime}=R$.}. As stated above, among these parameters, $\rm{min}(m_{i})$, $\theta_{13}$, ${\rm{sign}}(\D m^{2}_{23})$ and $\delta$ are expected to be determined in a near future, whereas the 11 remaining parameters will probably not. This is clearly a problem if one wants to make some predictions, hence the will to reduce this number. There are different ways to do so.\\
In the above discussion, we assumed that 3 right-handed neutrinos are added to the SM particle content. Actually, the minimal extension of the SM requires only 2 RHns to generate the 2 oscillation frequencies. In this model~\cite{2RHN}, the lightest LH neutrino is massless $\rm{min}(m_{i})=0$, hence the 6 yet free low-energy parameters are reduced to 4, since additionnaly the Majorana $CP$ violating phase associated to $\rm{min}(m_{i})$, $\phi_{1}$, can be rotated away. Furthermore, the high-energy sector reduces to  4 unknowns, although there is little experimental or theoretical support for this hypothesis.
%Another way to constrain the high-energy sector is to suppose textures for right-handed neutrinos~\cite{Textures}, which may be seen as an avatar of a grand-unified theory realisation, where ideally these parameters can be fully determined~\cite{hep-ph/0612021}.\\
%We will study flavoured leptogenesis in the type I seesaw model in chapter 5.
\subsection{Scalar triplets: type II seesaw}
This model consists in the addition of a scalar triplet $\D$ charged as $(\tbf{1},\tbf{3},\tbf{1})$ under $SU(3)\times SU(2)_{L}\times U(1)_{Y}$~\cite{typeII}. This field is a real scalar triplet under $SU(2)$, which lies in the adjoint representation. $\D$ can couple to a product of two left-handed fields. Writing
\bea
\D =\left(\begin{array}{cc} \D^{+} & \sqrt{2}\,\D^{++} \\ \sqrt{2}\,\D^{0} & -\D^{+} \end{array}\right)\, ,
\eea
and assuming that $\D$ takes a non-zero vev along its $U(1)_{em}$ singlet direction, $\left\langle\D^{0}\right\rangle=v_{L}\neq 0$, a Majorana mass term for light neutrinos can be generated. Indeed, since $Q_{Y}(\D)=-2\,Q_{Y}(\ell_{L})$, we have in the Lagrangian the additional term
\bea
\mathcal{L}\supset -\frac{1}{2}\,f_{ij}\,\ell_{L\,i}^{T}\,\D \,(\rm{i}\sigma_{2})\,C\,\ell_{L\,j}\, \stackrel{\left\langle\D^{0}\right\rangle =v_{L}}{\longrightarrow} -\frac{1}{2}\left(m_{\nu}^{II}\right)_{ij}\,\nu_{L\,i}^{T}\,C\,\nu_{L\,j} \, .
\eea
In this so-called type II seesaw, the light-neutrino mass is given by $m_{\nu}=f\,v_{L}$ and its suppression originates from the small vev of $\D^{0}$ $v_{L} \ll v$. This smallness can be understood from the other couplings of the triplet that are allowed in the Lagrangian, in particular its coupling with the SM Higgs doublet $\phi_{d}$:
\bea
\mathcal{L}_{\D}\supset -\mu_{\D}\,\phi_{d}^{T}\,(\rm{i}\sigma_{2})\,\D\,\phi_{d}-M_{\D}^{2}\,\rm{Tr}(\D^{\dagger}\,\D)+...
\eea
where the dots represent higher order couplings. When $\phi$ and $\D$ develop their vev, assuming $M_{\D}\gg v$, one obtains a seesaw like relation between $v_{L}$ and $v$:
\bea
v_{L}=-\frac{\mu_{\D}\,v^{2}}{M_{\D}^{2}} \ll v\, .
\eea
This ensures the smallness of $m_{\nu}^{II}$. This mechanism naturally occurs in GUT models, where scalar triplets are contained in the higher dimensional representions of Higgses used to break the GUT group down to the SM. In such a GUT framework, it is frequent that both type I and type II seesaw mechanisms occur.
\subsection{Fermion triplets: type III seesaw}
Finally, a third type of seesaw mechanism can occur, where fermion triplets $\Sigma$ are added to the SM particle content~\cite{typeIII}. These fermion triplets couple to leptons via additional couplings $\propto \la_{\Sigma}\,\ol{\ell}\,\phi_{u}\,\Sigma+h.c.$. Since these triplets reside in the adjoint representation of $SU(2)_{L}$, a Majorana mass term for $\Sigma$ is allowed $\propto M_{\Sigma}\,\ol{\Sigma}\,\Sigma^{c}$. In these scenarios, masses for the light neutrinos are generated according to:
\bea
m_{\nu}^{III}=-\frac{v^{2}}{2}\,\la_{\Sigma}^{T}\,M_{\Sigma}^{-1}\,\la_{\Sigma}\, ,
\eea
which is similar to the type I seesaw formula. %
\subsection*{Probing the seesaw ?}
All these mechanisms are built to account for a suppressed light neutrino mass. However, the scale at which $B-L$ is violated depends on the model and on its input parameters, naturally implying very different low-energy signatures~\cite{testseesaw}. It would be very interesting to determine which scenario is the most likely to occur, since this knowledge has important implications, notably in the building of a high energy model consistent with low-energy data.
\subsection*{Seesaw and leptogenesis}
The seesaw models share the interesting feature that the light neutrino mass is suppressed owing to the introduction of a intermediate scale -$M$, $M_{\D}$ or $M_{\Sigma}$- between the SM one and the GUT one, the seesaw scale, at which the lepton number is violated. This is a welcome by-product of seesaw, which is used in leptogenesis models. In chapters 3, 4 and 5, we will investigate leptogenesis in type I seesaw, while leptogenesis in type I+II seesaw is studied in chapter 6. We do not consider leptogenesis in the type III seesaw. In~\cite{LeptoTypeIII}, it has been found that the role played by $\Sigma$ in leptogenesis is similar to the one played by fermion singlets. 
\newpage
\chapter[The single flavour approximation]{Leptogenesis \\ in the single flavour approximation}
In the previous chapter we have seen that the seesaw mechanism, in addition to explaining neutrinos masses and mixings, implies the violation of lepton number. This $L$ violation is a welcome by-product of these models, and is the building-block of the Fukugita and Yanagida seminal paper~\cite{Fukugita}, where the foundations of leptogenesis were first laid.\\
In this chapter, we introduce leptogenesis in the so-called one-flavour approximation~\cite{lepto1saveurgenerale,towards,pedestrians}, which in most cases provides good estimates of the baryon asymmetry. Here we focus our discussion on the type I seesaw, the type II being latter addressed in chapter 6.\\
\section{Basics of leptogenesis}
In the type I seesaw model, the added right-handed neutrinos are very heavy particles, and only interact with leptons and the Higgs boson, via their Yukawa couplings. Being heavy, the right-handed neutrinos decouple early from the thermal bath. When they decouple, their decay will create leptons and antileptons, and if $CP$ is violated during these decays, an excess of leptons over antileptons can be obtained. Once right-handed neutrinos are completely frozen out, the lepton-antilepton asymmetry will evolve without being affected, as all other processes are $CP$ conserving. %carry on, as all other processes are CP conserving. 
Then the fast $B+L$ violating processes that are in-equilibrium partly convert this lepton asymmetry into a baryon asymmetry. This very rough picture highlights the three main features of leptogenesis: 
\begin{itemize}
\item A lepton-antilepton asymmetry is created due to $CP$-odd couplings/processes.
\item At the same time, leptons and antileptons are liable to inverse reactions that potentially wash-out the asymmetry.
\item Sphalerons partly convert the lepton asymmetry into a baryon asymmetry. 
\end{itemize}
We will discuss the different points in detail hereafter, but let us first give a qualitative picture of thermal leptogenesis.\\
%In the type I seesaw, 3 right-handed neutrinos (RHn) $N_{i}$ are added to the SM particle content.
The right-handed neutrinos are assumed to be Majorana spinors, and so can decay into both lepton and antilepton, due to the Yukawa coupling
\bea
\mathcal{L}\supset \la_{\al i}\ol{N}_{i}\,\ell_{\al}\,\phi+\la_{\al i}^{\dagger}\,N_{i}\,\ol{\ell}_{\al}\,\phi^{*}\,,
\eea
$\ell_{\al}$ being the lepton doublet of flavour $\al$, while $\phi$ is the SM Higgs doublet.\\
As we saw in the previous chapter, this coupling $\la$ is in all generality a $3\times3$ complex matrix, which, if complex, implies that $N_{i}$'s interactions distinguish between leptons and antileptons: a $CP$ asymmetry can be created in the decays of RHns. This $CP$ asymmetry is defined by the difference between the decay rate into leptons and antileptons
%, which couple to charged leptons and neutrinos via an extra Yukawa coupling $Y_{\nu}^{i \alpha}\ol{N_{i}} \ell_{\alpha} \phi +h.c.$. As seen in the previous chapter, this coupling is in all generality a $3\times3$ complex matrix, meaning that $N_{i}$'s interactions distinguish between fermions and anti-fermions: CP is violated. The CP asymmetry is defined as 
\bea
\epsilon_{CP}\equiv\frac{\Gamma(N\rightarrow \ell \phi)-\Gamma(N\rightarrow \ol{\ell} \ol{\phi})}{\Gamma(N\rightarrow \ell \phi)+\Gamma(N\rightarrow \ol{\ell} \ol{\phi})} \ .
\eea 
In the thermal scenario of leptogenesis, a population of right-handed neutrino is first created from the thermal bath via inverse decays $\ell \phi \rightarrow N$, $\ol{\ell} \ol{\phi} \rightarrow N$ and scatterings. From these inverse decays and scatterings a lepton asymmetry is produced, defined as:
\bea
Y_{L}\equiv \frac{n_{L}-n_{\ol{L}}}{s} \, ,
\eea
where $Y_{x}=n_{x}/s$ is the comoving number density, the number density to entropy density ratio. During the thermalisation of RHns, $Y_{L}$ is proportionnal to $- \epsilon_{CP} Y_{N}$.\\
Then, as the temperature of the Universe decreases, RHns begin to decouple and decay, creating an asymmetry $\propto  \epsilon_{CP} Y_{N}$, that potentially cancels out with the former one. However, during and after thermalisation of RHns, the lepton asymmetry undergoes washout processes and so is partly depleted. An exact cancellation is therefore avoided. The surviving asymmetry  is then $\propto \epsilon_{CP} \,Y_{N}\,(1-1/C_{wo})\simeq \epsilon_{CP}\,\eta \neq 0$, where $C_{wo}\gtrsim 1$ reflects the washout of the asymmetry. $\eta$ is called the efficiency factor; it reflects the competition between production and depletion of the asymmetry, as well as the ability of those process to thermalise the $N$'s. In thermal leptogenesis, $\eta \simeq \mathcal{O}(0.1)$. The overall factor is the equilibrium density of RHn at the beginning of the leptogenesis epoch, and is roughly $10^{-3}$.
% The overall proportionality constant is related to equilibrium thermodynamic and is $\simeq 10^{-3}$.
A baryon asymmetry is then generated owing to sphaleron processes that partly convert the $L$ asymmetry into a $B$ asymmetry, with a conversion factor that depends on the model and on the temperature, but is roughly $-1/2$.\\
The amount of baryon asymmetry produced by leptogenesis is then approximately
\bea
Y_{B}\simeq -\times 10^{-3}\,\eta \,\epsilon_{CP}\, ,
\eea
a value to be compared with the observed baryon asymmetry (cf appendix B for relating $Y_{X}$ to $n_{X}$):
\bea
Y_{B}^{obs.}=\frac{n_{b}-n_{\ol{b}}}{s}=\frac{\eta_{B}}{7.04}\simeq 8.7\times 10^{-11}\, .
\eea
It implies that the $CP$ asymmetry generated in RHn decays is required to be
\bea
\epsilon_{CP}\gtrsim \rm{few}\,\times 10^{-7} \ ,
\eea
which constrains the parameters of the model. Let us now discuss the different points in more detail.
\section{CP violation}
We have defined $\eps_{CP}$ as the difference between the decay rate into leptons and antileptons.
This $CP$ asymmetry emerges from the interference between the tree level decay rate and the 1-loop corrected one. Indeed, at tree level, for the decay of a right handed neutrino, and given that $\Gamma(N_{i}\Ra\ell_{\al} \phi)\propto \vert \la_{\al i} \vert^2$ and $\Gamma(N_{i}\Ra \ol{\ell_{\al}} \ol{\phi})\propto \vert \la_{\al i}^{\star} \vert^2$, trivially no CP asymmetry is possible.\\
If we define the tree level decay amplitude by $\mathcal{A}_{tree}=\alpha$, the CP conjugate process is $\ol{\mathcal{A}}_{tree}=\alpha^{\star}$, and their respective one-loop corrections are $\mathcal{A}_{1-loop}=\alpha \beta \mathcal{F}$ and $ \ol{\mathcal{A}}_{1-loop}=\alpha^{\star} \beta^{\star} \mathcal{F}$, where $\mathcal{F}$ is related to the loop function, and $\beta$ is some dimensionless coupling, then the $CP$ asymmetry is given by
\bea
\eps_{CP}&=&\frac{\int [\mathcal{D}]\left( \vert \al \left(1+\be \mathcal{F}\right) \vert ^2-\vert \al^{*}\left(1+\be^{*} \mathcal{F} \right)\vert ^2 \right)}{\int [\mathcal{D}]\left(\left \vert \al\left(1+\be \mathcal{F} \right)\vert ^2+\vert \al^{*}(1+\be^{*} \mathcal{F} \right) \vert ^2 \right)}\,, \nonumber \\
&\simeq &-2\,\IM{\be}\,\IM{ \int[\mathcal{D}] \mathcal{F}} \ ,
\eea
where $\int[\mathcal{D}]$ stands for phase-space integration.\\
We see that a non-vanishing $\eps_{CP}$ requires  both $\be$ and $\mathcal{F}$ to be complex, \ie the 1-loop corrections have to develop an absorptive part, that is, on-shell particles must be running in the loop. Moreover, the tree level and the loop corrections  have to differ by a relative phase which comes from $\be $.
\section*{Evaluation of $\eps_{CP}$}
\begin{figure}[h!]
\begin{center}
\includegraphics[trim =20mm 23cm 20mm 25mm, clip,scale=1]{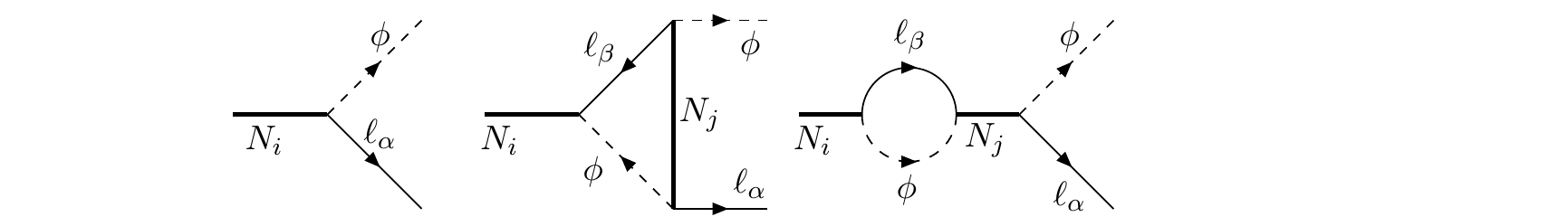} 
\end{center}
\caption{One-loop diagrams contributing to the asymmetry
from the $N_i$ decay.}
\label{figECP}
\end{figure}

In this model, the CP asymmetry arises from the interference between the tree level decay diagram and the one-loop corrections, whose Feynman diagrams are depicted in fig.\ref{figECP}.\\
%%%%%%%%%%%%%%%%%%%%%%%%%%%%%%%%%%%%%%%%%%%%%
%
\\
There are two types of corrections to the tree level diagram (a): one from the vertex correction (b) and another one from  the self-energy correction (c) ~\cite{CPdecays}-~\cite{CP}, with
\bea
\eps_{CP}=\eps_{CP}^{vertex}+\eps_{CP}^{self}\,. \nonumber
\eea
Since the decaying right-handed neutrinos are unstable particles, one cannot handle them in terms of asymptotic \textit{in-} and \textit{out-} states. One should rather deduce their properties from their on-shell contribution to lepton-lepton scatterings~\cite{CPplum}. The latter being stable, the usual formalism applies. Then, from the $S-$matrix elements of such scatterings, the $CP$ violating properties of RHn decays are inferred.\\
%However, due to the instability of right-handed neutrinos, one cannot define asymptotic $in$ and $out$ state for them. Instead one should look at their contribution to scatterings of stable particles, such as $\ell \phi \LRa \ell \phi$, hence determining right-handed neutrinos properties thanks to the S-matrix elements of such scatterings.
Nevertheless, it appears that in the case where the differences between RHn masses are large compared to their width differences, that is if $\vert M_{N_i}-M_{N_j}\vert \gg \vert \Gamma_{i}-\Gamma_{j} \vert$, the na\"ive calculation and the careful one give the same results. 
%We therefore evaluate in this naive way the vertex and self-energy diagram in the non-degenerate case, then discuss the resonance of the self-energy contribution when $M_{N_i}\simeq M_{N_j}$.
We first assume such non-degeneracy of RHn masses, and evaluate $\eps^{vertex}$ and $\eps^{self}$ in the na\"ive prescription. We then discuss the resonant regime, and how it affects the self-energy contribution.
\paragraph*{Evaluation of $\eps_{CP}^{vertex}$\\}
For the process $N_{i} \Ra \ell_{\al} H$, the tree level amplitude reads
\bea
\label{Atree}
\rm{i}\,\mathcal{M}^{0}_{i,\al}=\rm{i}\,\la_{\al i}^{*} \,\ol{u}_{\ell_{\al}}(p_{\ell})\,\rm{P_{R}}\,u_{N_{i}}(p_{N}) \ ,
\eea
where $\rm{P_{R(L)}}$ is the right(left) chirality projector $\rm{P_{R(L)}}=(1+(-)\gamma_{5})/2$.\\
The vertex correction reads
\bea
\rm{i}\,\mathcal{M}^{vertex}_{i,\al}&=& \sum_{j=1,2,3}\sum_{\be}(\rm{i})^{6}\,\la_{j\al}^{*}\la_{j\be}^{*}\,\la_{i\be}\,\ol{u}_{\ell_{\al}}(p_{\ell})\,\mathcal{F}\, u_{N_{i}}(p_{N})\,,
\eea
where
\bea
\mathcal{F}=\rm{P_{R}}\,\left(\int[\mathcal{D}]\,\frac{1}{\sla{k}-M_{j}}\,\rm{P_{R}}\,\frac{1}{\sla{q}-m_{\be}+\rm{i}\eps}\,\rm{P_{L}}\,\frac{1}{q^{2}_{H}-m_{H}^{2}+\rm{i}\eps}\right)\,.
\eea
The absorptive part of the loop-function comes from the fact that internal leptons $(\ell_{\be})$ and Higgses are on-shell. Moreover, we can neglect their masses, as $m_{\ell},m_{H}\ll M_{j}$\footnote{This assumptions holds at $T=0$. When taking temperature into account, particles acquire effective masses due to screening of the thermal plasma. While lepton masses can still be neglected, $m_{H}(T)\simeq 0.4 T$, and so at high temperatures $m_{H}$ can in fact be heavier than $M_{N_{i}}$.}.\\
In the loop function,  $k$, $q$ and $q_{H}$ are internal momenta, with $p_{N}=q+q_{H}$ and $k=q_{H}+p_{\ell}$. After a straightforward calculation, one obtains for the loop function:
\bea
\mathcal{F}=\frac{-\rm{i}}{16\, \pi^2}\frac{1}{M_{i}}\left(f_{1}\left(\frac{M_{j}^2}{M_{i}^2}\right)\,\sla{p_{N}}+f_{2}\left(\frac{M_{j}^2}{M_{i}^2}\right)\,\sla{p_{\ell}}\right).
\eea
As lepton masses are neglected, the part $\propto \sla{p_{\ell}}$ will not contribute and we neglect it in the following. In the expression above we introduce the function $f_{1}(x)$:
\bea
f_{1}(x)=\sqrt{x}\left(1-(1+x)\log{(1+\frac{1}{x})} \right)\left(\rm{i}\pi-(1+\log{x})\right)-(1+x)L_{2}\left(-\frac{1}{x}\right) \ ,
\eea
where $L_{2}(x)$ is the dilogarithm function.\\
The CP conjugated process $N_{i}\Ra \ol{\ell_{\al}}+H^{*}$ reads at tree level:
\bea
\rm{i} \,\ol{\mathcal{M}}_{i,\al}^{0}=\rm{i}\,\la_{\al i}\, v_{N_i}(p_{N})\,\rm{P_{L}}\,\ol{v}_{\ell_{\al}} \ ,
\eea
whereas the corresponding vertex correction reads:
\bea
\rm{i}\,\ol{\mathcal{M}}_{i,\al}^{vertex}=\sum_{j=1,2,3}\sum_{\be}(\rm{i})^{6}\,\la_{j\al}\, \la_{j\be} \,\la_{i\be}^{*}\, \frac{-\rm{i}}{16 \pi^2}\frac{1}{M_{i}} f_{1}\left(\frac{M_{j}^2}{M_{i}^2}\right)\, v_{N_{i}}(p_{N})\,\rm{P_{L}}\,\sla{p}_{N}\,\ol{v}_{\ell_{\al}}(p_{\ell}).
\eea
Then the evaluation of $\eps_{CP}^{vertex}$ is straightforward. The numerator simply reads
\bea
\rm{num}(\eps^{vertex}_{i,\al})=\frac{1}{16\, \pi^2}\frac{1}{M_{i}}\, 4 \,\left(\sum_{j=1,2,3}\IM{f_{1}\left(M_{j}^2/M_{i}^2\right)} \IM{\la_{\al i}^{*}\la_{\al j} \left(\la^{\dagger}\la\right)_{ij}}\right)\,\frac{1}{2}M_{i}^{3} \ ,
\eea
where we have detailed the different contributions, and used global four-momentum conservation $p_{N}=p_{\ell}+p_{H}$. The numerator is $\propto \la^{4}$. In this expression we write $\left(\la^{\dagger}\la\right)_{ij}=\sum_{\be}\la_{i\be}^{\dagger}\,\la_{\be j}$.\\
For the denominator one considers only the leading term:
\bea
\rm{den}(\eps^{vertex}_{i,\al})=M_{i}^{2}\,\left(\la^{\dagger}\la\right)_{ii} \ .
\eea
Thus we find the well known result concerning $CP$ violation coming from the vertex correction~\cite{CP}:
\bea
\eps_{i,\al}^{vertex}=\frac{1}{8 \pi}\sum_{j=2,3}\frac{\IM{\la_{\al i}^{*}\la_{\al j} \left(\la^{\dagger}\la\right)_{ij}}}{\left(\la^{\dagger}\la \right)_{ii}} f\left(\frac{M_{j}^2}{M_{i}^2}\right) \ .
\eea
Notice that the sum is made over $j\neq i$, since the contribution from $N_{i}$ running into the loop is the same for both $CP$ conjugated processes. In the expression above, the loop function $f(x)$ is the imaginary part of $f_{1}(x)$ :
\bea
f(x)=\sqrt{x}\left(1-(1+x)\log{(1+\frac{1}{x})} \right).
\eea

\paragraph*{Evaluation of $\eps_{CP}^{self}$\\}
Let us now turn to the evaluation of the the self-energy contribution to the CP asymmetry, still in the case of non-degenerate masses.\\
While the tree level contribution is given by eq.(\ref{Atree}), the one loop correction reads
\bea
\rm{i}\,\mathcal{M}^{self}_{i,\al}&=&-2\sum_{j=1,2,3}\la_{\al j}^{*}\,\la_{\be j}^{*}\,\la_{\be i}\nonumber \\
&\times & \ol{u}_{\ell_{\al}}(p_{\ell})\,\rm{P_{R}}\,\frac{1}{\sla{p_{j}}-M_{j}}\,\rm{P_{R}}\,\int[\mathcal{D}]\,\frac{1}{\sla{q}-m_{\be}+\rm{i}\eps}\,\frac{1}{q_{H}^{2}-m_{H}^{2}+\rm{i}\eps}\, \rm{P_{L}}\, u_{N_{i}}(p_{N}) \ ,
\eea
with obviously $p_{j}=p_{N}$, and $q_{H}=q-p_{N}$. The overall factor 2 comes from the fact that both components of the $SU(2)$ doublet can run into the loop. Analogous to the vertex correction, the contribution from $j=i$  cancels out with the $CP$ conjugated process.\\
Neglecting Higgs and lepton masses, the loop function is
\bea
\mathcal{F}=\int \frac{d^{4}q}{(2 \pi)^{4}} \frac{\sla{q}}{q^{2}+\rm{i}\eps}\frac{1}{(p_{N}-q)^{2}+\rm{i}\eps}\,,
\eea
and is clearly divergent. Using dimensional regularisation, and going on-shell $p_{N}^{2}=M_{i}^{2}$, one obtains
\bea
\mathcal{F}=\frac{\sla{p}_{N}}{(4\pi^2)}\left(\frac{1}{2}\left(\frac{2}{\eps}-\gamma+\rm{ln}(4\pi)-\rm{i}\pi \right)+1\right) +\mathcal{O}(\eps)\ ,
\eea
where $\eps=4-d$ and $\gamma$ is the Euler constant.\\
The evaluation of the self-energy correction then follows the same lines than the vertex correction. The $CP$ conjugate process $N_{i}\Ra \ol{\ell_{\al}} \phi^{*}$ presents the same divergence as above,  but the $CP$ asymmetry resulting in the difference of the two partial decay rates is finite:
\bea
\label{CPself}
\eps_{i,\al}^{self}=\frac{1}{8 \pi}\sum_{j\neq i}\frac{\IM{\la_{\al i}^{*}\la_{\al j} \left(\la^{\dagger}\la\right)_{ij}}}{\left( \la^{\dagger}\la \right)_{ii}} g\left(\frac{M_{j}^2}{M_{i}^2}\right) \ ,
\eea
with 
\bea
\label{loopself}
g(x)=\frac{\sqrt{x}}{1-x} \ .
\eea
Had we included lepton number conserving self-energy diagrams, we would have obtained an additional contribution
\bea
\eps_{i,\al}^{self^{\prime}}=\frac{1}{8 \pi}\sum_{j\neq i}\frac{\IM{\la_{\al i}^{*}\la_{\al j} \left(\la^{\dagger}\la\right)_{ji}}}{\left(\la^{\dagger}\la\right)_{ii}} g_{c}\left(\frac{M_{j}^2}{M_{i}^2}\right) \ ,
\eea
with $g_{c}(x)=x^{-1/2}\,g(x)$, which in the hierarchical limit is suppressed when compared to the lepton number violating one eq.(\ref{CPself}).\\
Adding the two contributions, and assuming that right-handed neutrinos are non-degenerate, the $CP$ asymmetry in the lepton flavour $\al$ produced by the decay of a right-handed neutrino $N_{i}$ is finally given by ~\cite{CP}:
\bea
\label{EQ325}
\eps_{i,\al}=\frac{1}{8\pi}\sum_{j\neq 1}\frac{\IM{\la_{\al i}^{*}\la_{\al j} \left(\la^{\dagger}\la\right)_{ij}}}{\left(\la^{\dagger}\la\right)_{ii}}\left(f(x_{j})+g(x_{j})\right) \
.
\eea
\paragraph*{A remark on the calculation of $\eps_{CP}^{self}$\\}
We see that while the vertex-correction is well defined, the self-energy correction is divergent for both $CP$ conjugated processes, although the difference is finite for non-degenerate RHn.\\
One expects the self-energy $CP$ asymmetry to vanish when the right-handed neutrinos are exactly degenerate, since then the $CP$ violating phases of the mixing matrix can be rotated away. We see that in the na\"ive  estimation above, the function $g(x)$ diverges when $x\Ra 1$, a clear proof that this treatment is incorrect. Actually, this comes from the fact that in the degenerate case, the widths of the RHns cannot be neglected, as will be shown below.\\
One way to deduce the $CP$ violating properties of $N$ decays is to look at Higgs-lepton two-body scatterings mediated by on-shell $N_{i}$ ~\cite{CPplum}-\cite{AnisimovETplum}. The unstable nature of right-handed neutrinos is taken into account by summing self-energy corrections in the propagator, which near the pole has a Breit-Wigner form:
\bea
S(q^2)\propto \frac{1}{q^{2}-M_{i}^2+\rm{i}\,M_{i}\,\Gamma_{i} }\ ,
\eea
where $\Gamma_{i}$ is the width of $N_{i}$, given by 
\bea
\Gamma_{i}=\frac{\left( \la^{\dagger}\la \right)_{ii}\,M_{i}}{8\,\pi} \ .
\eea
It has been found in ~\cite{AnisimovETplum}  that the self-energy correction reads (in the simplified $2$ RHn scheme):
\bea
g^{cor}(x)=-\frac{\sqrt{x}(x-1)}{(x-1)^{2}+(x\,a_{j}-a_{i})^{2}} \ ,
\eea
with $a_{k}=\Gamma_{k}/M_{i}$.\\
Then, in the non-degenerate case, for $\vert M_{N_i}-M_{N_j}\vert \gg \vert \Gamma_{i}-\Gamma_{j} \vert$, one recovers the function $g(x)$ of eq.(\ref{loopself}). When RHns are exactly degenerate, no divergence appears as $g^{cor}(1)=0$, as expected.\\
Having evaluated the $CP$ asymmetry produced by $N_{i}$ decays, let us now consider the mechanism that produces a lepton asymmetry.
\section{Leptogenesis in the single flavour approximation}
In the single flavour approximation, one assumes that only the processes involving the lightest RHn contribute to the production of a lepton asymmetry. In the case of a hierarchical spectrum, $M_{1} \ll M_{2,3}$, any asymmetry produced during $N_{2} (N_{3})$ leptogenesis, at $T\sim M_{2} (M_{3})$ is supposed be completely wiped out during the $N_{1}$ leptogenesis era, at $T\sim M_{1}$.\\
Another approximation concerns the decay products: in this simplified picture only the total lepton asymmetry is considered, without taking account the flavour content of the leptons produced. Hence $N_{1}$ decays into $L$ in a $CP$ violating way and the $CP$ asymmetry is obtained by summing over lepton flavours:
\bea
\label{EQ329}
\eps_{1}=\sum_{\al} \eps_{1,\al}=\frac{1}{8\pi}\sum_{j\neq 1}\frac{\IM{ \left(\la^{\dagger}\la\right)_{1\,j}^{2}}}{\left(\la^{\dagger}\la \right)_{11}}\left(f(x_{j})+g(x_{j})\right) \, .
\eea
Assuming $M_{1} \ll M_{2,3}$, and given the asymptotic behaviour of the loop function
\bea
f(x)+g(x)\stackrel{x\gg 1}{\longrightarrow}-\frac{3}{2\,\sqrt{x} } \, ,
\eea  
%the $CP$ asymmetry can be written
%\bea
%\eps_{1}&=&-\sum_{j=2,3}\frac{3}{16 \pi}\frac{M_{1}}{M_{j}}\frac{\IM{\left(\la^{\dagger}\la\right)_{1j}^{2}}}{\vert \la^{\dagger}\la \vert_{11}} \nonumber \\
%&=&-\frac{3\,M_{1}}{16\,\pi}\frac{\IM{\left(\la^{\dagger}\la\right)_{1j}^{2}}}{\vert \la^{\dagger}\la \vert_{11}} \ .
%\eea
and further using the seesaw formula for the light neutrino mass
\bea
m_{\nu}^{\al\be}=\la_{\al k}M_{k}^{-1}\la_{k \be}^{T} \ ,
\eea
the $CP$ asymmetry can be re-expressed as
\bea
\eps_{1}=\frac{3\,M_{1}}{16\,\pi}\,\frac{\IM{(\la^{T}.m_{\nu}^{*}.\la)_{11}}}{v^{2}\,\left( \la^{\dagger}\la \right)_{11}} \ .
\eea
Davidson and Ibarra have shown~\cite{DI} that the $CP$ asymmetry can be bounded from above by a function that only depends on right-handed and light neutrino masses. This bound reads
\bea
\label{EcpDI}
\vert \eps_{1} \vert \lesssim \eps^{DI}= \frac{3}{16 \pi}\frac{M_{1}(m_{\nu}^{\rm{max}}-m_{\nu}^{\rm{min}})}{v^{2}}\, ,
\eea
where $m_{\nu}^{\rm{max}} (m_{\nu}^{\rm{min}})$ is the largest (smallest) light neutrino mass. This bound was improved in~\cite{beta}:
\bea
\label{EcpDIimp}
\vert \eps_{1} \vert \lesssim \eps^{DI} \be(m_{\nu}^{\rm{min}},\tilde{m}_{1}) \, ,
\eea
where the function $\be$ is roughly given by:
\bea
\be(m_{\nu}^{\rm{min}},\tilde{m}_{1})\simeq  \left\{\begin{array}{ll}
1-m_{\nu}^{\rm{min}}/\tilde{m}_1 & \hbox{if $m_{\nu}^{\rm{min}}\ll m_{\nu}^{\rm{max}}$}\\
\sqrt{1- m_{\nu}^{\rm{min}\,2}/\tilde{m}_1^2}&
\hbox{if $m_{\nu}^{\rm{min}}\simeq m_{\nu}^{\rm{max}}$}
\end{array}\right. \,.
\eea
In the above equation, the smallest light neutrino mass is compared to the rescaled decay rate, 
\bea
\label{tildem}
\tilde{m}_{1}=8 \,\pi\,\frac{v^{2}}{M_{1}}\Gamma_{N_{1}}=\left(\la^{\dagger}\la \right)_{11}\,\frac{v^{2}}{M_{1}} .
\eea
The Davidson-Ibarra bound of eq.(\ref{EcpDI}) has important consequences for leptogenesis, since it fixes the mass scale of right-handed neutrinos.
% bound on $\eps_{1}$ has an important consequence that it translates into a lower bound on the mass of the decaying heavy neutrino.
Indeed, given that the baryon asymmetry produced by leptogenesis should at least match the observed one
\bea
Y_{B}\simeq 1.38\times 10^{-3}\,\eta \,\epsilon_{1}\gtrsim Y_{B}^{obs}\simeq 8.7 \times 10^{-11} \ ,
\eea
$M_{1}$ is lower bounded by:
\bea
M_{1}&\gtrsim& 6.5\times 10^{8} \GeV \,\left(\frac{1}{\eta}\right)\,\times\left(\frac{Y_{B}}{Y_{B}^{obs}}\right)\,\left(\frac{m_{\rm{atm}}}{m_{\nu}^{\rm{max}}-m_{\nu}^{\rm{min}}}\frac{1}{\be(m_{\nu}^{\rm{min}},\tilde{m}_{1})}\right)\,, \nonumber \\
&\simeq & 6.5\times 10^{8} \GeV \,\left(\frac{1}{\eta}\right) \ .
\eea
We will evaluate the efficiency factor $\eta$ later; for the moment it is enough to say that $\eta\lesssim 1$.
We thus see that the RHn mass scale has to be far beyond the electroweak scale, and this has many important  implications.\\
One of them is that the seesaw mechanism is hardly testable (if even possible), since such a high mass scale will (presumably) never be reached in laboratory experiments, and furthermore since the low-energy effects of RHns are highly suppressed.\\
Another caveat emerges when leptogenesis is embedded in a supersymmetric (SUSY) framework, and stems from the tension with gravitino over-production. We will discuss this issue in chapter 6, when studying leptogenesis in a SUSY-GUT framework.
%Such a high energy scale for leptogenesis is a potential source of problem for many reasons.\\
%The first one is that we will never be able to produce such a heavy particle in laboratory experiment, or any contribution of these right-handed neutrinos is highly suppressed hence extremely difficult (if possible) to observe.\\
%Another reason is the potential conflict with standard cosmology in supersymmetric versions of leptogenesis. Indeed, for the lepton asymmetry not to be washed out during reheating, leptogenesis has to occur at temperature lower than $T_{rh}$. Hence $T_{rh} \gtrsim 10^{8} \GeV$, and such a high reheating temperature may lead to unacceptable gravitinos or dangerous relics densities. However we discuss here only SM leptogenesis, for which no such tension exist. We discuss in more details the reheating problem in chapter 7 when dealing with a supersymmetric type II seesaw version of leptogenesis.
\section*{Basic framework of leptogenesis}
\begin{figure}[htb]
\begin{center}
  \includegraphics[scale=0.45]{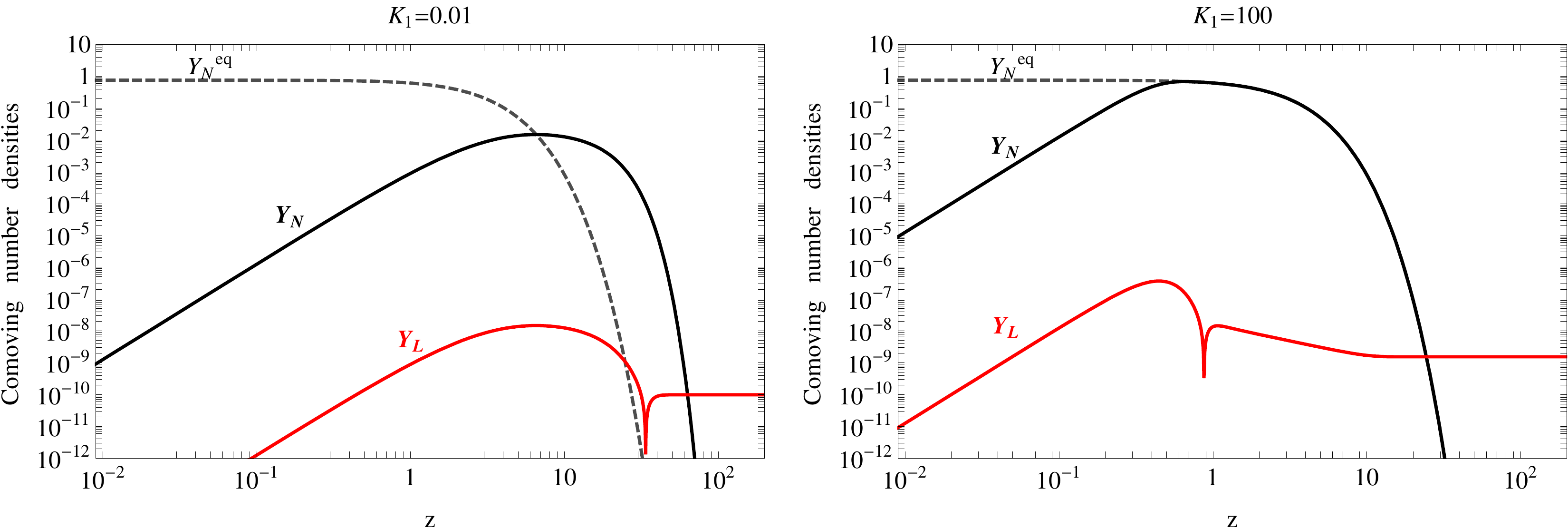}
    \caption{Evolution of comoving number densities, solutions of eqs.(\ref{eqBE1FLA}), as functions of $z=M_{1}/T$. The grey dashed line represents $Y_{N_1}^{eq}$, the number density of $N_{1}$ in thermal equilibrium, whereas the black line stands for $Y_{N}$, the solution of the Boltzmann equation. In red (light grey) is depicted the lepton asymmetry $Y_{L}$. For all these plots the $CP$ asymmetry is taken equal to $\eps_{1}=10^{-6}$, while the washout parameter is  $K_{1}=0.01 (100)$ on the left (right).} 
\label{Graphe1FLA}
\end{center}
\end{figure}

After having discussed its different building-blocks, let us see how leptogenesis works in the single flavour approximation~\footnote{We use the labelling convention of~\cite{pedestrians} ($D(z),K_{1},...$), upon which this section is inspired.}.\\
For the sake of illustration, we first consider a very simplified scheme, where we only include the leading terms at order $\mathcal{O}(\la^{4})$ that are decays and inverse-decays, and the on-shell contribution of $\D  L=2$ scatterings~\cite{towards}, included for consistency. We then evaluate the efficiency of leptogenesis in different washout regimes, and finally list the various processes which should be included in leptogenesis.\\
%We consider first a very simplified picture where we only include $\mathcal{O}(\la^{2})$ processes, that is decays and inverse decays, together with the resonant contribution of the $\Delta L=2$ scatterings for consistency. In this framework, the Boltzmann equations for the comoving number densities are given by:
The study of the evolution of the number densities $n_{x}(t)$ is found to be more tractable in terms of comoving number densities, $Y_{x}=n_{x}/s$, since both the number density $n_{x}$ and the entropy density $s$ scale as $T^{3}$.\\ The final lepton asymmetry is evaluated by solving the set of Boltzmann equations (BEs) for $Y_{L}$ and $Y_{N}$, which drive their evolution:
\bea
\label{eqBE1FLA}
\frac{d Y_{N}(z)}{dz}&=&-D(z)\,K_{1}\,(Y_{N}(z)-Y_{N}^{eq}(z))\,, \nonumber \\
\frac{d Y_{L}(z)}{dz}&=&\eps_{1}\,D(z)\,K_{1}\,(Y_{N}(z)-Y_{N}^{eq}(z))-W_{id}(z)\,K_{1}\,Y_{L}(z) \, ,
\eea
and compare the comoving number densities to their equilibrium value $Y_{X}^{eq}$. In these equations, the evolution parameter is $z=M_{1}/T$ and the decays $D(z)$ and inverse decays $W_{id}(z)$ reads:
\bea
\label{EQ514}
D(z)&=&z\,\frac{{\mathcal{K}}_{1}(z)}{{\mathcal{K}}_{2}(z)}\,, \nonumber \\
W_{id}(z)&=&\frac{1}{2}\frac{Y_{N}^{eq}}{Y_{\ell}^{eq}}\,D(z)=\frac{1}{4}\,z^{3}\,{\mathcal{K}}_{1}(z) \, ,
\eea
where ${\mathcal{K}}_{i}(z)$ are the modified Bessel function of the second kind. The lepton asymmetry production term is $\eps_{1}\,K_{1}\,D(z)$, where $\eps_{1}$ is defined in eq.(\ref{EQ329}), and the parameter $K_{1}$ reflects how fast are decays compared to the Hubble expansion rate:
%the washout parameter $K_{1}$ reflects the state of equilibrium of decays:
\bea
K_{1}=\frac{\Gamma(N_{1})}{H(M_{1})}=\frac{\tilde{m_{1}}}{m^{*}} \, ,
\eea
alternatively defined in terms of the ratio of the effective neutrino mass $\tilde{m}_{1}$, defined in eq.(\ref{tildem}) over the equilibrium neutrino mass $m^{*}$,
%and this state of equilibrium is alternatively given in term of the ratio of the effective neutrino mass
%\bea
%\tilde{m_{1}}=(\la \la^{\dagger})_{11}\,\frac{v^{2}}{M_{1}}
%\eea
%over the equilibrium neutrino mass
\bea
m^{*}=\frac{v^{2}}{M_{Pl}}\,\sqrt{\frac{256\,\pi^{5}\,g_{*}}{45}}\simeq 1.08\,\times 10^{-3} \eV \ ,
\eea
where $v\sim 174 \GeV$ is the SM Higgs vev, $M_{Pl}\sim 1.22\,\times 10^{19} \GeV$ is the Planck mass, and $g_{*}=106.75$ is the SM effective number of degrees of freedom at $T\gg T_{EW}$. For $K_{1}\gg 1$ or alternatively $\tilde{m}\gg m^{*}$, decays are in-equilibrium and lepton asymmetries are strongly washed out, while for $K_{1}\ll 1$ or $\tilde{m}\ll m^{*}$, the asymmetries are weakly washed out, as decays are out of equilibrium.\\
As for the equilibrium densities, even if  a Maxwell-Boltzmann distribution is used for the evaluation of the interaction rates, we correct them with a factor $3 \zeta(3)/4\simeq 0.9$, $\zeta$ being the Riemann Zeta function, to match the latter with the high energy behaviour of a Fermi-Dirac distribution~\cite{KolbWolfram}, so that $Y_{N}^{eq}$ reads:
\bea
Y_{N}^{eq}(z)=\frac{45}{2\,\pi^{4}\,g_{*}}\frac{3\zeta(3)}{4}\,z^{2}\,{\mathcal{K}}_{2}(z)\stackrel{z\ll 1}{\simeq} 3.9\times 10^{-3} \ .
\eea
We see how the Sakharov's conditions are fulfilled in this leptogenesis scenario:
\begin{itemize}
\item Baryon number violation: in this case, if $L$ was not violated during decays, we see that starting from a vanishing $Y_{L}$ no asymmetry could be generated.
\item $CP$ violation: $\eps_{1}\neq 0$ is mandatory.
\item Departure from thermal equilibrium: it is clearly seen above that, if right-handed neutrinos are in-equilibrium all along their evolution, the production term is null. Notice that this last point is only satisfied if the contribution of on-shell RHns to $\D L=2$ scatterings has been included.
\end{itemize}
Typical solutions of the set of eqs.(\ref{eqBE1FLA}) are depicted in fig.\ref{Graphe1FLA}, for various values of $K_{1}$, while the $CP$ asymmetry is arbitrarily set to $\eps_{1}=10^{-6}$. \\
\\
These solutions exhibit a very different behaviour. %Let us discuss them, so that we can evaluate the final lepton asymmetry.
%\paragraph*{thermalisation of $N_{1}$s\\}
%For high temperature, $z\ll 1$, the dominant processes are inverse decays $\ell \phi \Ra N$. While these processes  thermalise right-handed neutrinos, increasing its density, it creates in negative a lepton asymmetry:
%\bea
%Y_{N}(z)^{\prime}&\simeq & K_{1} \,D(z)\,Y_{N}^{eq}\simeq \frac{135 \zeta(3)}{8 \pi^{4} g_{*}}\,K_{1}\,z^{2} \nonumber \\
%Y_{L}^{\prime} &\simeq & -\eps_{1}\,Y_{N}(z)^{\prime} \ , \nonumber
%\eea
%During the early stage of thermalisation, for $z\ll 1$, given Bessel function asymptotic behaviour (cf appendix) one has
%\bea
%Y_{N}(z\ll 1)&\simeq &\frac{45 \zeta(3)}{8 \pi^{4} g_{*}}\,K_{1}\,z^{3} \nonumber \\
%Y_{L}(z\ll 1)&\simeq &-\eps_{1}\,\frac{45 \zeta(3)}{8 \pi^{4} g_{*}}\,K_{1}\,z^{3} \, .
%\eea
%thermalisation of $N_{1}$ occurs until they reach their equilibrium density. The more fast inverse decays are compare to the Hubble expansion rate, the more early are the $N_{1}$ thermalised.\\
%Once the equilibrium is reach, the evolution of the number densities depends on the strength of the washout.
\section{Evaluation of the lepton asymmetry}
This evaluation proceeds mainly in two parts, which correspond to the different contributions to the final asymmetry.\\
We work in the thermal scenario of leptogenesis, hence before decaying, right-handed neutrinos have to be produced. This thermalisation occurs due to scatterings/inverse decays of leptons present in the thermal bath. During this first stage, an (anti-)asymmetry is created.\\
If $K_{1}\gg 1$, since decays and inverse decays are far in-equilibrium, the thermalisation of RHns is fast and the latter reach thermal equilibrium at high temperatures. Conversely, if $K_{1}\ll 1$, thermal equilibrium  occurs late, as can be seen in fig.\ref{Graphe1FLA}. In both cases, during this first stage a lepton asymmetry is produced, which is $\propto \eps_{1}\,Y_{N}$.\\
Once thermal equilibrium is reached, the subsequent evolution of RHns and of the lepton asymmetry also depends on $K_{1}$. In the strong washout regime, $K_{1}\gg 1$, the fast decays maintain $Y_{N}$ close to its equilibrium value, whereas in the opposite weak washout regime $K_{1}\ll 1$, these decays occur late, when RHns are far out of equilibrium. During this second stage, a lepton asymmetry is produced, which may cancel out with the former one. This can be seen in fig.\ref{Graphe1FLA}, where the compensation between first and second stage asymmetry production is visible in the peak occuring at $z\sim 30 (1)$ for $K_{1}=0.01 (100)$.\\
Finally, as the temperature decreases, as RHns become too diluted, together with the fact that the different processes are freezing, the lepton asymmetry reaches its final value.\\
Let us evaluate this value in the two characteristic washout regimes.
\paragraph*{Strong washout regime: $K_{1}\gg1$\\}
%If $K_{1}\gtrsim 1$, decays and inverse decays are fast enough to keep $N$ close to thermal equilibrium, and one roughly has $Y_{N}\simeq Y_{N}^{eq}$ for temperature cooler than the equilibrium one, as can be seen in the right panel of fig.(\ref{Graphe1FLA}). In fact, after thermalisation, right-handed neutrino density is slightly bigger than the equilibrium one, hence $N$ decays will produce a lepton asymmetry of the opposite sign than the one resulting of thermalisation. If decays are fast enough, a significant amount of $Y_{L}$ is produced, that is washed out by lepton inverse decays. Finally,  $N$ become too diluted so that they can significantly interact and modify $Y_{L}$, and inverse decays become far out-of-equilibrium so that $Y_{L}$ gets its final value.\\
%We can evaluate this value quite easily. 
In the strong washout regime, the main contribution to the production of a lepton asymmetry comes from the second stage, after RHns are thermalised, because thermalisation is fast when inverse decays are far in-equilibrium. Thus, defining $\Delta N=Y_{N}-Y_{N}^{eq}$, the fact that the fast processes keep $N$ close to its equilibrium abundance implies that:
\bea
\Delta N^{\prime}(z)=-K_{1}\,D(z)\,\Delta N(z)-Y_{N}^{eq\,\prime}(z)\simeq 0 \, .
\eea
The formal solution of  eq.(\ref{eqBE1FLA}) for the lepton asymmetry then reads:
\bea
\label{formsol}
Y_{L}(z)&=&\eps_{1}\int_{0}^{z} dx\,K_{1}\,D(x)\,\Delta N(x)\,e^{-K_{1}\int_{x}^{z}W_{id}(y)\,dy}\,, \\
&=&-\eps_{1}\,\int_{0}^{z} dx\,Y_{N}^{eq\,\prime}(x)\,e^{-K_{1}\int_{x}^{z}W_{id}(y)\,dy} \ .
\eea
We can rewrite the integrand of eq.(\ref{formsol}) as $\int\,e^{-f}$, with 
\bea
f(x)=K_{1}\int_{x}^{z}dy\,W_{id}(y)-\rm{ln}(Y_{N}^{eq\, \prime}(x)) \, . \nonumber
\eea
This integral is evaluated using a saddle point approximation, and gets its maximum for
\bea
\frac{K_{1}}{4}\sqrt{\frac{\pi}{2}}\,\ol{x}^{5/2}\,e^{-\ol{x}}\simeq 1\, , \nonumber
\eea
that is, for
\bea
\ol{x}\simeq -\frac{5}{2}\,\rm{ProductLog}\left[\frac{-4\,K_{1}^{-2/5}}{5\,\pi^{1/5}}\right]\,, \nonumber 
\eea 
where $\rm{ProductLog}(z)$ is the Lambert $W(z)$ function\footnote{The Lambert function has an accurate asymptotic expansion for $z\gtrsim 3$: $W(z)=\rm{ln}(z)-\rm{ln(ln(z))}+...$~\cite{Math1}.}. For $K_{1}\gtrsim 1$, we find that the main contribution is for $z\gtrsim 1$, as $\ol{x}\propto \rm{ln}(K_{1})$. The lepton asymmetry in this strong washout regime is then accurately given by
\bea
\label{Strong}
Y_{L}\simeq \eps_{1}\left(\frac{0.4}{K_{1}^{1.16}}\right)Y_{N}^{eq}(z_{in}) \, ,
\eea
and the efficiency factor is therefore
\bea
\eta_{s}=0.4 /K_{1}^{1.16}\,.
\eea
\paragraph*{Weak washout regime: $K_{1}\ll1$\\}
The regime of weak washout is depicted in the left panel of fig.\ref{Graphe1FLA}. In this regime, since inverse decays are slow compared to the Hubble expansion rate, the right-handed neutrinos reach thermal equilibrium at $z_{eq}\gg 1$. Once thermalised, RHns do not stay in thermal equilibrium, but rather decay late.\\
These decays produce a lepton asymmetry that cancels out the one produced during thermalisation. However, since leptons have more time to participate in inverse reactions, this cancellation is only partial, preventing $Y_{L}=0$.\\
%As inverse decays are slowest than in the strong washout regime,  $N_{1}$s reach their thermal equilibrium abundance at lower temperature.\\
%Furthermore, decays and inverse decays are too slow for the $N_{1}$ to track equilibrium. Rather, once reach their maximum density, right-handed neutrinos maintain almost constant for a while, as decays are too slow compare to the expansion rate to significantly bring down $N_{1}$'s density.\\
%Nevertheless, during that time leptons have enough time to participate in inverse reaction, hence their number density is slightly washed out.\\
%Finally, as the temperature cools down decays become in-equilibrium, and the $N$ density abruptly decrease while the $N$ decay and produce a lepton asymmetry. This lepton asymmetry cancels with the one produced in hollow during thermalisation, and the only surviving asymmetry comes from the washout that happened before decays which prevent from a total cancellation.\\
Let us evaluate the different contributions.\\
Before the $N$s reach their thermal equilibrium abundance, $Y_{N}^{eq}$ dominates over $Y_{N}$ and so
%and one has
\bea
Y_{N}^{\prime}(z)&\simeq &K_{1}\,D(z)\,Y_{N}^{eq}(z)\,, \nonumber \\
Y_{L}^{\prime}(z)&\simeq &-\eps_{1}\,K_{1}\,D(z)\,Y_{N}^{eq}(z)-W_{id}(z)\,K_{1}\,Y_{L}(z) \, .
\eea
The lepton asymmetry is then given by
\bea
Y_{L}^{<}(z)\simeq -\eps_{1}\,K_{1}\,\int_{0}^{z}dx\,D(x)\,Y_{N}^{eq}(x)\,e^{-K_{1}\int_{x}^{z}dy\,W_{id}(y)} \ .
\eea
Defining for convenience $n_0=135 \zeta(3)/8\pi^{4}g_{*}$, such that $D(x)\,Y_{N}^{eq}(x)=4\,n_{0}\,W_{id}(x)$,  the lepton asymmetry produced at $z\lesssim z_{eq}$ is
\bea
Y_{L}^{<}(z)\simeq -\eps_{1}\,4\,n_{0}\left(1-e^{\frac{-3\pi}{8}\,K_{1}}\right) \ .
\eea
For temperatures below $T_{eq}$, $Y_{N}$ dominates over $Y_{N}^{eq}$, in such a way that
\bea
Y_{N}^{\prime}(z)=-K_{1}\,D(z)\,Y_{N}(z)\Ra Y_{N}^{>}(z)=Y_{N}^{eq}(z_{eq})\,e^{-\frac{K_{1}}{2}(z^{2}-z_{eq}^{2})}\ .
\eea
The equilibrium temperature is defined by $Y_{N}(z_{eq})=Y_{N}^{eq}(z_{eq})$. Given that
\bea
Y_{N}^{<}(z_{eq})\simeq K_{1}\int_{0}^{z_{eq}}dx\,D(x)\,Y_{N}^{eq}(x)\simeq n_{0}\,K_{1} \frac{\al\,z_{eq}^{3}}{3\,\al +z_{eq}^{3}}\,,
\eea
where $\al=\frac{3\pi}{2}$, and as $Y_{N}^{eq}(z_{eq})=n_{0}\,z_{eq}^{2}\,K_{2}(z_{eq})$
one immediately obtains $z_{eq}$:
\bea
z_{eq}=-\frac{3}{2}\rm{ProductLog}\left[-2^{2/3}\,\left(\frac{\pi}{3}\right)^{1/3}\,K_{1}^{2/3}\right]\propto \rm{ln}(1/K_{1})  \nonumber \, .
\eea
The equilibrium is reached when $Y_{N}=Y_{N}^{eq}$, at $z_{eq}\sim 10$ in fig.\ref{Graphe1FLA}.\\
We are now able to determine the final lepton asymmetry:
\bea
Y_{L}(z)&=&\eps_{1}\,\int_{0}^{z}dx\,D(x)\,\left(Y_{N}(x)-Y_{N}^{eq}(x)\right)\,e^{-K_{1}\int_{x}^{z}dy\,W_{id}(y) }\,,\nonumber \\
&=&\eps_{1}\left[-\int_{0}^{z_{eq}}dx\,D(x)\,Y_{N}^{eq}(x)e^{-K_{1}\int_{x}^{z}dy\,W_{id}(y)}+\int_{z_{eq}}^{z}dx\,Y_{N}^{\prime}(x) \right]\,,\nonumber \\
&=&\eps_{1}\left[-4\,n_{0}\,\left(1-e^{-\frac{\al}{4}K_{1}}\right)-n_{0}\,\al\,K_{1}\right] \nonumber \\
Y_{L}&\simeq &\eps_{1}\times 1.3\,K_{1}^{2}\,Y_{N}^{eq}(z_{in}) \, .
\eea
As expected, we find that the lepton asymmetry is non-zero due to the washout of the lepton asymmetry produced during thermalisation. The efficiency factor in the weak washout regime is then 
\bea
\eta_{w}=1.3\,K_{1}^{2}\, .
\eea
\paragraph*{Global parametrisation\\}
Finally, we can give a value of $\eta$ for all washouts from a simple interpolation ~\cite{towards}:
\bea
\eta\simeq \left(\eta_{w}^{-1}+\eta_{s}^{-1}\right)^{-1}\simeq\left(0.8 K_{1}^{-2}+2.5\,K_{1}^{-1.16}\right)^{-1}\,.
\eea
\paragraph*{Resulting baryon asymmetry\\}
As the Universe cools down, at $T\gtrsim M_{1}/100$, all processes are frozen and the lepton asymmetry remains constant, until being partly converted by sphalerons into a  baryon asymmetry. In the single flavour approximation, one has~\cite{BLconv}:
\bea
Y_{B}=-\frac{8\,n_{g}+4\,n_{H}}{14\,n_{g}+9\,n_{H}}\,Y_{L} \, ,
\eea
where $n_{g}=3$ is the number of fermion generations, and $n_{H}$ is the number of Higgs doublets in the considered model ($n_{H}=1$ in the SM). Hence, 
\bea
Y_{B}=-28/51\,Y_{L}\simeq -1.38\,\times 10^{-3}\,\eps_{1}\,\eta  \ .
\eea
This has to be compared with the observed value $Y_{B}^{obs}\simeq 8.7\,\times 10^{-11}$: assuming for example $\eps_{1}=10^{-6}$, $K_{1}$ should lie between $\sim 0.15$ and $\sim 15$.
\section{Dependence on the initial conditions}
\begin{figure}[h!]
\begin{center}
  \includegraphics[scale=0.45]{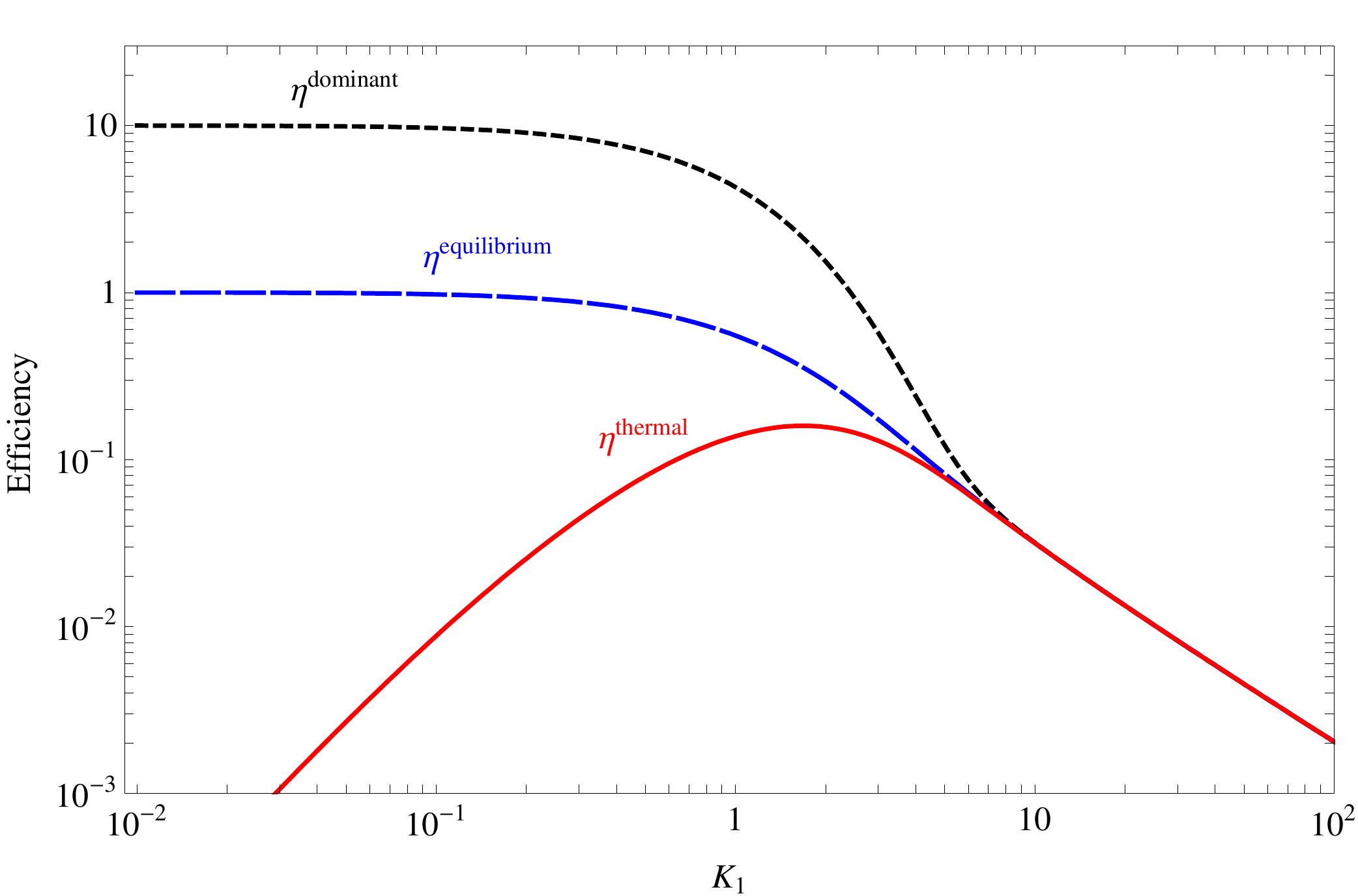}
    \caption{Efficiency factor of leptogenesis assuming different initial conditions. In red (solid line) the thermal scenario is plotted, while in long-dashed blue and short-dashed black, the equilibrium and dominant scenarios are respectively represented.} 
\label{GrapheINcond}
\end{center}
\end{figure}

In the thermal scenario, since right-handed neutrinos have to be produced by inverse decays, it is clear that a minimal amount of washout is required.\\
Actually, had we assumed that initially RHn were already thermalised, or even that their abundance was the dominant one, washout processes would not be required at all. These scenarios, which we may call equilibrium~\cite{pedestrians} or dominant~\cite{towards}, assume that before leptogenesis occurs, for $T\gg T_{lepto}$, some mechanism produces RHns. On the one hand these scenarios are model-dependent, but on the other hand the constraints derived for the thermal scenario are weakened.\\
To illustrate this, we plot in fig. \ref{GrapheINcond} the efficiency factor as a function of $K_{1}$, for the different scenarios, with $Y_{N}(z_{in})=0$ in the thermal case, $Y_{N}(z_{in})=Y_{N}^{eq}(z_{in})$ in the equilibrium one and $Y_{N}(z_{in})=10\,Y_{N}^{eq}(z_{in})$ in the dominant scenario.
\\
We observe that in the two latter cases, the efficiency is maximal for a vanishing washout, with maximum values $\eta^{eq}\simeq 1$ and $\eta^{dom}\simeq 10$, while in the thermal scenario a maximal efficiency is reached for $K_{1}\simeq 2$, for which $\eta^{th}\simeq 0.16$.\\
We further notice that for $K_{1} \gtrsim 10$, all scenarios yield the same efficiency: in a sense, the strong washout regime is more robust since it does not depend on the thermal history of the right-handed neutrinos~\cite{BdB}.
\section{Other processes}
\begin{figure}[h!]
\begin{center}
\includegraphics[trim =25mm 22cm 10mm 30mm, clip,scale=1]{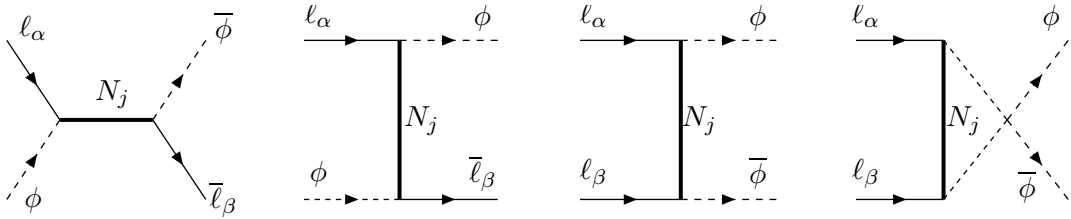} 
\end{center}
\caption{Feynman diagrams of $\Delta L=2$ scatterings mediated by $N_j, j=1,2,3$.
}
\label{DL2}
\end{figure}

\begin{figure}[h!]
\begin{center}
\includegraphics[trim =25mm 22cm 10mm 30mm, clip,scale=1]{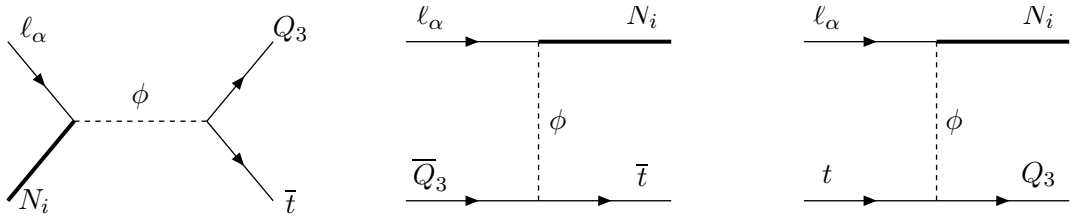}
\end{center}
\caption{Feynman diagrams of neutrino-top quark scatterings.}
\label{Topscat}
\end{figure}

\begin{figure}[htb]
\begin{center}
  \includegraphics[scale=0.45]{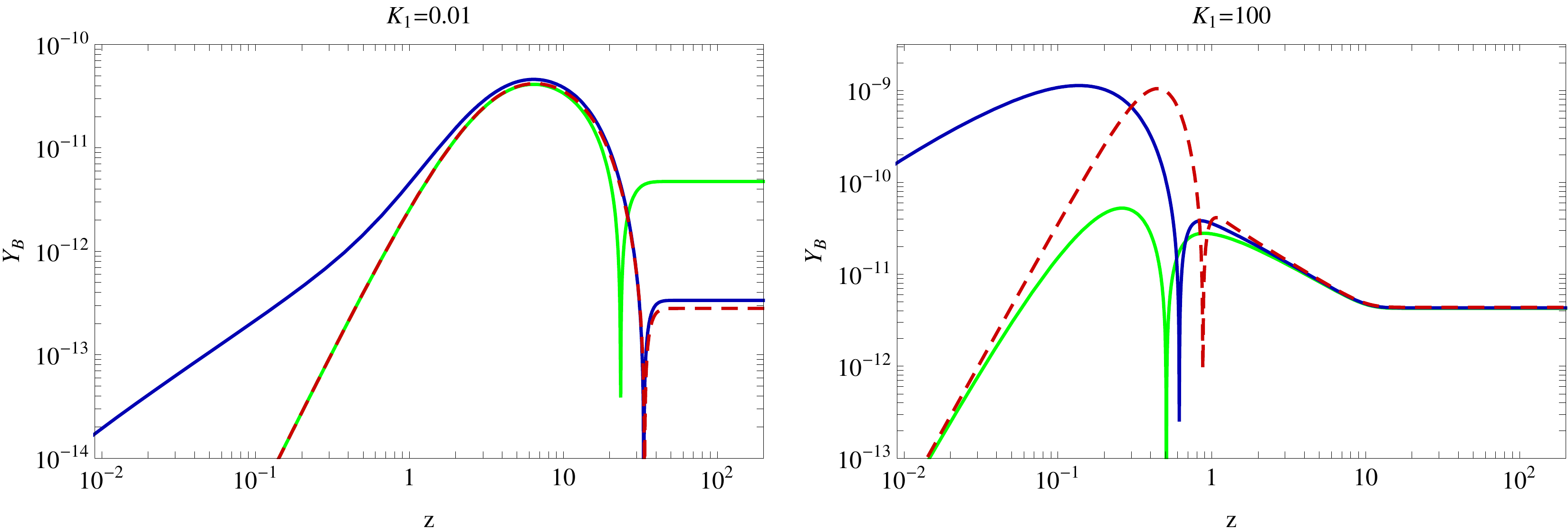}
    \caption{Influence of neutrino-top quark scatterings on the baryon asymmetry. The dashed-red line depicts the case in which scatterings are neglected, the green (light grey) line the case where scatterings are included only as damping term, while the blue (dark grey) curve represents the case where scatterings are added both as source and damping terms.} 
\label{GrapheYlProcesses}
\end{center}
\end{figure}

In the discussion above we only have included decays and inverse decays. But there are many other processes to take into account, that strongly modify the previous results. Let us discuss these terms.
\paragraph*{$\mathcal{O}(\la^{2})$ and $\mathcal{O}(\la^{4})$ processes\\}
Decays and inverse decays are $\mathcal{O}(\la^{2})$, but no $CP$ violation is possible at this order, and one needs to include $\mathcal{O}(\la^{4})$ as discussed in the beginning of the chapter. At this order, in addition to decays and inverse decays, $2-to-2$ scatterings mediated by $N$ exchange should be included.
\begin{itemize}
\item $s-$channel $N$ exchange: $\Delta L=0$ processes  $\ell_{\al}\phi \LRa \ell_{\be}\phi$ and $\Delta L=2$ processes  $\ell_{\al}\phi \LRa \ol{\ell_{\be}}\ol{\phi}$.
\item $t-$ and $u-$channel $N$ exchange: $\Delta L=0$ processes $\ol{\phi}\phi\LRa\ell_{\al}\ol{\ell_{\be}}$ and $\Delta L=2$ processes $\ol{\phi}\ol{\phi}\LRa\ell_{\al}\ell_{\be}$.
\end{itemize}
We represent in fig.\ref{DL2} the diagrams for the lepton number violating processes.\\
%For all these processes, we see that if $\al\neq\be$ lepton flavour is violated, but not necesserely the lepton number. We distinguish between the $s-$ and $t-,u-$ channel because neglecting thermal effect, only $s-$channel processes have an on-shell intermediate right-handed neutrino. However, we saw that a proper calculation of the $CP$ asymmetry coming from self-energy correction should be made by extracting the $CP$ violation due to on-shell $N$ in a $2\LRa2$ process. Hence the resonant contribution of the $s-$channel has already been taken into account, and one has to substract this contribution to the total scattering amplitude in order to avoid double counting. While the off-shell contribution can be safely neglected to some extent, the on-shell contribution to substract is $\gamma_{N,s}=1/4\,\gamma_{D}$ and has to be taken into account even at the order $\mathcal{O}(\la^{2})$, that we did. We refer to the Appendix C for a more detailled discussion about Real-Intermediate-State substracting.\\
\\
The $s-$channel receives contributions from on-shell RHns, with a corresponding rate $\gamma_{N}^{os}\propto \gamma_{D}(\la)^{4}$. As the production term of the lepton asymmetry is $\eps_{1}\,K_{1}\,D(x)\propto \gamma_{D}\,(\la)^{4}$, the resonant $s-$channel contribution has to be included for consistency at lowest order, as we did before. Then, when including $\D L=2$ scatterings, this should be carefully done in order to avoid double counting, and the on-shell part has to be subtracted from the total $\D L=2$ interaction rates~\cite{towards,pedestrians}.\\
When off-shell $\Delta L=2$ processes are in-equilibrium, their late scatterings can spoil a successful leptogenesis by dramatically washing out the lepton asymmetry.  
%We further illustrate in fig.(\ref{GrapheDeltaL2}) the effect of including those off-shell $\Delta L=2$ on the lepton asymmetry, by plotting ratios of $Y_{L}$ computed with and without those terms.
%\begin{figure}[htb]
%\begin{center}
%\includegraphics[scale=0.5]{GrapheDeltaL2.pdf}
%\caption{Effect of off-shell $\Delta L=2$ on the lepton asymmetry: the plot represents ratio of $Y_{L}$ computed with and without the off-shell $\Delta L=2$ mediated by $N_{1}$. The different lines represent different washout of the lepton asymmetry: in black is depicted a washout of $1\%$, in solid blue $10\%$ while in dashed blue the washout is of $20\%$. The red lines represent very important washout: the solid line stands for $Y_{L}^{with}/Y_{L}^{without}=50\%$ whereas the washout is of $99\%$ for the red dashed line.} 
%\label{GrapheDeltaL2}
%\end{center}
%\end{figure}
%We see that as long as
%\bea
%\frac{M_{1}}{10^{12} \GeV}\lesssim \left(\frac{100}{K_{1}}\right)^{2}
%\eea
%we can neglect the off-shell contribution of $\Delta L=2$ scatterings harmlessly.
The term that potentially washes out the lepton asymmetry is given at low temperature by~\cite{Buchmuller1}:
\bea
W_{\Delta L=2}(z)\simeq \frac{0.186}{z^{2}}\left(\frac{M_{1}}{10^{10}\GeV}\right)\left(\frac{ \overline{m}}{1\eV}\right)^{2} \ ,
\eea
where $\overline{m}^{2}=m_{1}^{2}+m_{2}^{2}+m_{3}^{3}$. However, for $z\gtrsim 15$ this term is small compared to inverse decays $ K_{1}\,W_{id}$, and can safely be neglected. Since at low temperature inverse decays read
\bea
K_{1}\,W_{id}(z\gg1)\simeq \frac{K_{1}}{4}\sqrt{\frac{\pi}{2}}\,z^{3/2}\,e^{-z} \ ,
\eea
this condition somehow constrains $M_{1}$ and $K_{1}$~\cite{matters}:
%We thus see that it effect is more important in the degenerate case, for which $m_{1}\simeq m_{2}\simeq m_{3} =m$. This term is however harmless if its contribution is negligible compare to inverse decays, that is if
\bea
\label{eqDL2}
\frac{M_{1}}{10^{14}\GeV}\lesssim 0.1 \times K_{1} \ .
\eea
If the relation above holds, we can neglect $\D L=2$ scatterings; on the other hand, if this constraint is not satisfied, to prevent the lepton asymmetry from being washed out by these processes implies to upper-constraint the light neutrino mass scale, as we will see at the end of the chapter.
%or else if $K_{1}\gtrsim 10^{-4}$. We will study washout above $10^{-2}$, hence we can safely neglect the off-shell $\Delta L=2$.
\paragraph*{$\Delta L=1$ processes $\mathcal{O}(\la^{2}h_{t}^{2})$ and $\mathcal{O}(\la^{4}h_{t}^{2})$\\}
Given the size of the top Yukawa coupling, processes involving $h_{t}Q_{3}H_{u}\,t_{R}^{c}$ should be included. At $\mathcal{O}(\la^{2}h_{t}^{2})$, these processes consist in $2\LRa 2$ Higgs mediated scatterings, both in $s-$channel $\ell\, N\LRa \ol{t}\, Q_{3}$ and $t-,u-$ channels $\ell \,\ol{Q}_{3} \LRa N\, \ol{t}$ and $\ell\, t\LRa N\, Q_{3}$. These processes are represented in fig.\ref{Topscat}.\\
At this order in the couplings, three body decays and inverse decays $N\LRa \ell\, \ol{Q}_{3}\,t$ should also be included. Actually, these processes are phase-space suppressed, and are not quantitatively relevant ~\cite{NardiCP}, hence we will neglect them.  The scattering processes violate lepton number by one unit, and act as source term for heavy neutrino production and as damping term for the lepton asymmetry.\\
The interesting feature of these scatterings is that at high temperature, contrary to decays and inverse decays that, scaling as $z^{2}$, are suppressed, the scatterings involving top quarks become constant at high temperature. Thus, for $z \ll 1$, thermalisation of RHns is significantly faster.\\
When evaluating the lepton asymmetry in the regime $K_{1}\ll1$, we noted that $Y_{L}\neq 0$ only owing to the partial washout due to inverse decays, and this dependence was illustrated in the fact that $Y_{L}\propto K^{2}$. Including $\Delta L=1$ scatterings strongly modifies the situation, and the lepton asymmetry now scales as $Y_{L}\propto K$.\\
However, going to order $\mathcal{O}(\la^{4}h_{t}^{2})$, $CP$ is violated in $\Delta L=1$ scatterings, which then also act as source terms for the lepton asymmetry. One then recovers the dependence of $Y_{L}\propto K^{2}$ in the weak washout regime, as source and damping terms now similarly act and cancel each other at order $K$.\\
It has been shown in~\cite{matters,resonantlepto} that the $CP$ asymmetry in scatterings is the same as in decays and inverse decays. For consistency, $2\LRa3$ scatterings should be included, but as for the three body decays they are phase-space suppressed and do not quantitatively contribute in leptogenesis.\\
We illustrate the effect of top-quark scatterings in fig.\ref{GrapheYlProcesses}, where we compare the lepton asymmetry computed with and without the $\Delta L=1$ processes (included both in washout and in source terms), in the regimes of weak and strong washout. When $\Delta L=1$ top scatterings are included, the Boltzmann equations for the comoving number densities are:
\bea
\label{eqBE1FLAs}
\frac{d Y_{N}(z)}{dz}&=&-P(z)\,K_{1}\,(Y_{N}(z)-Y_{N}^{eq}(z))\,, \nonumber \\
\frac{d Y_{L}(z)}{dz}&=&\eps_{1}\,P(z)\,K_{1}\,(Y_{N}(z)-Y_{N}^{eq}(z))-K_{1}\,W(z)\,Y_{L}(z) \, ,
\eea
where the production and washout terms are defined by ~\cite{pedestrians}:
\bea
P(z)&=&D(z)+2\,S_{s}(z)+4\,S_{t}(z) \, ,\nonumber \\
W(z)&=&W_{id}(z)\left(1+\frac{1}{D(z)}\left(2\,S_{s}(z)\frac{Y_{N}(z)}{Y_{N}^{eq}(z)}+4\,S_{t}(z)\right)\right)\, ,
\eea
where $S_{s}$ and $S_{t}$ represent the Higgs-mediated scatterings in the $s-$ and $t-$ channels. We refer the reader to appendix B where these scatterings are discussed in more detail.\\
\\
We clearly see the influence of scatterings in the plots: when one considers only the contribution of scatterings to the washout, they prevent the cancellation of the lepton asymmetry produced during thermalisation with the one generated in decays of the RHn, so that the final $Y_{L}$ is $\propto K_{1}$ instead of being $\propto K_{1}^{2}$. When the $CP$ asymmetry in scatterings is included, we see that $Y_{L}$ gets suppressed as $K_{1}^{2}$.\\
Among $\Delta L=1$ scatterings, one should take into account scatterings involving Higgs bosons-$U(1)$ or $SU(2)$ gauge bosons couplings~\cite{towards,resonantlepto}. These scatterings behave similarly to the top-quark scatterings, accelerating the thermalisation of right-handed neutrinos, and acting as a source and damping terms for the lepton asymmetry. We neglect these terms in this thesis, even if they are of the same order -or even larger than the top-quark scatterings-, as their inclusion does not drastically modify the picture here drawn.  
\section{Upper-bound on the light neutrino mass scale}
The requirement that leptogenesis should explain the observed baryon asymmetry of the Universe has led to a constraint on the light neutrino mass scale. Indeed, we have in the single flavour approximation the (improved) Davidson-Ibarra bound on the $CP$ asymmetry eqs.(\ref{EcpDI})-(\ref{EcpDIimp}):
\bea
\vert \eps_{1} \vert \lesssim \frac{3}{16 \pi}\frac{M_{1}(m_{\nu}^{\rm{max}}-m_{\nu}^{\rm{min}})}{v^{2}}\times \be(m_{\nu}^{\rm{min}},\tilde{m}_{1}) \, ,
\eea
which decreases as the light neutrinos become degenerate in mass. Indeed, for $m_{1}\sim m_{2}\sim m_{3}\sim m$, and assuming $M_{1}\ll M_{2},M_{3}$, one has:
\bea
\vert \eps_{1} \vert \lesssim \frac{3}{32 \pi}\frac{M_{1}\,\Delta m_{atm}^{2}}{m\,v^{2}}\times \be(m_{\nu}^{\rm{min}},\tilde{m}_{1}) \, .
\eea
In this quasi-degenerate case, $m \gtrsim m_{atm}\gg m_{*}$, and as $m_{1}\lesssim \tilde{m_{1}}\lesssim m_{3}$, we are in the strong washout regime. Thus we can apply eq.(\ref{Strong}) which tells us that $Y_{B}\propto 1/K$, hence
\bea
\label{wouldbe}
Y_{B}&\simeq &-1.38\times 10^{-3}\,\eps_{1}\,\eta(\tilde{m_{1}}) \propto \frac{M_{1}}{m^{2}} \ .
\eea
In order to keep $Y_{B}$ large enough, increasing the light neutrino mass scale could be compensated by increasing RHn mass. On the other hand, one cannot increase $M_{1}$ harmlessly.\\
There is a strong constraint coming from the requirement that off-shell $\Delta L=2$ scatterings do not wash-out the lepton asymmetry. These scatterings are potentially dangerous at low temperature, when other processes relevant in leptogenesis are already frozen. Their contribution is given in this case by \cite{Buchmuller1}
\bea
W_{\Delta L=2}(z)\simeq \frac{3\times 0.186}{z^{2}}\left(\frac{M_{1}}{10^{10}\GeV}\right)\left(\frac{ m}{1\eV}\right)^{2} .
\eea
This term implies a washout of the baryon (lepton) asymmetry, which then can be expressed as
\bea
Y_{B}=Y_{B}^{<}\,e^{-\frac{3\times0.186}{z_{B}}\,\left(\frac{M_{1}}{10^{10}\GeV}\right)\left(\frac{ m}{1\eV}\right)^{2}}\ ,
\eea
where $Y_{B}^{<}$ is the would-be baryon asymmetry without $\D L=2$ scatterings of eq.(\ref{wouldbe}).
Maximising $Y_{B}$ with respect to $M_{1}$ and to the effective neutrino mass $\tilde{m}$, and then requiring that the baryon asymmetry at least matches the observed value,  the authors of~\cite{MassBound,pedestrians,beta} have derived an upper-bound on the neutrino masses $m_{i},i=1,2,3$:
\bea
\label{condmass}
m_{i}\lesssim 0.12 \eV \, .
\eea
This bound on the light neutrino mass is certainly one of the most important phenomenological implication of successful leptogenesis in the one flavour approximation.\\
However, we will see in chapter 5 that this bound will no longer hold when lepton flavours are included.\\
Finally, let us remark that this strong bound only applies in the case the heavy neutrinos are hierarchical. Indeed, as shown in \cite{beta}, the DI bound, which is roughly
\bea
\label{eqbonus}
\eps^{DI}\sim \frac{M_{1} \Gamma_{j}}{M_{j}^{2}}\times \frac{m_{3}-m_{1}}{\tilde{m}_{j}}\,, 
\eea
gets suppressed for two reasons. On the one hand, the second factor comes from the orthogonality of the $R$ matrix and  explains the strong suppression of the $CP$ asymmetry in the limit of degenerate light neutrinos. On the other hand, the first term of eq.(\ref{eqbonus}) implies a suppression of the $CP$ asymmetry when heavy neutrinos are very hierarchical. However, since we are interested in the regime of degenerate light neutrinos, it might be more natural to suppose that the heavy neutrinos are degenerate too, unless postulating accidental compensation between right-handed neutrino masses and neutrino Yukawa couplings. When the RHns are close to the resonance, both suppressions do not occur. The first one gets enhanced by a resonance factor, while the second suppression does not occur due to the absence of an orthogonality relation. The author of \cite{beta} have thus shown that in the single flavour approximation, the constraint eq.(\ref{condmass}) can be avoided. Nevertheless, we will show in chapter 5 that assuming hierarchical RHns (even a very weak hierarchy), the inclusion of flavours relaxes the constraint eq.(\ref{condmass}).
\newpage
\chapter[Importance of lepton flavours]{On the importance of lepton flavours \\in leptogenesis}
%We have seen in the previous chapter that the thermal leptogenesis scenario can explain the observed baryon asymmetry, under few constraints: the right-handed neutrinos should heavy enough, the CP violation in the leptonic sector comes from high-energy phases and is above about $10^{-7}$ and that light neutrinos should not be too massive.
In the previous chapter we have studied leptogenesis in the one-flavour approximation, where one deals with the total lepton asymmetry, which is the sum over lepton flavours $L=\sum_{\al} L_{\al}$. Right-handed neutrinos decay into leptons and antileptons, and in this approximation one considers that the lepton asymmetry produced evolves coherently, until being converted into a baryon asymmetry through the $B+L$ violating sphalerons.\\
However, lepton doublets interact also via the Yukawa term $\mathcal{L} \supset -h_{\al}\,\ol{\ell}_{\al}\,\phi\, e_{R}^{\al}$, which gives mass to charged leptons after electroweak symmetry breaking. This interaction, if efficient, will decohere the lepton doublet, by projecting it onto a flavour basis, in which case the electron, muon and tau flavours can possibly be differentiated~\cite{barbieri99}-~\cite{issues,NardiNir2}.\\
If flavours are distinguishable, we should consider the asymmetry produced in each lepton flavour direction, instead of the collective asymmetry, for which accidental cancellations are always possible.\\
In this chapter, which is mostly based on the appendix of~\cite{issues}, we do not investigate the effects of lepton flavours in leptogenesis. This study will be led in chapter 5 and 6, in the framework of type I and type II seesaws, respectively. Here, we discuss how to handle flavoured lepton asymmetries: how should the Boltzmann equations be modified, what are the good quantities to consider, and especially under which conditions are flavours relevant in leptogenesis. 
\section{A toy model for flavoured Boltzmann equations}
We want to derive the set of Boltzmann equations for lepton flavours. We could infer, given the BE for the lepton asymmetry, the structure of these equations. However, in this approach, the flavour structure, and more importantly, the relevance of flavour, may appear rather ambiguous. We thus derive the BEs, looking at the evolution of the flavoured leptonic number density operator
\bea
\label{eq41}
f_{\Delta \ell}^{\al \be}=f_{\ell}^{\al\be}-f_{\bar{\ell}}^{\al \be}=a_{\ell_{\al}}^{\dagger}a_{\ell_{\be}}-a_{\bar{\ell}_{\be}}^{\dagger}a_{\bar{\ell}_{\al}} \ ,
\eea
where $a_{\ell_{\al}}^{\dagger}$($a_{\ell_{\al}}$) and $a_{\bar{\ell}_{\al}}^{\dagger}$($a_{\bar{\ell}_{\be}}$) are the particle and antiparticle number creation (annihilation) operators, respectively. Actually, this number density operator is not truly the leptonic one, but rather the leptonic doublet one, since lepton number is stored both in $SU(2)$ doublets and singlets. Moreover, as seen from eq.(\ref{eq41}), $f_{\Delta \ell}^{\al \be}$ is a matrix in flavour space~\cite{raffelt,barbieri99}.\\
A complete derivation of the BEs would require to also consider the Lorentz structure of the different fields. Here, we just give an heuristic derivation based on simple quantum mechanics, and reintroduce by hand at a later stage the Lorentzian structure. Nevertheless, such a treatment enables us to determine the flavour structure of the BE, and furthermore to determine explicitely which objects should be studied in flavoured leptogenesis.\\
The operators we introduce obey the same statistic that the fields they represent, namely anticommutation relations for the fermionic fields $\ell$:
\bea
\left\lbrace a_{\ell^{\al}},a_{\ell^{\be}}^{\dagger} \right\rbrace =\delta_{\al \be} \ , 
\left\lbrace a_{\ell^{\al}},a_{\ell^{\be}} \right\rbrace =0 \ ,
\left\lbrace a_{\ell^{\al}}^{\dagger},a_{\ell^{\be}}^{\dagger} \right\rbrace =0 \ ,
\eea
and similarly for the $e_{R}^{c}$ and $N_{1}$. The Higgs field obeys bosonic commutation relations:
\bea
\left[ a_{\phi},a_{\phi}^{\dagger} \right] =\delta_{\al \be} \ , 
\left[ a_{\phi},a_{\phi} \right] =0 \ ,
\left[ a_{\phi}^{\dagger},a_{\phi}^{\dagger} \right] =0 \ .
\eea
We first discuss the $N_{1}-\ell$ interaction. We only consider the interaction with the lightest right-handed neutrino $N_{1}$, droping the indice in $N_{1}$. The interaction Hamiltonian can be written as
\bea
H^{N}=\la_{\rho 1} \aNd \aphi \aell{\rho}+\la^{*}_{\rho  1}\aN \aphid \aelld{\rho} \ .
\eea
The evolution of the leptonic doublet number density is given by the perturbative expansion of the Heisenberg equations of motion (we use here the expectation value instead of the operator):
\bea
\frac{\partial \fell{\al \be}}{\partial t}&=&-\left[H^{N} \left[H^{N},f_{\ell}^{\al \be}\right] \right]\,, \nonumber \\
&=&\la_{\al 1}\la_{1 \rho}^{\dagger}\left(f_{N}(1+f_{\phi})(\delta_{\rho \be}-f_{\ell}^{\rho \be})-(1-f_{N})f_{\phi}f_{\ell}^{\rho \be} \right) \nonumber \\
&+& \left( f_{N}(1+f_{\phi})(\delta_{\al \rho}-\fell{\al \rho})-(1-f_{N})f_{\phi} f_{\ell}^{\al \rho} \right) \la_{\rho 1}\la_{1 \be}^{\dagger}\,.
\eea
In this expression, we define the RHn and Higgs distribution function as $f_{N}=\aNd\aN$ and $f_{\phi}=\aphid\aphi$.\\
The first and third terms of this expression represent decays of right-handed neutrinos whereas the second and fourth terms stand for the inverse decays. If we neglect Pauli blocking and Bose enhancement factors $(1\pm f)$, we can rewrite this expression as
\bea
\frac{\partial \fell{\al \be}}{\partial t}=2 f_{N} \left( \la \la^{\dagger} \right)_{\al \be}-f_{\phi} \left\lbrace f_{\ell},\left( \la \la^{\dagger} \right) \right\rbrace _{\al \be} \ .
\eea
The evolution of the antileptonic number density  is directly deduced from the previous one:
\bea
\frac{\partial \fbell{\al \be}}{\partial t}=2 f_{N} \left( \la \la^{\dagger} \right)_{\al\be}^{\dagger}-f_{\phi} \left\lbrace f_{\bar{\ell}},\left( \la \la^{\dagger} \right)^{\dagger} \right\rbrace _{\al \be} \ ,
\eea
with $\left( \la \la^{\dagger} \right)^{\al \be}=\left( \la_{\al 1} \la^{\dagger}_{1 \be} \right)$.
For simplicity, we have taken the Higgs field to be real\footnote{Had we considered the Higgs to be a complex scalar field, we would have obtained an asymmetry stored in the Higgs degrees of freedom, which could be expressed in terms of the $B/3-L_{\al}$ one, via a conversion matrix, $C_{\phi}$ in~\cite{NardiNir1,NardiNir2}.}. We therefore obtain the evolution of the leptonic doublet asymmetry
\bea
\label{eq48}
\frac{\partial f_{\Delta \ell}^{\al \be}}{\partial t}&=&2 f_{N}\left( (\la \la^{\dagger})-(\la \la^{\dagger})^{\dagger} \right)_{\al \be} \nonumber \\
&-& f_{\phi}\left[\left\lbrace f_{\Delta \ell} , \frac{( \la \la^{\dagger})+( \la \la^{\dagger})^{\dagger}}{2} \right\rbrace ^{\al \be}+2 \left\lbrace f_{\ell}^{eq} ,\frac{( \la \la^{\dagger})-( \la \la^{\dagger})^{\dagger}}{2} \right\rbrace^{\al \be} \right] \ ,
\eea
where we define the "equilibrium" leptonic number density 
\bea
\label{eqNumeq}
f_{\ell}^{\al \be}+\fbell{\al \be}=2 f_{\ell}^{eq\,{\al \be}} \ .
\eea
The first and third terms of eq.\ref{eq48} are nul at this order of expansion in the couplings. Forgetting about it for a while, eq.\ref{eq48} can be further simplified if we assume that the Higgs fields are in thermal equilibrium $f_{\phi}=f_{\phi}^{eq}$, and use the equilibrium condition for decays $f_{N}^{eq}=f_{\phi}^{eq}\,f_{\ell}$, that, once symmetrised, reads 
\bea
\label{eqcond}
f_{N}^{eq} A_{\al \be}=f_{\phi}^{eq}\, \left\lbrace f_{\ell}^{eq} ,\frac{A}{2} \right\rbrace^{\al \be} .
\eea
We thus obtain
\bea
\frac{\partial f_{\Delta \ell}^{\al \be}}{\partial t}&=&2\left( f_{N}-f_{N}^{eq}\right) \left( (\la \la^{\dagger})-(\la \la^{\dagger})^{\dagger} \right)_{\al \be} - f_{\phi}^{eq}\left\lbrace f_{\Delta \ell} , \frac{( \la \la^{\dagger})+( \la \la^{\dagger})^{\dagger}}{2} \right\rbrace ^{\al \be}\ .
\eea
The first term represents the production of a lepton asymmetry in the decays of $N$. Having only made the expansion to second-order, no $CP$ asymmetry can emerge, and the production term is nul. However, if we introduce in the interaction Hamiltonian an effective coupling mimicking the one obtained when calculating $\eps_{CP}$, cf. eq.(\ref{EQ329}), that is 
\bea
H^{N,eff}\sim \sum_{j, \be} \la_{ j \al}^{*} \la_{j \be}^{*} \la_{1 \be}\times g(x_{j}) \aN \aphid \aelld{\al}
+\sum_{j, \be} \la_{ j \al} \la_{j \be} \la_{1 \be}^{*}\times g(x_{j}) \aNd \aphi \aell{\al} \ ,
\eea
we can obtain non-zero $CP$ violation. Assuming this to be the case, and defining the washout factor $K_{1}= \sum_{\al} \vert \la_{\al 1} \vert^{2}$ and the flavoured washout factors $\kappa_{\al \be}= (\la \la^{\dagger})^{\al \be}$, we rewrite the evolution equation as:
\bea
\frac{\partial f_{\Delta \ell}^{\al \be}}{\partial t}=2\left( f_{N}-f_{N}^{eq}\right) K_{1} \eps_{\al \be}-f_{\phi}^{eq}\left\lbrace f_{\Delta \ell},\kappa \right\rbrace^{\al \be} .
\eea
For the moment, we have only considered the neutrino Yukawa couplings, but we also have to include the charged lepton Yukawa couplings, among which we will only consider the interaction involving the tau Yukawa. Assuming that the process $\phi \LRa \ell_{\tau} \tau_{R}^{c}$ is in-equilibrium, it contributes to the interaction Hamiltonian as
\bea
H^{\tau}=h_{\rho \tau} \aphid \atau \aell{\rho}+h_{\rho \tau}^{*}\aphi \ataud \aelld{\rho} \,.
\eea
In the basis where the charged Yukawa couplings are diagonal, $h_{\rho\tau}=h_{\tau}\delta_{\rho\,\tau}$, their contribution to the lepton doublet asymmetry evolution can be written as:
\bea
\label{LeptoOp}
\frac{\partial f_{\Delta \ell}^{\al \be}}{\partial t} &=&2\left( f_{N}-f_{N}^{eq}\right) K_{1} \eps_{\al \be}-f_{\phi}^{eq}\left\lbrace f_{\Delta \ell},\kappa \right\rbrace^{\al \be} \nonumber \\
&-&\vert h_{\tau} \vert^{2} \left(f_{\Delta \tau^{c}}f_{\ell}^{eq}+\,f_{\tau^{c}}^{eq}f_{\Delta \ell} \right)^{\al \be}\left(\delta_{\al \tau}+\delta_{\be \tau}\right) \ ,
\eea
with $f_{\Delta \tau^{c}}=f_{\tau^{c}}-f_{\bar{\tau}^{c}}$ and $f_{\tau^{c}}^{eq}=(f_{\tau^{c}}+f_{\bar{\tau}^{c}})/2$.\\
Before moving further on, it is instructive to look at the equation for the equilibrium density defined eq.(\ref{eqNumeq}). Neglecting the contribution from the interaction Hamiltonian $H^{N}$, we find
\bea
\label{LeptoOp2}
\frac{\partial f_{\ell}^{eq\,\al \be}}{\partial t}=2\vert h_{\tau} \vert^{2}\,\delta_{\al\tau}\delta_{\be\tau}\,f_{\phi}^{eq}-\vert h_{\tau} \vert^{2}f_{\tau^{c}}^{eq}\left(\delta_{\al\tau}\,f_{\ell}^{eq\,\tau \be}+f_{\ell}^{eq\,\al \tau}\,\delta_{\tau\be}\right) \ .
\eea
Here we took two flavours into account, which are labelled $\tau$ and $\mu$ for simplicity. We see that neither the $\mu\mu$ component, nor the $\tau \tau$ one are affected by the charged Yukawa interactions, owing to the equilibrium condition of the interaction rate, which is similar to eq.(\ref{eqcond}). On the other hand, for the non-diagonal entries, we have
\bea
\frac{\partial f_{\ell}^{eq\,\mu \tau}}{\partial t}=-\vert h_{\tau} \vert^{2}f_{\tau^{c}}^{eq}\,f_{\ell}^{eq\,\mu \tau} \ ,
\eea
and similarly for $f_{\ell}^{eq\,\tau \mu}$.\\
We see that when the charged Yukawas are in-equilibrium, "equilibrium" densities of quantum-correlations are exponentially damped~\cite{issues}, and do not affect the flavour-eigenstate "equilibrium" densities. Therefore, we now write $f_{\ell}^{eq\,\al \be}=f_{\ell}^{eq}\,\delta_{\al \be}$.\\
\section{Flavoured Boltzmann Equations}
Eq.(\ref{LeptoOp}) above clearly shows the flavour structure that the Boltzmann equations should have. Before discussing this, we make a few transformations in order to extrapolate this equation to the comoving number density asymmetry, through the following "rough" procedure~\footnote{Had we kept the spinorial structure, all the steps indicated would have been rigorous. Since we are confident on this point, we use a simple matching procedure for our equations. Following the different steps indicated does allow us to identify the different terms without ambiguity.}:
\begin{itemize}
\item we factorise the equilibrium distribution function, using for the last two terms the equilibrium conditions $f_{\phi}^{eq}f_{\ell}^{eq}=f_{N}^{eq}$ and $f_{\tau^{c}}^{eq}f_{\ell^{\tau}}^{eq}=f_{\phi}^{eq}$,
\item we "sort of" integrate over the phase-space to work with number densities,
\item we  reintroduce the expansion of the Universe through
\bea
\frac{\partial }{\partial t}\Ra \frac{d}{dt}-3H(T) \ ,\nonumber
\eea
and change time to $z=M_{1}/T$ and number densities to comoving number densities.
At this point we get,
\bea
\frac{d Y_{\Delta \ell}^{\al \be}}{dz}&=&\frac{1}{z\,s\,H(T)}\left(\eps_{\al \be} K_{1} \left[\frac{n_{N}^{eq}}{Y_{N}^{eq}}\right](Y_{N}-Y_{N}^{eq})-\left[\frac{n_{N}^{eq}}{Y_{N}^{eq}}\frac{Y_{N}^{eq}}{Y_{\ell}^{eq}}\right]\frac{1}{2}\left\lbrace Y_{\Delta \ell},\kappa \right\rbrace^{\al \be} \right)\nonumber \\
&-&\frac{1}{z\,s\,H(T)}\left(\left[\frac{n_{\phi}^{eq}}{Y_{\ell}^{eq}}\right]\,\vert h_{\tau} \vert^2 \left(Y_{\Delta \tau^{c}}\,\delta_{\al \be}+\frac{Y_{\tau^{c}}^{eq}}{Y_{\ell}^{eq}}\,Y_{\Delta \ell}^{\al \be}\right) \left(\delta_{\al \tau}+\delta_{\be \tau}\right)\right) \ . \nonumber 
\eea
\end{itemize}
The last step consists in "sort of" redefining the couplings in order to reintroduce the usual interaction rates. In an accurate calculation, the interaction rates naturally appear when integrating over the phase-space.\\
Further using the fact that $Y_{\tau^{c}}^{eq}=Y_{\ell}^{eq}/2$, we obtain the flavoured Boltzmann equations for the comoving number densities:
\bea
\label{flavBE}
\frac{d Y_{\Delta \ell}^{\al \be}}{dz}&=&\eps_{\al \be} K_{1} D(z)(Y_{N}-Y_{N}^{eq})-W_{id}(z)\frac{1}{2}\left\lbrace Y_{\Delta \ell},\kappa \right\rbrace^{\al \be} \nonumber \\
&-&\frac{z}{s H(M_1)}\frac{\gamma_{\tau}}{Y_{\ell}^{eq}}\left(2\,Y_{\Delta \tau^{c}}\,\delta_{\al\be}+Y_{\Delta \ell}^{\al \be}\right) \left(\frac{\delta_{\al \tau}+\delta_{\be \tau}}{2}\right)  \, ,
\eea
%where $D(z)$ stands for decays and $W_{id}$ for the inverse decays that wash-out the leptonic doublet asymmetry.
The last term represents the depletion of the lepton $SU(2)$ doublet asymmetries by the charged Yukawa interaction, with~\cite{leptorept},\cite{SingletAsymmetry}:
\bea
\label{quotedrate}
\frac{\gamma_{\al}}{n_{\ell}^{eq}}\simeq 10^{-2} \vert h_{\al} \vert ^2 T\ .
\eea
When extrapolating from eq.(\ref{LeptoOp}) to the BE for comoving number densities, the matching with the $\tau^{c}$ interaction term may seem arbitrary. However, we can derive the evolution equation for the $\tau^{c}$ asymmetry~\cite{SingletAsymmetry}, by looking at the evolution of the operator 
\bea
f_{\Delta \tau^{c}}=\ataud\atau -\abtaud\abtau\,.
\eea
Doing so, and using the same matching procedure as before, we find
\bea
\label{BEtau}
\frac{d Y_{\Delta \tau^{c}}}{dz}&=&-\frac{z}{s H(M_1)}2 \vert h_{\rho} \vert^2 \left(2\,Y_{\Delta \tau^{c}}+Y_{\Delta \ell}^{\tau \tau}\right) \nonumber \\
&=&-\frac{z}{s H(M_1)}\frac{\gamma_{\tau}}{Y_{\ell}^{eq}}\left(2\,Y_{\Delta \tau^{c}}+Y_{\Delta \ell}^{\tau \tau}\right)\, ,
\eea
hence justifying the quoted rate eq.(\ref{quotedrate}).\\
Given the expression above we can now deduce the temperature regime for which lepton flavours are relevant in leptogenesis.\\
We see that the interaction projecting lepton doublets onto the flavour space is in-equilibrium if~\cite{issues,NardiNir2}
\bea
z \,\gamma_{\al} \gtrsim H(M_{1})\,,
\eea
which provides a bound on the decaying right-handed neutrino mass. Evaluating the equilibrium condition at the temperature for which the asymmetry will be mostly produced, $T\sim M_{1}$, this bound reads:
\bea
\label{Equilcond}
M_{1} \lesssim 7\times 10^{15} h_{\al}^{2} \GeV \,,
\eea
%while this bound reads in the MSSM
%\bea
%M_{1} \lesssim 5\times 10^{15} h_{\al}^{2}\left(1+\tan{(\be)}^2\right) \GeV .
%\eea
Therefore:
\begin{itemize}
\item $\tau$-Yukawas are in-equilibrium if $M_{1} \lesssim 10^{12}$ GeV,
\item muon-Yukawas are in-equilibrium for $M_{1}\lesssim 3\times 10^{9}$ GeV,
\item electron-Yukawas are in-equilibrium for $M_{1}\lesssim 6\times 10^{4}$ GeV.
\end{itemize}
However, thermal equilibrium does not guarantee that the lepton doublet is to be projected onto flavour space~\cite{Zeno}. Indeed, we see that the terms involving $N-\ell$ couplings in eq.(\ref{flavBE}) are independent of the choice of the flavour basis, which is obviously not the case for the charged Yukawa interactions. The condition for the flavour basis to be relevant is actually more stringent than the equilibrium condition eq.(\ref{Equilcond}). The processes decohering the lepton doublets have to be faster than the processes which recohere lepton doublets at the time the asymmetry is produced, that is~\cite{Zeno,FlavourOsc}
\bea
\label{CondFlav}
\frac{z \gamma_{\al}}{H(M_1)} \gtrsim W_{1}(z) \, .
\eea 
Considering only inverse decays, and evaluating them at their maxima, $z\simeq 2$, one finds 
\bea
M_{1} &\lesssim & 5\times 10^{15} \GeV \frac{h_{\al}^{2}}{\kappa_{\al \al}} \ ,
%M_{1} &\lesssim & 5\times 10^{15} \GeV \frac{h_{\al}^{2}}{\kappa_{1 \al}}\left(1+\tan{(\be)}^2\right) \, ,
\eea
where $\kappa_{\al \al}$ parametrises the individual washout of the flavour $\al$.
For the flavour $\tau$, this constraint reads:
\bea
\label{Fullcond}
M_{1}\lesssim \frac{10^{12} \GeV}{\kappa_{\tau \tau}}\, .
\eea
Under this constraint, flavours are fully relevant in leptogenesis. The situation where charged Yukawas are in-equilibrium but not faster than recohering processes is more involved, and we postpone its discussion until the end of the chapter.
\section{Flavour structure}\label{FlavStruc}
\subsection{One flavour scheme}
This case is encountered if $M_1 \gtrsim 10^{12} \GeV$, and in this case none of the charged Yukawas are in-equilibrium. It has been shown in~\cite{FlavourOsc} that by a rotation of lepton doublet states, one can recover the BE of the one flavour case studied in the previous chapter, that is
\bea
\frac{d Y_{N}(z)}{dz}&=&-D(z) K_{1} \left(Y_{N_1}(z)-Y_{N_1}^{eq}(z)\right) \ , \nonumber \\
\frac{d Y_{L}(z)}{dz}&=&\eps_{1} D(z) K_{1} \left(Y_{N_1}(z)-Y_{N_1}^{eq}(z)\right)-W_{id}(z)K_{1} Y_{L} \ ,
\eea
with $\eps_{1}=Tr[\eps]$. For $M_1 \gtrsim 10^{12} \GeV$, the results of the previous chapter hold.
\subsection{Two flavour scheme}
\begin{figure}[htb]
\begin{center}
\includegraphics[scale=0.5]{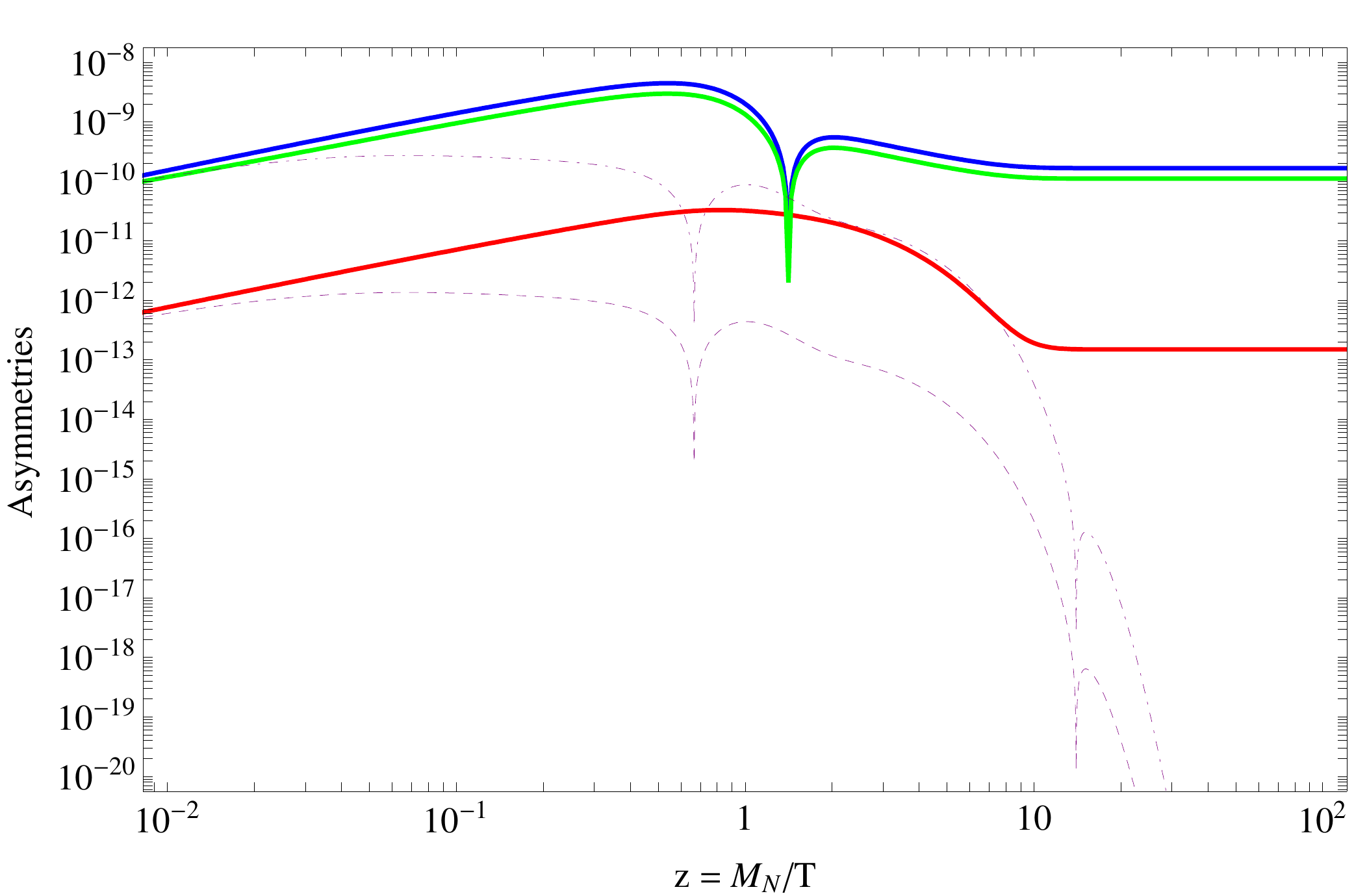} 
\caption{Evolution of comoving number densities. The thick lines represent diagonal terms of the flavoured Boltzmann equation, with $Y_{L}^{\tau \tau}$ in blue (upper curve), $Y_{\Delta \ell}^{\tau \tau}$ in green (light grey)  and $Y_{\Delta \ell}^{\perp \perp}$ in red (lower curve), respectively. The thin purple lines represent the quantum-correlations, \ie the off-diagonal terms when the charged Yukawas are in-equilibrium. The dashed line stands for $Y_{\Delta \ell}^{\perp \tau}$ whereas the dot-dashed one for $Y_{\Delta \ell}^{\tau \perp}$. The input parameters have been chosen, so that the components are distinguishable, with $\kappa_{\perp,\perp}=\kappa_{\perp,\tau}=0.1$, $\kappa_{\tau,\perp}=\kappa_{\tau,\tau}=20$ for the washout parameters and $\eps_{\perp,\perp}=\eps_{\perp,\tau}=0.1\times 10^{-6}$, $\eps_{\tau,\perp}=\eps_{\tau,\tau}=20\times 10^{-6}$ for the $CP$ asymmetries.}
\label{GrapheQuanCor}
\end{center}
\end{figure}

Let us now consider  the case where only the tau Yukawas are in equilibrium, that is $3\times 10^{9} \GeV \lesssim M_1 \lesssim 10^{12} \GeV$.\\
In this case, the tau flavour is distinguishable, but neither the muon nor the electron flavours are. Hence, the lepton produced in $N_1$ decays will be projected onto the tau direction, as well as onto an orthogonal direction to that of the tau. This second direction is composed of the muon and electron flavour directions that coherently mix, and which we will label $L_{\perp}=L_{e}+L_{\mu}$. We have a two flavour regime $(L_{\tau},L_{\perp})$ and in this case eq.(\ref{flavBE}) holds, with $\al=\tau$ and $\be=\perp$:
\bea
\frac{d Y_{N}(z)}{dz}&=&-D(z) K_{1} \left(Y_{N_1}(z)-Y_{N_1}^{eq}(z)\right) \ , \nonumber \\
\frac{d Y_{\Delta \ell}^{\perp\perp}(z)}{dz}&=&\eps_{\perp \perp} K_{1} D(z)(Y_{N}(z)-Y_{N}^{eq}(z))-W_{id}(z) \kappa_{\perp\perp}Y_{\Delta \ell}^{\perp\perp}(z)\nonumber \\
&-&W_{id}(z)\frac{1}{2}\left(\kappa_{\perp\tau}Y_{\Delta \ell}^{\tau\perp}(z)+Y_{\Delta \ell}^{\perp\tau}(z)\kappa_{\tau\perp}\right) \ , \nonumber \\
\frac{d Y_{\Delta \ell}^{\perp\tau}(z)}{dz}&=&\eps_{\perp\tau} K_{1} D(z)\,(Y_{N}(z)-Y_{N}^{eq}(z))-W_{id}(z)\,Y_{\Delta \ell}^{\perp\tau}(z)\,\left(\frac{\kappa_{\perp\perp}+\kappa_{\tau\tau}}{2}\right) \nonumber \\
&-&\kappa_{\perp \tau}\,\left(\frac{Y_{\Delta \ell}^{\perp\perp}(z)+Y_{\Delta \ell}^{\tau\tau}(z)}{2}\right)-\frac{D_{\tau}}{2} Y_{\Delta \ell}^{\perp \tau}(z) \ , \nonumber \\
\frac{d Y_{\Delta \ell}^{\tau\perp}(z)}{dz}&=&\frac{d Y_{\Delta \ell}^{\perp\tau}(z)}{dz}(\perp \LRa \tau) \ ,\nonumber \\
\frac{d Y_{\Delta \ell}^{\tau\tau}(z)}{dz}&=&\frac{d Y_{\Delta \ell}^{\perp \perp}(z)}{dz}(\perp \LRa \tau)-D_{\tau}\left(2 Y_{\Delta \tau^{c}}(z)+Y_{\Delta \ell}^{\tau \tau}(z)\right) \ . 
\label{TwoFla}
\eea
In these equations, the $CP$ asymmetries $\eps_{\perp \perp}$ and $\eps_{\perp \tau}$ are defined by
\bea
\eps_{\perp \perp}\equiv\eps_{e e}+\eps_{\mu \mu} \, ,\nonumber \\
\eps_{\perp \tau}\equiv\eps_{e \tau}+\eps_{\mu \tau} \ , \nonumber 
\eea
and similar relations hold for the washout parameters $\kappa$. We also define for convenience
\bea
D_{\tau}=\frac{z}{H(M_1)}\frac{\gamma_{\tau}}{Y_{\ell}^{eq}}\simeq \frac{7\times 10^{15} \GeV}{M_1}\times h_{\tau}^{2}\simeq  \frac{7\times 10^{11} \GeV}{M_1}.
\eea
Notice that, according to eq.(\ref{BEtau}), the Boltzmann equations for the $\tau$ diagonal entry can be written:
\bea
\frac{d Y_{\Delta \ell}^{\tau\tau}(z)}{dz}&=&\frac{d Y_{\Delta \ell}^{\perp \perp}(z)}{dz}(\perp \LRa \tau)+\frac{d Y_{\Delta \tau^{c}}}{dz} \ .
\eea
Since the leptonic $\tau$ doublet $\ell^{\tau\tau}$  interacts with right-handed neutrinos, its evolution equation receives contributions from the interaction Hamiltonian $H^{N}$, corresponding to an asymmetry production and depletion. However, since $\ell^{\tau\tau}$ also interacts with the right-handed singlet $\tau^{c}$, it receives a further contribution from $H^{\tau}$.\\
This is quite satisfactory, since we do not want to study the asymmetry stored in lepton doublets, but the lepton asymmetry 
\bea
L_{\al}=L_{\ell^{\al}}+L_{e_{R}^{\al}}=L_{\ell^{\al}}-L_{e_{R}^{\al\,c}} \ ,
\eea
which is stored in lepton doublets as well as in singlets~\cite{SingletAsymmetry}.\\
The muon-Yukawas being out of equilibrium, the leptonic asymmetry $Y_{\ell^{\perp}}$ is not influenced by any effect of this type: muon (or electron) isosinglet density is too far from thermal equilibrium to disrupt the $\ell_{\perp}$ propagation.\\
The off-diagonal terms  are  however sensitive to the $\tau^{c}$ population, which acts as a damping term. Furthermore, the term $\propto D_{\tau} Y_{\Delta \ell}^{\al \be}$  exponentially drives the quantum-correlations to zero: when the interactions involving charged Yukawa couplings of a given flavour are in-equilibrium, the quantum correlations between this flavour and the orthogonal one(s) are exponentially damped, and the flavours, being distinguishable, evolve separately. Notice that since we did not include the free Hamiltonian $H^{0}\propto \omega_{\tau}\,\aelld{\tau}\aell{\tau}+w_{\perp}\,\aelld{\perp}\aell{\perp}$, the number densities are not liable to flavour oscillations~\cite{issues}, which have been studied in detail in~\cite{FlavourOsc}. Actually these oscillations only affect the quantum correlations by inducing a faster damping~\cite{FlavourOsc}.\\
We illustrate this damping in fig.\ref{GrapheQuanCor}, were we depict the evolution of the different asymmetries according to the set of Boltzmann equations in eqs.(\ref{BEtau}-\ref{TwoFla}).
\\
In these equations, we have chosen the washout parameters and $CP$ asymmetries in order to distinguish the different components. The effect of charged Yukawa interactions in damping the quantum correlations is clear, hence in the temperature regime where these interactions are in-equilibrium the off-diagonal terms can be safely neglected.\\
Furthermore, we plotted in this graph the leptonic doublet asymmetry $Y_{\Delta \ell}^{\tau \tau}$ (green-light grey) and the lepton asymmetry $Y_{\Delta L}^{\tau \tau}=Y_{\Delta \ell}^{\tau \tau}-Y_{\Delta \tau^{c}}$ (blue-upper curve). We see that the singlet contribution to the leptonic asymmetry accounts only for ${\mathcal{O}}(1)$, and thus, for now, we will neglect it on the evolution of the asymmetries.
\subsection{Three flavour scheme}
If $M_{1}\lesssim 3 \times10^{9} \GeV$, then both muon and tau Yukawas are in equilibrium. These interactions decohere the lepton doublet by projecting it onto a three flavour basis $(L_{\tau},L_{\mu},L_{e})$. The muon and electron flavours being now distinguishable, muonic and electronic quantum-correlations are exponentially damped, and one is left with the set of Boltzmann equations for the lepton number densities:
\bea
\frac{d Y_{L}^{\al \al}(z)}{dz}=\eps_{\al \al}\,K_{1}\,D(z)\,\left(Y_{N}(z)-Y_{N}^{eq}(z)\right)-W_{id}(z)\,\kappa_{\al\al}\,Y_{L}^{\al\,\al}(z) \ ,
\eea
with $\al=e,\mu,\tau$.
\section{$B-L$ conversion and the Baryon asymmetry}
In the previous section we have derived the Boltzmann equations for the lepton asymmetries. However, as explained in appendix A, sphalerons conserve $B/3-L_{\al}$ asymmetries~\cite{BLconv}, but not the leptonic ones. Therefore, it is preferable to work with 
\bea
Y_{\Delta \al}\equiv \frac{1}{3}Y_{\Delta B}-Y_{\Delta L^{\al}} \ ,
\eea
computed from the leptonic doublet asymmetries by
\bea
Y_{\Delta \al}=\sum_{\be}A_{\al \be} Y_{\Delta \ell^{\be}} \ ,
\eea
where the sum is made over all flavours that were distinguishable during leptogenesis. Details about the derivation of the entries of  the conversion matrix $A_{\al\be}$ are given in the appendix B. This relation is obtained by expressing the chemical potentials of the $B/3-L_{\al}$ asymmetries in terms of the leptonic doublet ones, and it depends on the different processes that are in-equilibrium at the temperature leptogenesis occurs~\cite{HarveyTurner,equilcond}.\\
Since we do not consider $B$ violating processes, the Boltzmann equations for $Y_{N}$ and for the $B/3-L_{\al}$ asymmetries are finally given by (dropping the redundant double indices):
\bea
\label{EQBEFLAV}
\frac{d Y_{N}(z)}{dz}&=&-D(z) K_{1} \left(Y_{N_1}(z)-Y_{N_1}^{eq}(z)\right) \ , \nonumber \\
\frac{d Y_{\Delta \al}(z)}{dz}&=&-\eps_{\al}\,K_{1}\,D(z)\left(Y_{N}(z)-Y_{N}^{eq}(z)\right)-W(z)\,\kappa_{\al}\sum_{\be}A_{\al\be}\,Y_{\Delta \be}(z) \ .
\eea 
This set of equations is written here in all generality, and the expressions of the production term $D(z)$ and of the washout one $W(z)$ depend on the model, as well as the parameters which govern these equations, $\kappa_{\al}$, $K_{1}$ and $\eps_{\al}$.\\
Finally, $B/3-L_{\al}$ asymmetries are conserved, until they are converted by sphalerons. The baryon asymmetry is then related to the individual asymmetries by:
\bea
Y_{B}=a_{sph}\times \sum_{\al} Y_{\Delta \al} \ ,
\eea
the sum being over the flavour that were distinguishable when leptogenesis occurs. The conversion factor is $a_{sph}\sim 1/3$, and its value also depends on the model, and on whether the sphalerons freeze before or after the electroweak-phase transition~\cite{HarveyTurner,equilcond}.
\section{A remark on the full flavour regime}
\begin{figure}[htb]
\begin{center}
\includegraphics[scale=0.6]{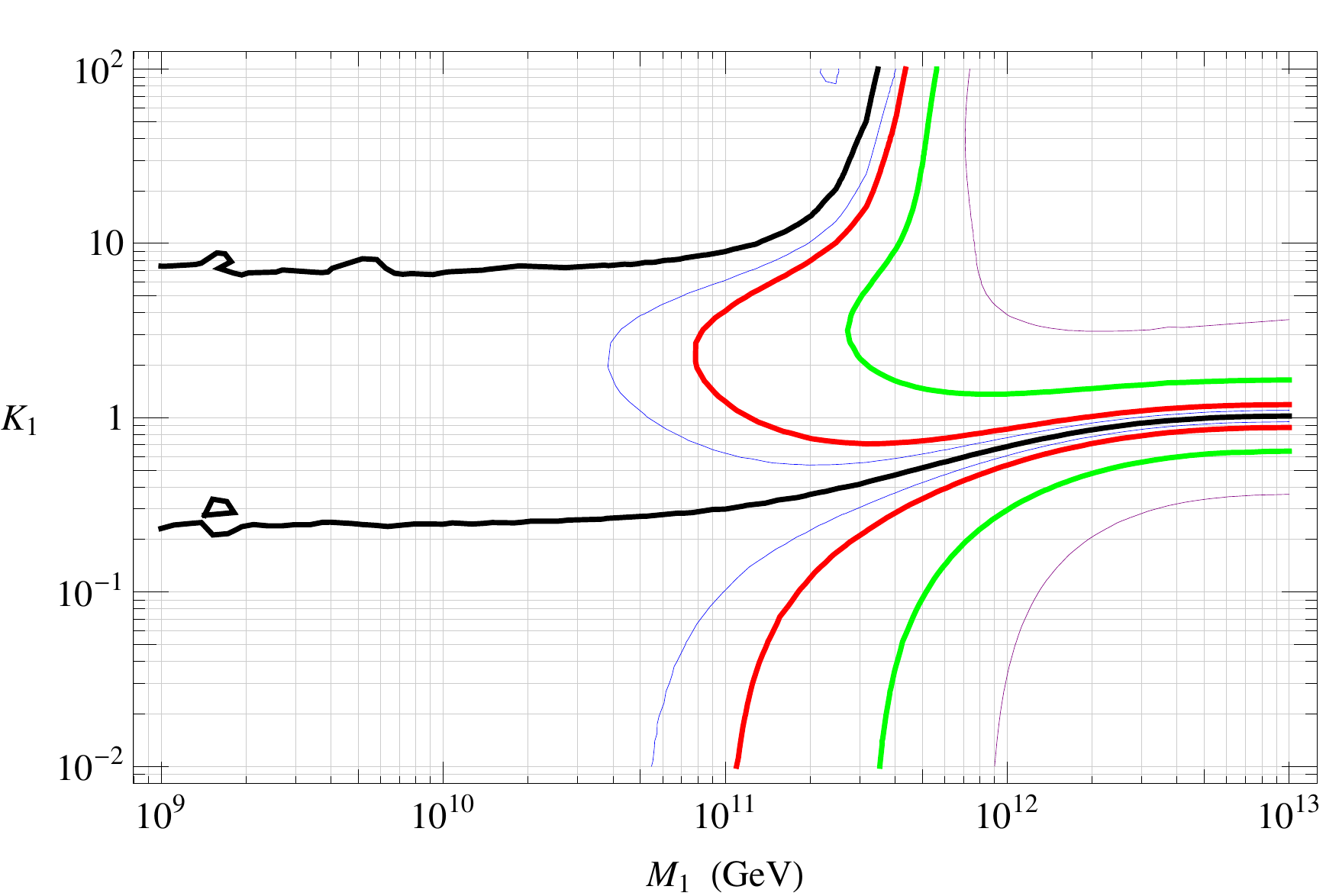} 
\caption{Contour plot of the ratio between the lepton doublet asymmetries computed with and without quantum correlations. The black line stands for $Y_{\Delta \ell^{\al \al}}^{with}/Y_{\Delta \ell^{\al \al}}^{without}=1$, whereas the blue, red, green and purple lines represents deviation (from left to right) of $5\%$, $10\%$, $30\%$ and $60\%$, respectively.}
\label{QuantCorrCont}
\end{center}
\end{figure}

In section \ref{FlavStruc}, when we discussed the flavour structure of the BEs, the criterion used for the relevance of a given flavour in leptogenesis was presented eq.(\ref{Equilcond}) and translates into having the interactions involving charged lepton Yukawa couplings in-equilibrium, $\Gamma_{\al}\gtrsim H(M_N)$.\\
However, as explained around eqs.(\ref{CondFlav})-(\ref{Fullcond}), thermal equilibrium does not guarantee that the lepton doublets will be projected onto the flavour basis. A more accurate constraint comes from the requirement that the decohering rates $\gamma_{\al}$ are faster than any recohering interaction, either the production term $D(z)$ or the washout one $W(z)$~\cite{Zeno}.
Considering only inverse decays as depletion reactions, and requiring that at least the interaction involving the charged tau Yukawa couplings are in-equilibrium, a bound on $M_1$ can be derived:
\bea
\label{condflav}
M_{1}\lesssim \frac{10^{12} \GeV}{\kappa_{\al}} \left(\frac{h_{\al}}{h_{\tau}}\right)^{2}\, .
\eea
If this constraint is satisfied, then the flavours are fully relevant in leptogenesis.\\
Given the equilibrium condition for sphalerons, $M_{1}\lesssim 10^{12} \GeV$, which is roughly the same than the equilibrium condition for tau-Yukawas, we see that only in the strong washout regime is the constraint of eq.(\ref{condflav}) more stringent than the equilibrium condition eq.(\ref{Equilcond}).\\
Let us consider a lepton doublet (asymmetry) produced in decays of $N_1$ at a temperature $T\sim M_{1}$, in the case where the tau-Yukawas are in-equilibrium. If during the caracteristic time for two $\ell-H-\tau^{c}$ interactions to happen, inverse decays have time to occur, the lepton doublet will not be projected onto the flavour basis, and the faster the decays/inverse decays, the more coherently the lepton doublet will evolve.\\
In such case, the lepton asymmetry should be computed by solving the set of BEs including off-diagonal terms. As already stated, ref.~\cite{FlavourOsc} shows that by an appropriate redefinition of the lepton doublets, corresponding to a rotation in flavour space (allowed since the $N$-interactions are the $\ell$-labelling ones), one recovers the single flavour picture.\\
However, through this redefinition, couplings are also modified, and so the differences between including or neglecting quantum correlations are less transparent. In order to study the validity of the full flavour regime, we solve the sets of eqs.(\ref{TwoFla}) with and without quantum correlations. We depict in fig. \ref{QuantCorrCont} the effect of these correlations by showing a contour plot of the ratio of the lepton doublet asymmetry for a given flavour computed with and without these terms, $Y_{\Delta \ell^{\al \al}}^{with}/Y_{\Delta \ell^{\al \al}}^{without}$, allowing $M_{1}$ to vary around the central value $M_{1}=10^{12} \GeV$ and the washout parameter around $K_{1}=1$. For simplicity, we have taken the $CP$ asymmetries and the washout parameters of the different flavours to be equal. One expects that if the condition of eq.(\ref{condflav}) is not satisfied, the damping of the quantum correlations will be inefficient and these off-diagonal terms will significantly contribute to the lepton asymmetry (in fact, given our choice of parameters, in an equal amount). On the other hand, if eq.(\ref{condflav}) holds, the quantum correlations will be damped and the ratio is expected to approach unity.
\\
We see in fig.\ref{QuantCorrCont} that for $M_{1}\lesssim 5\times10^{11} \GeV$, the inclusion of quantum correlations modifies the results up to $30\%$, while for $M_{1}\lesssim 10^{11} \GeV$, the approximate eq.(\ref{EQBEFLAV}) gives very accurate results, and so in that case quantum correlations can be safely neglected.\\
To conclude on the validity of eqs.(\ref{EQBEFLAV}) regarding a precise determination of the lepton asymmetry, we can say that for $M_{1}\lesssim 10^{11} \GeV$, the quantum correlations are effectively damped and one can use eqs.(\ref{EQBEFLAV}). For $5\times10^{11} \GeV\gtrsim M_{1}\gtrsim 10^{11} \GeV$, the results are valid only up to $10\%-30\%$, while for $M_{1}\gtrsim 5\times 10^{11} \GeV$, a precise computation does indeed require the inclusion of quantum correlations. For heavier $M_{1}$, the one flavour approximation is valid.\\
However, we stress that the results here presented have been derived in the Standard Model (plus RHns). Had we considered a supersymmetric extension, the discussion would have been dramatically different. Indeed, in SUSY extensions, the charged Yukawa couplings is now corrected by the ratio of Higgs vevs, $h_{\tau}^{2}\times(1+\rm{tan}^{2}(\be))$, where $\rm{tan}(\be)=v_{u}/v_{d}$ is the ratio of the SUSY Higgs vevs. Hence for  $\rm{tan}^{2}(\be)) \gtrsim 3 (10)$, the results are valid up to $10\%$ for $M_{1}\lesssim 10^{12 (14)} \GeV$, so that for $M_{1}\lesssim 10^{12} \GeV$, as required for sphalerons to be in-equilibrium, flavours are always fully relevant in supersymmetric leptogenesis.\\
In the following, we will neglect quantum correlations.
\newpage
\chapter[Flavours in type I leptogenesis]{On the role of lepton flavours\\ in type I leptogenesis}
In this chapter we are going to study leptogenesis in the type I seesaw mechanism, 
%In this scenario, 3 right-handed neutrinos are added to the SM model particle content. These neutrinos, being singlet under SM gauge group, can have a bare mass term in the potential. As this mass does not arise from the vev of the SM Higgs field, its value is not dictated by the scale of electroweak-symmetry breaking. By pushing this mass up to high energy scale, one obtains after diagonalisation of the neutrino mass term two eigenstates (per family). a very massive one, usually labelled $N$ and mainly composed of the right-handed Majorana neutrino, and a very light state, mainly composed of left-handed field, which is implied in low energy experiments. The mass of the light state is given by:
%In this scenario, 3 right-handed neutrinos are added to the SM particle content. These neutrinos are assumed to be Majorana spinors, and are singlets under the SM gauge symmetries. They nevertheless couple to light leptons and SM Higgs via an extra Yukawa coupling 
were three right-handed neutrinos are added to the SM model particle content in order to explain neutrino masses. The Lagrangian contains the extra terms
\bea
\mathcal{L}\supset -\la_{\al i}\ol{N}_{i}\,\ell_{\al}\,\phi-\frac{1}{2}\ol{N}_{i}M_{i}N_{i}^{c}+h.c.\,,%+\la_{\al i}^{\dagger}\,N_{i}\,\ol{\ell}_{\al}\,\phi^{*}\, .
\eea
%a mass for light neutrinos is obtained
%\bea
%m_{\nu}^{ij}=v^{2}\la_{ik}.M_{k}^{-1}.\la_{kj}^{T} \ ,
%\eea 
%which is suppressed compared to the electroweak scale thanks to the heavy Majorana mass of right-handed neutrinos. This is the famous type I relation~\cite{Seesaw}.\\
which provide masses to neutrinos from the seesaw mechanism. %The mass $m_{\nu}$  is a complex $3\times 3$ matrix and is diagonalized by the $PMNS$ unitary matrix $U$:
%\bea
%d_{m}=U^{T}.m_{\nu}.U \ 
%\eea
%where $m_{1,2,3}$ are the mass implied in low-energy experiments.
These masses are not yet measured, but rather the differences of squared masses are deduced from the oscillations of neutrinos.
%, together with the mixing angles $\theta_{12}$, $\theta_{23}$, the last angle $\theta_{13}$ being only upper-constrained. The mass ordering, as well as the $CP$ violating phase $\delta$, may be determined in future oscillation experiments. The Majorana $CP$-violating phases, which are also not yet know, might be determined in neutrinoless double-beta decay experiments.\\
However, should we believe that leptogenesis is indeed the mechanism responsible for the observed BAU, an upper bound on the light neutrino mass can be derived~\cite{MassBound,pedestrians,beta}, $m_{i} \lesssim 0.12 \eV$, a bound that may be soon tested.\\
%On the other hand, nothing is known about the heavy right-handed neutrinos. Neither their existence, which is postulated, nor their masses and mixings. This is one of the gap of leptogenesis that may be never filled in.\\
%Nevertheless, as we saw in chapter 4, if leptogenesis is the mechanism responsible for the observed baryon asymmetry,
On the other hand, the right-handed sector is experimentally unconstrained. This is one of the gaps of leptogenesis that may be never filled in. Nevertheless, the Davidson-Ibarra bound~\cite{DI} constrains the right-handed neutrino mass scale to be above $M_{1}\gtrsim 10^{9} \GeV$ . Moreover, $CP$ violation in the leptonic sector -related to the $CP$ phases of $U$ and $R$- should be quantitatively large enough so that $CP$ asymmetry is larger than $\eps_{CP}\gtrsim 10^{-7}$.\\
All these constraints have been derived in the single flavour picture, where one considers that the lepton involved in the decays of the RHn evolves coherently after being produced. Nonetheless, we saw in the previous chapter that for $M_{1}\lesssim 10^{11-12} \GeV$, not only are the interactions involving charged lepton Yukawa couplings in-equilibrium, but their rate can also be faster than recohering processes, the decays and inverse decays, in which case the flavours are relevant and should be included for a precise computation of the baryon asymmetry obtained by leptogenesis.\\
This chapter, which is mostly based on \cite{matters} and \cite{StudyOf}, is organised as follows: firstly, we set up the framework of flavoured leptogenesis and we qualitatively discuss how lepton flavours affect the key-parameters of leptogenesis, namely the $CP$ asymmetries and the washout factors. Secondly, we evaluate the resulting baryon asymmetry produced by leptogenesis, and show how it differs from the one-flavour picture provided in chapter 3. Finally, we consider the constraints derived for the low and high-energy parameters. 
\section{Flavoured leptogenesis}
We first recall the various definitions and the framework of Boltzmann equations.
We consider a hierarchical spectrum for the right-handed neutrinos $M_{1}\ll M_{2},M_{3}$, and thus neglect contributions from the heavier neutrinos in the lepton asymmetry production. Indeed, since in this case the leptons  produced by $N_{2,3}$ leptogenesis  participate via inverse decays in the thermalisation of $N_{1}$, one can consider that  the lepton asymmetry produced during $N_{2,3}$ leptogenesis epoch will be washed out. However, this assumption, if valid in the single approximation~\cite{neglectN2}, is not strictly justified when flavours are taken into account~\cite{Vives}, as the lepton flavour content may differ for the lepton involved in $N_{2,3}$ leptogenesis and the one involved in $N_{1}$ leptogenesis. This point will be discussed in the following chapter; presently, we only consider the  processes that involve $N_{1}$.\\
The BEs for right-handed neutrinos $N\equiv N_{1}$ and $\D_{\al}\equiv B/3-L_{\al}$ asymmetry comoving number densities are~\cite{matters}:
\bea
\label{eqBE}
\frac{d Y_{N}^{eq}(z)}{dz}&=&-K_{1}\,P(z)\,\left(Y_{N}(z)-Y_{N}^{eq}(z)\right) \ , \nonumber \\
\frac{d Y_{\Delta \al}(z)}{dz}&=&-\eps_{\al}\,K_{1}\,P(z)\,\left(Y_{N}(z)-Y_{N}^{eq}(z)\right)+W(z)\,\kappa_{\al}\sum_{\be}A_{\al\be}\,Y_{\Delta \be}(z) \, ,
\eea
%where the equilibrium (comoving) number density is given by
%\bea
%Y_{N_1}^{eq}(z)\simeq\frac{45 \, \zeta(3)}{ 2\pi^4 g_{\star}}\frac{3}{4} \, \,z^{2} K_{2}(z)\ . 
%\eea
In these equations, individual washout parameters are given by
\bea
\kappa_{\al}=\frac{\Gamma(N\Ra\ell \phi)}{H(M_{1})}=\vert \la_{1 \al} \vert^{2} \frac{v^2}{M_{1}\,m^{*}}=\frac{\tilde{m}_{\al}}{m^{*}} \ ,
\eea
with the effective neutrino mass $\tilde{m}_{\al}$ being given by
\bea
\tilde{m}_{\al}&=&\vert \la_{1 \al} \vert^{2} \frac{v^2}{M_{1}}\,, 
%m^{*}&=&\frac{8 \pi v^{2}}{M_{Pl}}\sqrt{\frac{8 \pi^{3} g_{*}}{90}}\simeq 1.08\times 10^{-3} \eV \ .
\eea
and $K_{1}=\sum_{\al}\kappa_{\al}$.
The other quantities have been defined in chapter 3, section 3.3. We nevertherless recall the BE for the lepton asymmetry in the single flavour picture:
\bea
\frac{d Y_{L}(z)}{dz}&=&\eps_{1}\,K_{1}\,P(z)\,\left(Y_{N}(z)-Y_{N}^{eq}(z)\right)-W(z)\,K_{1}\,Y_{L}(z) \, .
\eea
%
%The total washout is simply the sum over the individual washouts $K_{1}=\sum_{\al} \kappa_{\al}$.\\
The individual $CP$ asymmetries $\eps_{\al}$ are given by eq.(\ref{EQ325}):
\bea
\label{ECPflav}
\eps_{\alpha}  %&  %\equiv & \frac{\Gamma(N1\Ra \ell_{\alpha})-\Gamma(N1\Ra\bar{\ell}_{\alpha})}{\sum_{\alpha}\left(\Gamma(N1\Ra \ell_{\alpha})+\Gamma(N1\Ra\bar{\ell}_{\alpha})\right) } \nonumber \\ 
%&=&\frac{1}{8\pi}\sum_{j\neq 1}\frac{\IM{ \left(\la^{\dagger}\la\right)_{1\,j}^{2}}}{\left(\la^{\dagger}\la \right)_{11}}\left(f(x_{j})\right) \, \\ 
&=&\frac{1}{8\pi}\sum_{j\neq 1}\frac{\IM{\la_{\al i}^{*}\la_{\al j} \left(\la^{\dagger}\la\right)_{ij}}}{\left( \la^{\dagger}\la \right)_{ii}}\,\left(f(x_{j})+g(x_{j})\right),
\eea
where
\bea
f(x)+g(x)=\sqrt{x}\left(\frac{2-x}{1-x}-(1+x)\log{(1+\frac{1}{x})} \right).
\eea
The total $CP$ asymmetry is the sum of the individual ones $\eps_{1}=\sum_{\al}\eps_{\al}$. \\
The set of equations (\ref{eqBE}) governs the $B-L_{\al}$ asymmetry evolutions. As explained in the previous chapter, these $B-L_{\al}$ asymmetries are deduced from the leptonic doublet asymmetries, by relating the chemical potentials of the $B-L_{\al}$ asymmetry to the ones of $\ell_{\al}$ according  to the different processes that are in-equilibrium at the temperature at which leptogenesis occurs.\\
We saw that if $M_{1}\lesssim 10^{9} \GeV$, interactions involving muon and tau charged lepton Yukawa couplings are in-equilibrium, and are potentially faster than decays and inverse decays, thus projecting the lepton asymmetry onto a three flavour basis $(L_{e},L_{\mu},L_{\tau})$. Given that for $T\lesssim 10^{9}\GeV$, interactions involving charm, bottom and top quark Yukawa couplings are also in-equilibrium, the conversion matrix $A$ reads~\cite{matters}:
\begin{eqnarray}
A= \left( \begin{array}{ccc}
-151/179 & 20/179 &  20/179 \\ 
 25/358 &  -344/537 &  14/537 \\ 
 25/358 & 14/537 & -344/537
\end{array} \right) \ .
\end{eqnarray}
One then has to solve the coupled BEs for those three distinguishable flavours. The BEs are governed by individual $CP$ asymmetries $\eps_{\al}$, $\al=e,\mu,\tau$ and individual washouts $\kappa_{\al}$, $\al=e,\mu,\tau$.\\
If the decaying RHn is heavier, with $10^{9}\GeV \lesssim M_{1} \lesssim 10^{12} \GeV$, then only tau Yukawas are in-equilibrium (together with $b, c, t$ quarks), so that the lepton asymmetry is projected onto the tau direction $(L_{\tau},\eps_{\tau},\kappa_{\tau})$ and onto an orthogonal direction in flavour space, which coherently mixes electron and muon flavours $(L_{o},\eps_{o},\kappa_{o})=(L_{e}+L_{\mu},\eps_{e}+\eps_{\mu},\kappa_{e}+\kappa_{\mu})$.\\
In this two flavours scheme the $A$-matrix reads:
\bea
A=\left( \begin{array}{cc}
-417/589 &  120/589 \\ 
 30/589 &  -390/589  \\ 
\end{array} \right)\ .
\eea
For $M_{1} \gtrsim 10^{12} \GeV$, none of the charged lepton Yukawas are in-equilibrium, thus the one flavour approximation holds and one solves the BE for $(Y_{B-L},\eps_{1},K_{1})$.\\
Finally, the baryon asymmetry produced by sphalerons conversion is~\cite{BLconv}:
\bea
Y_{B}=\frac{12}{37} \sum_{\al} Y_{\Delta \al} \, ,
\eea
where the factor $12/37$ is obtained in the SM, and the sum is made over the flavours which were distinguishable when the asymmetries were produced.\\
\\
In this chapter, similarly to chapter 3, we consider $\Delta L=1$ two-body decays and inverse decays which act as source and damping terms, and whose expressions are given by eq.(\ref{EQ514}).
%\bea
%D(z)&=&z\,\frac{K_{1}(z)}{K_{2}(z)} \, , \nonumber \\
%W_{id}(z)&=&\frac{1}{2}\,D(z)\,\frac{Y_{N}^{eq}}{Y_{\ell}^{eq}}\simeq \frac{1}{4}\,z^{3}\,K_{1}(z) \ .
%\eea
We further include $\Delta L=1$ 2-to-2 scatterings involving the third generation of quark doublets and top quark singlets,
% in the s-channel $t^{c}\,Q_{3} \LRa N \,\ell_{\al}$ and in the t (and u) channels $Q_{3}\, N \LRa \ol{t^{c}}\,\ell_{\al}$,$Q_{3}\,\ol{\ell}_{\al} \LRa \ol{t^{c}}\,N$.
whose contribution to the washout term is given by~\cite{pedestrians}
\bea
W_{scat}(z)=2\,S_{s}(z)\left(\frac{Y_{N}(z)}{Y_{N}^{eq}(z)}\right)+4\,S_{t}(z) \ ,
\eea
where the precise expression of these scatterings is given in appendix B. Finally the total washout reads:
\bea
W(z)=W_{id}(z)+W_{scat}(z).
\eea
Neutrino-top scatterings also contribute to the production of the lepton asymmetry, the $CP$ asymmetry in scatterings being equal to $CP$ asymmetry in decays~\cite{resonantlepto,issues}. The production term is thus:
\bea
P(z)=D(z)+2\gamma_{s}(z)+4\gamma_{t}(z) \ .
\eea
Furthermore, as we want to study the impact of lepton flavours in leptogenesis, $M_{1}$ should not be too heavy, $M_{1} \leq 10^{12} \GeV$. Therefore, as explained in chapter 3, for values of $K_{1}$ between $0.01$ and $100$, we can safely neglect off-shell $\Delta L=2$ while the on-shell contribution is included.
%Finally, we neglect $\Delta L=1$ gauge boson scatterings, as their contribution is qualitatively the same as neutrino-top scatterings, as well as other higher order effect which are discussed in chapter 3.
\section{Effect of lepton flavours in the type I seesaw}
The key parameters for leptogenesis, which are the couplings of light neutrinos to the heavy states $\la$ and the mass of the  right-handed neutrinos, are all we need to know to evaluate $Y_{B}$. Going in the basis where charged leptons and RHns mass matrices are diagonal, we use the Casas-Ibarra parametrisation~\cite{CasasIbarra} for the Yukawa couplings:
\bea
\label{EQ518}
\la_{i \al}=\left(\sqrt{M}.R.\sqrt{m}.U^{\dagger}\right)_{i\al} \ ,
\eea
where the PMNS matrix $U$ and the matrix $R$ are defined in chapter 2, section 2.5.1.
%where $M=\rm{diag}(M_{1},M_{2},M_{3})$ is the heavy neutrino mass matrix, and $m=\rm{diag}(m_{1},m_{2},m_{3})$ is the light neutrino one. The PMNS matrix is parametrised in terms of three mixing angles and 1+2 $CP$ violating phases, while the matrix $R$ is a $3\times3$ orthogonal matrix that depends on three complex angles:
%\bea
%\label{Rmatrix}
%R=\left[\begin{array}{ccc}
%c_{3}c_{2} & c_{3}s_{2} & s_{3} \\
%-c_{1}s_{2}-s_{1}s_{3}c_{2} & c_{1}c_{2}-s_{1}s_{3}s_{2} & s_{1}c_{3} \\
%s_{1}s_{2}-c_{1}s_{3}c_{2} & -s_{1}c_{2}-c_{1}s_{3}s_{2} & c_{1}c_{3} 
%\end{array}\right] \, .
%\eea
%that the low energy sector depends on 3 masses, 3 angles and 3 phases. Among these parameters, only 4 are yet measured (2 mixing angles and 2 squared mass differences), while $\theta_{13}$ is only upper-constrained. The best constraint on the neutrino mass scale ($m_{1}$ in our hierarchical ordering) comes from cosmology~\cite{WMAP5b}, and is $m_{1}\lesssim 0.6 \eV$. Hence we have 3+1+1(+1) free parameters (phases+$\theta_{13}$+$m_{min}$(+light neutrino mass ordering)) coming from the low energy sector. On the other hand, the  parameters of the $R$ matrix are free, as well as RHn masses.\\
We recall that the type I seesaw yet contains $15$ independent degrees of freedom to fit a single observable: the baryon asymmetry. Even though we cannot hope to determine the would-be high-energy masses and mixings, we can nevertheless try to shed some light on the optimal regions of the parameter space. Such a work has been done in the context of flavoured leptogenesis in~\cite{Towardsbis}, and we do not carry here such a detailled analysis.\\
Using the above parametrisation, let us see how the inclusion of lepton flavours influence $CP$ asymmetries and washout factors.
\subsection{$CP$ asymmetry}
\begin{figure}[h!]
\begin{center}
\includegraphics[scale=0.45]{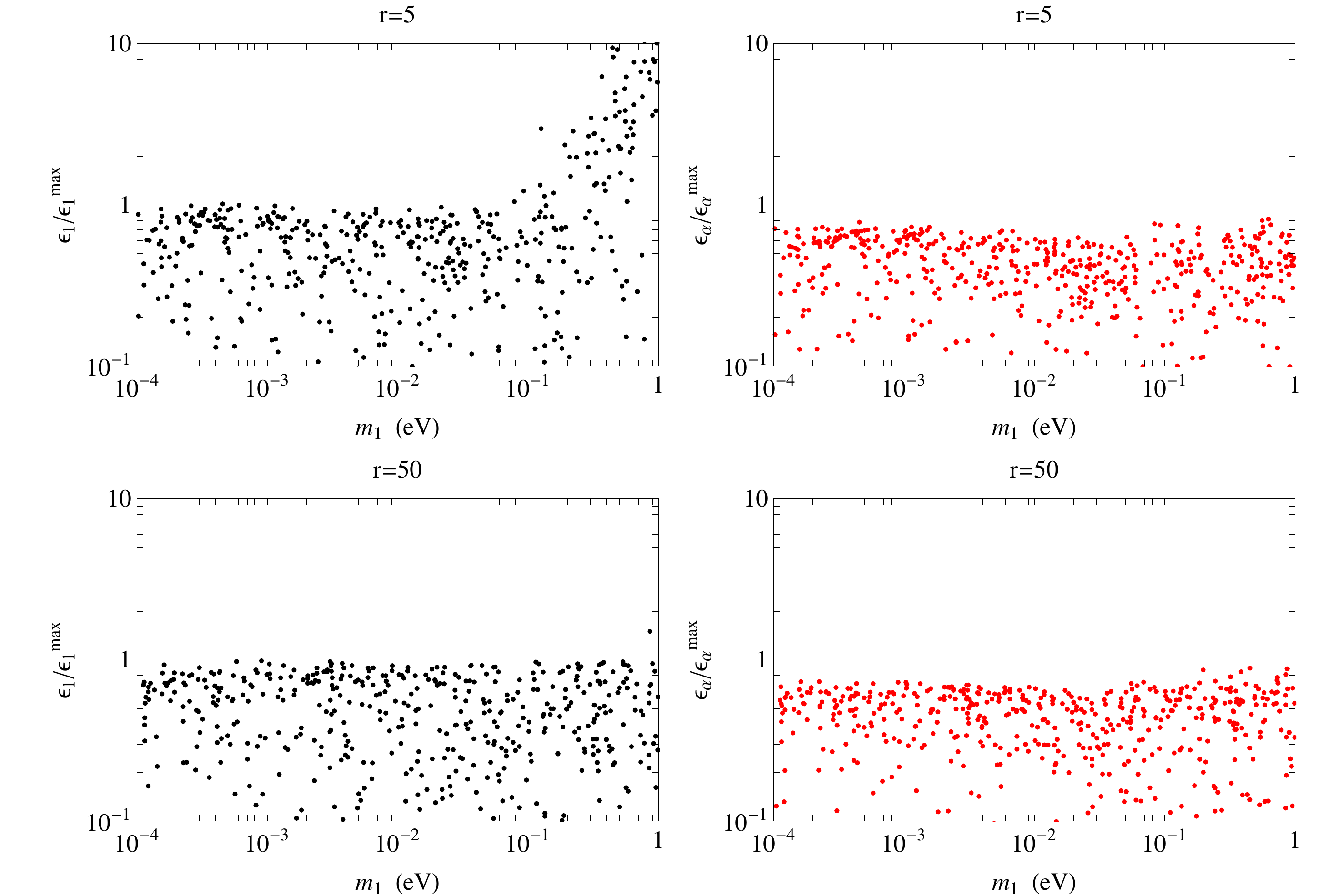} 
 \caption{Scatter plot of $\eps_{\al}/\eps_{\al}^{max}$ as a function of $m_{1}$, for different ratios of heavy neutrino masses. These plots have been obtained by a random scan of the parameter space, imposing the perturbativity constraint $\la_{ij}< \sqrt{4\pi}$. $M_{1}$ ranges from $10^{9} \GeV$ to $10^{11} \GeV$, and $m_{1}$ from $10^{-4} \eV$ to $1\eV$, even if the upper limit is already above the cosmological bound.} 
\label{GrapheEcpFla}
\end{center}
\end{figure}

\begin{figure}[h!!]
\begin{center}
\includegraphics[scale=0.45]{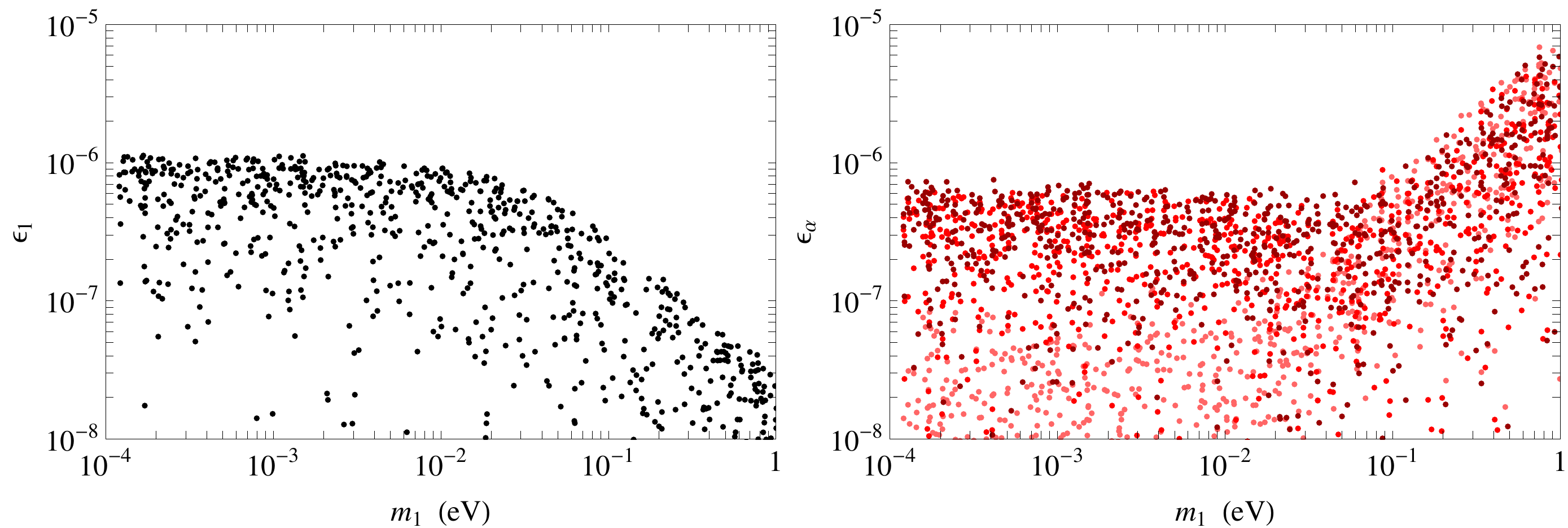} 
 \caption{Scatter plot of $\eps_{1}$ (left panel) and $\eps_{\al}$ (right panel) as a function of the $m_{1}$. $M_{1}$ is set at $10^{10} \GeV$, and we take $r=100$.} 
\label{EcpCompFla}
\end{center}
\end{figure}

Assuming a strong hierarchy in the heavy neutrino sector, and using the parametrisation of eq.(\ref{EQ518}), eq.(\ref{ECPflav}) can be recast into
\bea
\label{EQ520}
\eps_{\al}=-\frac{3}{16\pi}\frac{M_{1}}{v^{2}}\frac{\IM{\sum_{\be,\rho}m_{\be}^{1/2}\,m_{\rho}^{3/2} U_{\al\be}^{*}\,U_{\al\rho}\,R_{1\be}R_{1\rho}}}{\sum_{\be} m_{\be}\vert R_{1\be}\vert^{2}} \, .
\eea
Summing over the lepton flavours, the total $CP$ asymmetry in decays is found to be
\bea
\eps_{1}=-\frac{3}{16\pi}\frac{M_{1}}{v^{2}}\frac{\IM{\sum_{\be}m_{\be}^{2}\,R_{1\be}^{2}}}{\sum_{\be} m_{\be}\vert R_{1\be}\vert^{2}} \, ,
\eea
no longer dependent on the low-energy phases of the PMNS matrix.%: there is more source of $CP$ asymmetry in flavoured leptogenesis.\\
%However,%given our parametrisation of the PMNS matrix, 
However, these phases appear with a factor $\propto m_{1,2}^{1/2}$ or $ m_{1,2}^{3/2}$, and so are suppressed compared to the dominant term $\propto m_{3}^{2}$. Thus we expect them to play a subdominant role in the general case of a complex $R$ matrix, or if light neutrinos are very hierarchical. On the contrary, if $CP$ is violated only in the low-energy sector, \ie if $R$ is real, we see that whereas $\eps_{1}$ vanishes, individual $CP$ asymmetries do not~\cite{matters}. However it has been found that relying only on low-energy phases, a viable leptogenesis requires a sizable amount of fine-tuning~\cite{RiottoPetcov}.\\
In general, the $CP$ asymmetries will receive contribution from both high and low-energy sectors\footnote{cf. \eg~\cite{Molinaro:2008rg}.}.\\
An important point to notice is the modification of the Davidson-Ibarra bound on the $CP$ asymmetry. 
Indeed, in the one flavour approximation, when summing over lepton flavours  and assuming a strong hierarchy between right-handed neutrinos, the $CP$ asymmetry is bounded by
\bea
\eps_{1}^{max}=\frac{3}{16 \pi}\frac{M_{1}(m_{3}-m_{1})}{v^{2}}\times \beta(m_{1},\tilde{m}_{1})\, .
\eea
Hence, for degenerate neutrinos, $m_{1}\simeq m_{3}$, and thus the $CP$ asymmetry scales as $m_{1}^{-1}$.\\
When lepton flavours are included, the bound on individual $CP$ asymmetries is modified. Indeed, in the limit of hierarchical RHns, eq.(\ref{ECPflav}) reads
\bea
\eps_{\al}\simeq \frac{3\,M_{1}}{16\,\pi\,v^{2}}\,\sum_{\be}\frac{\IM{\la_{\al\,1}\la_{\be\,1}\left(m^{\dagger}\right)_{\al\,\be}}}{\left(\la^{\dagger}\,\la\right)_{11}} \, .
\eea
Then, writting $\la_{\al\,1}=\Phi_{\al}\,\tilde{\la}_{\al}$, where $\tilde{\la}=\vert \la\vert$ and $\Phi$ stands for the phase of $\la$, the above expression reads
\bea
\eps_{\al}\simeq \frac{3\,M_{1}}{16\,\pi\,v^{2}}\frac{1}{\sum_{\rho}\tilde{\la}^{2}_{\rho}}\sum_{\be}\,\IM{\Phi_{\al}\Phi_{\be}\,\left(m^{\dagger}\right)_{\al\,\be}}\tilde{\la}_{\al} \,\tilde{\la}_{\be}\, ,
\eea
The washout factors are given by
\bea
K_{1}=\sum_{\al}\kappa_{\al}=\sum_{\al}\frac{\tilde{\la}_{\al}^{2}\,v^{2}}{M_{1}}\,.
\eea
Using $m_{\al\be}=\left(U^{*}d_{m}U^{\dagger}\right)_{\al\be}$, and since
%\sum m \tilde{\la} Im{phase}\less \sum m \tilde{\la}\less m_{max}\sum \tilde{\la}=m_{\max}\sum \tilde{\la}/\sqrt{\sum \tilde{\la}^{2}} \times  \sqrt{\sum \tilde{\la}^{2}}\less m_{max} \sqrt{\sum \tilde{\la}^{2}} car \tilde{\la}>0
\bea
\sum_{k,\be}m_{k}\,\tilde{\la}_{\be}\,\IM{\Phi_{\al}\Phi_{\be}U_{\al\,k}U_{\be\,k}}\lesssim \rm{max}(m)\,\sqrt{\sum_{\be}\tilde{\la}_{\be}^{2}}\, ,
\eea
we obtain that the individual $CP$ asymmetries are bounded by~\cite{issues}:
\bea
\label{DIfla}
\eps_{\al}\lesssim \frac{3}{16\,\pi}\frac{M_{1}\,m_{3}}{v^{2}}\times\sqrt\frac{\kappa_{\al}}{K_{1}} \,. 
\eea
This is an important result, since in the limit of degenerate light neutrinos, instead of scaling as $m^{-1}$ like the total $CP$ asymmetry, individual $CP$ asymmetries scale as $m$ hence are not suppressed. We study in fig.\ref{GrapheEcpFla} the validity of the bound of eq.(\ref{DIfla}) by plotting $\eps_{\al}/\eps_{\al}^{\max}$ as a function of $m_{1}$. As this bound is derived assuming a strong hierarchy for the right-handed neutrino masses, we plot $\eps_{\al}/\eps_{\al}^{\max}$ for different ratios $M_{i}/M_{j}=r=5,50$. We see that even for  $r=5$, the bound on $\eps_{\al}$ can be safely used.
\\
The different behaviour in the degenerate regime between the flavoured case and the single flavour approach can be seen in fig.\ref{EcpCompFla}, where we display $\eps_{1}$ and $\eps_{\al}$ as a function of $m_{1}$.
\\
In the degenerate regime when $m_{1}\gtrsim m_{atm}\simeq 5\times 10^{-2}\eV$, we clearly see that while the total $CP$ asymmetry is suppressed and scales as $m_{1}^{-1}$, such a suppression does not occur for individual $CP$ asymmetries. On the contrary, since in the degenerate regime $\eps_{\al}\propto m_{3}\simeq m_{1}$, the flavoured asymmetries  indeed increase with $m_{1}$.
\subsection{Washout factors}
Using the Casas-Ibarra parametrisation, we can rewrite the washout factors as
\bea
K_{1}&=&\sum_{\be}\frac{m_{\be}}{m_{*}}\vert R_{1\be}\vert^{2}\, , \nonumber \\
\kappa_{\al}&=&\sum_{\be}\frac{m_{\be}}{m_{*}}\vert R_{1\be}\,U_{\al\be}^{*}\vert^{2} \, .
\eea
Then, while the total washout parameter reads 
\bea
K_{1}=k_{1} \vert c_{3}c_{2}\vert^{2}+k_{2} \vert c_{3}s_{2}\vert^{2}+k_{3} \vert s_{3}\vert^{2}\, ,\,k_{i}=m_{i}/m_{*}\, .
\eea
The dependence of $K_{1}$ on $m_{3}$ being $k_{3} \vert s_{3} \vert^{2}$, the dependence of, for instance, the electron-flavour washout $\kappa_{e}$ on $m_{3}$ is given by $k_{3}\,s_{13}^{2}\vert s_{3}e^{i\delta} \vert^{2}$ and is very suppressed given the bound $\theta_{13}\lesssim 0.05$.\\
Then, besides the fact that the individual washouts are smaller than their sum $K_{1}$, one has in general an interesting non-democracy of washout strengths. Indeed, it is possible that one flavour is only weakly washed out while still having $K_{1}\gg 1$. Furthermore, due to the orthogonality of the matrix $R$, we see that $\tilde{m}_{1}$ is bounded from below by $m_{1}$. In general such bounds on $\tilde{m}_{\al}$ do not exist. Then one can have $\tilde{m}_{\al}\lesssim m_{1}$, which is particularly interesting in the strong washout/degenerate regime.\\
In this sense, one somehow evades the constraint of the one-flavour picture, where the total washout is upper bounded by $K_{1}\lesssim 20\times (\eps_{1}/5\,\times 10^{-6 })^{0.86}$.\\
% We illustrate this point in fig.(\ref{GrapheRapK}), which is a scatter plot of the washout $\kappa_{e+\mu}$ vs. $\kappa_{\tau}$. We have chosen $M_{1}$ to vary between $10^{9} \GeV$ to $10^{11} \GeV$ hence the two flavours case applies. 
%\begin{figure}[h]
%\begin{center}
%\includegraphics[scale=0.7]{GrapheRapK.pdf} 
%    \caption{Scatter plot of $\kappa_{e+\mu}$ vs. $\kappa_{\tau}$. These plots have been obtained by a random scan of the parameter-space parameter, with the perturbative constraint $\la_{ij}< \sqrt{4\pi}$. $M_{1}$ is varying from $10^{9} \GeV$ to $10^{11} \GeV$, and $m_{1}$ from $10^{-6} \eV$ to $1\eV$.} 
%\end{center}
%\label{GrapheRapK}
%\end{figure}
%This picture shows how the washout differs from one flavour to the other. We see however that a strong discrepancy between the different washout such as $\kappa_{\al}/\kappa_{\be} \gtrsim 100$, even if possible is hardly obtained in our scan, meaning that such domination of one flavour on the other results from some fine tuning of the high-energy phases.\\ 

\subsection{Qualitative picture}
\begin{figure}[!h]
\begin{center}
\includegraphics[scale=0.7]{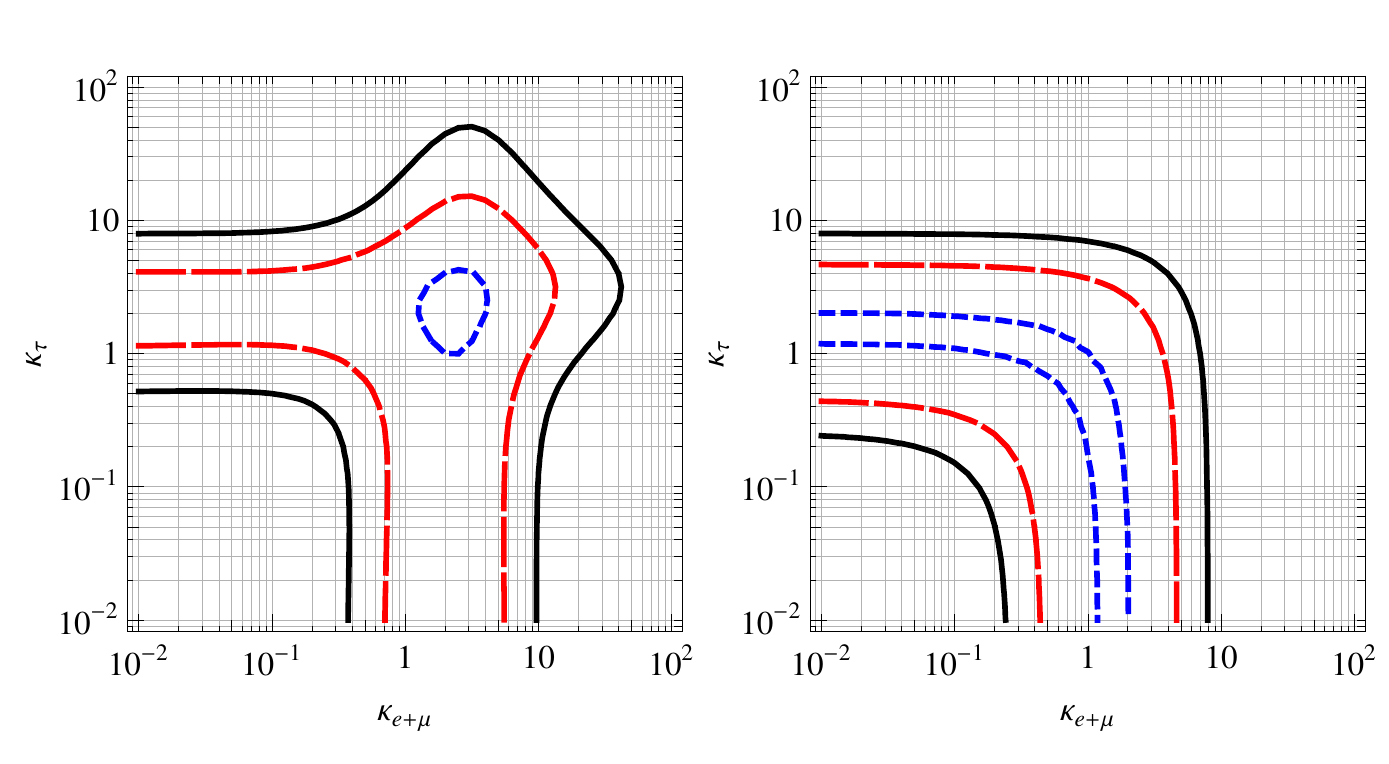} 
    \caption{Contour plot of $Y_{B}^{num}/Y_{B}^{obs}$, for $\kappa_{e+\mu}$ and $\kappa_{\tau}$ varying from $10^{-2}$ to $10^{2}$. The left (right) panel represents the case with (without) flavours. The different contours stand for a ratio of $1, 2$ and $4$ and are respectively depicted in black, long-dashed red and dashed blue.} 
\label{GrapheQualFla}
\end{center}
\end{figure}

The rough picture of flavoured leptogenesis follows the same lines as the single flavour approximation. First, a thermalisation period during which a lepton asymmetry is produced in hollow. Then, after RHns reach thermal equilibrium, decay processes  wash-out the lepton asymmetry.\\
When the Universe has cooled down to about $M_{1}/100$, the different processes are frozen, and the lepton asymmetry remains constant until being converted by sphalerons.\\
The difference with respect to the single flavour approximation concerns how the different lepton flavours are coupled to the decaying RHn, that is the rate at which they are produced or washed out. We illustrate this difference in fig.\ref{GrapheQualFla}, by plotting contours of the baryon asymmetry obtained by solving the set of BEs in eqs.(\ref{eqBE}), normalised to the observed value $Y_{B}^{obs}=8.7\times10^{-11}$, in the case where  flavours are included (left panel) or not (right panel). Leading to this plot we did not use the seesaw formula for the Yukawa couplings, since then we do not control the different washout parameters. Instead we uses the $\kappa_{\al}$s as input parameters, setting the  $CP$ asymmetries to their maximal values. We further choose $M_{1}=10^{10} \GeV$, so that we work in a two flavour scheme, and set for convenience $m_{1}=10^{-5} \eV$\footnote{Recall that $\tilde{m}_{1} \geq m_{1}$, hence for $K_{1}=10^{-2}$, $m_{1}$ has to be lighter than the quoted value.}.
\\
In this figure we clearly see the influence of having included flavours. While in the single flavour case the total washout is required to be $0.1\lesssim K_{1}\lesssim 10$, in the flavoured case we see that a total washout $K_{1}=\kappa_{e+\mu}+\kappa_{\tau}$ as large as $10-50$ is still possible, if one of the individual washouts is close to its optimal value $\kappa_{\al}\simeq 3$.
\section{Evaluation of the baryon asymmetry}
To begin with, we neglect the effects of the flavour conversion due to sphalerons~\cite{StudyOf}, typically $\mathcal{O}(0.1)$, which thus have a subdominant effect. These terms will be studied later on. The BEs for the number density then reads:
\bea
\label{eqBEf}
\frac{d Y_{N}(z)}{dz}&=&-K_{1}\,P(z)\,\D N(z) \ , \nonumber \\
\frac{d Y_{\Delta \al}(z)}{dz}&=&-\eps_{\al}\,K_{1}\,P(z)\,\D N(z)-W(z)\,\tilde{\kappa}_{\al}\,Y_{\Delta \al}(z) \, ,
\eea
where $\D N(x)=Y_{N}(x)-Y_{N}^{eq}(x)$ and  we introduce  $\tilde{\kappa}_{\al}=-A_{\al\al}\,\kappa_{\al}$ (recall that diagonal entries of the conversion matrix $A$ are negatives).
The formal solution of this equation for the $\D\al$ asymmetry is
\bea
Y_{\D \al}(z)=-\eps_{\al}K_{1}\int_{z_{in}}^{z}dx\,P(x)\,\D N(x)\,e^{-\tk{\al}\int_{x}^{z}dy\,W(y)} \, .
\eea
As in the single flavour case, the $\D\al$ asymmetry evolves differently according to the specific washout regime.
\subsubsection*{All flavours in the strong washout regime}
In the case where $\kappa_{\al}\gg 1$,  $Y_{N}(z)\simeq Y_{N}^{eq}(z)$, so that $\D N(x)= -Y_{N}^{eq\,\prime}/K_{1}\,P(x)$ and 
\bea
Y_{\D \al}\simeq \eps_{\al}\int_{z_{in}}^{z}dx\,Y_{N}^{eq\,\prime}(x)\,e^{-\tk{\al}\int_{x}^{z}dy\,W(y)} \, .
\eea
In this integral, since the main contribution to the asymmetry production comes from $z\gtrsim 1$, the scattering term in the washout can be neglected, and therefore we recover the result of the one-flavour approximation, namely
%We can evaluate this integral using a saddle point approximation, and find a similar result as in the one-flavour case for the final $B-L$ asymmetry:
\bea
\label{Strong}
Y_{\D \al}\simeq -\eps_{\al}\times \left(\frac{0.4}{\tk{\al}^{1.16}}\right)\,Y_{N}^{eq}(z_{in}) \, ,
\eea
with $Y_{N}^{eq}(z_{in})\simeq 3.9\times 10^{-3}$. The efficiency factor for the flavour $\alpha$ in the strong washout regime is thus given by:
\bea
\eta_{\al}^{s}\simeq \frac{0.4}{\tk{\al}^{1.16}}\, .
\eea
By summing over the flavours, the baryon asymmetry is found to be
\bea
Y_{B}=-\frac{12}{37}\sum_{\al} \eps_{\al}\left(\frac{0.4}{\tk{\al}^{1.16}}\right)\,\,Y_{N}^{eq}(z_{in}) \, .
\eea
This shows the difference with respect to the one-flavour approximation, for which the baryon asymmetry in the strong washout regime is
\bea
Y_{B}^{1f}\propto \frac{\sum_{\al}\eps_{\al}}{\left(\sum_{\al}\kappa_{\al}\right)^{1.16}} \ .
\eea
If, for example, $\kappa_{1}=3$ and $\kappa_{2}=7$ ($K_{1}=10$), then assuming equal $CP$ asymmetries, the baryon asymmetry in the flavoured case is $\sim 2.5$ times the unflavoured one; for instance if $\kappa_{1}=1$ and $\kappa_{2}=14$ ($K_{1}=15$), then the enhancement is of one order of magnitude.
\subsubsection{All flavours in the weak washout regime}
In the case where $\kappa_{\al}\ll 1$ for all distinguishable flavours,  when $\Delta L=1$ scatterings are included in both source and damping terms, one expects the $B-L$ asymmetry to scale as $K_{1}\,\kappa_{\al}$. Indeed, during the first stage of RHn thermalisation, an (anti-)asymmetry is produced, which is proportional to $\eps_{\al} K_{1}$, and is latter washed out by decays of the RHn. However, since this asymmetry undergoes washout processes that are $\propto \kappa_{\al}$, the surviving part is $\propto K_{1} \kappa_{\al}$.\\ The evaluation of $Y_{\D \al}$ follows the same lines as in the single flavour case, where one distinguishes anti-asymmetry production for $z\lesssim z_{eq}$ during thermalisation of the $N$, and the asymmetry production at $z\gtrsim z_{eq}$ that cancels out with the former one:
\bea
Y_{\D \al}(z)&=&-\eps_{\al}\int_{0}^{z}dx\,K_{1}\,P(x)\,\D N(x)e^{-\tk{\al}\int_{x}^{z}dy W(y)} \nonumber \\
&\simeq &\eps_{\al}\left(K_{1}\int_{0}^{z_{eq}}dx\,P(x)Y_{N}^{eq}(x)e^{-\tk{\al}\int_{x}^{z}dy W(y)}+\int_{z_{eq}}^{z}dx\,Y_{N}^{\prime}(x)e^{-\tk{\al}\int_{x}^{z}dy W(y)} \right)\,.
\eea
The first term represents the asymmetry produced in hollow during $N$ thermalisation. Its evaluation is simplified using the approximate expression for the source term~\cite{pedestrians}:
\bea
P(z)\simeq \frac{K_{s}}{K_{1}}+z\,,\quad K_{s}=\frac{9 m_{t}^{2}}{8 \pi^{2}\,v^{2}}K_{1}
\eea
where $K_{s}\simeq 0.1 K_{1}$ parametrises the strength of the $\D L=1$ scatterings. In this weak washout case, at high temperatures RHn are far below their thermal equilibrium, $Y_{N}(z\lesssim~z_{eq})\ll Y_{N}^{eq}(z\lesssim~z_{eq})$, and so a similar expression can be found for the damping term
\bea
W(z)=W_{id}(z)\left(1+\frac{\be}{z}\right) \, , \be \simeq \frac{2\,K_{s}}{3\,K_{1}}+2 \ .
\eea
The first term can be written as
\bea
\label{eqYwa}
Y_{\D \al}\simeq \eps_{\al}\,\frac{K_{1}}{\tk{\al}}\,4\,n_{0}\,\int_{0}^{z_{eq}}dx\left(\tk{\al} W_{id}(x)\right)\left(\frac{K_{s}}{K_{1}}+x\right)e^{-\tk{\al}\int_{x}^{z_{eq}}W_{id}(y)\left(1+\frac{\be}{y}\right)dy} \ ,
\eea
where we note that
\bea
n_{0}=\frac{1}{2}\,Y_{N}^{eq}(z\ll 1)=\frac{135\zeta(3)}{8\pi^{4}\,g_{*}} \, .
\eea
The expression in eq.(\ref{eqYwa}) can be further simplified since the leading contribution comes from $z\sim 1$, such that the source term in this weak washout regime reads $D(z)+S(z)\simeq K_{s}/K_{1}$. Given that 
\bea
\int_{x}^{\infty}dy\,y^{2}K_{1}(y)=x^{2}K_{2}(x) 
\eea
with $x^{2}K_{2}(x)\simeq 2$ for $x\lesssim 1$, and using the fact that
\bea
\int_{0}^{\infty}dy\,y^{3}K_{1}(y)=\frac{3\pi}{2}\equiv \tilde{\al} \ , 
\eea
one obtains the antisymmetry production during thermalisation:
\bea
Y_{\D \al}^{<}\simeq \eps_{\al}\frac{K_{1}}{\tk{\al}}\,4n_{0}\,\frac{K_{s}}{K_{1}}\left(1-e^{-\frac{\tk{\al}}{4}\tilde{\al}}\right)\,e^{-\frac{\tk{\al}}{4}\,2\be} \ .
\eea
For $z\gtrsim z_{eq}$,  when RHns dominate over their equilibrium abundance, and since the washout terms can be neglected, the asymmetry produced in the decay can be writen as:
\bea
Y_{\D \al}^{>}&=&-\eps_{\al} \int_{z_{eq}}^{z}\,K_{1}\,P(x)\,Y_{N}(x) \nonumber \\
&\simeq &\eps_{\al}\int_{z_{eq}}^{z} Y_{N}^{\prime}(x)\simeq-\eps_{\al}\,n_{0}\,\frac{K_{s}}{K_{1}}\tilde{\al} \ .
\eea
Adding the two contributions, the $B-L$ asymmetry is given in the weak washout regime by
\bea
\label{Weak}
Y_{\D \al}&=& Y_{\D \al}^{<}+Y_{\D \al}^{>} \nonumber \\
&\simeq & -\eps_{\al}\times \left(3.7 K_{1}\,\tk{\al}\right) Y_{N}^{eq}(z_{in}) \, .
\eea
In this weak washout regime, the efficiency factor is therefore found to be:
\bea
\eta_{\al}^{w}\simeq 3.7 K_{1}\,\tk{\al} \, .
\eea
Finally, the baryon asymmetry in this case is 
\bea
Y_{B}=-\frac{12}{37}\sum_{\al} \eps_{\al}\times \left(3.7 K_{1}\,\tk{\al}\right) Y_{N}^{eq}(z_{in}) \ ,
\eea
clearly differing from the baryon asymmetry evaluated in the one flavour case:
\bea
Y_{B}^{1f}\propto -\left(\sum_{\al} \eps_{\al}\right)\times K_{1}\left(\sum_{\al}\kappa_{\al}\right)Y_{N}^{eq}(z_{in}) \ .
\eea
In the regime of weak washout, the inclusion of lepton flavours is found to lower the baryon asymmetry when compared to the unflavoured case.
\subsubsection{Mixed regime: some flavours  weakly washed out, but $K_{1}\gg 1$}
This case does not correspond to any of the cases encountered in the single flavour picture.\\
We assume that one flavour, labelled $\be$, is only weakly washed out with $\kappa_{\be}\ll 1$, while the total washout is strong since all other $\kappa_{\al}\gg1, \al\neq \be$.\\
This case has the interesting feature of combining the fast RHn thermalisation, owing to strong inverse decays or scatterings, and which typically ensure a copious asymmetry production, with the   fact that the flavour $\be$, being only weakly washed out, is somewhat protected from the strong washouts of the fast RHn decays. Let us see this in detail.\\
During the first stage of thermalisation, one has
\bea
Y_{\D \be}^{<\,\prime}(z) \simeq \eps_{\be}\,Y_{N}^{\prime}(z)-\tk{\be}\,W(z)\,Y_{\D\be}(z) \, .
\eea
This gives a contribution
\bea
Y_{\D\be}^{<}\simeq \eps_{\be}\int_{0}^{z_{eq}}dx\,Y_{N}^{\prime}(x)\,e^{-\tk{\be}\int_{x}^{z_{eq}}dy\, W(y)} \, .
\eea
As $\tk{\be}\ll 1$ and the RHns reach thermal equilibrium at high temperatures, in this integral we can approximate  $e^{-\tk{\be}\int W}\simeq 1-\tk{\be}\int W$, such that
\bea
Y_{\D\be}^{<}\simeq \eps_{\be}Y_{N}(z_{w})-\eps_{\be}\,\tk{\be}\,\left(\int_{0}^{z_{w}}dx\,Y_{N}(x)\int_{x}^{z_{w}}dy\,W(y)\right) \, ,
\eea
where $z_{w}$ reflects the temperature below which the washout cannot be neglected with $z_{w}\gtrsim 1$. Therefore, in a second stage, for $z\gtrsim z_{w}$, one has
\bea
Y_{\D\be}^{\prime}(z)\simeq \eps_{\be}\,Y_{N}^{\prime}(z) \Ra Y_{\D\be}^{>}\simeq -\eps_{\be}\,Y_{N}(z_{w})\, .
\eea
Adding the two contributions gives:
\bea
Y_{\D\be}=Y_{\D\be}^{<}+Y_{\D\be}^{>}\simeq -\eps_{\be}\,\tk{\be}\,\left(\int_{0}^{z_{w}}dx\,Y_{N}(x)\int_{x}^{z_{w}}dy\,W(y)\right) \, .
\eea
The washouts freeze after they reach their maxima, which is evaluated to occur at $z_{w}\simeq 2.4$, and for which the above integrand is $\simeq 0.4 Y_{N}^{eq}(z_{in})$.
%This gives a contribution to the $B-L$ asymmetry $Y_{\D \be}^{<}(z) \simeq -\eps_{\be} Y_{N}(z)$.
%The RHns rapidly reach their equilibrium abundance, which they closely follow since $K_{1}\gg1$. For $z\gtrsim z_{eq}$, in first order we can write the evolution of $Y_{\D\be}$ during this second stage as:
%\bea
%Y_{\D \be}^{>\,\prime}(z)\simeq -\eps_{\be}\,Y_{N}^{\prime}(z)-\tk{\be}\,W(z)\,Y_{\D \be}(z) \ ,
%\eea
%such that 
%\bea
%Y_{\D \al}^{>}\simeq \eps_{\be}\int_{z_{w}}^{z}dx\,Y_{N}^{eq\,\prime}(x)e^{-\tk{\be}\int_{x}^{z}dy\,W(y)} \ .
%\eea
%The individual washout $\tk{\be}$ being $\ll 1$, $e^{-\tk{\be}\int W}\simeq 1-\tk{\be}\int W$. Furthermore, given that $Y_{N}^{eq\,\prime}(x)\simeq -x^{2}K_{1}(x)$, after straitforward integration one obtains
%\bea
%Y_{\D \be}^{>}\simeq -\eps_{\be}\,Y_{N}^{eq}(z_{w})\,(1-\tk{\be}) \ .
%\eea
%Finally, adding the two contribution gives $Y_{\D \be}\simeq \eps_{\be}\,\kappa_{\be}\,Y_{N}^{eq}(z_{w})$.
%The damping term becomes non-negligible for $z_{w}\sim 2$, so that one has
Therefore, the $B-L$ asymmetry in the flavour $\be$ is given by
\bea
\label{Int}
Y_{\D \be} \simeq -\eps_{\be}\,0.4\,\tk{\be}\,Y_{N}^{eq}(z_{in})\, ,
\eea
hence $\eta_{\be}^{m}\simeq 0.4\,\tk{\be}$. For the flavour(s) $\al$ which is (are) strongly washed out, the result eq.(\ref{Strong}) still holds.\\
\subsubsection{Global parametrisation}
\begin{figure}[h!!]
\begin{center}
\includegraphics[scale=0.4]{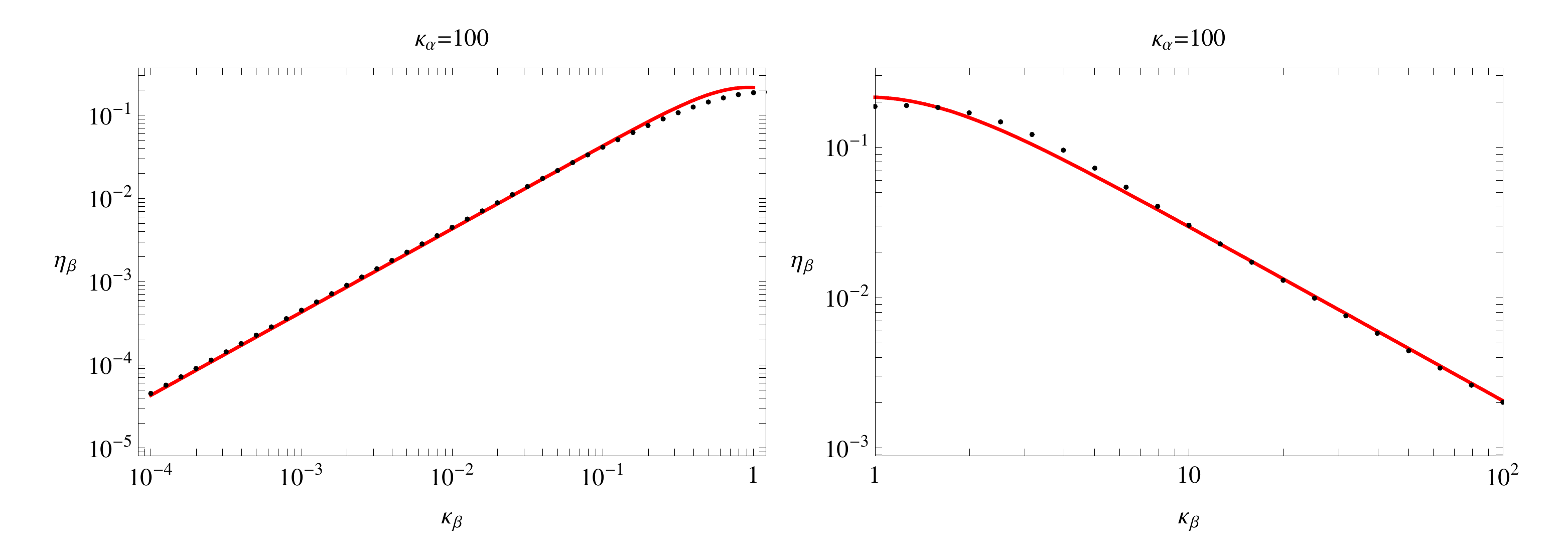} 
    \caption{Individual efficiency factor $\eta_{\be}$ as a function of the corresponding washout $\kappa_{\be}$. We impose that the other flavour is strongly washed-out, with $\kappa_{\al}=100$. The black points represent the numerical results, while the red line stands for the approximate formulae given in eqs.(\ref{Strong},\ref{Int}).} 
\label{EffDiagA}
\end{center}
\end{figure}

\begin{figure}[!h]
\begin{center}
\includegraphics[scale=0.4]{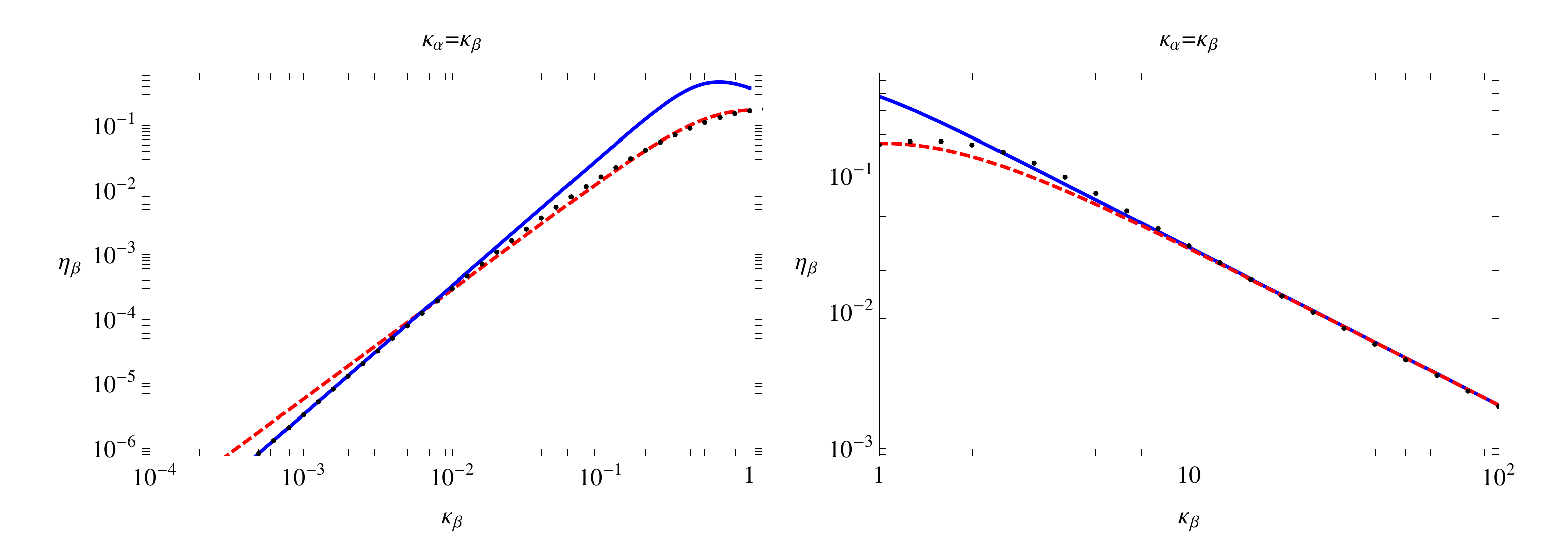} 
    \caption{Individual efficiency factor $\eta_{\be}$ as a function of the corresponding washout $\kappa_{\be}$, in the case where the washout of the other flavours also varies, with $\kappa_{\al}=\kappa_{\be}$. The black points represent the numerical results, while the blue line stands for the approximate formulae given in eqs.(\ref{Strong},\ref{Weak}). The dashed red line represents the fit of eq.(\ref{WeakFit}).} 
\label{EffDiagB}
\end{center}
\end{figure}

Similarly to the global parametrisation in the single flavour picture, by means of a simple interpolation we can infer the efficiency factor for all washout regimes
\bea
\eta \simeq \left(\eta_{w}^{-1}+\eta_{s}^{-1}\right)^{-1}\, ,
\eea
which provides a good idea of the result, even if some fine-tuning is still required in order to best fit the numerical results. As an illustrative example, we consider a two-flavour case, $Y_{\D\al}-Y_{\D\be}$. We plot in fig.\ref{EffDiagA} the efficiency factor for  flavour $\be$, whose washout parameter varies from $10^{-4}$ to $10^{2}$, while fixing $\kappa_{\al}=100$. Hence, the total washout is strong, and eqs.(\ref{Strong},\ref{Int})  apply.
\\
We observe a good agreement between the formulae derived and the numerical results. However if we allow the washout of  flavour $\al$ to vary, setting for example $\kappa_{\al}=\kappa_{\be}$, we see in fig.\ref{EffDiagB} that the formula of eq.(\ref{Weak}), which is depicted in blue, is accurate only for $\kappa \lesssim 0.05$. For $\kappa \gtrsim 0.01$, some fine-tuning of eq.(\ref{Weak}) is preferable to best fit the result. In the plot of fig.\ref{EffDiagB}, we represent in a red dashed line the formula
\bea
\label{WeakFit}
\eta_{\be}\simeq \frac{1}{\left((1.3\,K_{1}\,\kappa_{\be}^{0.7})^{-0.4}+(0.4\,\kappa_{\be}^{-1.16})^{-0.4}\right)^{1/0.4}}\, ,
\eea
which gives an accurate result for $\kappa_{\be}\gtrsim 0.01$.
However, this is an important fine-tuning of the parameters, and can be thus disregarded.
\subsection{Thermal initial population}
\begin{figure}[!h]
\begin{center}
\includegraphics[scale=0.5]{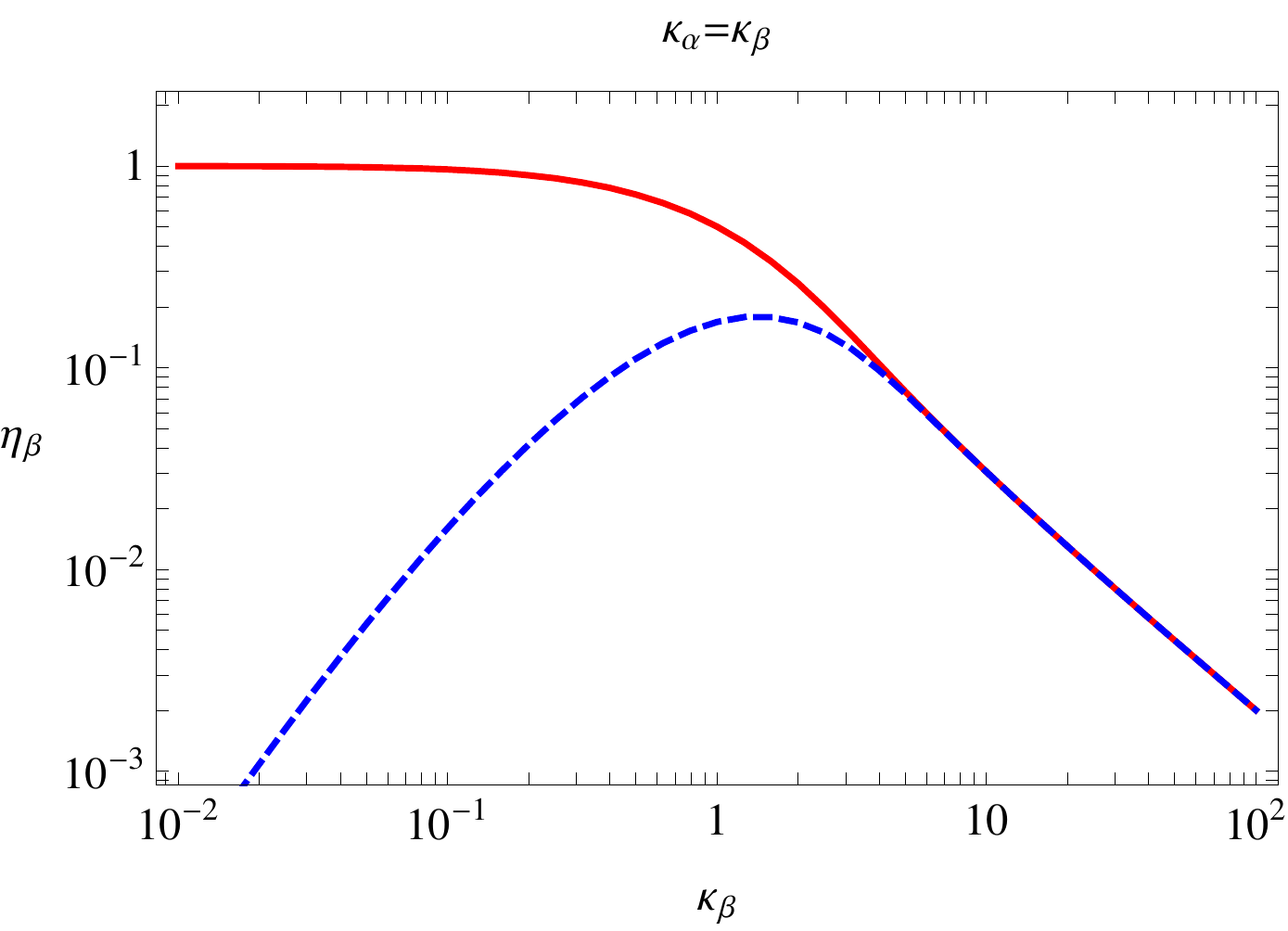} 
    \caption{Efficiency factor of a given flavour $\be$, in the thermal case (dashed blue) and in the case where the $N$ are initially in thermal equilibrium (red).} 
\label{GrapheEffTherm}
\end{center}
\end{figure}

The evaluation of the efficiency factor in the equilibrium scenario is very similar to the one flavour case~\cite{BdB}. Indeed, we saw in chapter 3 that in this scenario, washout is not necessary to produce a population of RHns.
%It is interesting to look at the efficiency factor in the case where initially right-handed neutrinos are in thermal equilibrium. Since in this case, washout are no longer necessery to create $N$, one expects a maximal efficiency in the small washout regime. Conversely, in the strong washout regime, we do not expect difference with the thermal scenario, as in the latter case thermalisation of the $N$ rapidly occurs.\\
Nevertheless, for the sake of illustration, we represent in fig.\ref{GrapheEffTherm} the eficiency factor for a given flavour, in the thermal case (dashed-blue) and in the case where the $N$s are initially in thermal equilibrium.
\\
In the case where all flavours are weakly washed out,
\bea
Y_{\D \al}(z)=\eps_{\al}\int_{z_{in}}^{z}dx\,Y_{N}^{\prime}(x)e^{-\tk{\al}\int_{x}^{z}dy\,W(y)}\,,
\eea 
the exponential term can be neglected, and given that $Y_{N}(z_{in})=Y_{N}^{eq}(z_{in})$ one simply has
\bea
Y_{\D \al}\simeq -\eps_{\al}\,Y_{N}^{eq}(z_{in}) \ , 
\eea
resulting in a maximal efficiency factor $\eta=1$.\\
On the other hand, when all flavours are strongly washed-out, decays maintain RHns in thermal equilibrium. Therefore, similarly to the thermal case, one can neglect high temperature contribution, and so
\bea
Y_{\D \al}(z)&=&\eps_{\al}\int_{z_{in}}^{z}dx\,Y_{N}^{\prime}(x)\,e^{-\tk{\al}\int_{x}^{z}dy\,W(y)} \nonumber \\
&\simeq &\eps_{\al}\int_{z_{in}}^{z}dx\,Y_{N}^{eq\prime}(x\,)e^{-\tk{\al}\int_{x}^{z}dy\,W(y)} \nonumber \\
&\simeq &\eps_{\al}\,\frac{n_{0}\,4}{\tk{\al}}\int_{z_{in}}^{z}dx\,\tk{\al}\,W(x)\,e^{-\tk{\al}\int_{x}^{z}dy\,W(y)} \ .
\eea
The $B-L$ asymmetry is equal to the one derived for the thermal case: 
\bea
Y_{\D \al}\simeq -\eps_{\al}\,\frac{0.4}{\tk{\al}^{1.16}}\times Y_{N}^{eq}(z_{in}) \ .
\eea
We see that the strong washout regime provides a similar dependence on the washout factor, regardless of the thermal history of the RHns. This, together with the fact that the atmospheric and solar neutrino masses  point towards the strong washout regime, 
\bea
\tilde{m}&\simeq & m_{sol} \Ra \kappa \simeq 8 \, ,\nonumber\\
\tilde{m}&\simeq & m_{atm} \Ra \kappa \simeq 45 \, ,
\eea
indicates that the strong washout regime is a more robust and reliable regime.
\subsection{Influence of the off-diagonal terms of the $A$ matrix}
\begin{figure}[h!]
\begin{center}
\includegraphics[scale=0.4]{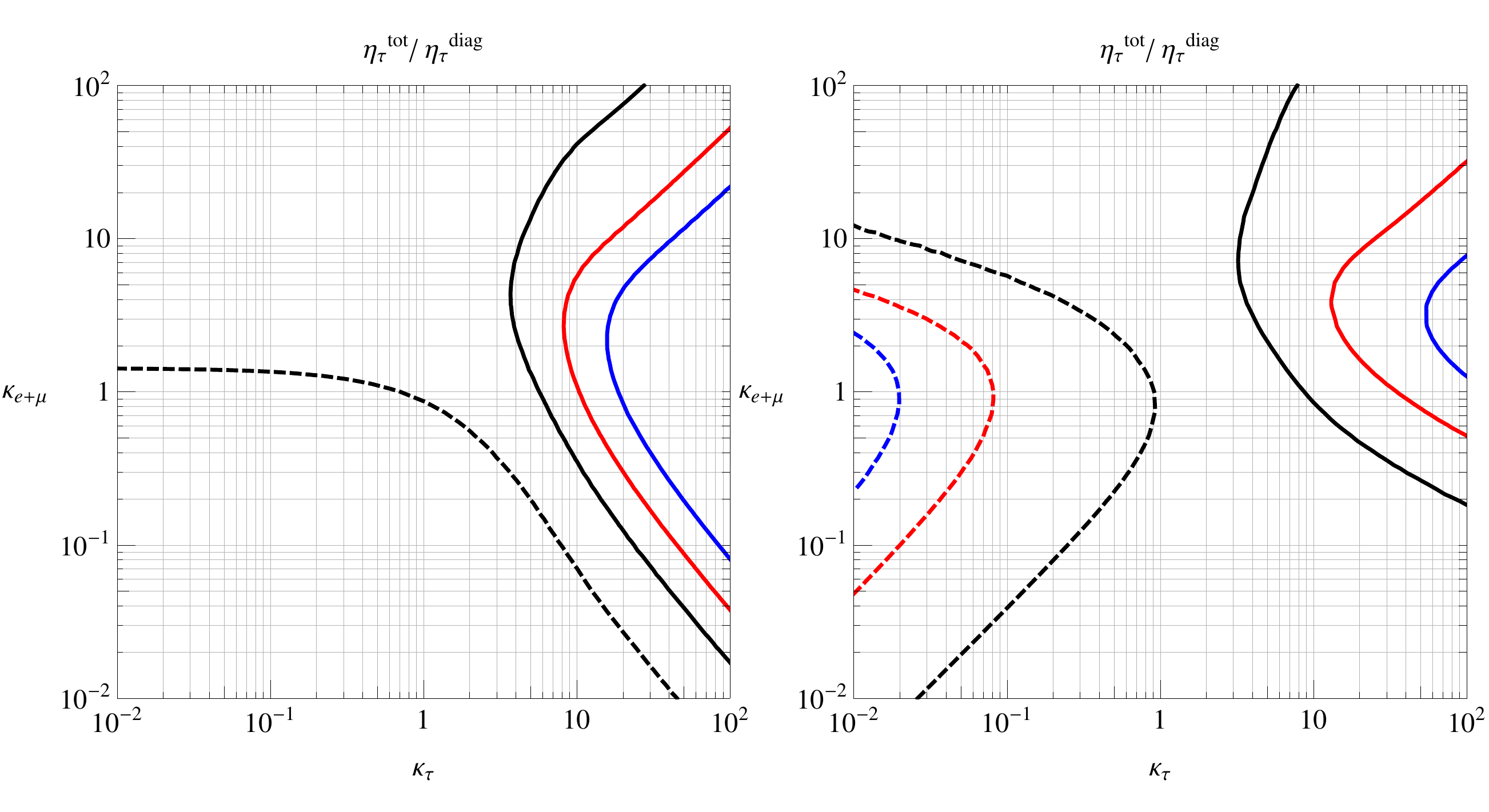} \\
\includegraphics[scale=0.4]{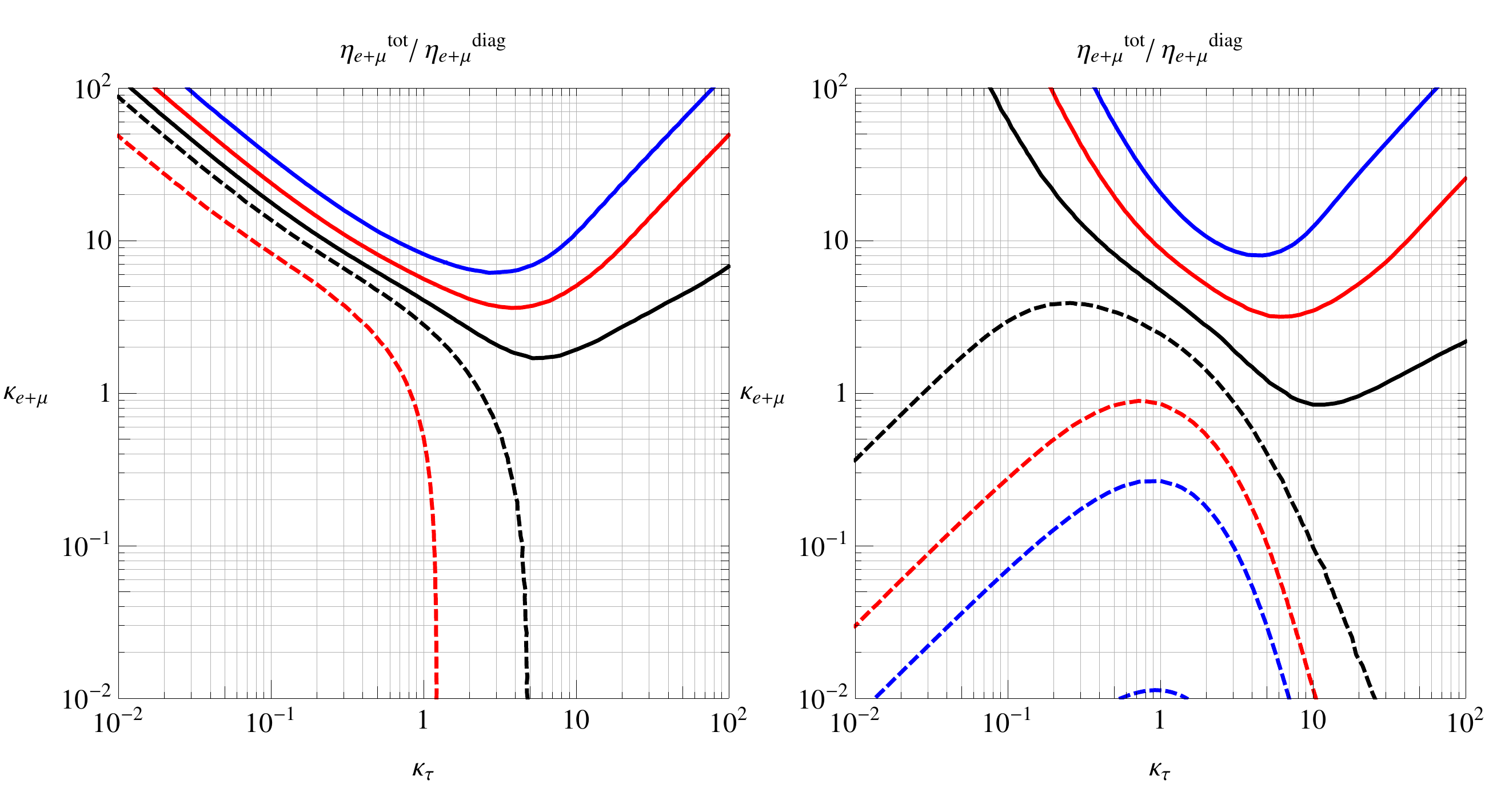}
    \caption{Influence of the flavour conversion on the asymmetry in lepton flavours.  The influence on the flavour $\tau$ ($e+\mu$) is displayed in the up (down) panel. The contours represent different ratios of $\eta_{\al}^{total}/\eta_{\al}^{diag}$, with either equal $CP$ asymmetries $\eps_{\tau}=\eps_{e\mu}$ (left panel) or else $CP$ asymmetries set to their maxima $\eps_{\al}=\eps_{\al}^{max}$ (right panel) . An increase in the efficiency is represented by solid lines, whereas  the dashed lines represent a reduction of $\eta$, with a colour code: black $(\pm 5 \%)$, red $(\pm 20 \%)$ and blue $(\pm 50 \%)$.} 
\label{GrapheEffOD}
\end{center}
\end{figure}

\begin{figure}[h!]
\begin{center}
\includegraphics[scale=0.6]{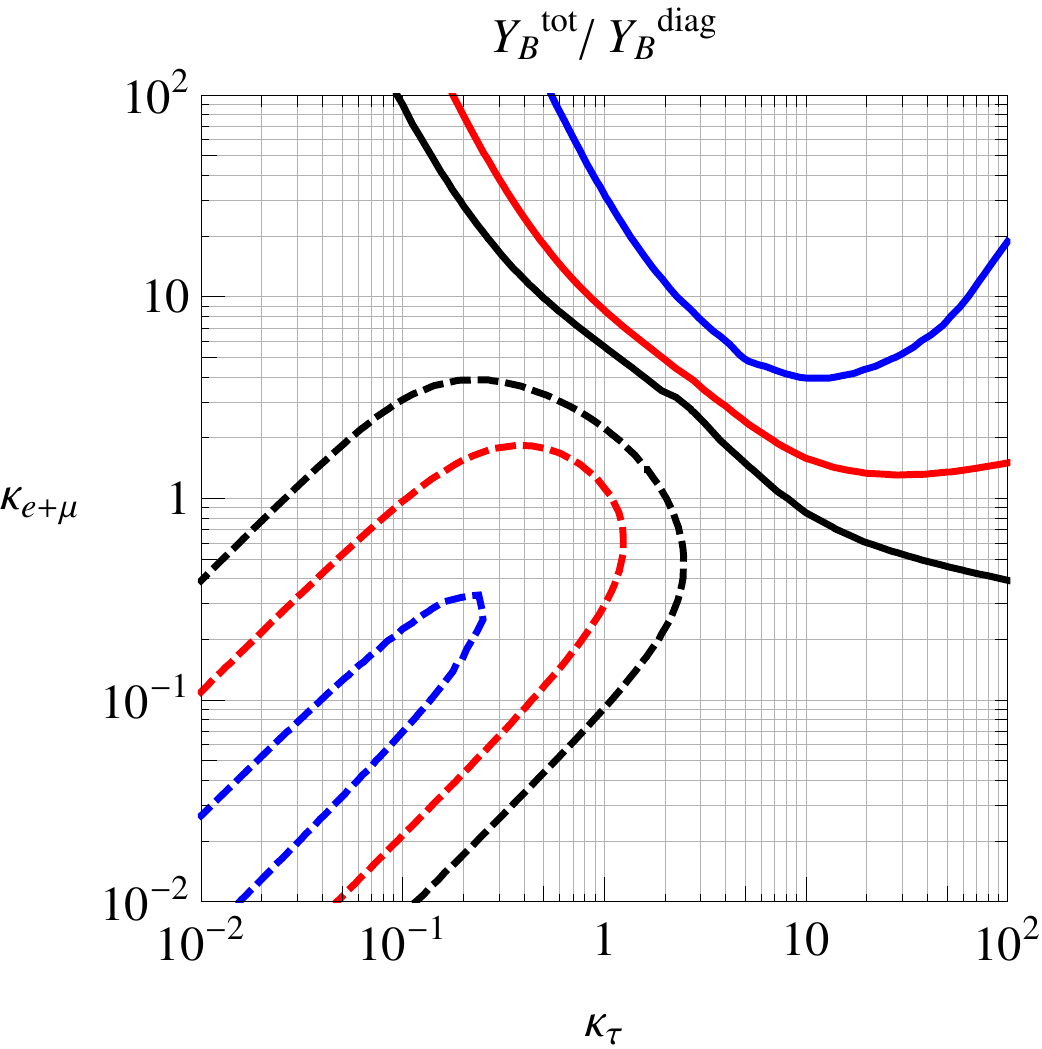} 
    \caption{Influence of the flavour conversion on the baryon asymmetry. The contours represent different values of the ratio $Y_{B}^{total}/Y_{B}^{diag}$, the enhancement being depicted by solid lines and the reduction by dashed lines, with the colour code: black $(\pm 5 \%)$, red $(\pm 10 \%)$, blue $(\pm 20 \%)$.} 
\label{GrapheEffODyb}
\end{center}
\end{figure}

So far we have only included the dominant terms which are the diagonal entries of the matrix $A$, since the off-diagonal terms are $\mathcal{O}(0.1)$. Nonetheless, it is worth studying them.\\
When considering the dominant order, we saw that each flavour could be washed out very differently. As the off-diagonal terms couple flavours among themselves, it is likely that a flavour, being suppressed by either a very weak or a very strong washout, may nontheless provide a non-negligible contribution to the final baryon asymmetry, having been pushed up by another flavour.\\
The evaluation of this off-diagonal term has been done in~\cite{StudyOf}, where we showed that while preventing a complete washout of flavours in particular cases, the effect of the diagonal terms on the baryon asymmetry is less than $20 \%$. Here we illustrate these results.\\
The off-diagonal term contributes to the $B-L$ asymmetry as:
\bea
Y_{\D \al}^{od}(z)=\kappa_{\al}\sum_{\be\neq \al}A_{\al\be}\int_{z_{in}}^{z}dx\,W(x)\,Y_{\D\be}(x)\,e^{-\tk{\al}\int_{x}^{z}dy\,W(y)} \ ,
\eea
and so does not depend on the initial condition of the right-handed neutrinos.
The evaluation of the  above  integral is made difficult by the fact that it implies knowing the dynamic of the individual flavours.\\
We thus make a few approximations. First, we neglect the off-diagonal terms entering in the BE for the flavour $\be$. This is quite justified, since its effect on the flavour $\al$ is $A_{\al \be}^{2}\ll 1$. Then we assume that $Y_{\D \be}(z)\simeq Y_{\D \be}(\infty)$ so that $Y_{\D\be}$ is factorised out of the integral.
We then obtain, dropping out the sum on $\be$:
\bea
Y_{\D \al}^{od}&\simeq &\kappa_{\al}A_{\al\be}\,Y_{\D\be}\,\int_{z_{in}}^{\infty}dx\,W(x)\,e^{-\tk{\al}\int_{x}^{\infty}dy\,W(y)} \nonumber \\
&\simeq &\frac{A_{\al\be}}{A_{\al\al}}\,Y_{\D\be}\left(1-e^{\tk{\al} \ol{\al}}\right)
\eea
where 
\bea
\ol{\al}=\int_{z_{in}}^{\infty}dx\,W(x)\simeq 1.3 \ .
\eea
Had we only included inverse decays, we would have obtained $\ol{\al}=3\pi/8$.\\
The relative factors $A_{\al \be}/A_{\al\al}$ strongly differ from one flavour to the other. For example, in a two-flavour scheme, the relative factors are
\bea
Y_{\D e\mu}^{nd}&\simeq &-0.3 Y_{\D \tau}\left(1-e^{-\tk{e\mu} \ol{\al}}\right) \nonumber \\
Y_{\D \tau}^{nd}&\simeq &-0.08 Y_{\D e\mu}\left(1-e^{-\tk{\tau} \ol{\al}}\right) \ .
\eea
If $\kappa_{\al}\gg 1$, the off-diagonal term is about $\sim   \left(A_{\al \be}/A_{\al\al}\right)\,Y_{\D \be}$, while if $\kappa_{\al}\ll 1$, this term is roughly $\sim  \left(A_{\al \be}/A_{\al\al}\right)\,Y_{\D \be} \times \tk{\al} \ol{\al}$.\\
The $B-L$ flavour conversion is expected to have a significant impact in the case where one flavour is very strongly or very weakly washed-out. Indeed, flavour conversion acts as a source term for the asymmetry. Therefore in the former case, it partly compensates  the washout from RHns decays, and so we expect a small increase in the asymmetry. In the weak washout case, this extra source term further cancels with the asymmetry produced during thermalisation. So, in the weak washout case, one expects a depletion of the asymmetry.\\
We illustrate in fig.\ref{GrapheEffOD} the effect of including the off-diagonal terms on the efficiency factor. We consider a two-flavour regime, with washout varying from $10^{-2}$ to $10^{2}$ \footnote{For simplicity, we set $\kappa_{e}=0$ so that the washout of $Y_{\D e+\mu}$ is $\kappa_{e+\mu}=\kappa_{\mu}$.}. On the left hand side, the $CP$ asymmetries of the different flavours are chosen equal and arbitrary, while on the right hand side the $CP$ asymmetries are set to their maximum values, cf. eq.(\ref{DIfla}).
\\
Looking only at washout effects (left panels), we see that, as expected,  the flavour conversion affects $\D \al$ asymmetries mostly when $\kappa_{\al} \gg 1$ or $\ll 1$. The fact that the flavour $\D e\mu$ is (much) more affected comes from the relative size of the  matrix elements of $A$: $A_{e+\mu,\tau}\simeq 0.2=4\, A_{\tau,e+\mu}$. \\
When realistic values are used for the $CP$ asymmetries (right panels), we see that the global behaviour remains unchanged, albeit this further restricts the influence of the off-diagonal terms to the case where $\sqrt{\kappa_{\al}/K_{1}}$ is not strongly suppressed.\\
%\begin{figure}[!h]
%\begin{center}
%\includegraphics[scale=0.5]{GrapheOD.pdf} 
%    \caption{} 
%\label{GrapheOD}
%\end{center}
%\end{figure}
What about the baryon asymmetry?\\
Individual asymmetries are affected by $\pm 50 \%$, and given that $Y_{B}=12/37 \sum_{\al} Y_{\D \al}$, the baryon asymmetry is affected up to $20 \%$, $Y_{B}$ being enhanced in the strong washout regime, and depleted in the weak washout regime. This can be seen in fig.\ref{GrapheEffODyb}.
\\
We see that the flavour conversion does not significantly modify the baryon asymmetry. However we notice that  for some specific configurations, it prevents $Y_{B}$ from vanishing: in the case where $\eps_{e\mu}+\eps_{\tau}=0$, and $\tk{e\mu}=\tk{\tau}$, without flavour conversion the baryon asymmetry is zero, while with flavour conversion one has 
\bea
Y_{B}& \simeq &\frac{12}{37}\left(Y_{\D e \mu}+Y_{\D \tau}\right) \nonumber \\
%& \simeq &\frac{12}{37}Y_{\D \tau}^{d}\left(\frac{A_{ \tau,e\mu}}{A_{\tau ,\tau}}-\frac{A_{ e\mu,\tau}}{A_{e\mu,e\mu}}\right)(1-e^{-\tk{e\mu}\,\ol{\al}})
&\simeq & 7\times 10^{-2}\,Y_{\D \tau}^{d} \ ,
\eea
which is non zero, even if compatibility with the observed baryon asymmetry may be hardly obtained.
\section{Constraints from a successful leptogenesis}
We saw in chapter 3 that requiring a successful leptogenesis in the single flavour picture somewhat constrains light and heavy neutrinos. It is mandatory that the decaying RHn is heavy enough, $M_{1} \gtrsim 2\times 10^{9} \GeV$, and that light neutrino masses are lighter than $\sim 0.12 \eV$. Concerning the different mixing angles and $CP$ violating phases, the only generic constraint derived is the necessity that the matrix $R$ contains $CP$-odd phases, otherwise no $CP$ asymmetry can be generated.\\Given the importance of this constraint, and the fact that the inclusion of lepton flavours strongly modifies the key parameters of leptogenesis, it is interesting to investigate the possible influence of lepton flavours on the above constraints.
\subsection{Lower bound on heavy neutrino mass}
\begin{figure}[h!]
\begin{center}
\includegraphics[scale=0.6]{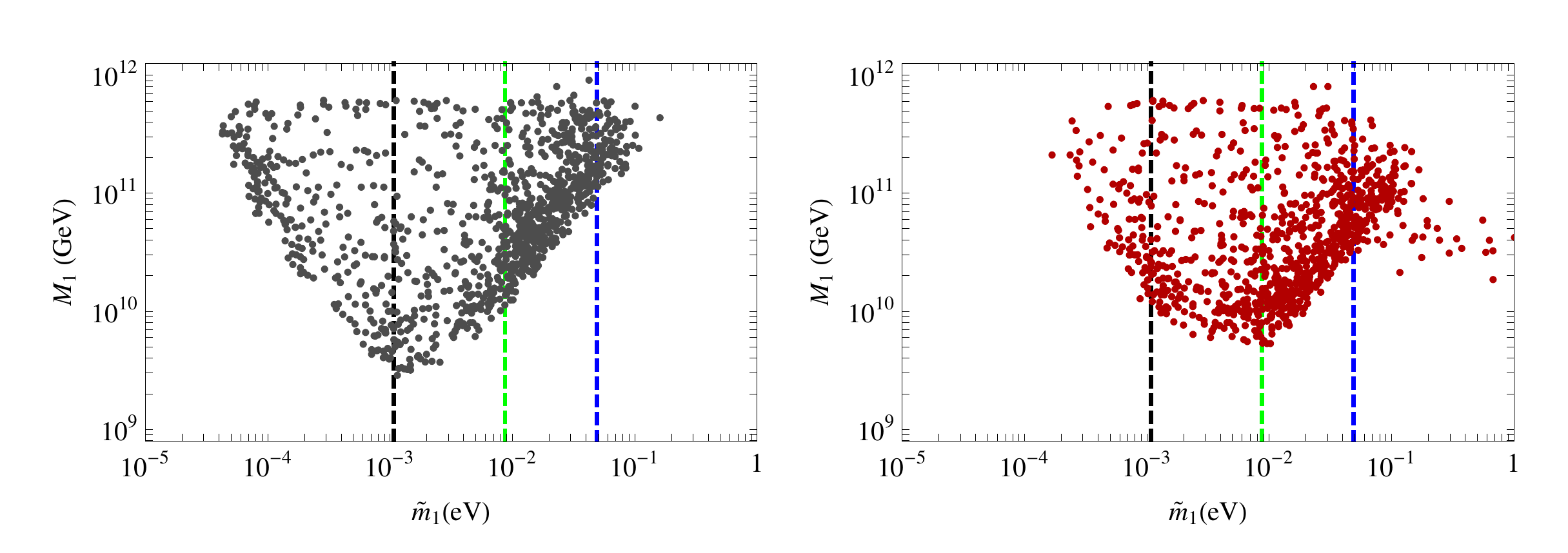} \\
\includegraphics[scale=0.6]{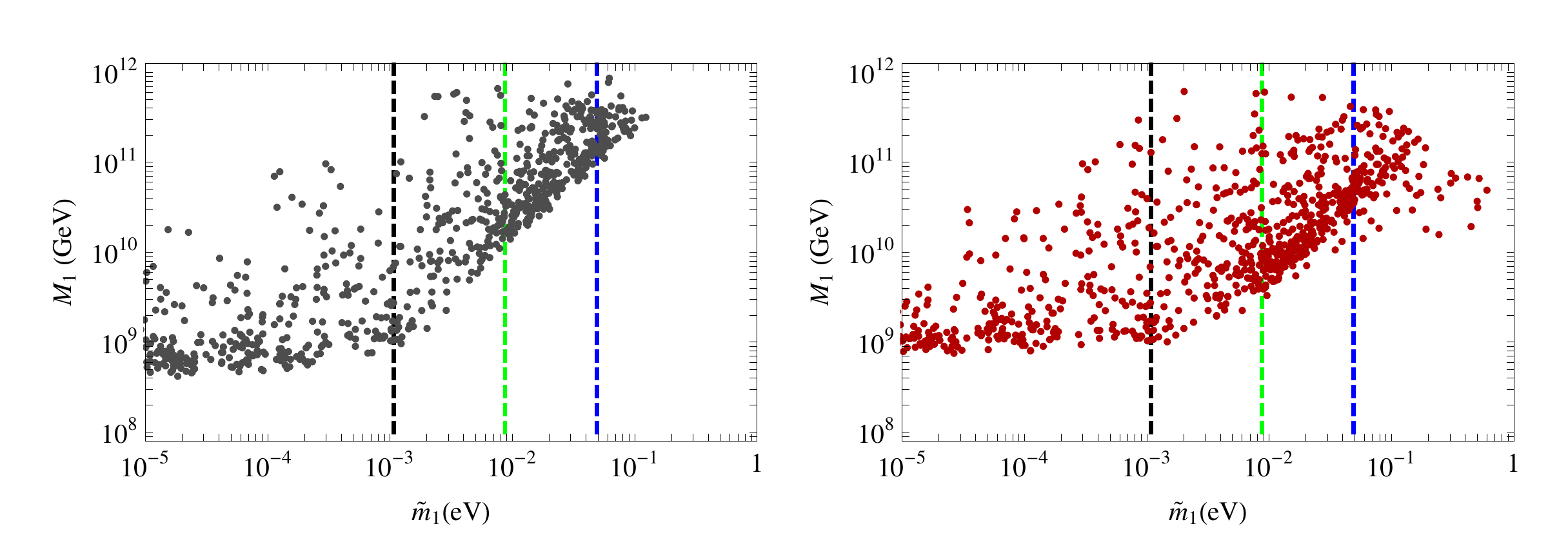}
 \caption{$M_{1}-\tilde{m}_{1}$ parameter-space compatible with successful leptogenesis:  bound on $M_1$,  in the case of a zero (thermal) initial $N_1$ abundance for the up (down) panels. The right (left) panels show the allowed ($M_{1}$-$\tilde{m}_1$) parameter space  in the case the lepton flavours are included (neglected). The vertical lines represent the equilibrium mass $m_{*}\simeq 1.08\,\times 10^{-3} \eV$ (dashed-black),  the solar mass $\sqrt{\Delta m^{2}_{\rm{sol}}}\simeq \sqrt{7.6\,\times 10^{-5}} \eV$ (in green) and the atmospheric mass $\sqrt{\Delta m^{2}_{\rm{atm}}}\simeq \sqrt{2.4\,\times 10^{-3}} \eV$ (in blue).} 
\label{GrapheM1tm1}
\end{center}
\end{figure}

Concerning the lower bound on RHn masses, the inclusion of lepton flavours is a mixed blessing. On the one hand, when light neutrinos are (very) hierarchical, since the upper bound on the $CP$ asymmetry is roughly the same, we do not expect any important modifications to the lower bound on $M_{1}$. Actually, since individual $CP$ asymmetries are roughly $1/3$ of the total one, we even expect in general a slight increase of $M_{1}^{min}$. On the other hand, assuming hierarchical RHns, when light neutrinos are (quasi) degenerate, we see that while $\eps_{1}$ becomes suppressed, the  $CP$ asymmetries in each lepton flavour are in fact enhanced. In this degenerate limit, we expect the lower bound on $M_{1}$ to decrease. Let us see this in detail, for both the thermal and the equilibrium RHn abundance scenarios.\\
From the bound on each individual CP asymmetry~~\cite{issues},
\begin{eqnarray}
\eps_{\alpha}\lesssim \frac{3 M_{1} m_{\rm{max}}}{16 \pi v^{2}} \sqrt{\frac{\kappa_{\alpha}}{K_{1}}}\ ,
\end{eqnarray}
 one has
\begin{eqnarray}
\vert Y_{\cal B} \vert &\simeq & 1.26\times 10^{-3}\sum_{\alpha}\,\eps_{\alpha}\:\eta_{\alpha} \nonumber \\
&\lesssim& 1.26\times10^{-3}\,\frac{3\,M_{N_{1}}\,m_\text{max}}{16 \pi v^{2}}\sum_{\alpha} \sqrt{\frac{\kappa_{\alpha}}{\kappa}}\:\eta_{\alpha}\, ,
\end{eqnarray}
from which a lower bound on $M_{N_1}$ is derived,  
\begin{eqnarray}
\label{boundM1}
%M_{N_1}&\gtrsim &\frac{16 \pi }{3\times 1.26\times10^{-3}}\frac{v^2}{m_\text{max}} \frac{\vert Y_{\cal B} \vert }{\sum_{\alpha} \sqrt{\frac{\kappa_{\alpha}}{\kappa}}\:\eta_{\alpha}} \\
M_{N_1} &\gtrsim & 7.1\times10^{8} \GeV \left( \frac{m_{\rm{atm}}}{m_{\rm{max}}} \right)\,\left| \frac{Y_{\cal B}}{Y_{\cal B}^{obs}} \right|  \frac{1}{\sum_{\alpha} \sqrt{\frac{\kappa_{\alpha}}{\kappa}}\:\eta_{\alpha}}\, .
\end{eqnarray}
Since the lower bound on $M_{1}$ is inversely proportional to the efficiency $\eta_{\alpha}$, it will therefore depend on the thermal history of the decaying right-handed neutrino. In the case where  $N_1$ are produced by scatterings, the efficiency is maximised for a washout $\kappa_{\alpha}\simeq 1$, where $\eta_{\alpha}\simeq 0.2$. In the case where  $N_1$ are non-thermally produced, the efficiency reaches its mamximal value of $1$ for a very weak washout $\kappa_{\alpha} \ll 1$. The lower bound also depends on the alignment of flavours, and in the case of democratic washouts one has~\cite{StudyOf}: 
\begin{eqnarray}
\label{bound}
M_{N_1}\gtrsim \left\lbrace 
\begin{array}{l}
4.1\times10^{8} \GeV\: \ {\rm{ in\: the\: thermal\: case}}\\
2.5\times10^{9} \GeV \:\ {\rm{ in\: the\: dynamical\: case}} \, .
\end{array} \right .
\end{eqnarray}
This bound is close to the one derived in the one-flavour approximation, where $M_{N_1} \gtrsim 4.2\times10^{8} \GeV$ in the thermal case and $M_{N_1} \gtrsim 2.1\times10^{9} \GeV$ in the dynamical one~~\cite{towards}.\\
% Besides flavour effects, the difference between the lower bounds of the flavoured and unflavoured cases comes from a different  factor in the  $B-L\leftrightarrow B$ conversion. Indeed, in the one flavour dominance, the Davidson-Ibarra bound reads~\cite{DI}: 
%\bea
%\eps \leq \frac{3}{16 \pi} \frac{M_{N_{1}} (m_{\rm{max}}-m_{\rm{min}})}{v^{2}}\,\be(\tilde{m}_{1},m_{\rm{min}}) \ ,
%\eea
%and the conversion from sphalerons is $ Y_{B}=28/51 Y_{L} $. Therefore the lower bound on $M_{N_{1}}$ in the unflavoured case is $\sim 12/37\times 51/28 \sim 0.6$ times the bound encountered for flavoured case.\\
The parameter space $M_{1}-\tilde{m}_{1}$ compatible with successful leptogenesis is shown in fig.\ref{GrapheM1tm1}, where we represent the thermal scenario (up panels) and the case where the $N$s are in thermal equilibrium at high temperature (down panels), both in the flavoured case (right panels) and in the single flavour approximation (left panels).
\\
These scatter plots have been obtained by numerically solving the set of BEs, using the Casas-Ibarra parametrisation for the neutrino Yukawa couplings. We choose a normal mass ordering for both low and high energy sectors, setting the solar and atmospheric mass differences, together with the corresponding mixing angles, to their best fit values, while $\theta_{13} \lesssim 13^{\circ}$. For the light neutrino mass, we impose the constraint $m_{1}\lesssim 1 \eV$. We use as perturbative limit $\la \lesssim 1$, hence the upper-bound on $M_{1}$. The hierarchy for heavy neutrinos is imposed $M_{3}=r M_{2}=r^{3} M_{1}$, with $r=10$. The low-energy $CP$ violating phases, as well as those of $R$, are not constrained.\\
We scan over $33000$ random sets, and select the solutions satisfying $8.7\times 10^{-11} \lesssim Y_{B} \lesssim 2\times8.7\times 10^{-11}$.  When scanning over the different parameters, we tried to target the lower part of the curve, with unequal success, hence the lower bound numerically differs from analytical estimates, by a factor $2-3$.\\
We clearly see in these plots the different behaviour regarding the washout.\\
In both scenarios, the transition between strong and weak washout rapidly occurs in the one flavour case at $\tilde{m}_{1}\simeq m_{*}$, while this transition occurs at slightly higher $\tilde{m}_{1}$ in the flavoured case, due to the $A< 1$ matrix elements. Furthermore, we clearly see that washouts are necessary in the thermal scenario, with $\tilde{m}\gtrsim 10^{-4} \eV$ or equivalently $K_{1}\gtrsim 0.1$, while this is not the case in the equilibrium scenario.\\
Moreover, we see the important impact of lepton flavours in the limit of degenerate light neutrinos, with a  re-opening of the parameter space. Accordingly, the lower bound on $M_{1}$ is turned down in this regime, with for example a diminution of one order of magnitude when $\tilde{m}_{1}\simeq 0.1\eV$.
%Successful leptogenesis in the degenerate case require however some fine-tuning of the parameters, as can be seen from the faint point-density in this region. 
\subsubsection{Influence of $\Delta L=2$ and the "full flavour regime"}
An important point to notice is that the above plot have been obtained neglecting $\D L=2$ scatterings, and assuming that the flavours were fully relevant in leptogenesis. These assumptions are justified as long as
\bea
\label{del2}
M_{1}/10^{14}\GeV &\lesssim & c_{1}\, 0.1\,\kappa_{\al} \\
\label{fulfla}
M_{1}/ 10^{12}\GeV &\lesssim & c_{2}\,\kappa_{\al}^{-1} \ ,
\eea
respectively. Here $c_{1,2}$ are numerical factors that are $\mathcal{O}(1)$.\\
If satisfied, eq.(\ref{fulfla}) implies that lepton flavours are fully relevant in leptogenesis as explained in chapter 4, while  eq.(\ref{del2}) tells us that $\D L=2$ lepton violating scatterings can be safely neglected.\\
We wonder about the robustness of the above discussion regarding these two constraints. On the one hand, strengthening the constraint of eq.(\ref{del2}) reduces the $M_{1}-\tilde{m}_{1}$ parameter-space by increasing the minimal  allowed washout, whereas on the other hand strengthening eq.(\ref{fulfla}) lowers the upper bound on $M_{1}$.
\begin{figure}[htb]
\begin{center}
\includegraphics[scale=0.7]{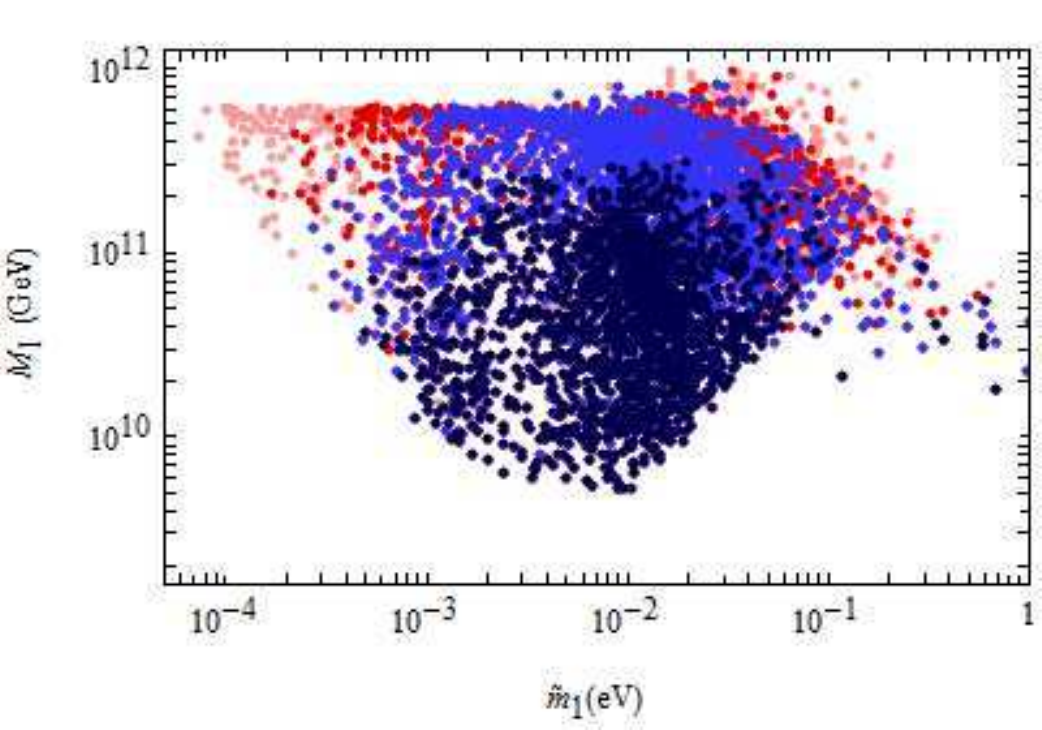} \quad \includegraphics[scale=0.7]{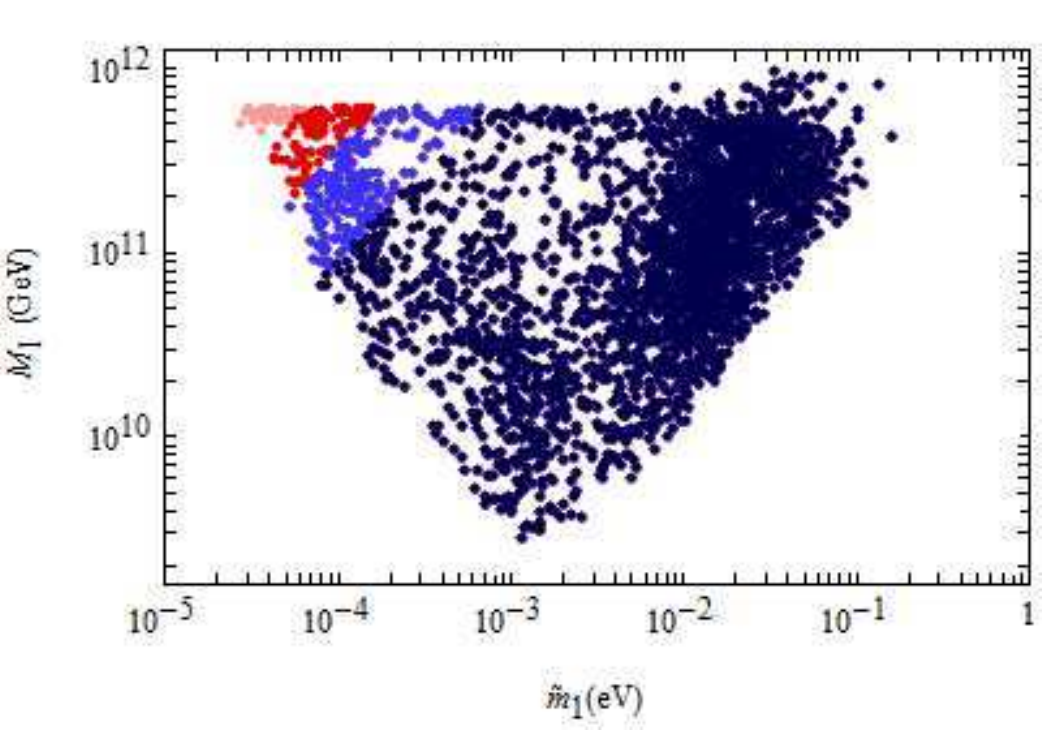}
    \caption{Influence of $\Delta L=2$ and of recohering processes, in the flavoured case (left) and unflavoured case (right). In light red is depicted the case $c=10$, whereas in dark red, light blue and dark blue  are represented the case $c_{1}=c_{2}=c$, with $c=1$, $c=0.6$ and $c=0.1$, respectively.} 
\label{GrapheM1adef}
\end{center}
\end{figure}\\
In order to quantify this point, we illustrate in fig.\ref{GrapheM1adef} the  $M_{1}-\tilde{m}_{1}$ parameter-space which is allowed by leptogenesis when the above constraints are included. These constraints are strengthened/weakened by modifying the factor $c_{1,2}$: we thus vary $c_{1}$ and $c_{2}$ for the flavoured case, and only $c_{1}$ in the single flavour case. For simplicity, we took $c_{1}=c_{2}=c$, with $c=10$ depicted in light red, which would correspond to the case eqs.(\ref{del2}-\ref{fulfla}) are satisfied. In dark red is depicted the case $c=1$, in light blue $c=0.6$ and in dark blue $c=0.1$. We see how the allowed parameter-space dramatically reduces when we strengthen the constraint.\\
Since the plots of figs.\ref{GrapheM1tm1},\ref{GrapheM1m1} have been obtained assuming $c=0.8$, we can say that the conclusion of the previous (and following) section applies.
\subsection{Upper bound on light neutrino mass}
\begin{figure}[htb]
\begin{center}
\includegraphics[scale=0.6]{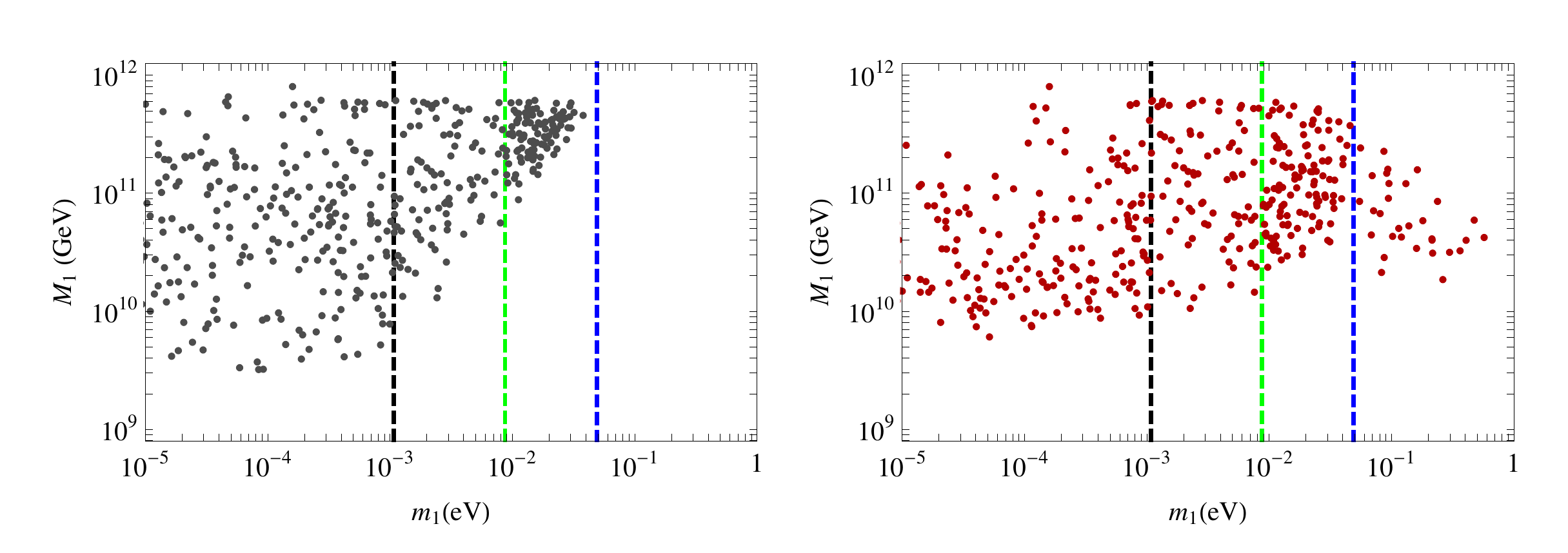} \\
\includegraphics[scale=0.6]{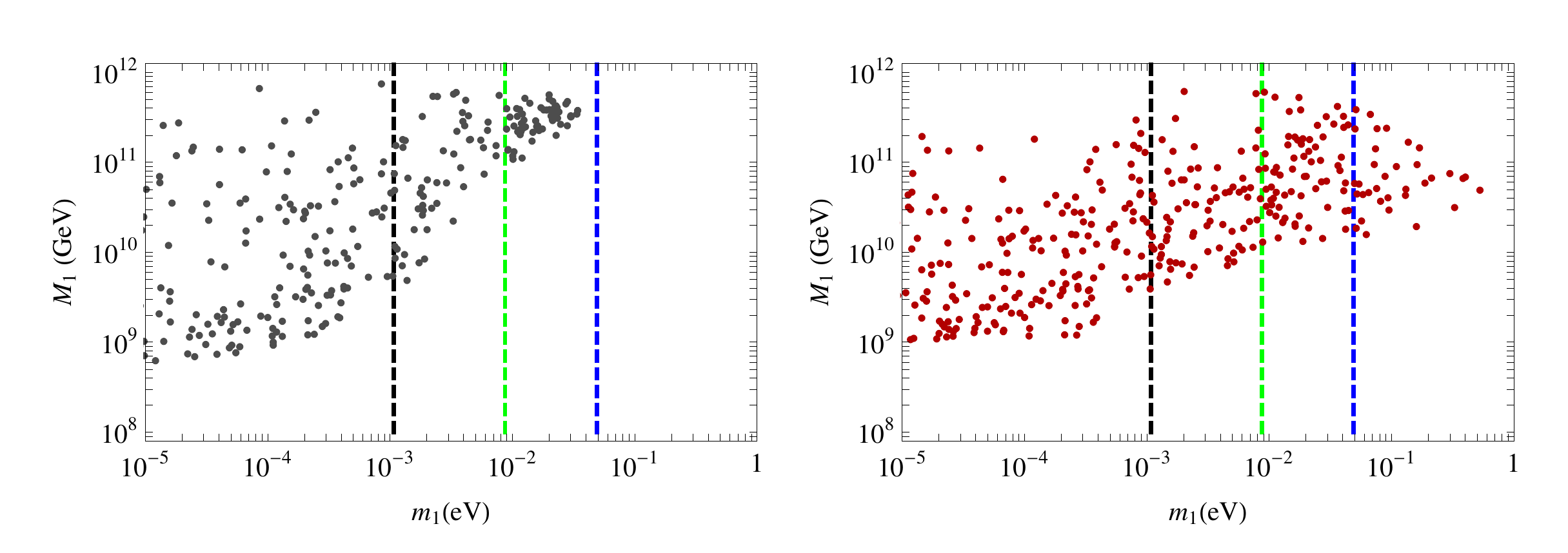} 
    \caption{
Successful leptogenesis:  $M_1$-$m_{1}$ parameter-space, in the case of a vanishing (equilibrium) initial RHn abundance for the up (down) panels. The right (left) panels show the allowed ($M_{1}$-$m_{1}$) parameter space  for the flavoured (unflavoured) case. The vertical lines represent the equilibrium mass $m_{*}\simeq 1.08\,\times 10^{-3} \eV$ (dashed-black),  the solar mass $\sqrt{\Delta m^{2}_{\rm{sol}}}\simeq \sqrt{7.6\,\times 10^{-5}} \eV$ (in green) and the atmospheric mass $\sqrt{\Delta m^{2}_{\rm{atm}}}\simeq \sqrt{2.4\,\times 10^{-3}} \eV$ (in blue).} 
\label{GrapheM1m1}
\end{center}
\end{figure}

The reason why the light neutrino mass is bounded from above in the single flavour approximation  is twofold.
First, in the one flavour approximation, when summing over lepton flavours  and assuming a strong hierarchy for the right-handed neutrinos, the $CP$ asymmetry is bounded by
\bea
\eps_{1}^{max}=\frac{3}{16 \pi}\frac{M_{1}(m_{3}-m_{1})}{v^{2}}\times \beta(m_{1},\tilde{m}_{1})\, .
\eea
Hence, for degenerate neutrinos, $m_{1}\simeq m_{3}$ and the $CP$ asymmetry scales as $m_{1}^{-1}$ and is therefore suppressed. This suppression could be compensated by increasing the heavy neutrino mass scale.\\
However, $M_{1}$ cannot be increased  harmlessly: $\D L=2$ washouts that are $\propto M_{1}$ must be out of equilibrium, which means that  $M_{1}$ is upper-constrained  and so is $m_{min}$.\\
When lepton flavours are included in leptogenesis, the situation is drastically different, since the flavoured $CP$ asymmetries are no longer suppressed. Indeed, the upper-bound on the $CP$ asymmetry in a lepton flavour $\al$ reads
\bea
\label{DIfla}
\eps_{\al}\lesssim \frac{3}{16\,\pi}\frac{M_{1}\,m_{3}}{v^{2}}\times\sqrt\frac{\kappa_{\al}}{K_{1}} \,, 
\eea
no longer suppressed for degenerate light neutrinos. Therefore, the bound
\bea
m_{1}\lesssim 0.12 \eV
\eea
no longer holds. We can see this in the scatter plots of fig.\ref{GrapheM1m1}, which have been obtained in a similar way as the plots of fig.\ref{GrapheM1tm1}. It is clear how the inclusion of lepton flavours re-opens the parameter space in the degenerate/strong-washout regime.
\\
Working  in the flavoured framework, the authors of~\cite{FlavourOsc} have derived an upper-bound on $m_{1}$, under the constraint that flavours are fully relevant in leptogenesis. However this bound roughly reads $\ol{m}\lesssim 2 \eV$, which is already above the cosmological constraint on neutrino mass. 
\subsection{Constraints on mixing angles and $CP$ violating phases}
Constraining the seesaw parameters from leptogenesis is an ambitious task that requires a systematic analysis of the different high energy phases. Such work has been attempted, both in the the single flavour approximation \eg in~\cite{CP1saveur}, as well as in flavoured picture~\cite{Towardsbis,RiottoPetcov,CPsaveur}. Due to the degeneracy among the parameters, even if favourable configurations have been identified, it has not been possible to clearly constraint the parameters.
Hence the necessity to tie the constraints  coming from leptogenesis with other (non-)observations, such as lepton-flavour violating processes~\cite{Tie}. But even then, high energy parameters would remain unconstrained, and only most favourable parameter space regions could be determined.
\section{Summary}
In this chapter, we have studied the effects of lepton flavours in the type I seesaw model of leptogenesis. Actually, the conclusions to which we are led are not particular to the type I seesaw, but rather they are generic conclusions which may apply in all leptogenesis models, under the constraint that at the time the asymmetry is created, the interactions involving the charged lepton Yukawa couplings are in-equilibrium/faster than the coherent processes.\\
However, we made this study in the type I seesaw, and so let us summarise the results we derived in this case.\\
In the flavoured treatment of leptogenesis, the individual asymmetries $Y_{\D\al}$ are independently produced and washed out. For example, the washout factor $\propto \la^{2}$ depends only on the coupling of the lepton flavour to the right-handed neutrino. This coupling being in general different for  the distinct flavours, the latter are differently washed out. This non-alignment of lepton flavours is particularly interesting in the strong washout regime $K_{1}\gg1$, since it is possible to have flavours which are only weakly washed out, allowing to somewhat escape the usual constraint that $K_{1}$ should not be larger than $\sim 15$. A similar effect is found for the $CP$ asymmetries.\\
We showed that in the relevant temperature regime, the $CP$ asymmetries in each lepton flavour have to be considered, instead of the total $CP$ asymmetry (summed over the flavours). This has mainly two consequences. The first one comes from the fact that the $\eps_{\al}$s depend on the low-energy $CP$ violating phases of the PMNS mixing matrix. Therefore, one could ask, relying only on this low-energy phases, whether or not leptogenesis could work. Such a study has been done in~\cite{RiottoPetcov}, where it was shown that while leptogenesis is indeed possible in that case, there is however no clear constraint on the seesaw parameters, owing to the unknown $R$-matrix elements. It has been further shown, in~\cite{Insensivity}, that leptogenesis is actually insensitive to these low-energy phases, in the sense that for any values of the PMNS matrix, one can always find a configuration of the $R$-matrix for which leptogenesis works.\\
The other consequences, which have a direct implication for the low energy phenomenology, stem from the modification of the upper-bound on the $\eps_{\al}$s for hierarchical  heavy neutrinos. Since the $CP$ asymmetries in lepton flavours are no longer suppressed in the limit of (quasi) degenerate light-neutrinos, the upper-bound on this mass scale, which was derived in the single flavour picture, does not hold in the flavoured context.\\
We studied the lower-bound on the lightest right-handed neutrino mass, and showed that the bound derived in the single flavour case is only slightly modified, and still sets the scale of leptogenesis to be $M_{1}\gtrsim (2-3)\,\times 10^{9}\GeV$.\\
Finally, we can say that while the inclusion of lepton flavours somewhat weakens the constraints derived in the single flavour picture, it only significantly  affects leptogenesis in the strong wash-out regime. Furthermore, the inclusion of flavours does not provide any significant constraints on either the high-energy $CP$ violating sector, or the low-energy one. This comes from our complete ignorance of the neutrino Yukawa couplings, which are not constrained in the SM type I seesaw. However,  this is not always the case, as we will see in the next chapter, where we consider a grand-unified framework, where the latter couplings are theoretically predicted.
\newpage
\chapter[Leptogenesis in the type II seesaw]{Leptogenesis in the type II seesaw}
We saw in the previous chapter that leptogenesis in the type I seesaw can explain the observed baryon asymmetry under the main constraint that the lightest right-handed neutrino is heavier than $\sim 4\times 10^{9} \GeV$. However, we also observed that while the low-energy sector is somehow constrained by low-energy observations, we have failed to constrain the high-energy sector. This lack of predictivity originates from our ignorance of the neutrino Yukawa couplings $\la$ and the masses and mixings of right-handed neutrinos. Indeed, in the SM, these masses are bare mass terms, which are thus unconstrained.\\
In $SO(10)$-based grand unified theories, the RHns are no longer gauge singlets, and thus the Majorana mass term can only arise from the dynamical breaking of an underlying symmetry. Moreover, due to the unifying picture, the neutrino Yukawa couplings are related to the Yukawa couplings of other field(s), and therefore one could think of a predictive GUT framework in which to embed the type I seesaw.\\
However in $SO(10)$-GUT scenarios, the typical mass of the lightest RHn predicted by a type I seesaw is $M_{1}\lesssim 10^{7} \GeV$, and so hardly compatible with thermal leptogenesis~\cite{LowM1}.\\
Nevertheless, we saw in chapter 2 that different realisations of the seesaw mechanism are possible. 
%In the type I seesaw, one has 
%\bea
%m_{\nu} \simeq -v^{2}\la M^{-1} \la^{T} \ ,
%\eea
%the smallness of $m_{\nu}$ coming from the high scale of $M$. 
In the (pure) type II seesaw~\cite{typeII}, where one advocates the exchange of scalar $SU(2)_{L}$ triplets, a mass term 
\bea
m_{\nu}\sim v_{L} f_{L}
\eea
is obtained, the smallness of $m_{\nu}$ being explained by the small vev of an $SU(2)_{L}$ triplet,  $v_{L} \ll v$. Such a small vev could appear quite unnatural. On the contrary, for GUT models that are  based on $SU(2)_{L}\times SU(2)_{R}$~\cite{PS,LR}, such a suppression appears very naturally, due to a seesaw-like relation among the vevs of the different Higgs representations that the model contains.\\
In the type II seesaw\footnote{Here we label pure type II seesaw the case where only scalar triplets are added, and type II seesaw the case where fermion singlets and scalar triplets are added.}, light neutrino mass emerges from the neutrino couplings with fermion singlets (the RHn) and $SU(2)_{L}$ triplets. Since these models contain two sources of lepton number violation, it is interesting to look at the feasibility of leptogenesis.\\
In this chapter, we investigate the viability of leptogenesis in a supersymmetric GUT model, based on a Left-Right symmetric type II seesaw model. This study is based on \cite{AHJML}, where we complete the previous study of \cite{HLS} by the inclusion of lepton flavours and heavier neutrino effects. We first introduce the GUT framework in which this study is conducted, untill the type II seesaw relation is obtained. Inverting this relation will allow us in a second time to infer the high-energy parameters (right-handed neutrino masses and couplings...), using a reconstruction procedure developped in \cite{HLS}. The knowledge of neutrino Yukawa couplings and RHn masses enables us to discuss the thermal scenario of leptogenesis in this SUSY-GUT framework. This scenario has \textit{a priori} important differences with respect to the type I scenario discussed in the previous chapter, mainly coming from the gauging of $SU(2)_{R}$, the inclusion of a scalar triplet and the fact that we now work in a supersymmetric extension of the SM, with its inherent problems. We finally conduct a numerical study of leptogenesis in this framework, discussing in particular the influence of the key parameters of our model, and if the requirement of a viable leptogenesis yields constraints for the parameter-space.
\section{Framework}
This supersymmetric GUT model is based on $G_{3221}=SU(3)\times SU(2)_{L}\times SU(2)_{R}\times U(1)_{B-L}$. Our model consists of three $\bf{16}$ representation, which account for all SM fermions plus 3 right-handed neutrinos. This $\bf{16}$, which is spinorial, is the smallest complex representation of $SO(10)$.\\
Given the product decomposition under $SO(10)$, 
\bea
\bf{16}\times\bf{16}=\bf{10}_{s}+\bf{120}_{a}+\bf{126}_{s} \, ,
\eea
fermion bilinears can be formed by coupling this product to the appropriate Higgs representations, either the symmetric $\bf{10}_{s}$ and $\bf{126}_{s}$ or the antisymmetric $\bf{120}_{a}$. Another solution  consists in using non-renormalisable operators, which couple fermion bilinears to products of Higgses, whose decompositions contain one of the representations above.\\
We consider here only symmetric representations, which consist in two fundamental $\bf{10}$'s and a rank 5 (complex anti-self dual) $\bf{\ol{126}}$. The inclusion of a (self-dual) $\bf{126}$ is then necessary in order to prevent supersymmetry breaking at the GUT scale by a non-vanishing $D$-term. Therefore, the superpotential contains the following Yukawa couplings:
\bea
\label{Yuk}
W_{Y}=Y_{ij}^{u}\,\tbf{16}_{i}\,\tbf{16}_{j}\,\tbf{10}_{u}+Y_{ij}^{d}\,\tbf{16}_{i}\,\tbf{16}_{j}\,\tbf{10}_{d}+f_{ij}\,\tbf{16}_{i}\,\tbf{16}_{j}\,\tbf{\ol{126}} \ .
\eea
Due to the symmetry of the chosen Higgs representation, the couplings $Y^{a,b}$ and $f$ are complex symmetric matrices. Assuming that only the $SU(2)_{L}$ doublets of the $\bf{10}$s acquire a vev, the relation above tells us that at the GUT scale
\bea
m_{u}=m_{D}\, , \quad m_{d}=m_{e} \ .
\eea
The GUT framework therefore constrains the neutrino Dirac type mass $m_{D}$ to be equal to the up quark type mass $m_{u}$. On the other hand, the relation $m_{d}=m_{e}$ among down-quark and charged lepton masses is clearly not satisfactory~\cite{MeMd}, and we will correct it later on.\\
Under the Pati-Salam group $SU(4)\times SU(2)_{L}\times SU(2)_{R}$~\cite{PS}, the $\bf{\ol{126}}$ decomposes as follows~\footnote{For the different decomposition, see, for example, ref.~\cite{SO10modelbuilding}.}:
\bea
\bf{\ol{126}}=(\bf{6},\bf{1},\bf{1})+(\bf{\ol{10}},\bf{3},\bf{1})+(\bf{10},\bf{1},\bf{3})+(\bf{15},\bf{2},\bf{2}) \, .
\eea
%If we suppose that the $(\bf{10},\bf{1},\bf{3})$ takes its vev along the SM gauge singlet direction, the last term of eq.(\ref{Yuk}) gives us the Majorana mass of RHn of the pure type I seesaw.\\
In the type II seesaw, we assume that the two scalar triplets of $\bf{\ol{126}}$, $\D _{L}=(\bf{1},\bf{3},\bf{1},\bf{1}) $ and $\D _{R}=(\bf{1},\bf{1},\bf{3},-\bf{1}) $ under $G_{3221}$, respectively couple to light and heavy neutrinos, according to:
\bea
{\mathcal{L}}_{m}=\frac{1}{2}f_{L}^{ij}\,\D_{L}\,L_{i}\,L_{j}+\frac{1}{2}\,f_{R}^{ij}\,\D_{R} N^{c}_{i}\,N^{c}_{j}+Y_{ij}^{u}L_{i}H_{u}\,N^{c}_{j}\, .
\eea
A non-zero vev of the $SU(2)_{R}$ triplet then implies a Majorana mass for right-handed neutrinos, whereas a vev for $SU(2)_{L}$ triplet implies a Majorana mass for the left-handed neutrinos.\\
We suppose here that the breaking of $G_{3221}$ occurs due to $\D_{R}$, whose vev $v_{R}$ lies at an intermediate scale between the GUT and the electroweak scales. This breaking leads to a Majorana mass term of RHns, which violates $B-L$ by two units.\\
%\bea
%\mathcal{L}_{\sla{L}}\supset -\frac{1}{2}m_{\nu\,ij}^{II} \nu_{L\,i}^{T}\,C\,\nu_{L\,j}-\frac{1}{2}\ol{N}_{i}\,M_{i}\,N^{c}_{i} \ .
%\eea
After electroweak symmetry breaking, the light neutrino masses receive two contributions, one coming from the type I seesaw~\cite{Seesaw}:
\bea
m_{\nu i j}^{I}=-v_{u}^{2}\la_{ik} M_{k}^{-1} \la_{kj}^{T}=-\frac{v_{u}^{2}}{v_{R}}\,\la_{ik}\,f_{R\,k}^{-1}\,\la_{jk} \, ,
\eea
where $v_{u}$ is the vev of the up-type Higgs doublet, $v_{u}=v\,\rm{sin}(\be)$, and one coming from the pure type II seesaw~\cite{typeII}:
\bea
m_{\nu\,ij}^{II}=v_{L}\,f_{L}^{ij} \, .
\eea
The smallness of the type I mass is obvious, since in this model the breaking of $B-L$ occurs far above the electroweak scale, $v_{R} \gg v_{u}$. Conversely, the smallness of pure type II contribution comes from the suppression of the $SU(2)_{L}$ triplet vev.\\
For instance, this seesaw-like relation for $v_{L}$ can be obtained through the addition of a $\bf{54}$ dimensional Higgs that couples both to the $\bf{10}$s and to the $\bf{\ol{126}}$. The Higgs sector of the superpotential contains
\bea
W_{H}\supset \rho_{1} \tbf{54}\,\tbf{\ol{126}}\,\tbf{\ol{126}}+\rho_{2}\tbf{10}\,\tbf{10}\,\tbf{54} + M_{126}\tbf{\ol{126}}\,\tbf{126} +...,
\eea
where $\rho_{1,2}$ are model-dependent couplings, the knowledge of which is beyond the scope of this chapter. Since under $G_{3221}$, the $\bf{54}$ contains a bi-triplet $\tbf{54}\supset \D_{54}=(\tbf{1},\tbf{3},\tbf{3},\tbf{0})$, it can couple to the $SU(2)_{L,R}$ triplets of $\bf{\ol{126}}$ and to the the product of two) Higgs bidoublets $\bf{10}\supset \Phi=(\bf{1},\bf{2},\bf{2},0)$, $\bf{10}\times\bf{10} \supset (\bf{1},\bf{3},\bf{3},0)$, according to
\bea
W_{H}\supset \rho_{1}\,\D_{54}\,\D_{R}\,\D_{L} +\rho_{2} \Phi\,\Phi\,\D_{54}\ +M_{\D}\D_{L}\D_{R}.
\eea
Minimising the scalar potential with respect to $v_{L}$, one obtains the seesaw relation for the $SU(2)_{L}$ triplet vev:
\bea
v_{L}\simeq -\frac{\rho_{1}\rho_{2} v^{2} v_{R}}{M_{\D}^{2}} \ll v \, ,
\eea
which ensures that the (pure) type II contribution to light neutrino mass is indeed suppressed. In general, the superpotential contains many other terms (as indicated by the dots), which can contribute to the above relation. Therefore, in general $v_{L}$ depends on many other couplings. Nevertheless, we always have
\bea
v_{L}=\frac{v_{u}^{2}}{M_{\D_{L}}}\, ,
\eea
$M_{\D_{L}}$ being defined as a function of the relevant couplings.\\
Moreover, we assume here that $G_{3221}$ is Left-Right (LR) symmetric~\cite{LR}, so that the Lagrangian invariant under $G_{3221}$ is also invariant under the $SU(2)_{L}\LRa SU(2)_{R}$ exchange
\bea
\left(
\begin{array}{c}
\nu_{L} \\ e_{L}
\end{array} \right) \LRa \left(
\begin{array}{c}
N^{c} \\ e^{c}
\end{array} \right)\,,\quad \Phi \LRa \Phi^{T} \,,\quad \D_{L}\LRa \D_{R} \, .
\eea
This LR symmetry enforces $f_{L}=f_{R}\equiv f$, leading to two important consequences:
\begin{itemize}
\item Firstly, the seesaw relation now reads~\cite{typeII}
\bea
\label{SSII}
m_{\nu}=\frac{v_{u}^{2}}{M_{\D}}\,f-\frac{v_{L}^{2}}{v_{R}} \la\,f^{-1}\,\la \ .
\eea
Given the low-energy constraints on $m_{\nu}$, and the GUT relation between the neutrino and the up-quark Yukawa couplings, we can extract $f$.
\item Secondly, the right-handed neutrino masses simply read
\bea
M=f\,v_{R}\, ,
\eea
and so the knowledge of $f$ from the seesaw relation in turn predicts the RHn spectrum, a clear gain when compared to the non-GUT seesaw, where the RHn masses are put by hand.
\end{itemize}
\section{Inferring the unknown parameters}
\subsection{Inverting the seesaw}
Starting from the seesaw relation of eq.(\ref{SSII}), the authors of~\cite{HLS,Akhmedov:2005np} have developed a procedure which allows to  determine the properties of the complex symmetric matrix $f$, assuming the knowledge of $m_{\nu}$ and $\la$. Here we summarise their results, which are the building-blocks of our study~\cite{AHJML}.
Given the symmetry of $\la$, it is possible to find $\la^{1/2}$ such that $\la=\la^{1/2}.\la^{1/2\,T}$. Hence eq.(\ref{SSII}) can be written
\bea
\label{eqrec}
Y=\al\,X+\be\,X^{-1}
\eea
where
\bea
Y=\la^{-1/2}.\,m_{\nu}.\,\left(\la^{-1/2}\right)^{T}\;,\; X=\la^{-1/2}.\,f.\,\left(\la^{-1/2}\right)^{T}\;,
\eea
and
\bea
\al =v_{L}=\frac{v_{u}^{2}}{M_{\D}}\;,\; \be= \frac{v_{u}^{2}}{v_{R}} \, .
\eea
The complex symmetric matrix $Y$ is then diagonalised by an orthogonal matrix $O_{Y}$, which also diagonalises  $X$:
\bea
\rm{diag}(y_{1},y_{2},y_{3})&=&O_{Y}^{T}.Y.O_{Y} \nonumber \\
\rm{diag}(x_{1},x_{2},x_{3})&=&O_{Y}^{T}.X.O_{Y} \, .
\eea
The equation (\ref{eqrec}) then leads to a second degree equation for the eigenvalues $y_{i}=\al x_{i}+\be x_{i}^{-1}$, which is easily solved:
\bea
x_{i}^{\pm}=\frac{1}{2\al}\left(y_{i}\pm\rm{sign}(\rm{Re}(y_{i}))\sqrt{y_{i}^{2}+4\,\al\,\be}\right) \ .
\eea
The couplings $f$ are then given by:
\bea
f=\la^{1/2}.X.\left(\la^{-1/2}\right)^{T}=\la^{1/2}.O_{Y}.\rm{diag}(x_{1},x_{2},x_{3}).O_{M}^{T}.\left(\la^{-1/2}\right)^{T} \, .
\eea
Finally, the masses of the RHns are obtained after the diagonalisation of $f$, through an unitary matrix $U_{f}$
\bea
M_{i}=f_{i}\,v_{R}\;,\;f_{i}=\left(U_{f}^{\dagger}.f.U_{f}^{*}\right)_{ii} \ ,
\eea
where the $f_{i}$s are  positive, and by convention $f_{1}\leq f_{2}\leq f_{3}$. The unitary matrix $U_{f}$ relates the basis where the right-handed neutrino mass matrix is diagonal to the basis where the coupling $\la$ is symmetric. We place ourselves in the former basis, which implies a redefinition of $\la^{new}=U_{F}^{\dagger} \la^{old}$.\\
Since there are two possible choices for each eigenvalue $x_{i}$, there are $8$ different solutions for the matrix $f$, that is for the right-handed neutrino spectrum. These $8$ solutions ($2^{n}$ for $n$ generations) constitute the "eight-fold" ambiguity \cite{HLS} of the LR symmetric seesaw mechanism: for one low-energy spectrum, we have eight possible solutions, which can be distinguished through their high-energy effects.
\subsection*{Behaviour of the different solutions}
In the above relations, the contributions from the different seesaws are clear: $\be$ stands for the type I contribution whereas $\al$ denotes the pure type II seesaw.\\
It is instructive to consider the asymptotic value  $4 \al\,\be \ll \vert y_{i}\vert^{2}$, since in this case the $\pm$ solutions exhibit the different seesaw contributions:
\bea
x_{i}^{+}\simeq \frac{y_{i}}{\al}\;,\;x_{i}^{-}\simeq -\frac{\be}{y_{i}} \, .
\eea
The $"+"$ solution corresponds to the dominance of the pure type II  while $"-"$ solutions correspond to a dominance of the type I.\\
The different solutions for $f$ are labelled according to the constitutive $x_{i}^{\pm}$: for example $(+,+,-)$ refers to the solution built from $(x_{1}^{+},x_{2}^{+},x_{3}^{-})$. For the $f_{i}$s, in general one cannot clearly assess which seesaw is the dominant one, apart from the $f^{(+,+,+)}$ which clearly tends, in the $4 \al\,\be \ll \vert y_{i}\vert^{2}$ limit, towards a pure type II solution. Similarly, in this limit the $f^{(-,-,-)}$ solution approaches the type I case.\\
However, we notice that from the asymptotic behaviour of the $x_{i}$s we can infer the evolution of right-handed neutrino masses as a function of the $B-L$ breaking scale $v_{R}$.\\
Indeed, for the $"+"$ solution, as $x_{i}^{+}\simeq y_{i}/\al \propto v_{R}$, one expects at high $v_{R}$ that the RHn mass  will continuously increase with $v_{R}$. For the $"-"$ solutions, since $x_{i}^{-} \simeq -\be /y_{i} \propto 1/v_{R}$, $v_{R}\,f_{i}\sim$ constant, and one expects that at high $B-L$ breaking scale the $"-"$ RHn masses will also become constant. The value of this constant is of major importance for leptogenesis.\\
In the opposite asymptotic limit, $4 \al\,\be \gg \vert y_{i}\vert^{2}$, we have
\bea
x_{i}^{\pm}\simeq \pm \rm{sign}(\rm{Re}(y_{i}))\sqrt{\be/\al}
\eea
which shows that the type I and the pure type II somehow cancel each other in $m_{\nu}$. Indeed, in this case, one roughly has $f\sim \la\,\sqrt{\be/\al}$ and 
\bea
m_{\nu}^{II}\simeq \sqrt{\al \be} \la \sim m_{\nu}^{I} \ .
\eea
In the intermediate regime, type I and II contributions are of the same order.\\
We are interested in the dependency of the different parameters on the $B-L$ breaking scale. Therefore, in the following, we fix the ratio 
\bea
\frac{\be}{\al}=\frac{v_{u}^{2}}{v_{R}\,v_{L}}=\frac{M_{\D}}{v_{R}}\simeq 0.1 \ .
\eea
In fact, depending on the different couplings, $\be/\al (M_{\D})$ could be either larger or smaller than 1$(v_{R})$. Its value roughly sets the relative importance of the different seesaw contributions to the light neutrino masses. Modifying this value only accounts for a translation of the solution on the $v_{R}$ axis. We choose $\be/\al=0.1$ due to numerical instabilities encountered with $\be/\al=1$ in the high $v_{R}$ regime.
\subsection*{Fixing parameters}
In the type I seesaw, we saw that there are 18 parameters among which only 4 are known. Let us see how many parameters we have in our type II framework.\\
We place ourselves in the basis where charged lepton (and down-type quark) and RHn masses ($f$) are diagonal. Then, using the GUT relation $m_{D}=m_{u}$, we have
\bea
\la & =& U_{q}^{T}.\rm{diag}(y_{u},y_{c},y_{t}).U_{q} \\
m_{\nu} &=& U_{\ell}^{*}.\rm{diag}(m_{1},m_{2},m_{3}).U_{\ell}^{\dagger} \, ,
\eea 
where $y_{q}$~\footnote{In general, the Yukawa couplings we use are linear combinations of those of eq.(\ref{Yuk}), and their precise relation is beyond the scope of this chapter.} is the Yukawa coupling for the $q$-quark, and $m_{i}$ are light neutrino masses. The unitary matrices $U_{q}$ and $U_{\ell}$ are related to the CKM~\cite{CKM} and PMNS~\cite{PMNS} mixing matrices. They respectively read:
\bea
U_{q}& =& P_{u}.V_{CKM}.P_{d}\,, \\
U_{\ell}&=& P_{e}.U_{PMNS}.P_{\nu} \, .
\eea
At the GUT scale, since quarks and leptons belong to the same $\bf{16}$ representation, they cannot be independently rotated, hence the matrices $P_{k}$ ($k=u,\nu,e,d$).\\
However, at low energy  $P_{e,d,u}$ can be rotated away, while the matrix $P_{\nu}$ contains the usual Majorana $CP$ violating phases. All these matrices contain 3 phases, among which two global phases can be re-absorbed. We parametrise these matrices as follows:
\bea
P_{k}&=&\rm{diag}(e^{\rm{i} \phi_{1}^{k}},e^{\rm{i} \phi_{2}^{k}},e^{\rm{i} \phi_{3}^{k}})\;,k=u,\nu \\
P_{l}&=&\rm{diag}(e^{\rm{i} \phi_{1}^{l}},e^{\rm{i} \phi_{2}^{l}},1)\;,k=e,d \;. 
\eea
Hence, at the GUT scale, we have 7 additional $CP$ violating phases when compared to the type I seesaw. This is a welcome feature regarding leptogenesis, even if these extra phases induce a loss of the predictivity for our model.\\
The input values for the quark masses and for the CKM matrix at $M_Z$ have been taken
from Refs.~\cite{Jamin06} and~\cite{PDG}, respectively, and subsequently
evolved to the GUT scale using the Mathematica package REAP~\cite{REAP}
with an effective SUSY threshold $M_{SUSY} = 1$ TeV and $\tan \beta = 10$.
%yielding:
%
%\bea
%  m_u\, (M_{GUT})\ =\ 0.70\, \mbox{MeV}\, , \quad
%  m_c\, (M_{GUT})\ =\ 250\, \mbox{MeV}\, , \quad
%  m_t\, (M_{GUT})\ =\ 95\, \mbox{GeV}\, .
%&  m_u = 0.7\, \mbox{MeV}\, , \quad
%  m_c = 250\, \mbox{MeV}\, , \quad
%  m_t = 95\, \mbox{GeV}\, ,  \\
%&  V_{us} = 0.227\, ,  \quad  |V_{ub}| = 0.0037\, ,  \quad
%  V_{cb} = 0.040\, ,  \quad  \delta = 56^\circ\, .
% COMMENT (SL): for CKM, I used the values given by Ross and Serna
% for tan beta = 10. It remains to be checked whether they agree with REAP.
%\eea
%
%The CKM matrix was taken from the fit of Ref.~~\cite{CKMfitter}, and
%the renormalisation group effects were taken into account by
%using $A (M_{GUT}) = 0.7$.
The light neutrino mass spectrum has been assumed to be hierarchical, with $m_1 = 10^{-3}$ eV, and the
oscillation parameters have been set to the best fit values of~\cite{neutrinos_fit}, while keeping $\theta_{13}$ only upper-constrained $\theta_{13}\lesssim 13^{\circ}$.
%for the oscillation parameters (the parametrisation of the PMNS matrix
Renormalisation group effects induce a multiplicative factor for the different Yukawa couplings: for the first two generations of quarks, renormalisation induces a factor $\sim 0.41$, the top quark Yukawa factor being $\sim 0.56$, while the neutrino sector is multiplied by a factor $\sim 1.08$.\\
All in all, we are left with 17 "free" parameters: the light neutrino mass scale, the ordering of the neutrino spectrum and the mixing angle $\theta_{13}$; we also have 10 free $CP$ phases. The $B-L$ breaking scale can vary from $\sim 10^{12} \GeV$ to $10^{16} \GeV$; however perturbativity of the couplings $f$ restricts this range: for the value $\al /\be=0.1$ that we choose, from  $\sim 10^{12}\GeV$ to $\sim 5\times10^{14} \GeV$. Finally, there is another constraint coming from the reheating temperature, which we will discuss later and which gives us another parameter, $T_{RH}$ which is allowed to lie between $10^{9}\GeV$ and $10^{11} \GeV$. Working in a supersymmetric framework, we have to fix the ratio of the low-energy Higgs doublet vev $v_{u}/v_{d}=\tan(\be)$. As already said, we choose here $\tan(\be)=10$ as a representative value.\\
Considering the large number of parameters, and our limited computation ability, it is not feasible to scan over the full parameter range, but we instead fixed some of the parameters. We list in the following table the usual values for the inputs that we used, distinguishing the parameters which we usually vary from those whose value is fixed (NMO stands for neutrino mass ordering, which is chosen normal ordered).
\bea
\begin{array}{|c|c|c|c|c|c|c|}
\hline
v_{R} (\GeV) &  T_{RH} (\GeV) & m_{\nu 1} (\eV)&  \theta_{13}& \phi_{2}^{u,\nu} \\ 
\hline 
10^{12}-5\times 10^{14}  & 5\times 10^{9}-10^{11} \GeV & 10^{-5}-10^{-1} & 0-13^{\circ} & 0,\pi/4    \\ 
\hline
\end{array} 
\eea
\bea
\begin{array}{|c|c|c|c|c|c|c|c|c|c|}
\hline
 \rm{NMO} &\be/\al  & \rm{tan}(\be) & \phi_{1,3}^{u,\nu} &\phi_{1}^{e,d} &\phi_{2}^{e,d}  \\ 
\hline 
\tbf{NH} & 0.1 & 10 & 0 & 0 & 0 \\ 
\hline
\end{array} 
\eea
\subsection*{Correcting the GUT relation $m_{e}=m_{d}$}
Since the only source for the Dirac mass term is the $SU(2)_{L}$ doublet of the two $\bf{10}$ Higgses, we have  $m_{u}=m_{D}$ and $m_{e}=m_{d}$. The latter relation is, strictly speaking, not a problem since the Dirac-type mass of light neutrinos is unknown. However, the relation $m_{e}=m_{d}$ is clearly wrong, and must be improved. \\
A usual correction is to assume that the $SU(2)_{L}$ doublets of the $\bf{\ol{126}}$ also acquire a vev at the electroweak scale. Doing this, one thus obtains~\cite{MeMd}: 
\bea
m_{d}&=&v_{10}^{d}\,\la^{d}+v_{126}^{d}\,f\,, \nonumber \\
m_{e}&=& v_{10}^{d}\,\la^{d}-3\,v_{126}^{d}\,f\,, \nonumber \\
m_{u}&=&v_{10}^{u}\,\la^{u}+v_{126}^{u}\,f\,, \nonumber \\
m_{\nu}&=& v_{10}^{u}\,\la^{u}-3\,v_{126}^{u}\,f\,,
\eea
such that the GUT relations for the masses are
\bea
m_{D}&=&m_{u}-4\,v_{126}^{u}\,f\,, \nonumber \\
m_{e}&=&m_{d}-4\,v_{126}^{d}\,f\,.
\eea
Inserting the first relation in the seesaw formula, one could in principle invert it in the same way as previously explained. However, the fact that $m_{e}\propto f$ renders the extraction of $f$ a really hard task.\\
We could assume that an anti-symmetric $\bf{120}$ does the job, but our model does not contain such a representation. The remaining possibility is to invoke adjoint $\bf{45}$ representations which couple to the $\bf{10}$ and create an effective $\bf{120}$, as can be seen from the following decomposition:
\bea
\bf{10}\times\bf{45}=\bf{10}\times\bf{120}\times\bf{320}\, .
\eea
Neglecting the higher dimensional representation, we see that both the $\bf{10}$ and the $\bf{120}$ can contribute to the fermion masses. However, the $\bf{10}$  equally contributes to leptons and quarks, and so we  neglect it (should it take a vev, a redefinition of $\la^{d}$ would absorb it). Therefore, we need the $\bf{120}$ to acquire a vev. Among the different possibilities, the only products which leave $G_{SM}$ invariant at the GUT scale after the $\bf{45}$ develops a vev are
\bea
\mathcal{L}&\supset&\frac{Y_{1}}{\Lambda}\,(\tbf{16}\times \tbf{16})_{\vert (\tbf{1},\tbf{2},\tbf{2})}\times\left[(\tbf{1},\tbf{2},\tbf{2})_{10}\times (\tbf{1},\tbf{1},\tbf{3})_{45}\right]\nonumber \\
&+&\frac{Y_{2}}{\Lambda}\,(\tbf{16}\times \tbf{16})_{\vert (\tbf{15},\tbf{2},\tbf{2})}\times\left[(\tbf{1},\tbf{2},\tbf{2})_{10}\times (\tbf{15},\tbf{1},\tbf{1})_{45}\right] \, ,
\eea
where we wrote the decomposition under the Pati-Salam  group $SU(4)\times SU(2)_{L}\times SU(2)_{R}$.
We see that the two terms contribute differently: the first one is $\propto (\tbf{1},\tbf{1},\tbf{3})_{45}$, and so a vev in this direction does not distinguish  quarks from leptons. On the contrary, the second term, which is $\propto (\tbf{15},\tbf{1},\tbf{1})_{45}$, clearly does.\\
Considering that only the latter term acquires a vev, at the GUT scale the relation $m_{e}=m_{d}$ can be corrected as:
\bea
m_{e}&=&v_{d}\left(\la_{d}-\frac{3 \hat{v_{15}}}{\Lambda} Y_{2}\right)\,, \\
m_{d}&=&v_{d}\left(\la_{d}+\frac{ \hat{v_{15}}}{\Lambda} Y_{2}\right)\,,
\eea
where $\hat{v_{15}}\,v_{d}=\left\langle (\tbf{15},\tbf{1},\tbf{1})_{45} \right\rangle$. 
Once corrected, the masses of charged leptons and down type quarks at the GUT scale are found to be:
\begin{equation}
\begin{array}{ccc}
m_d=0.94 \mbox{ MeV} & m_s=17 \mbox{ MeV} & m_b=0.98 \mbox{ GeV}\ ,  \\
m_e=0.346 \mbox{ MeV} & m_\mu =73.0 \mbox{ MeV} & m_\tau =1.25 \mbox{ GeV}\ .
\end{array}
\end{equation}
However, for a consistent correction of the relation $m_{e}=m_{d}$, since we only include one $\tbf{45}$, we should obtain that $m_{\tau}^2-m_{b}^2\sim m_{\mu}^2$, which is not the case with the values above.\\
Nevertheless, supersymmetric threshold effects arising from the SUSY breaking sector (see, \eg~\cite{threshold} and references therein) allow to solve this problem. Indeed, at the tree level, the down quarks couple only to one Higgs doublet, $H_{d}$, which is not the case for their scalar partner, given that $W \supset\mu\,H_{u}.H_{d}+Y_{d}\,Q.H_{d}\,D^{c}$. Once SUSY breaking terms are included, this will provide a correction to the down quark masses, according to~\cite{threshold}:
\bea
m_{b}\Ra m_{b}\left(1+\delta_{m_{b}}\rm{tan}(\be)\right) \, ,
\eea
where 
\bea
\delta_{m}\simeq \frac{2\al_{s}}{3\pi}\,\frac{\mu\,M_{3}}{m_{\tilde{b_{R}}}^{2}}\, ,
\eea
which is evaluated to be $\delta_{m} \lesssim 2\%$. Once these corrections are added, the bottom quark mass is $m_b(M_{GUT})\simeq 1.17 \GeV$, which is just enough to fit $m_{e}$ and $m_{d}$ with the procedure discussed before.\\
The two mass matrices differ $m_{e}\neq m_{d}$, and so we introduce a unitary matrix $U_{m}$, the matrix rotating from the basis where $m_{e}$ is diagonal to the basis where $m_{d}$ is diagonal:
\bea
m_{e}=\hat{m_{e}}\Ra m_{d}=U_{m}^{T}.\hat{m_{d}}.U_{m}\, .
\eea
The precise expression of $U_{m}$ is found after fitting the charged lepton and down quark masses, and is given by

\begin{equation}
U_m = e^{i\phi_g^m}\left(
\begin{array}{ccc}
e^{i\phi_1^m} & 0 & 0\\
0 & e^{i\phi_2^m} & 0\\
0 & 0 & 1
\end{array}
\right) V(\theta_{12}^m,\theta_{13}^m,\theta_{23}^m,\delta^m) \left(
\begin{array}{ccc}
e^{i\phi_3^m} & 0 & 0\\
0 & e^{i\phi_4^m} & 0\\
0 & 0 & 1
\end{array}
\right) \ .
\end{equation}
$V$ is a CKM-like mixing matrix, with three real angles and one complex phase. The different values we will use for the phases of $U_{m}$ are listed below, with the corresponding non-zero phases of the matrices $P_{k}$.
\begin{center}
\begin{tabular}{|c|c|c|c|c|c|c|c|c|c|c|}
\hline Sets & $\theta_{12}^m$ & $\theta_{13}^m$ & $\theta_{23}$ & $\delta^m$ & $\phi_g^m$ & $\phi_1^m$ & $\phi_2^m$ & $\phi_3^m$ & $\phi_4^m$ & $\pi/4$ \\
\hline 1 & 1.07 & 0.22 & 0.21 & 5.80 & 3.21 & 4.37 & 5.86 & 0.87 & 6.16 & $\phi_2^u$ \\
\hline 2 & 1.07 & 0.22 & 0.21 & 5.80 & 3.21 & 4.37 & 5.86 & 0.87 & 6.16 & $\phi_2^\nu$ \\
\hline 3 & 0.17 & 0.066 & 0.29 & 0.23 & 3.14 & 0.54 & 0.015 & 6.27 & 0.0032 & $\phi^u_2$ \\
\hline 4 & 0.28 & 0.089 & 0.37 & 0.062 & 3.15 & 3.12 & 6.03 & 2.94 & 6.19 & $\phi_2^u$ \\
\hline
\end{tabular}
\end{center}
\subsection*{Perturbativity of the couplings}
\begin{figure}[!h]
\begin{center}
\includegraphics[scale=0.5]{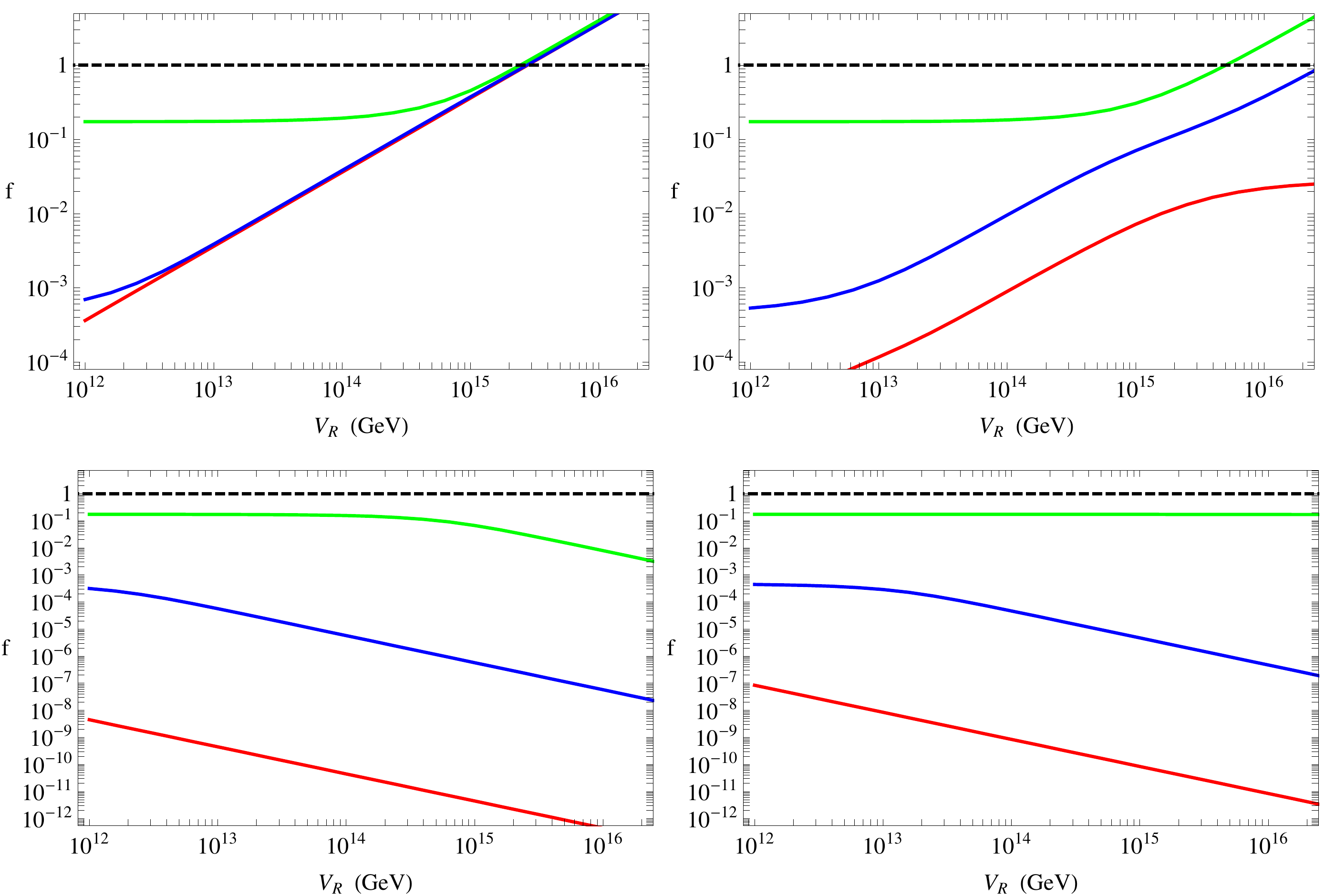} 
 \caption{Evolution of the coupling $f$ as a function of the $B-L$ breaking scale $v_{R}$ for the case $(+,+,+)$ (up panels) and $(-,-,-)$ (down panels), for different light-neutrino masses $m_{1}$, with $m_{1}=0.1 \eV, $ and $10^{-5} \eV$ for the left and right panels, respectively. In green (upper-curves) we depict $f_{3}$, in blue $f_{2}$ and in red $f_{1}$.} 
\label{GrapheFpert}
\end{center}
\end{figure}

Now that we have been able to reconstruct the couplings $f$, let us see how they depend on the $B-L$ breaking scale.\\
First of all, given the asymptotic behaviour of the $x_{i}$s, we can say that among the different solutions, the case $(+,+,+)$  will reach the highest values of $f$, and so of $M$, while on the contrary the case $(-,-,-)$   will reach the lowest ones.\\
We thus plot in fig.\ref{GrapheFpert} the evolution of the coupling $f_{i}$ for these two solutions, as a function of the $B-L$ breaking scale, also considering different values of the smallest light neutrino mass. We assume here a normal mass ordering and put all phases $\phi_{i}^{k}$ and $\theta_{13}$ to zero; furthermore, we  neglect $U_{m}$ for the moment.
\\
We observe that the coupling $f$ becomes non-perturbative  for $v_{R}\gtrsim 10^{15}\GeV$. From now, we therefore restrict ourselves to  $v_{R}\lesssim 10^{15} \GeV$.
\subsection{Right-handed neutrino spectrum}
Let us study the right-handed neutrino spectra for the different solutions. 
\begin{figure}[h!]
\begin{center}
\includegraphics[scale=0.45]{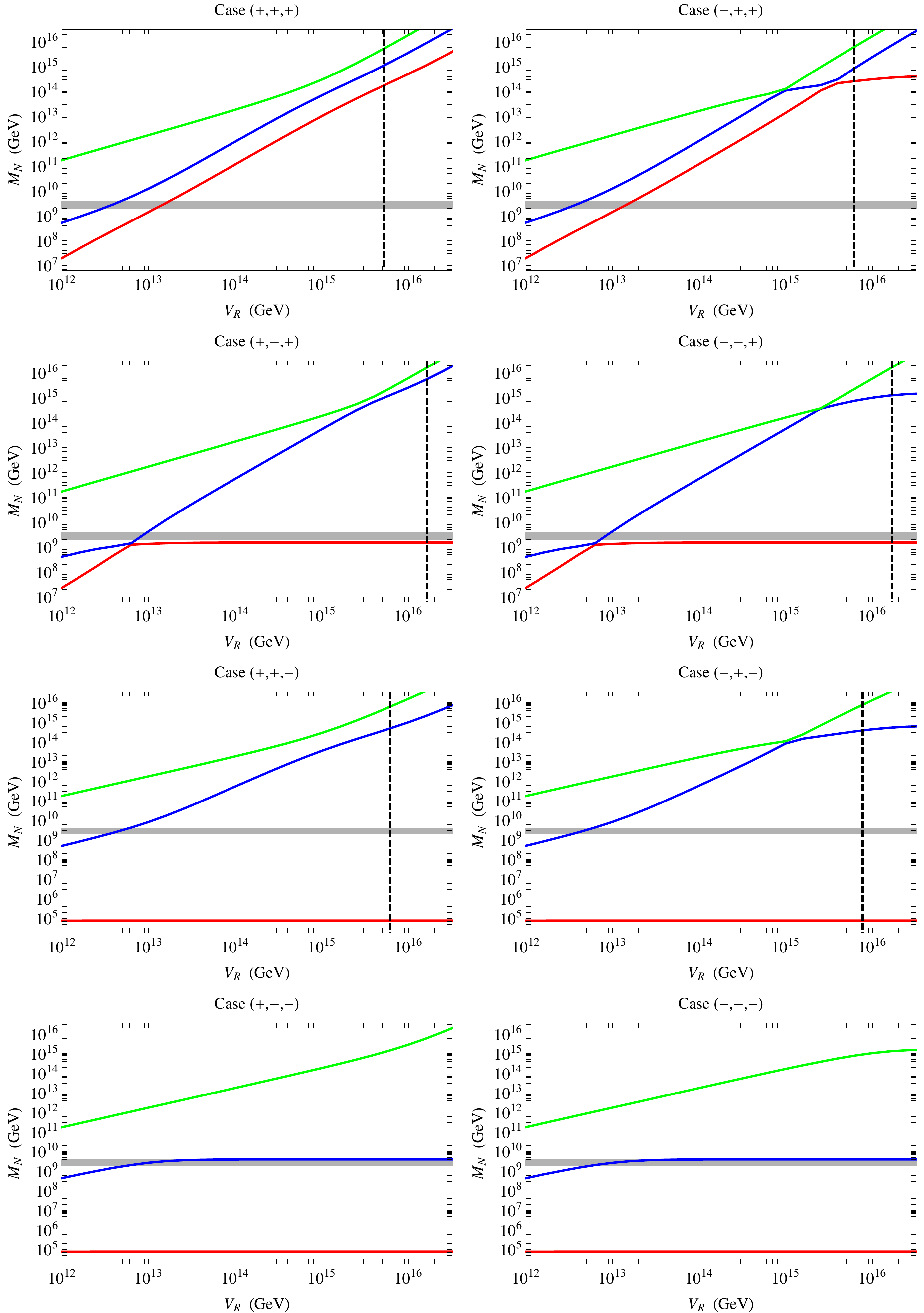} 
 \caption{Spectra of right-handed neutrinos resulting from the different solutions for $f$, as a function of $v_{R}$. The colour code is similar to fig.\ref{GrapheFpert}. The vertical dashed line represents the perturbative limit for $f$, while the horizontal gray band denotes the lower bound on $M$, as derived in the type I seesaw (see previous chapter).} 
\label{GrapheMsol}
\end{center}
\end{figure}\\
In fig.\ref{GrapheMsol}, we plot the masses $M_{i},i=1,2,3$ for the 8 solutions $(\pm,\pm,\pm)$ as a function of the $B-L$ breaking scale $v_{R}$. We choose $m_{1}=10^{-3}\eV$, the other parameters being fixed as before. Qualitatively, we see that the different solutions behave as expected: $(+,+,+)$ masses continuously increase with  $v_{R}$, whereas $(-,-,-)$ masses must reach a plateau.\\
Moreover, a $"-"$ solution for $x_{i}$ entails that one of the masses reaches such a plateau. The precise relation between $x_{i}^{-}$ and the mass $M_{j}$ which stabilises can be found in the appendix B of~\cite{HLS}.\\ 
Finally, we can classify the different solutions using the plateau argument, and group them in four pairs, $(\pm,+,+)$, $(\pm,-,+)$, $(\pm,+,-)$ and $(\pm,-,-)$ as displayed in fig.\ref{GrapheMsol}. 
\begin{itemize}
\item In the first pair $(\pm,+,+)$, right-handed neutrino masses keep growing with increasing $v_{R}$, at least until $f\sim 1$ is reached. For $v_{R}\lesssim 10^{14}\GeV$, we have $M_{1}\leq M_{2}\ll M_{3}$. A priori, one could think of $N_{1}$ and $N_{2}$ as potential candidate for leptogenesis.
\item In the second case $(\pm,-,+)$, the lowest RHn mass rapidly reaches a plateau, for $v_{R}\gtrsim 10^{13} \GeV$. The constant value is $M_{1}\simeq 10^{9} \GeV$, closed to but still slightly below the type I leptogenesis bound $M_{1} \lesssim M_{1}^{\rm{min,I}}\sim (2-3)\,\times 10^{9} \GeV$. We notice that $M_{1}\sim M_{2}$ at low $v_{R}$. Here also, $N_{1}$ and $N_{2}$ are "leptogenesis-friendly", albeit $M_{1}$ seems too light.
\item The third pair $(\pm,+,-)$ exhibits a very different mass pattern: above the $B-L$ breaking scale considered, $M_{1}\simeq 10^{5} \GeV$, while $M_{2}$ varies from $10^{9}\GeV$ to $10^{11}\GeV$ for $10^{12}\GeV\lesssim v_{R} \lesssim 10^{14} \GeV$. Thus $N_{1}$ cannot be of any help for leptogenesis, while $N_{2}$ clearly is.
\item In the fourth type of solutions, $(\pm,-,-)$ cases, both $M_{1}$ and $M_{2}$ reach a plateau. $M_{1}$ stabilises around $10^{5} \GeV$, and thus $N_{1}$ is too light for leptogenesis, while $N_{2}$, reaching a constant mass value $M_{2}\simeq M_{1}^{\rm{min,I}}$ provides an interesting possibility.
\end{itemize}
Therefore, in the following, we will focus only on four typical solutions: $(+,+,+)$ which corresponds to the pure type II in the high $v_{R}$ limit; $(-,-,-)$ which conversely corresponds to the type I solution in the high $v_{R}$ limit; solutions $(+,+,-)$ and $(+,-,+)$, which are interesting mixed type I-pure type II solutions.
\section{Leptogenesis}
Now let us proceed to the analysis of leptogenesis in this scenario, which has already been widely studied in the single flavour approximation~\cite{HLS},~\cite{SU2R}-~\cite{LR2}.\\
The sources of lepton number violation consist in interactions involving right-handed neutrinos and scalar triplets. The former are found similar to the type I seesaw, except for the fact that the RHns are no longer singlets, since they belong to an $SU(2)_{R}$ doublet. Thus they interact with the additional gauge bosons, which may have consequences for leptogenesis.\\
Furthermore, the inclusion of $SU(2)_{L}$ $\D_{L}$ triplets induces additional $CP$ asymmetries, as we will see below.\\
%Finally, we work in a supersymmetric extension.\\
Before examining these points, a remark is in order, regarding the different right-handed neutrino spectra discussed above.\\
In leptogenesis, it is customary to assume that the contributions of the heavier neutrinos are negligible, since the asymmetries produced during $N_{2,3}$ decays will be erased by the inverse decays which produce $N_{1}$. In our case, since $M_{3}\gg M_{2},M_{1}$, it is true that the asymmetry produced during $N_{3}$ decays will be washed-out by $N_{1,2}$ processes. However, we see that in the $(\pm,\pm,-)$ cases, $M_{1}$ is too small for leptogenesis, while $M_{2}$ has potentially  the good order of magnitude. Hence when evaluating the produced baryon asymmetry, we must consider both $N_{1}$ and $N_{2}$ processes, while $N_{3}$ ones can be disregarded. 
\subsection{Right-handed neutrinos as part of an $SU(2)_{R}$ doublet}
Having gauged $SU(2)_{R}\times U(1)_{B-L}$, our model contains four extra gauge bosons $\sim B^{'},W_{R}^{0},W_{R}^{1,2}$, which acquire mass when the $SU(2)_{R}$ Higgs triplet $\D_{R}$ develops a vev. Right-handed neutrinos are charged under $G_{3221}$, $N^{c}=(1,1/2,1,1)$, and so they interact with the extra gauge bosons through different processes~\cite{SU2R,Cosme}:
\begin{itemize}
\item Decays of $N$ into $W_{R}$, either with on-shell $W_{R}$ via two-body decays when $M_{1}\gtrsim M_{W_{R}}$, or into off-shell $W_{R}$ via three-body decays, when $M_{1}\lesssim M_{W_{R}}$.
\begin{figure}[h!]
\begin{center}
\includegraphics[trim=20mm 232mm 13cm 27mm, clip, scale=1]{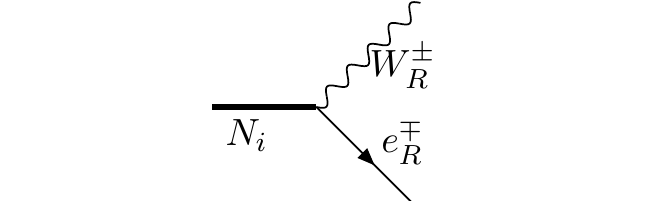} 
\caption{Two-body RHn decays.}
\end{center}
\label{GrapheEcpD}
\end{figure}
\item Gauge boson mediated scatterings: either (a) right-handed quark-lepton $W_{R}$ mediated scatterings , through  $s-$ channel $N\,e_{R}\LRa \ol{u_{R}}\,d_{R}$, and $t-$, $u-$ channels $N \,d_{R}\LRa e_{R} \,u_{R}$ and $N \,\ol{u_{R}}\LRa e_{R}\, \ol{d_{R}}$, or  (b) lepton-lepton scatterings involving two $N$s, via  $Z^{\prime}$  mediated scatterings in the $s-$ channel $N\,N^{c}\LRa e_{R}\,\ol{e_{R}}$, and $W_{R}$ $t-,u-$ channels $e_{R}\,\ol{e_{R}}\LRa N\,\ol{N}$ and $e_{R}\,N_{i}\LRa N_{j}\,e_{R}$. 
\begin{figure}[h!]
\begin{center}
\includegraphics[trim=37mm 185mm 20mm 20mm, clip, scale=1]{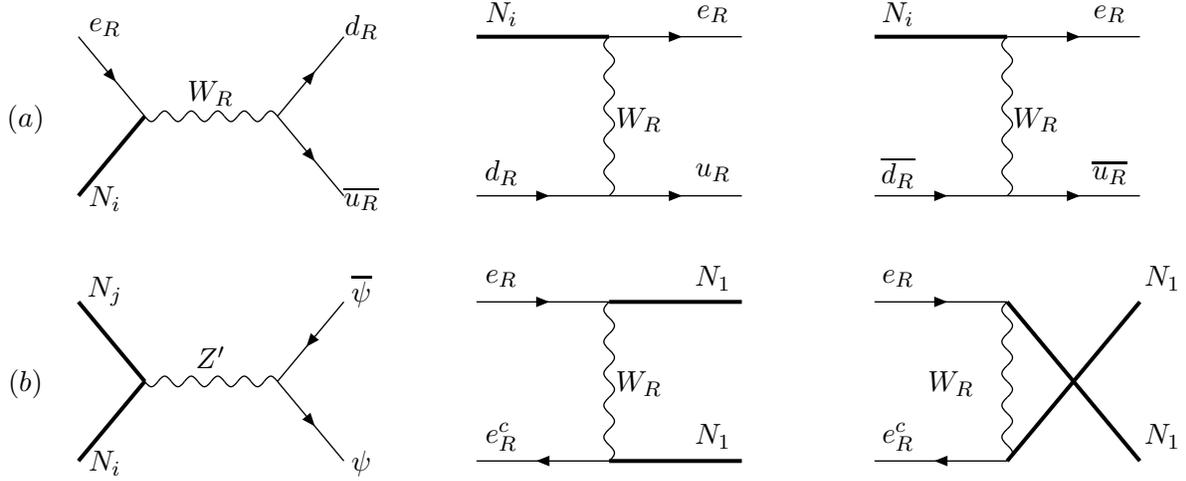} 
\caption{Gauge bosons scatterings.}
\end{center}
\label{GrapheEcpD}

\end{figure}\\
\end{itemize}
The decay channel of $N$ into $W_{R}$ may affect leptogenesis by diluting the $CP$ asymmetry generated in RHn decays into leptons~\cite{gaugedilution}. However, as $M_{\D}\gg M_{1}$ we can neglect this dilution.
The potential danger associated to these new interactions comes from scatterings, which tend to keep right-handed neutrinos in-equilibrium with the thermal bath. Two cases may happen: either $M_{i}\gtrsim M_{W_{R}}$ or $M_{1}\lesssim M_{W_{R}}$. It has been shown in~\cite{SU2R} that in the former case no lepton asymmetry can be generated, unless $M_{i}\gtrsim 10^{16} \GeV$ when the scatterings $e_{R}\,W_{R}^{+}\Ra N \Ra \ol{e_{R}}\,W_{R}^{-}$ become out of equilibrium. In the  considered case, under the constraint $v_{R}\lesssim 10^{15} \GeV$, we do not reach such high values for the $M_{i}$. However, since $M_{W_{R}}\simeq g_{R}\,v_{R}/2\simeq v_{R}/3$ and $M_{i}=f_{i} v_{R}$, this case may be only encountered in the $(+,+,+)$ solution for the heaviest right-handed neutrino, for $v_{R}\simeq 10^{15} \GeV$.\\
Thus the danger may only arise from the other case $M_{i}\lesssim M_{W_{R}}$, when scatterings mediated by the extra gauge bosons force the RHns to be in thermal equilibrium, and so prevent the third Sakharov condition from being satisfied. The condition for these scatterings to be out of equilibrium is~\cite{SU2R}
\bea
M_{i} \lesssim 9\,\times 10^{12}\GeV \times \left(v_{R}/10^{14} \GeV \right)^{4/3} \, ,
\eea
while the condition for the scatterings to be slower than $\D L=1$ scatterings is
\bea
M_{i}\lesssim 2\times 10^{13}\,K_{i}\times\left(\frac{v_{R}}{10^{14}\GeV}\right)^{2}\times\left(0.06\left(\frac{v_{R}}{M_{i}}\right)^{2}+1.15\right) \ ,
\eea
where $K_{i}$ is the washout factor of the RHn $N_{i}$ (see below).\\
These two conditions are fulfilled for both $N_{1}$ and $N_{2}$, for the $B-L$ breaking scale we considered $v_{R}\lesssim 5\times10^{14}\GeV$ and for light neutrino masses $m_{1}\lesssim 0.3 \eV$. Actually, this constraint may only apply for heavy RHns, $M_{i}\gtrsim 10^{11} \GeV$, for which one encounters an even stronger constraint which stems from the gravitino problem, as we will see hereafter. For that reason, we can safely neglect the effect of the $SU(2)_{R}\times U(1)_{B-L}$ gauge bosons.
\subsection{Scalar triplets in leptogenesis}
\subsubsection{Additional $CP$ asymmetries}
The first thing to notice is that $\D_{L}$ contributes to the leptonic $CP$ asymmetry in two different ways~\cite{Hambye2}.
Since $\D_{L}$ couples to  leptons and to the Higgs doublet, a $CP$ asymmetry can be generated during $\D_{L}$ decays through the interference between the tree level $\D_{L}\Ra L_{i} L_{j}$ and the vertex correction involving right-handed neutrinos, depicted in fig.\ref{GrapheEcpD}.
%%%%%%%%%%%%%%%%%%%%%%%%%%%%%%%%%%%%%%%%%%%%%
%
\begin{figure}[h!]
\begin{center}
\includegraphics[trim=20mm 230mm 13cm 27mm, clip, scale=1]{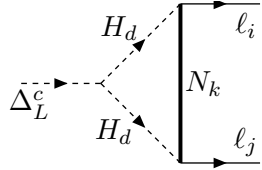} 
\end{center}
\caption{Vertex correction for $\Delta_L$ decays.}
\label{GrapheEcpD}
\end{figure}\\
These decays violate lepton number by two units. However, since we have $M_{\D_{L}}\simeq v_{R}/3$, the triplet is much heavier than the lightest RHn, $M_{\D_{L}}\gg M_{1}$. Therefore the asymmetry produced during these decays will be washed out by the subsequent processes involving $N_{2}$ and $N_{1}$, and so we can neglect all processes with on-shell $\D_{L}$s.\\
The other possible contribution comes from the vertex correction of the right-handed neutrino decay diagram, with an off-shell $\D_{L}$ running in the loop:
%%%%%%%%%%%%%%%%%%%%%%%%%%%%%%%%%%%%%%%%%%%%%
%
\begin{figure}[h!]
\begin{center}
\includegraphics[trim=20mm 230mm 13cm 27mm, clip, scale=1]{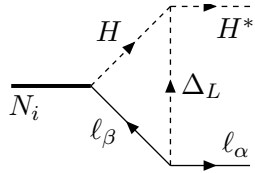} 
\end{center}
\caption{Additional $CP$ asymmetry for the type II seesaw.}
\label{figVertexD}
\end{figure}\\
%%%%%%%%%%%%%%%%%%%%%%%%%%%%%%%%%%%%%%%%%%%%%
As the couplings of $\D_{L}$ to lepton and Higgses are both complex, they induce a new interference term with the tree level diagram $N\Ra \ell \phi$. The corresponding $CP$ asymmetry in lepton flavours reads~\cite{boundII,FlavTypeII}: 
\bea
\eps_{i\al}^{II}=\frac{3}{8 \pi}  \sum_{k=1,3}\frac{\IM{\lam_{i k}\,(m_{\nu^{II}}^{*})_{k \al}\,\lam^{*}_{i \al}}}{(\lam \lam^{\dagger})_{i i}} \frac{M_{i}}{v_{u}}\,f^{II}\left(\left(\frac{M_{\D}}{M_{i}}\right)^{2}\right) \ ,
\eea
where the loop function is given in the MSSM by~\cite{Hambye2,boundII}:
\bea
f^{II}(y)&=& x \log{\left(\frac{1+x}{x}\right)} \stackrel{x\gg 1}{\longrightarrow} 1 \ .
\eea
\subsection{Supersymmetric leptogenesis}
The supersymmetric version of thermal leptogenesis is not very different from the SM one~\cite{Plumachersusy,towards}. Indeed, given the energy scale considered, $\sim M_{i}\gg 1 $TeV, the soft SUSY breaking terms are negligible, and the masses of fermions and sfermions are equal $M_{N}=M_{\tilde{N}}$. Furthermore, from the Yukawa couplings in the  superpotential 
\bea
W\supset \la_{i \al}L_{\al}.H_{u}N^{c}_{i} \ ,
\eea\beq
  \Gamma (N_i \rightarrow \ell_\alpha H_u)\,
    =\, \Gamma (N_i \rightarrow \bar \ell_\alpha H^*_u)\,
    =\, \Gamma (N_i \rightarrow \tilde \ell_\alpha \tilde H_u)\,
    =\, \Gamma (N_i \rightarrow \tilde \ell^*_\alpha \bar{\tilde H}_u)\,
    =\, \frac{M_i}{16 \pi}\, |\lambda_{i \alpha}|^2\ ,
\eeq
\beq
  \Gamma (\widetilde{N^c_i} \rightarrow \bar \ell_\alpha \bar {\tilde H}_u)\,
    =\,  \Gamma (\widetilde{N^c_i} \rightarrow \tilde \ell_\alpha H_u)\,
    =\, \frac{M_i}{8 \pi}\, |\lambda_{i \alpha}|^2\ .
\eeq
Supersymmetry implies the equality of the total decay widths,
$\Gamma_{N_i} = \Gamma_{\widetilde{N^c_i}}
= M_i (\lambda \lambda^\dagger)_{ii} / 4 \pi$.
Hence a lepton asymmetry is created through decays of right-handed neutrinos $N$ and (RH) sneutrinos $\tilde{N}$ in an amount equal to that of the slepton asymmetry
\bea
\eps_{N_{i},L_{\al}}=\eps_{\tilde{N^{c}}_{i},L_{\al}}=\eps_{N_{i},\tilde{L}_{\al}}=\eps_{\tilde{N^{c}}_{i},\tilde{L}_{\al}} \ .
\eea
The $CP$ asymmetry generated in $N_{i}$ decays is~\cite{CP}:
\bea
\label{ECPI}
\e_{i\al}^{I}=\frac{1}{8 \pi} \sum_{j\neq i} \frac{\IM{\lam_{i \al} (\lam \lam^{\dagger})_{i j} \lam^{*}_{j \al}}}{(\lam \lam^{\dagger})_{i i}} f_{I}\left(\left(\frac{M_{j}}{M_{i}}\right)^{2} \right)\,.
\eea
Since we have an additional scalar trilinear coupling $\tilde{L}H_{u}\tilde{N}^{c}$, the vertex correction is modified when compared to the SM case, as well as the self-energy correction, since twice more particles are running in the loop. The type I correction thus reads in the MSSM~\cite{CP}:
\bea
f_{I}(x)&=&\sqrt{x}\left(\frac{2}{1-x}-\log{\left(\frac{1+x}{x}\right)}\right) \stackrel{x\gg 1}{\longrightarrow} -\frac{3}{\sqrt{x}}  \ .
\eea
As we saw in chapter 3, when RH neutrinos are almost degenerate, the self-energy correction should be treated with care. As we can see in fig.\ref{GrapheMsol}, for example in the $(+,-,+)$ at low $v_{R}$, we have such a (partial) degeneracy $M_{1}\simeq M_{2}$. We therefore use for the $CP$ asymmetry the following formula~\cite{resonantlepto}:
\bea
\eps_{i \al}^{I}=\frac{1}{8 \pi} \frac{1}{(\lam \lam^{\dagger})_{i i} } \sum_{j\neq i} \IM{ \lam_{i \al}\lam^{*}_{j \al} \left[(\lam \lam^{\dagger})_{i j} ( C_{v}^{i j}+2 C_{s,a}^{i j})+2\,(\lam \lam^{\dagger})_{j i}\,C_{s,b}^{i j} \right] } \ ,
\eea
where $C_v^{i j}$ is the  vertex correction 
\bea
C_{v}^{i j}=-\sqrt{x} \log{\left(\frac{1+x}{x}\right)} \, , \sqrt{x}=\frac{M_{j}}{M_{i}} \ ,
\eea
the other term standing for the  self-energy correction 
\bea\label{csb}
C_{s,a}^{i j}=\sqrt{x}\, C_{s,b}^{i j}=\frac{\sqrt{x}(1-x)}{(1-x)^{2}+x\left(\lam \lam^{\dagger}\right)^{2}_{i j}/16\pi^{2} } \ .  
\eea
In the limit of hierarchical RHns, $x\gg1$ and the terms $C_{s,b}$ can be neglected, so that we recover the non-resonant formula eq.(\ref{ECPI}).\\
Another difference in the supersymmetric version of leptogenesis, is the relation among chemical potentials, from which we deduce the conversion between the lepton doublet asymmetry and the $B-L_{\al}$ asymmetries,
\bea
Y_{L_{\al}}(z)=\sum_{\be} A_{\al \be} Y_{\D \be}\ .
\eea
For $M_{1} \lesssim 10^{9} \GeV \times (1+\tan{\be}^{2})$ , tau- and muon-Yukawa couplings  are in equilibrium, thus the three flavoured asymmetries  $Y_{\D e}$,$Y_{\D \mu}$,$Y_{\D \tau}$ are distinguishable. In the  MSSM, the $A$ matrix is given by~\cite{RiottoPetcov}:
\bea
A=\left( \begin{array}{ccc}
-93/110 & 6/55 &  6/55 \\ 
 3/40 &  -19/30 &  1/30 \\ 
 3/40 & 1/30 & -19/30
\end{array} \right).  \ 
\label{bigA}
\eea
For $M_{1}$ between $10^{9} \GeV \times (1+\tan{\be}^{2})$ and $10^{12} \GeV \times (1+\tan{\be}^{2})$, only the tau-Yukawas are in equilibrium, thus the lepton asymmetry is projected onto a 2 flavour-space $(Y_{\D e+\mu}=Y_{\D e+\D \mu}, Y_{\D \tau})$.  The conversion $L \leftrightarrow B-L$ now reads:  
\bea
A=\left( \begin{array}{cc}
-541/761 & 152/761  \\ 
46/761 &  -494/761  
\end{array} \right).  \
\label{smallA} 
\eea
Finally, if $M_{i}$ is above $10^{12}\times (1+\tan{\be}^{2})$ GeV, none of the interactions involving charged lepton Yukawa couplings are in equilibrium, and we recover the flavour-independent treatment of leptogenesis with $A=-\rm{diag}(1,1,1)$.\\
Finally, with two light Higgs doublets, the baryon asymmetry resulting from the fast $B+L$ violating processes  is~\cite{HarveyTurner}:
\bea
Y_{B}=\frac{10}{31} \sum_{\al} Y_{\D \al} \ .
\eea
With our choice of $\tan{\be}=10$, the 3 flavour regime will generally apply, since in most of the solutions  ($(-,-,-)$, $(+,+,+)$, etc...) the RH neutrino masses are below $10^{11} \GeV$.\\
Let us turn to the discussion of the major problem a supersymmetric scenario of leptogenesis faces: the  constraint on the reheating temperature of the Universe, $T_{RH}$.
\subsection{Reheating temperature and the gravitino problem}
\begin{figure}[h!]
\begin{center}
\includegraphics[scale=0.4]{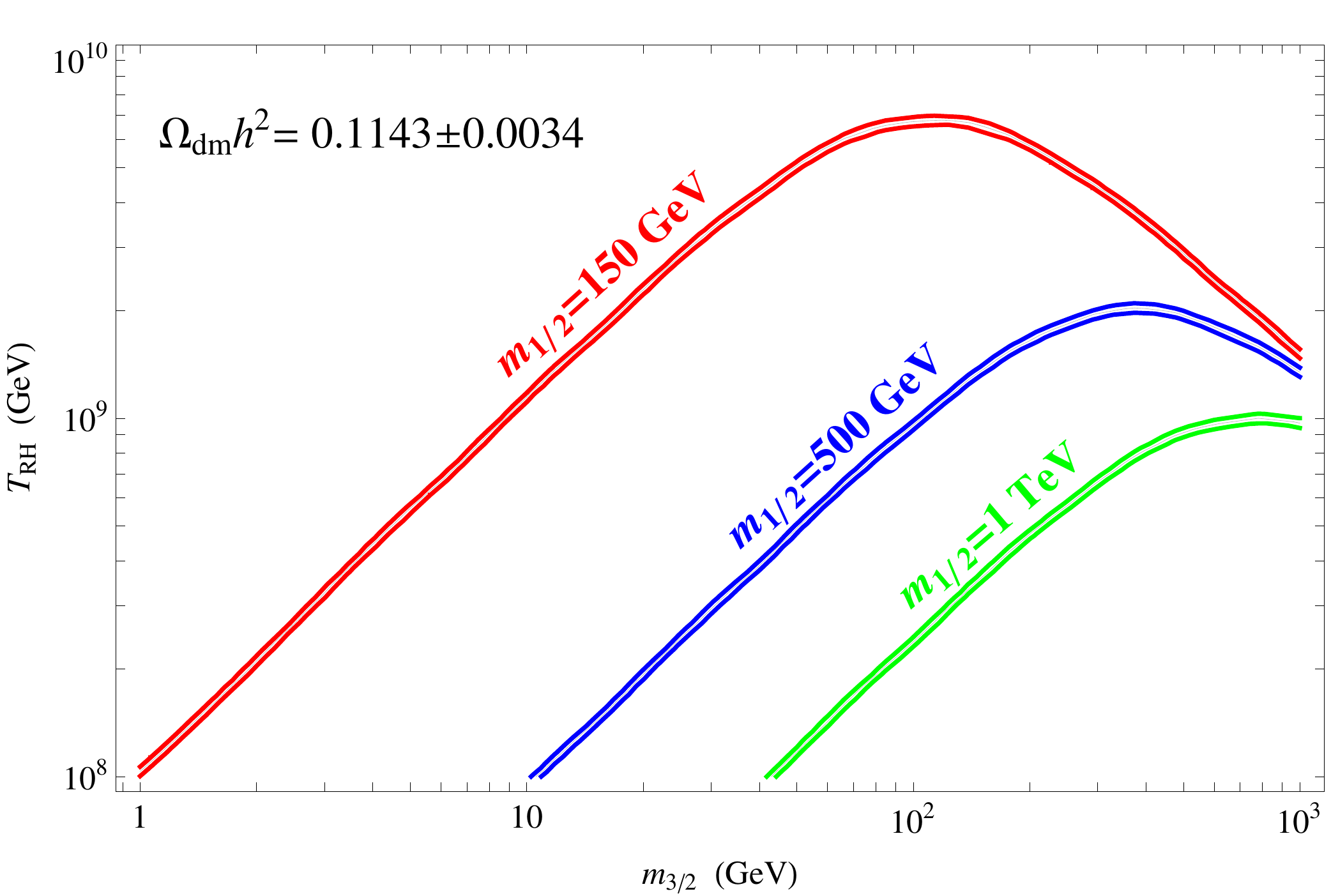} 
 \caption{Constraint on the reheating temperature from gravitino abundance, assuming $\Omega_{dm}=\Omega_{3/2}$. The plot has been obtained assuming universal boundary conditions for the gaugino masses, with $m_{1/2}=1$ TeV (green), $m_{1/2}=500\GeV$ (blue) and $m_{1/2}=150\GeV$ (red). For each case, upper and lower curves represent $\Omega_{3/2}=0.11435+0.0034$ and $\Omega_{3/2}=0.11435-0.0034$ respectively, while the central value is indicated by a grey line. Plot adapted from refs.\cite{StrumiaGrav}. } 
\label{GrapheGrav}
\end{center}
\end{figure}

The standard scenario of cosmology relies on the existence of an inflationary period~\cite{KolbTurner,Liddle:1999mq}, which explains, among other issues, the observed homogeneity and isotropy of the Universe. The inflationary epoch lasts until the field responsible for driving inflation, the inflaton, reaches the minimum of its potential. At this time, the energy density of the Universe is dominated by the vacuum energy of the inflaton. As this field oscillates around its minimum, it reheats the Universe by a release of its potential energy. Then, when $T\simeq T_{RH}$, the inflaton field decays, creating degrees of freedom, and so the temperature drops down.\\
Therefore all species existing before the reheating of the Universe will be completely diluted by the entropy production. Similarly, any pre-existing lepton asymmetry will be washed-out, so that it is mandatory that leptogenesis takes place after reheating $T_{\rm{lepto}}\lesssim T_{RH}$. As $T_{\rm{lepto}}\sim M_{1}\sim  10^{9}\GeV$  thermal leptogenesis requires $T_{RH}\gtrsim 10^{9}\GeV$.\\
On the other hand, such high temperatures can be problematic when one comes to local supersymmetry~\cite{Sugra}. Indeed, in such a case, the SUSY partner of the graviton field, the spin $3/2$ gravitino, imposes stringent constraints on the reheating temperature~\cite{GravitinoPb}. The gravitino only interacts gravitationaly, thus very weakly. Essentially two situations can occur. The lifetime of the gravitinos can be larger than the age of the Universe, in which case they might either overclose the Universe or their relic density may give an overabundant dark matter density~\cite{GravDarkMatter}. In the opposite case, gravitinos can be unstable, and their decays can either spoil the success of Big Bang Nucleosynthesis~\cite{BBNgrav}, or their decay products, if stable, can also provide an overabundant dark matter relic density.\\
The constraints on the reheating temperature come from the fact that the thermal production of gravitinos, which happens during or just after the reheating stage, is more efficient for high reheating temperatures. Indeed, the Boltzmann equation for the gravitino number density $n_{3/2}$, in a bath dominated by the inflaton field and radiation, reads~\cite{gravitinoprod,StrumiaGrav}:
\bea
\label{BEgrav}
&\frac{d \rho_{\phi}}{dt}&+3H(T)\,\rho_{\phi} = -\Gamma_{\phi}\,\rho_{\phi}\,, \nonumber \\
&\frac{d \rho_{R}}{dt}&+4H(T)\,\rho_{R} = \Gamma_{\phi}\,\rho_{\phi}\,, \nonumber \\
&\frac{d n_{3/2}}{dt}&+3H(T)\,n_{3/2} = \gamma(T)\,, \nonumber
\eea
where $\rho_{R}=\pi^{2}g_{*}\,T^{4}/30$ is the radiation energy density, and $\rho_{\phi}$ the energy density of the inflaton scalar field. The decay width of the inflaton field is related to the reheating temperature by:
\bea
\Gamma_{\phi}=\frac{T_{RH}}{M_{pl}}\sqrt{\frac{8\pi^{3}}{90}\,g_{*}}\ .
\eea
where in the MSSM the effective number of degrees of freedom is $g_{*}=228.75$. The thermal production rate of gravitino is roughly~\cite{StrumiaGrav} $\gamma(T)\simeq 10\,T^{6}/M_{pl}^{2}$.\\
Since gravitinos interact only gravitationnaly, one can consider that the density after reheating remains unchanged until gravitinos decay (if ever), such that one has
\bea
Y_{3/2}\simeq 6\times 10^{-12}\frac{T_{RH}}{10^{10}\GeV}\,.
\eea
Clearly, thermal production of gravitinos is more efficient for high reheating temperatures.\\
Now, if gravitinos are stable, they might either overclose the Universe, which leads to the constraint $T_{RH}\lesssim 2\times 10^{10}\GeV$, or if they provide the dominant contribution to dark matter, their relic density might be greater than the inferred density $\Omega_{dm}h^{2}=0.1143\,\pm0.0034$~\cite{WMAP5b}. Assuming that gravitinos provide the sole contribution to dark matter, one has 
\bea
\Omega_{dm}h^{2}=\Omega_{3/2}h^{2}\simeq 2.78\times 10^{10}\,Y_{3/2}\,\frac{m_{3/2}}{100 \GeV} .
\eea
This relation strongly constraints the reheating temperature, as can be seen in fig.\ref{GrapheGrav}, where we plot the relic density of gravitinos (assumed to correspond to the total dark matter abundance). In this plot, we further assume that gravitinos are only produced via thermal scatterings~\cite{StrumiaGrav}, which may not always be the case. Indeed, assuming that gravitinos are the lightest supersymmetric particles (LSP), decays of the next-to-LSP may give rise to a non-thermal production of gravitinos. However, we see that a stable gravitino implies $T_{RH}\lesssim 10^{10}\GeV$, which is slightly above the leptogenesis bound.
\\
As already stated, it is also possible to have unstable gravitinos. Nonetheless they can be long-lived particles and decay late. If these decays happen during or after BBN, $T\lesssim T_{BBN}$, the entropy released might jeopardise the successful predictions of BBN~\cite{BBNgrav}. As the decay rate of gravitinos is roughly
\bea
\Gamma_{3/2}\simeq\frac{m_{3/2}^{3}}{M_{Pl}^{2}}\ ,
\eea 
gravitinos decay after BBN if $\Gamma_{3/2}\lesssim H(T_{BBN})$. Assuming this is the case, the gravitino abundance is required to be small enough to not affect BBN, implying an upper-bound on the reheating temperature. For $m_{3/2}\lesssim 1$ TeV, one requires $T_{RH}\lesssim 10^{5-6} \GeV$.%, otherwise gravitino decays will photo-dissociate deuterium
This value is clearly incompatible with thermal leptogenesis. If one has $1 \TeV\lesssim m_{3/2}\lesssim 50 \TeV$, then the constraint roughly reads $T_{RH}\lesssim 10^{9}\GeV$.\\
%comes from photodissociation of $\hbox{$^{4}He$}$, and roughly reads $T_{RH}\lesssim 10^{9}\GeV$.
Assuming even heavier gravitinos, one evades the BBN constraint, but then gravitino decays will result in a non-thermal production of the LSP, which is assumed to be stable and the dominant (if not the only) contribution to the observed dark matter abundance. Considering only this non-thermal production for the LSP, its number density is related to the gravitino one by
\bea
\Omega_{LSP}h^{2}\simeq 2.78\times 10^{10}\times Y_{3/2}\,\left(\frac{m_{\chi_{1}^{0}}}{100\GeV}\right)\, ,
\eea
where $\chi_{1}^{0}$ is the lightest neutralino (in this case the LSP). Under the constraint that $\Omega_{LSP}\lesssim \Omega_{dm}$, reheating temperatures below $3\times 10^{10}\GeV$ are required.\\

To summarise this discussion, we can say that even if the gravitino abundance puts stringent constraints on the reheating temperature, leading in some models to a clear incompatibility with the scenario of thermal  leptogenesis, there are however possibilities in which the upper-bound on the reheating temperature is sufficiently high to allow for a successful leptogenesis.\\
Therefore, even if aware of this caveat, in the following we study the thermal scenario of leptogenesis, assuming that $T_{RH}=10^{11}\GeV$. This high reheating temperature will already allow us to distinguish the  solutions which provide enough baryon asymmetry from those which do not. Then, in a later stage, we will investigate more realistic values for $T_{RH}$.
\subsection{Boltzmann equations}
As before, the evolution of the comoving number densities is obtained by solving the set of Boltzmann equations.
For the RH (s)neutrino one has: 
\begin{eqnarray}
\label{BEYN}
Y_{N_i}^{\prime}(z)&=& - 2\,K_i\ \left( D_{i}(z)+S_{i}(z) \right)\ \left(Y_{N_i}(z)-Y_{N_i}^{eq}(z)\right) \ , \nonumber \\
Y_{\tilde{N_i}}^{\prime}(z)&=& - 2\,K_i\ \left( D_{i}(z)+S_{i}(z) \right)\ \left(Y_{\tilde{N_i}}(z)-Y_{\tilde{N_i}}^{eq}(z)\right) \ ,
\end{eqnarray}
where $z=M_{1}/T$. In this equation, $Y_{N_i}$ is compared to the number density in thermal equilibrium~\footnote{Assuming a Maxwell-Boltzmann (MB) distribution for the equilibrium abundance gives $Y_{N_{i}}^{eq}=Y_{\tilde{N_{i}}}^{eq}$. Taking Fermi-Dirac (FD) or Bose-Einstein (BE) distributions, the abundances differ at high temperatures by a spin factor~\cite{KolbWolfram}
\bea
n_{_{FD}}(T\gg M)=\frac{3}{4} \, n_{_{BE}}(T\gg M)=\frac{3 \zeta(3)}{4}\,n_{MB}(T\gg M) \ , \nonumber
\eea
and are equal at low temperatures. So even if we assume a MB statistic for the distribution functions, we  nevertheless correct the abundances by the high temperature expansion, and therefore we take $Y_{N_i}^{eq}=\frac{3}{4}  Y_{\tilde{N_{i}}}^{eq}$, with  $Y_{N_i}^{eq}$ given by eq. (\ref{Yneq}).} $Y_{N_{i}}^{eq}$\,, which is given by the following equation:
\bea
\label{Yneq}
Y_{N_i}^{eq}(z)\simeq \frac{135 \zeta(3)}{8 \pi^{4} g_{*} }R_{i}^{2}z^{2}K_{2}(R_{i} z) \stackrel{T\gg M_{i}}{\longrightarrow} \frac{135 \zeta(3)}{4 \pi^{4} g_{*} }\simeq 1.8\times 10^{-3} \ ,
\eea
where we have introduced  $R_i = M_{i}/M_{1}$.\\
In the MSSM, the number of degrees of freedom in the thermal bath is $g_{*}=228.75$ (without RH neutrinos or triplets). The individual $B-L$ asymmetries $Y_{\D \al}$ are driven by: 
\begin{eqnarray}
\label{beflav}
Y_{\D \al}^{\prime}(z)&=&-2\sum_{i=1,2}\e_{i \al}\, K_i \, \left( D_{i}(z)+S_{i}(z) \right).\left(Y_{N_i}(z)-Y_{N_i}^{eq}(z) + \left(Y_{ \tilde{N_i} }(z)-Y_{\tilde{N_i}}^{eq}(z)\right)\right) \nonumber \\ 
 &+& 2\sum_{i=1,2} \kappa_{i \al} \sum_{\be} W_{i}(z)\,A_{\al \be}\,Y_{\D \be}(z)  \ .
\end{eqnarray}
The washout parameters $\kappa_{i \al}$ are as usual given by:
\bea
\kappa_{i \al}\equiv \frac{\Gamma(N_i \rightarrow \ell_{\al} H_{d})}{H(M_{i})}=\frac{\tilde{m}_{i \al}}{m_{*}} \ ,
\eea
but with a slight modification of $\tilde{m}$ and $m_{*}$:
\bea
\tilde{m}_{i \al}&=&\frac{\lambda_{i \al} \lambda_{\al i}^{\dagger} v_{u}^{2}}{M_{i}}\,,\nonumber \\
m_{*}&=&\frac{16 \pi^{5/2} \sqrt{g^{*}}}{3 \sqrt{5}}\frac{v_{u}^{2}}{M_{Pl}}\simeq 1.56\times 10^{-3}\rm eV \ .
\eea
The total washout parameter $K_i$ is obtained by summing over flavour indices: 
\begin{eqnarray}
K_i=\sum_{\al} K_{i \al}=\frac{\tilde{m}_i }{m_*}\simeq \frac{\tilde{m}_i }{1.56\times10^{-3} \rm eV} \ .
\end{eqnarray}
The thermally averaged decay rate $D_i$ are given by:
\bea
D_{i}(z)=z\,R_{i}^{2}\,\frac{{\mathcal{K}}_{1}(R_{i} z)}{{\mathcal{K}}_{2}(R_{i} z)} \ .
\eea
The $\D L=1$ scatterings are Higgs-mediated  processes involving top quarks and antiquarks, and receive contributions from the $s-$ and $t-$channels: 
\bea
S_{i}(z)=2\gamma_{s}^{i}(z)+4\gamma_{t}^{i}(z) \, .
\eea
%We can approximate
%\bea
%1+\frac{S_{i}(z)}{D_{i}(z)}\simeq 1+\frac{2}{10} \frac{1}{R_{i}^{2} z^{2}} \ .
%\eea
The washout term $W_{i}(z)=W_{i}^{id}(z)+W_{i}^{s}(z)$ results from the contribution from inverse decays 
\bea
W_{i}^{id}(z)=\frac{1}{4}\,R_{i}^{4} \, z^{3} \,{\mathcal{K}}_{1}(R_{i} z) \ ,
\eea
and $\D L=1$ scatterings
\bea
W_{i}^{s}(z)=\frac{W_{i}^{id}(z)}{D_{i}(z)}\left(2\gamma_{s}^{i}(z)\left(\frac{Y_{N_i}(z)}{Y_{N_i}^{eq}(z)}+\frac{Y_{\tilde{N}_i}(z)}{Y_{\tilde{N}_i}^{eq}(z)}\right)+8\gamma_{t}^{i}(z)\right)\ .
\eea
A careful study of the reheating problem should be done by considering the influence of the inflaton field on the expansion rate of the Universe, similarly to eqs.(\ref{BEgrav}), as done in~\cite{towards,Towardsbis,PlumiReheat}. We do not carry such an analysis, but take the reheating into account in an effective way. We consider that the inflaton field decays instantaneously when the temperature drops below  $T_{RH}$, and thus we consider that the temperature at which leptogenesis starts $T_{in}$ coincides with $T_{RH}$. For masses $M_{i}\lesssim T_{RH}$, this is a rather good approximation. On the other hand, for heavier masses, thermalisation processes of $N_{i}$ are Boltzmann suppressed  and for $M_{i}\gtrsim (2-3)\times T_{RH}$ the number density of the created  RHns is negligible.\\
Had we properly included the contribution of the inflaton field, such a suppression of the thermalisation processes would not have occurred; however, as for $T\gtrsim T_{RH}$ the Universe is dominated by the inflaton energy density, the large Hubble expansion rate $\propto \sqrt{\rho_{\phi}+\rho_{R}+\rho_{N}}\sim \sqrt{\rho_{\phi}}$ would have diluted processes involving RHns. The difference between the na\"ive reheating we consider, and the proper treatment similar to eq.(\ref{BEgrav}) has been done, for example, in~\cite{StrumiaGrav} for the gravitino abundance, where it has been found that the na\"ive prescription, albeit twice larger, is of the correct order of magnitude.\\
As already stated, in a first stage we will use $T_{in}=10^{11}\GeV$ to discriminate among solutions, and then we will lower this value to more acceptable values.
\section{Results for the four typical solutions}
Now that the framework of our model has been set up, we can examine the viability of thermal leptogenesis for the four characteristic solutions we will focus on:
\begin{itemize}
\item Solution $(+,+,+)$ for which the pure type II seesaw dominates at large $v_{R}$, and where RHn masses continuously grow.
\item Solution $(+,-,+)$, an interesting type I+II mixed solution where, contrary to $M_{2}$ (which increases with $v_{R}$), $M_{1}$ reaches a plateau close to the type I bound.
\item Solution $(+,+,-)$, which is also a mixed solution, with $M_{1}$ reaching a constant value, but with $M_{1}\simeq 10^{5} \GeV$, far below the type I bound.
\item Solution $(-,-,-)$, which tends to the type I in the high $v_{R}$ limit. As in the previous case, $M_{1}\ll 10^{9}\GeV$.
\end{itemize}
From the above spectra, we can infer that for $(+,+,+)$, the lightest RH neutrino will contribute dominantly to the lepton asymmetry production, while for $(+,-,+)$, $N_{2}$ might play a role since $M_{1}$ could be excessively light. On the other hand, for the remaining solutions $(\pm,\pm,-)$, it is obvious that $N_{2}$ is the (ultra) dominant contribution.\\
We first assume that $m_{d}=m_{e}$, with a high reheating temperature $T_{RH}=10^{11}\GeV$, and then relax these assumptions in order to study their influence. In the next section, we will study the dependence on low-energy parameters that are the light neutrino mass, the $CP$ violating phase $\delta_{PMNS}$ and the mixing angle $\theta_{13}$, but for the moment, we fix $m_{1}=10^{-3} \eV$, $\theta_{13}=0$, put $\delta_{PMNS}$ to zero as well as all the other phases, except for $\phi_{2}^{u}=\pi/4$.
\subsection{Influence of flavours}
\begin{figure}[h!]
\hspace{-0.7cm}
\includegraphics[scale=0.6]{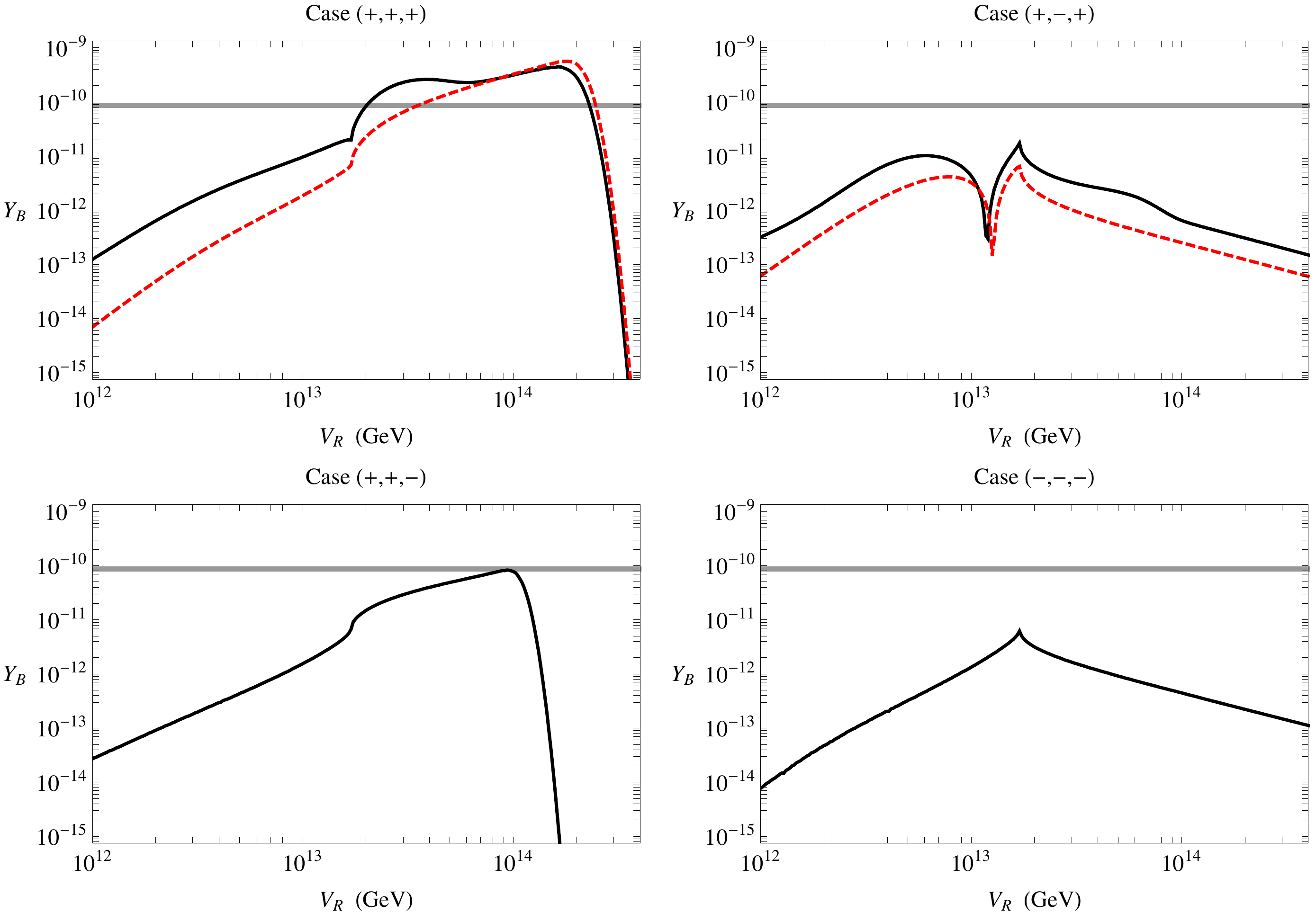}
 \caption{Evolution of the baryon asymmetry as a function of the $B-L$ breaking scale $v_{R}$, for the four characteristic solutions. In these plots we compared the flavoured case, depicted in solid-black lines, and the unflavoured case, in dashed-red lines. } 
\label{GrapheYBALL}
\end{figure}

\begin{figure}[h!]
\begin{center}
\includegraphics[scale=0.45]{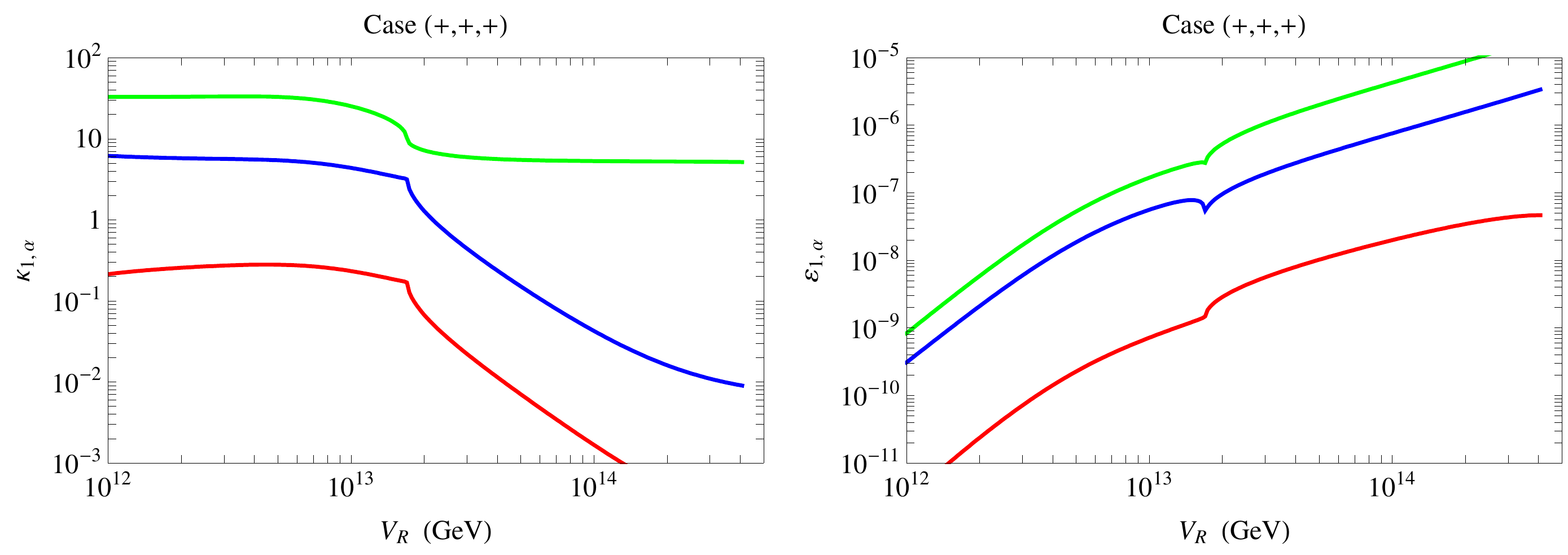}\\
 \includegraphics[scale=0.45]{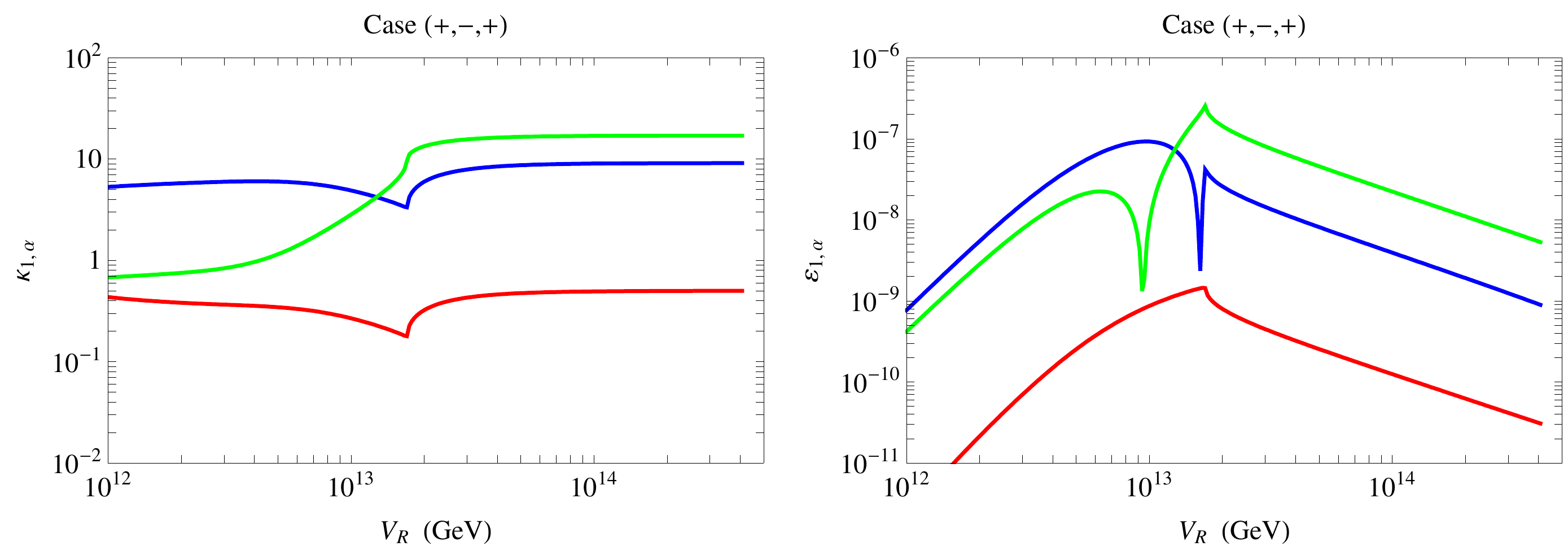}\\
\includegraphics[scale=0.45]{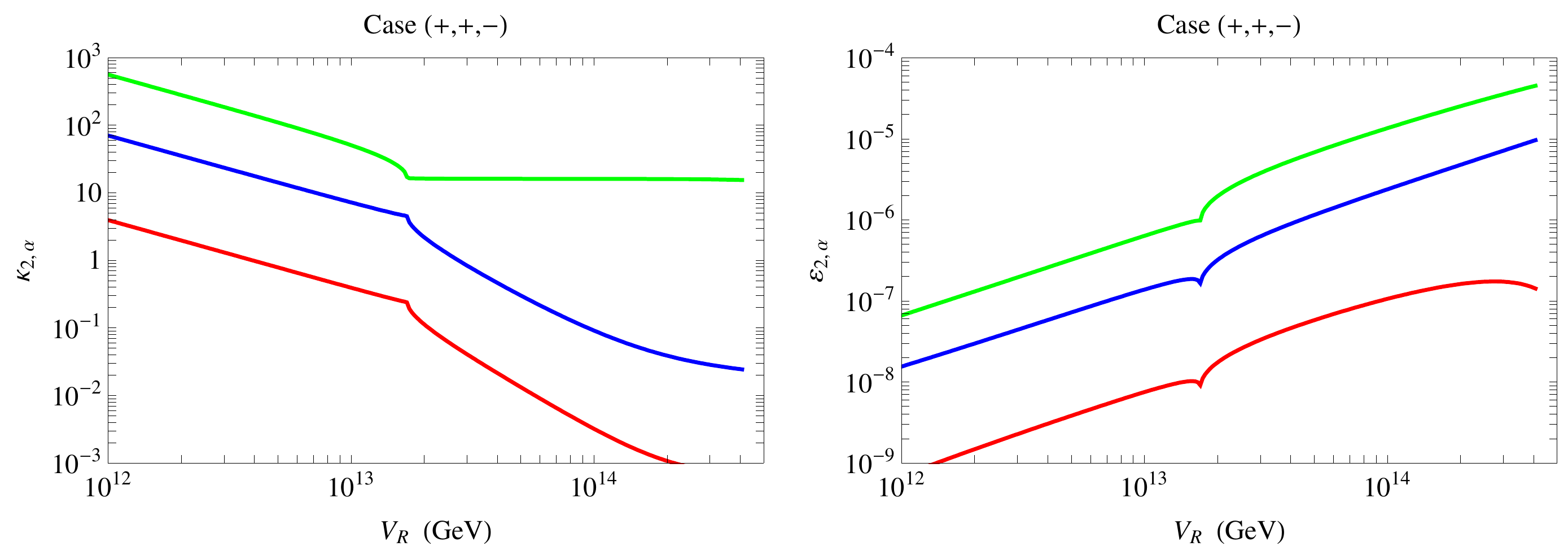}\\
\includegraphics[scale=0.45]{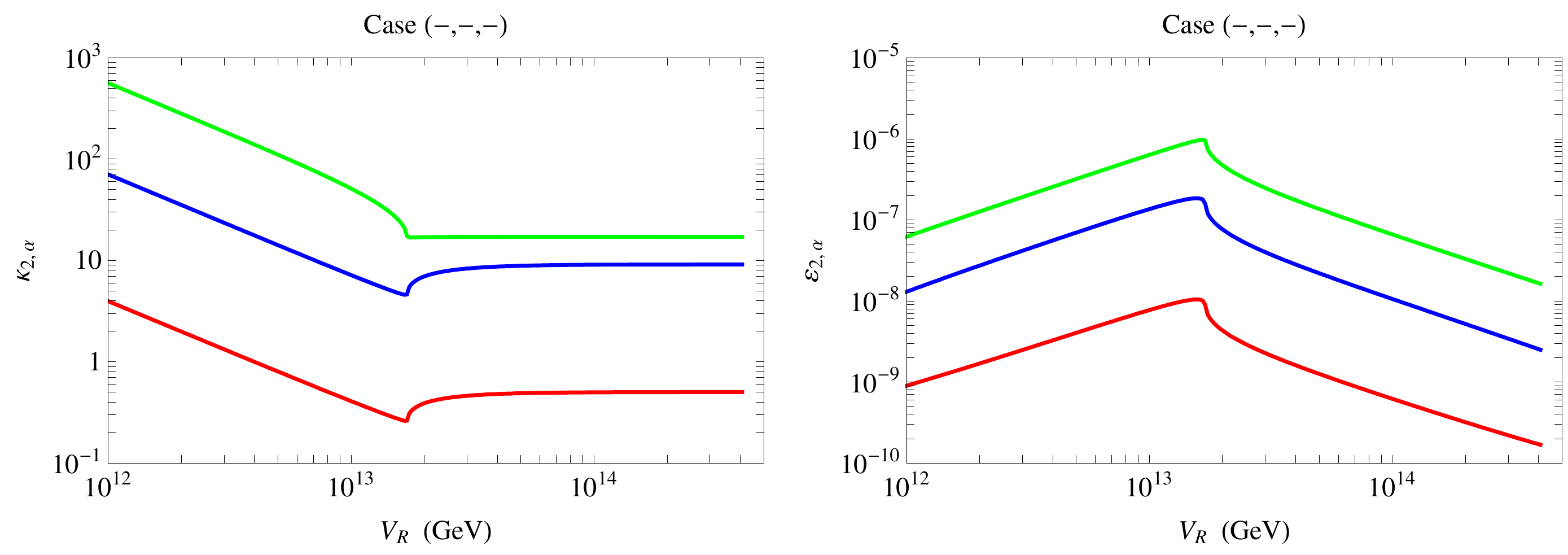}
 \caption{Evolution of individual $CP$ asymmetries $\eps_{i,\al}$ and washout factors $\kappa_{i\al}$ as a function of $v_{R}$ for the four characteristic solutions. Working in a three flavour scheme, we depicted in green the tau flavour, in blue the muon flavour and in red the electron one. We only plot the $\eps_{i,\al}$s and $\kappa_{i,\al}$s corresponding to the RHn $N_{i}$ that provides the dominant contribution to leptogenesis.} 
\label{Grapheek}
\end{center}
\end{figure}

We first investigate the influence of flavours on our four characteristic solutions by considering fig.\ref{GrapheYBALL}, where we plot the baryon asymmetry as a function of the $B-L$ breaking scale, including lepton flavours (solid black line) or not (red dashed line).
\\
The different behaviours are more easily understood by looking at fig.\ref{Grapheek}, where we plot the $CP$ asymmetries and washout factors involving the dominant RHn for the various solutions (see the caption for the colour code).
\\
We first notice that for the present choice of parameters, in the flavoured case (black lines)  $(+,+,+)$ can easilly saturate the observed baryon asymmetry, while $(+,+,-)$ is only marginally compatible. This is easily understood in fig.\ref{Grapheek}. These two solutions have in common that the dominant RHn strongly couples to the $\tau$  flavour (in green). Hence an important $CP$ asymmetry is created in this direction, even if somehow compensated by the strong washout. For $(+,+,+)$, we see that around $v_{R}\sim (2-3)\times10^{13} \GeV$, $\eps_{1,\tau}\gtrsim 10^{-6}$ while at the same time the washout decreases, dropping from $\sim 10^{2}$ down to $\sim 5$. Hence the baryon asymmetry is enhanced, as can be seen from the bump in $Y_{B}$ around $v_{R}\sim (2-3) \times10^{13} \GeV$.\\
Why is then $(+,+,-)$ comparatively so small?\\
It is because the $\tau$ flavour, even if copiously produced, is also strongly washed-out: for $v_{R}\gtrsim (2-3)\times10^{13}\GeV$, $\kappa_{2,\tau}\simeq 20$. Hence, a washout four times larger than for the $(+,+,+)$ case requires a $CP$ asymmetry four times greater, and such a value for the $CP$ asymmetry can only be reached for higher $v_{R}$.\\
Finally, as $v_{R}$ increases, $M_{1}$ or $M_{2}$ approach $T_{in}=10^{11}\GeV$, and so $N_{1}$ or $N_{2}$ are no longer efficiently produced by inverse decays, thus the dramatic fall of $Y_{B}$.\\
We observe a huge difference when we consider the unflavoured case (dashed red line): while for $(+,+,+)$ the unflavoured case is close to the flavoured one, particularly in the high $v_{R}$-$\eps_{CP}$ regime, the $(+,+,-)$ is highly suppressed\footnote{Actually, in this case, as in the $(-,-,-)$ one, the unflavoured picture gives $Y_{B}\simeq 10^{-18}$; but due to the numerical precision required to generate these plots, some instabilities arise, so we preferred to impose a cut around $Y_{B}\sim 10^{-15}$. See the next section for a plot with an increased precision to avoid the instabilities in $(\pm,\pm,-)$ solutions.}. In the next section, we will concentrate on this feature, which corresponds to the case discussed in \cite{Vives,N2lepto}.\\
Let us examine the remaining $(\pm,-,\pm)$ solutions. For the parameters chosen, the observed value $Y_{B}^{obs}$ is not reached. Still, the qualitative description carried for the previous case holds, up to the different behaviours of the $CP$ asymmetries. Notice also the strong suppression of the unflavoured $Y_{B}$ in the $(-,-,-)$ case, which also nicely illustrates the survival of the asymmetry produced by $N_{2}$ due to flavour effects~\cite{Vives} that we now discuss in greater detail.

\subsubsection{Survival of $N_2$ leptogenesis due to lepton flavour effects}
\begin{figure}[htb]
\hspace{-1cm}
\includegraphics[width=9cm]{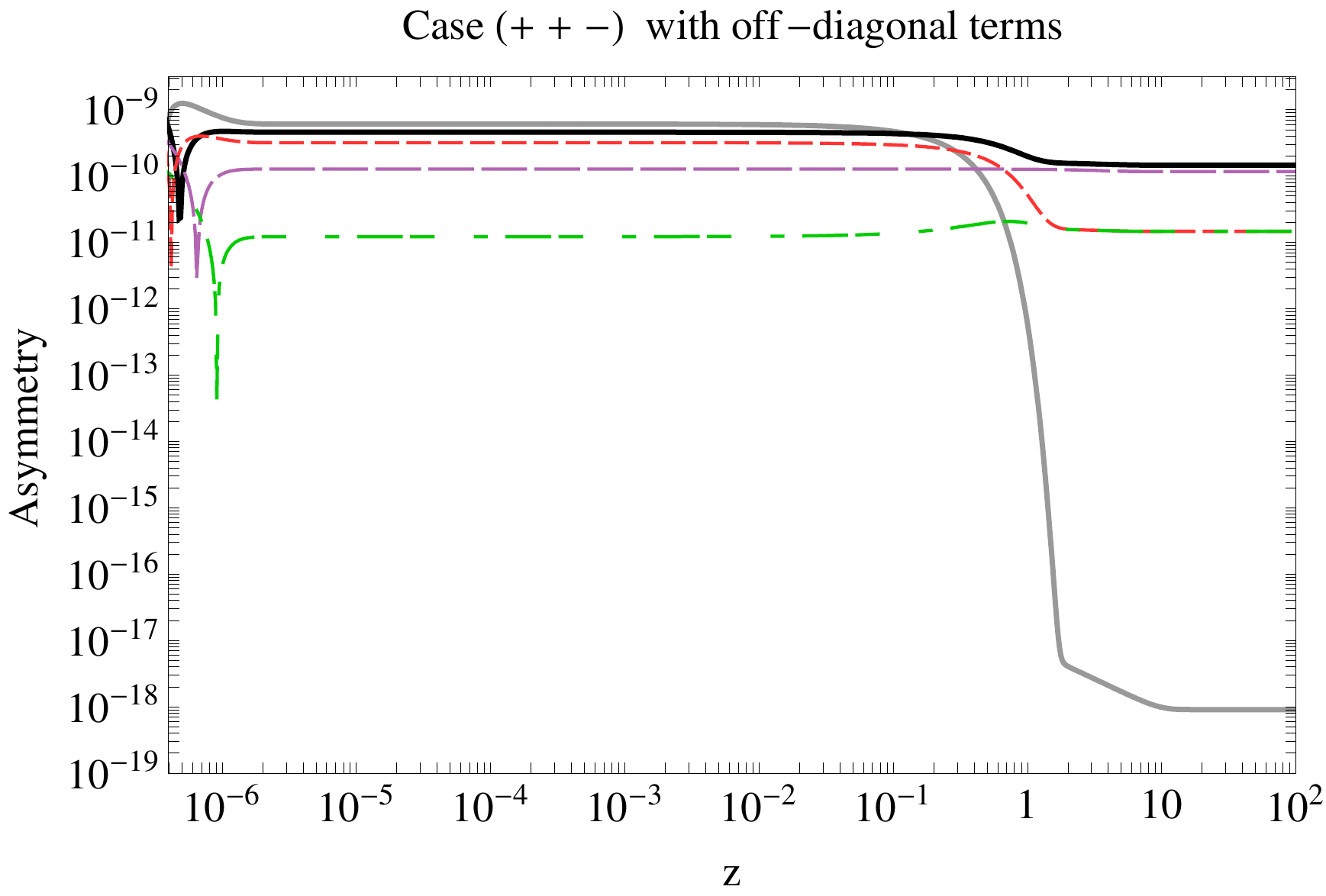}\quad
\includegraphics[width=9cm]{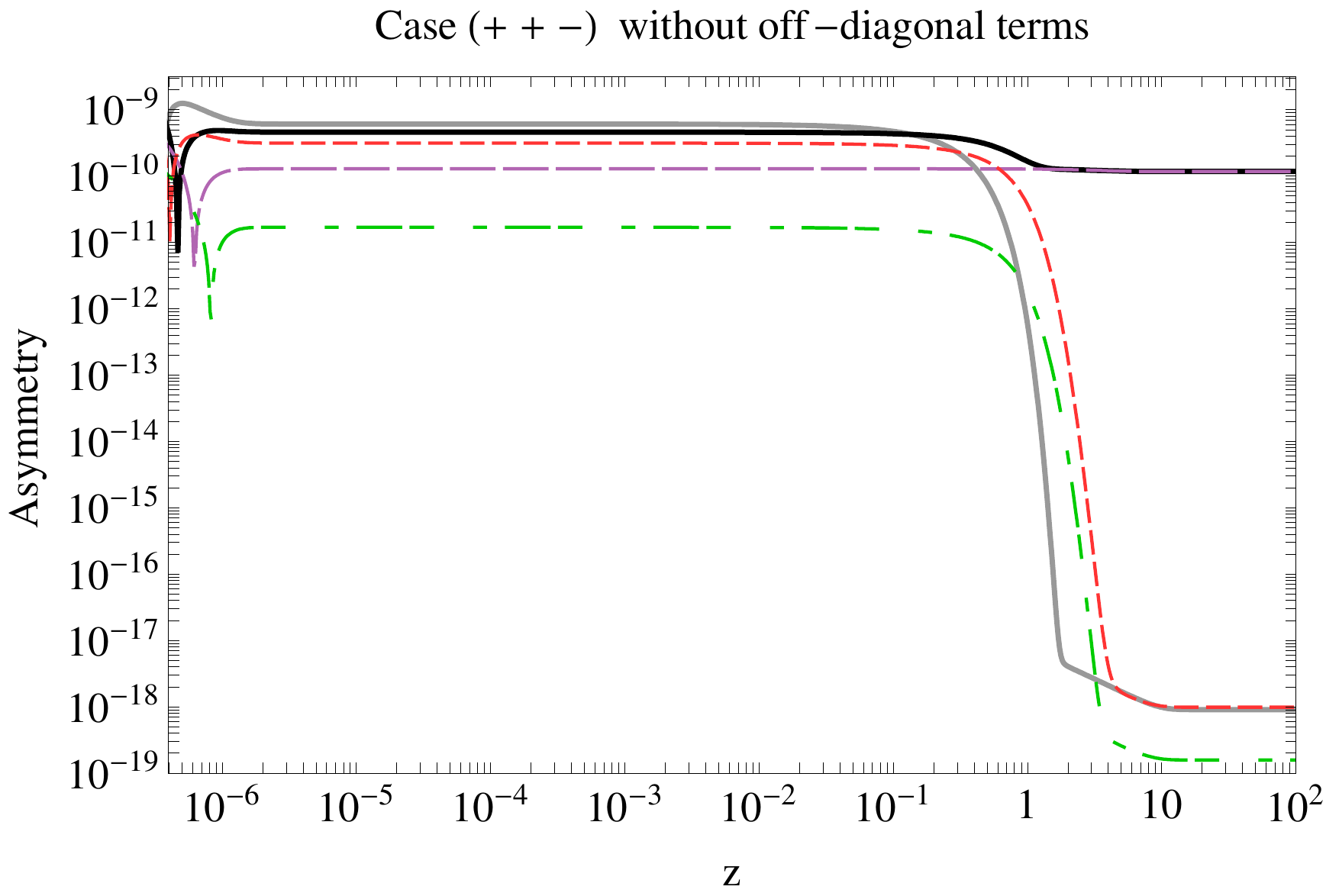}
\caption{Evolution of the asymmetries in the (+ + -) case, as a function of $z=M_1/T$. The thick black and grey lines represent the baryon asymmetry computed with and without flavour effect, respectively, highlighting the very effect of lepton flavours in leptogenesis. The thin lines represent the contribution to the baryon asymmetry of the different flavours:  electron (purple), muon (red) and tau (green). On the left panel we include  off-diagonal terms when solving the BE, whereas in comparison those terms are not included in the right panel. This clearly illustrates the muon and tau survival due to their non-diagonal couplings to the electron flavour.}
\label{GrapheN2leptocas112}
\end{figure}

As we just saw, for solutions $(+,+,-)$ and $(-,-,-)$ a huge difference exists between the flavoured and the unflavoured picture.\\
Assuming that all high energy phases and angles are set to zero, except for $\Phi_{2}^{u}=\pi/4$\footnote{The case considered here corresponds to a non-trivial $U_{m}$. For convenience, we rather choose the set 1 of $U_{m}$, defined in section 6.2.1. This choice only affects the quantitative result, and not the discussion. For example, while we obtain $Y_{B}\simeq 1.25\times 10^{-10}$, had we taken $U_{m}=1$  we would have obtained $Y_{B}\simeq 7.8\,\times 10^{-11}$, slightly below the observed value.}, we have for example the following $CP$ asymmetries and washout parameters at $v_R \simeq 10^{14} \GeV$: 
\begin{center}
\begin{tabular}{|c|c|c|c|c|c|c|}
\hline $(-,-,-)$ & $N_{1}, e$ & $N_{1}, \mu$ & $N_{1}, \tau$ & $N_{2}, e$ & $N_{2}, \mu$ & $N_{2}, \tau$ \\ 
\hline $\e_{i\al}$ & $1.1 \times 10^{-16}$ & $9.6 \times 10^{-15}$ & $5.8 \times 10^{-14}$ & -$1.2 \times 10^{-7}$ & -$6.4 \times 10^{-8}$ & -$3.4 \times 10^{-7}$ \\ 
\hline $\kappa_{i\al}$ & 0.04 & 17.2 & 16.2 & 2.3 & 0.7 & 2.7 \\ 
\hline \hline $(+,+,-)$ & $N_{1}, e$ & $N_{1}, \mu$ & $N_{1}, \tau$ & $N_{2}, e$ & $N_{2}, \mu$ & $N_{2}, \tau$ \\  
\hline $\e_{i\al}$ & $1.2 \times 10^{-16}$ & $9.7 \times 10^{-15}$ & $5.7 \times 10^{-14}$ & $7.0 \times 10^{-7 \,} $& $2.0 \times 10^{-7 \,}$ & $2.6 \times 10^{-6 \,}$ \\ 
\hline $\kappa_{i\al}$ & 0.04 & 17.2 & 16.2 & 0.5 & 0.2 & 3.5 \\ 
\hline 
\end{tabular} 
\end{center}
%We see that for both solutions, the inputs that govern leptogenesis are very similar: in both case, the $CP$ asymmetry in $N_{1}$ decays is very small, hence the contribution of $N_1$ to asymmetry production is negligible.
We see that in both solutions the washout parameters and the $CP$ asymmetries exhibit similar patterns. In both cases, the $CP$ asymmetry from $N_{1}$ decays is very small, and one can neglect the lepton asymmetry produced by $N_{1}$.
%Moreover, the values of the washout coming from $N_1$ interactions are quite similar and feature a crucial difference between an electron flavour very weakly washed out ($\kappa_{1e}\simeq 4\times 10^{-2}$) and $\mu-\tau$ flavours strongly washed out ($\kappa_{1\mu,\tau}\simeq 15$).
Moreover, in both cases $N_{1}$ interactions present important differences regarding  the washout of lepton flavours: while the electron flavour is very weakly washed out ($\kappa_{1e}\simeq 4\times 10^{-2}$), the  $\mu-\tau$ flavours are strongly washed out ($\kappa_{1\mu,\tau}\simeq 15$).\\
On the other hand, $e$, $\mu$ and $\tau$ couplings to $N_2$ are of the same order.\\
Hence, once $N_2$ has completely decayed, the three asymmetries $(L_e,L_\mu,L_\tau)$ are comparable, until $z\sim 1$, when $L_\mu$ and $L_\tau$ are strongly suppressed by $N_1$ washouts.
%Only the surviving $L_{e}$ will essentially contribute to the baryonic asymmetry, as well as prevent $\mu$ and $\tau$ asymmetries to be completely depleted, thanks the non-diagonal $(e-{\mu,\tau})$ elements of the $A_{\alpha\beta}$ matrix displayed in eq. \ref{bigA}.
However, since the electron flavour is not affected by these washouts, the baryon asymmetry generated by $N_{2}$ processes is not erased by  $N_{1}$ leptogenesis, opposed to what occurs in the single flavour approximation. 
The effects of $(L_{e},N_{1})$ orthogonality can be seen without ambiguity in figs.\ref{GrapheN2leptocas112}. Notice also the dramatic effect of flavour conversion on the muon and tau flavours, albeit this effect is only sub-leading for the baryon asymmetry.
\\
%Following the same reasoning as in~\cite{Vives}, one can easily evaluate quantitatively those effects. Given that right-handed masses are very hierarchical, we can treat $N_2$ and $N_1$ leptogenesis independantly. Here for illustration we consider only the $(+ + -)$ solution. The production of lepton asymmetries by $N_2$ leptogenesis, neglecting off-diagonal terms, is evaluated to be
Following~\cite{Vives}, we evaluate those effects in the illustrative $(+ + -)$ solution. Given that right-handed neutrino masses are very hierarchical, we can address $N_2$ and $N_1$ leptogenesis independently. Neglecting off-diagonal terms of the matrix $A$, after $N_{2}$ leptogenesis the asymmetries are given by:
\bea
\begin{array}{cc}
Y_{{\Delta e}_{\vert N_2}}\simeq -4\times 10^{-10}\ , &Y_{{\Delta \mu}_{\vert N_2}}\simeq -4\times 10^{-11} \ ,\\
Y_{{\Delta \tau}_{\vert N_2}}\simeq -1\times 10^{-9}\ ,&Y_{{L}_{\vert N_2}}\simeq 6\times 10^{-10} \ .\\
\end{array}
\nonumber
\eea
%\bea
%Y_{{\Delta e}_{\vert N_2}}&=& -\epsilon_{2,e}\eta_{2,e}\simeq -3.9 \times 10^{-3}\epsilon_{2,e}\times 0.3\,\kappa_{2,e}\simeq -3.8\times 10^{-10}&\quad&
%Y_{{\Delta \mu}_{\vert N_2}}&=& -\epsilon_{2,\mu}\eta_{2,\mu}\simeq -3.9 \times 10^{-3}\epsilon_{2,\mu}\times 0.3\,\kappa_{2,\mu}\simeq  -4.5\times 10^{-11} \nonumber \\
%Y_{{\Delta \tau}_{\vert N_2}}&=& -\epsilon_{2,\tau}\eta_{2,\tau}\simeq -3.9 \times 10^{-3}\epsilon_{2,\tau}\times 3.5\left(\frac{1}{6\kappa_{2,\tau}}\right)^{1.16}\simeq  -1\times 10^{-9} &\quad &
%Y_{{L}_{\vert N_2}}&=& \sum_{i=e,\mu,\tau}\epsilon_{2,i}\times\eta_{2}\simeq 2.1 \times 10^{-3}\epsilon_{2}\times 3.5\left(\frac{1}{6\kappa_{2}}\right)^{1.16}\simeq 6.4\times 10^{-10} \ .\nonumber 
%\eea
As the Universe cools down, the comoving number densities remain constant until $z\sim 1$, where $N_{1}$ washout processes  come into equilibrium, damping the asymmetries according to:
%$N_1$ processes become in-equilibrium at $z\sim 1 $.
%Given the smallness of the $CP$ asymmetries produced in $N_1$ decays, we can neglect their contribution and consider only the washout terms in the BE's:
\bea
Y_{\Delta \al}^{\prime}(z)=2\kappa_{1\al} A_{\al \al}W_{1}(z)Y_{\Delta \al}(z)+2\kappa_{1\al} \sum_{\be\neq\al}A_{\al \be}W_{1}(z)Y_{\Delta \be}(z) \ ,
\eea
%that can easily be solved. For the flavour $\alpha$, one finds:
and yielding the  formal solution
\bea
Y_{{\Delta \alpha}_{\vert N_1}}(z)&\simeq & Y_{{\Delta \alpha}_{\vert N_2}}\times e^{2\,A^{\al \al} \kappa_{1\alpha}\int_{zin}^{z}dx\  W_{1}(x)} \nonumber \\
&+&2\,\kappa_{1\alpha}\sum_{\be}A_{\al \be}\int_{z_{in}}^{z} dx\,W_{1}(x)Y_{\Delta \be}(x)\,e^{2\,A^{\al \al} \kappa_{1\alpha}\int_{x}^{z}dy \ W_{1}(y)} \ .
\eea
The first term corresponds to the depletion of the $Y_{\Delta \al}$ from $N_1$ washouts, whereas the second term represents  the effect of non-diagonal couplings between flavours. The contribution of the former to the final asymmetry turns out to be:
\bea
Y_{\Delta \alpha}^{d}\simeq Y_{{\Delta \alpha}_{\vert N_2}}\times e^{\frac{3 \pi}{4}\,A^{\al \al} \kappa_{1\alpha}} \ ,
\eea
and the second term, evaluated in~\cite{StudyOf}, gives a contribution which is:
\bea
Y_{\Delta \alpha}^{od}\simeq \sum_{\be\neq \al} Y_{\Delta \be}^{d}\times A_{\al \be} \frac{\kappa_{1\al}}{1+0.8 (-A_{\al \al} \kappa_{1\al})^{1.17}} \ .
\eea
The above formulae show that only the electron flavour survives $N_1$ washouts, while muon and tau flavours are suppressed by a factor $\simeq 10^{-6}$, illustrated on the right panel of fig.\ref{GrapheN2leptocas112}. 
The baryon asymmetry, produced from the electron flavour, is:
\bea
Y_{B}\simeq\frac{10}{31}\sum_{\al} Y_{\Delta \al}\simeq \frac{10}{31} \,1.19\,Y_{{\Delta e}_{\vert N_2}}\simeq -1.25\times 10^{-10} \ ,
\eea
\noindent
which is slightly above the observed value (notice the excellent agreement with the numerical result $Y_{B}\simeq-1.45\times 10^{-10}$).\\
In the one flavour approximation, $N_{1}$ washouts exponentially suppress  the total lepton asymmetry, as seen in fig.\ref{GrapheN2leptocas112}:
\bea
Y_{L}\simeq Y_{L_{\vert N_2}}\times e^{-\frac{3 \pi}{4}\kappa_{1}} .
\eea
\noindent \\
We further notice the effect of the off-diagonal couplings 
which prevent the total depletion of the $\mu,\tau$ asymmetries:
\bea
Y_{\Delta \mu (\tau)}^{od}\simeq 0.12\,Y_{\Delta e}\simeq 0.12 Y_{{\Delta e}_{\vert N_2}}\,e^{\frac{3 \pi}{4}A_{ee}\kappa_{1e}}\simeq -4.4\times 10^{-11} \ .
\eea
This effect is spectacular for the individual lepton asymmetries, but only marginal for the baryon asymmetry.
\\
This example clearly illustrates how the orthogonality of the leptonic directions and the decay of $N_1$ and $N_2$ in flavour space  allow $N_2$ leptogenesis to survive the washout of $N_1$.\\
In the following, we do not pursue the discussion of the unflavoured picture.
\subsection{Influence of mass corrections}
\begin{figure}[h!]
\hspace{-0.7cm}
\includegraphics[scale=0.6]{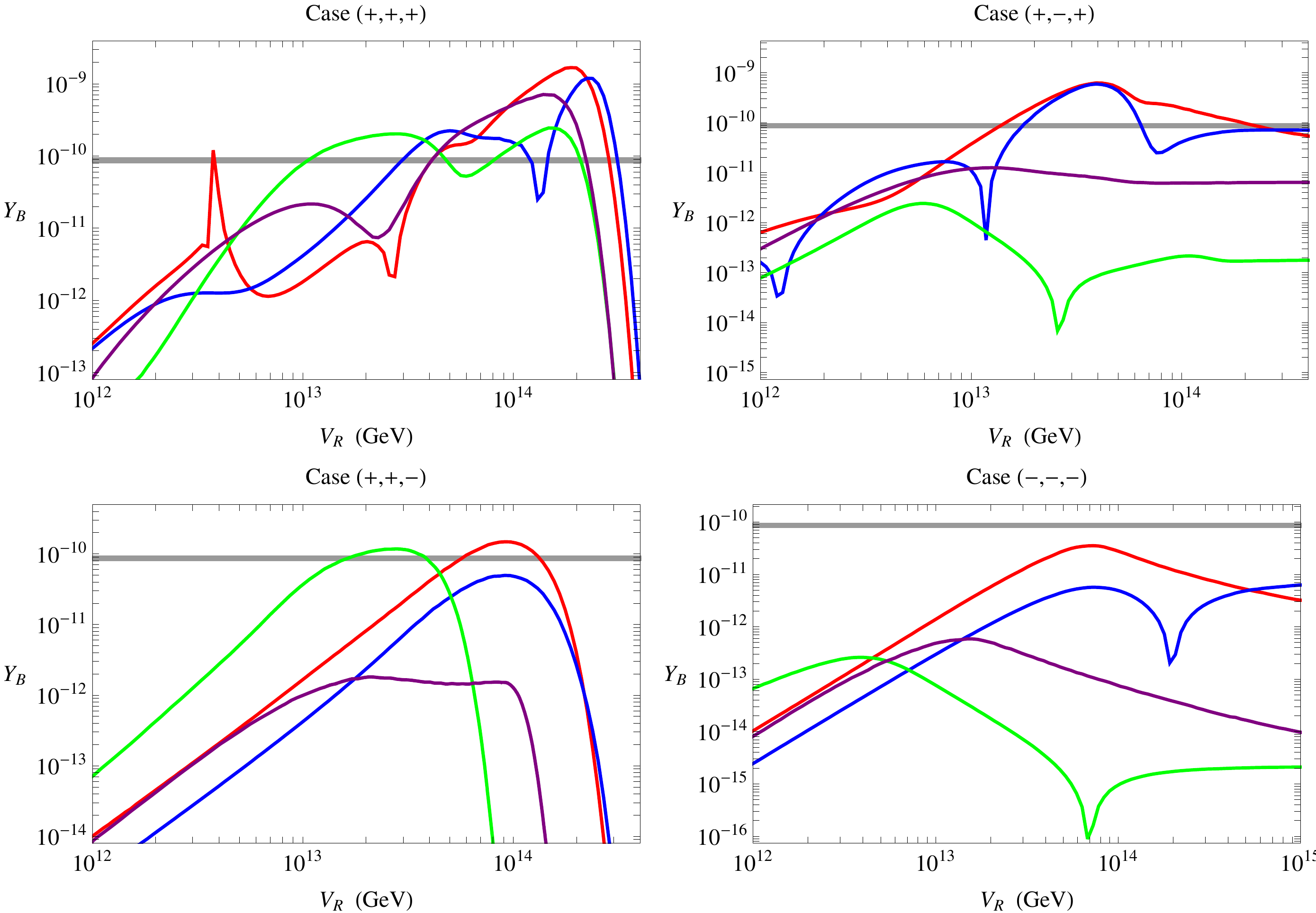}
 \caption{Evolution of the baryon asymmetry as a function of the $B-L$ breaking scale $v_{R}$, for the four characteristic solutions, and for the four sets of matrix $U_{m}$ defined in section 6.2.1. The different colours represent the different sets, with sets 1,2,3 and 4 depicted in red, blue, green and purple respectively.} 
\label{GrapheUMYBALL}
\end{figure}

\begin{figure}[h!]
\hspace{-0.7cm}
\includegraphics[scale=0.6]{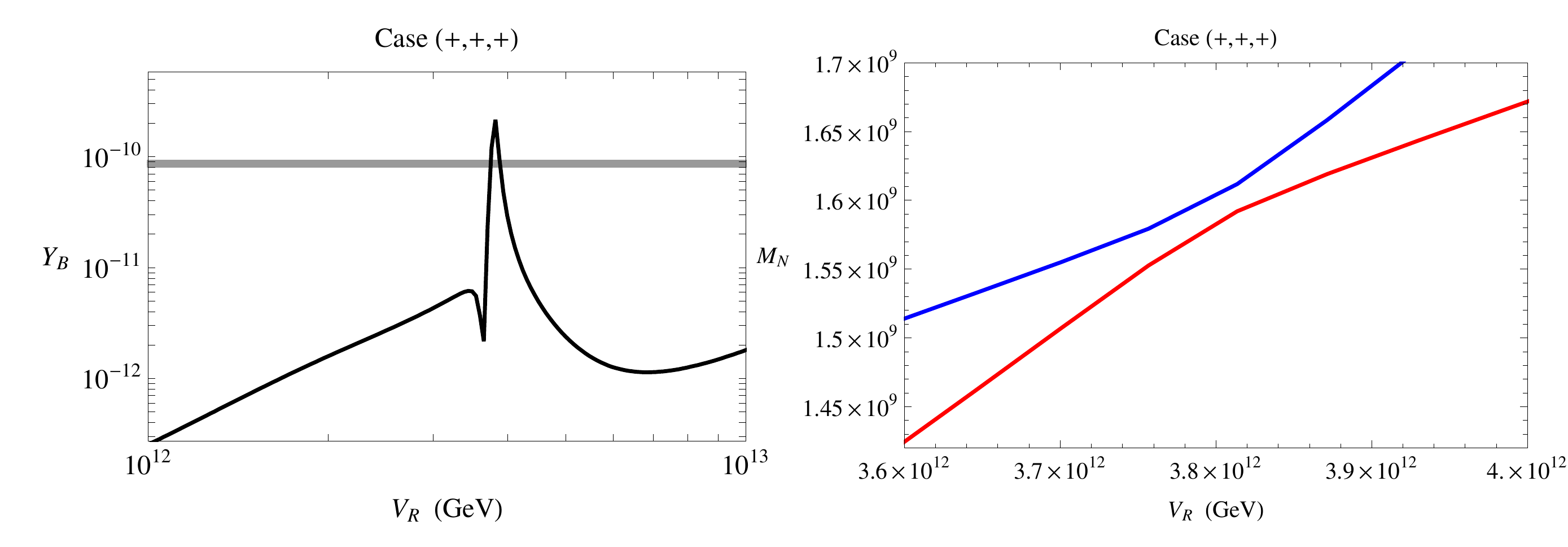}
 \caption{Zoom on the quasi-degenerate region of the $(+,+,+)$ solution depicted in red in fig.\ref{GrapheUMYBALL}. On the right-hand side, the plot shows the evolution of $M_{1}$ and $M_{2}$, which become quasi-degenerate around $v_{R}\simeq 3.8\times10^{12}\GeV$. This quasi-degeneracy implies a resonance of the baryon asymmetry, as shown on the left-hand side.} 
\label{GrapheRes111}
\end{figure}

\begin{figure}[h!]
\hspace{-0.7cm}
\includegraphics[scale=0.6]{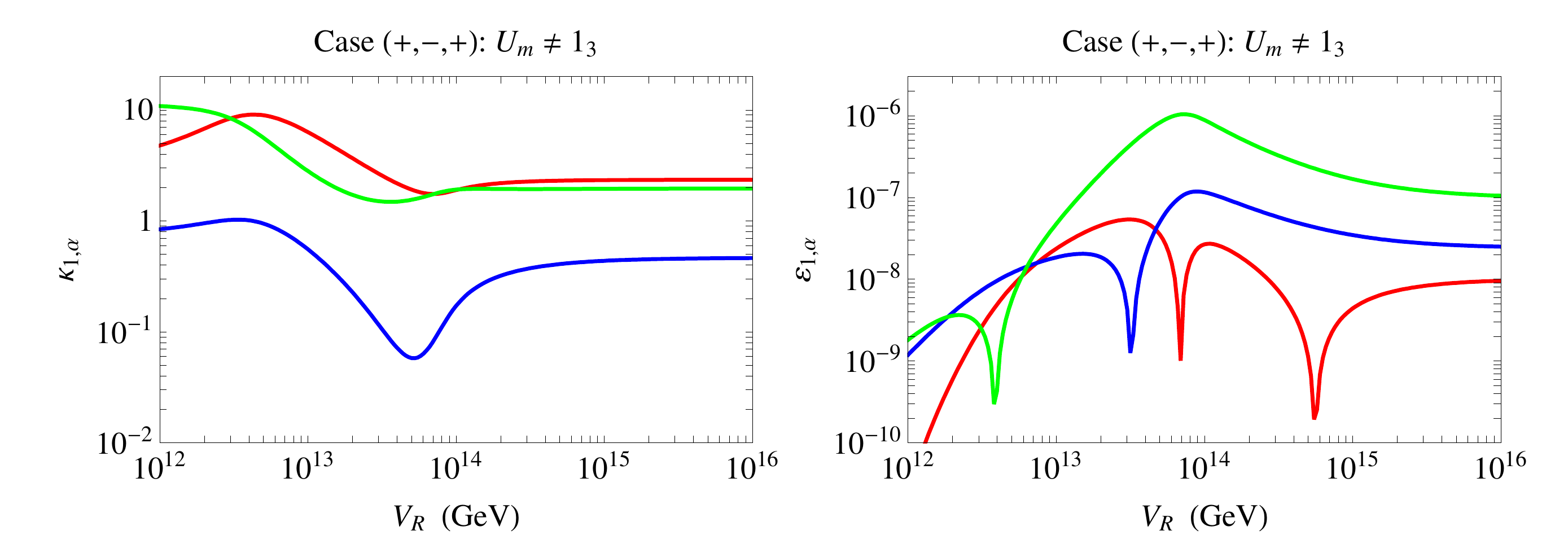}
 \caption{$CP$ asymmetries and washout factors for the $(+,-,+)$ solution depicted in red fig.\ref{GrapheUMYBALL}. Colour code as in fig.\ref{Grapheek}. } 
\label{GrapheUM121EK}
\end{figure}

Let us now examine the influence of the correction to the relation $m_{e}=m_{d}$. This correction entails a redefinition of the down-quark Yukawa coupling $\la_{d}\Ra U_{m}^{*}\,\la_{d}\,U_{m}^{\dagger}$ through the matrix $U_{m}$ which diagonalises the charged lepton masses. Going to the basis where down-quark masses are diagonal, the neutrino Yukawa couplings are affected by $U_{m}$, as is the right-handed neutrino spectrum, as well as the $CP$ asymmetries and washout factors.
Using the four sets of $U_{m}$ defined in section 6.2.1, we plot in fig.\ref{GrapheUMYBALL} the baryon asymmetry as a function of $v_{R}$. We do not analyse the different $U_{m}$ separately, but just remark their main features.
\\
The first thing to notice is that for solutions  $(+,+,\pm)$, the qualitative picture remains unchanged: $Y_{B}$ increases with $v_{R}$, until the right-handed neutrino becomes too heavy and thus its production is suppressed, accounting for the fall of $Y_{B}$ at high $v_{R}$. If we look more carefully, we see that these two solutions are enhanced, especially for the set 1 (in red), for which both two solutions reach the observed value.\\
Interestingly, for the $(+,+,+)$ case, $N_{1}$ and $N_{2}$ become partially degenerate around $v_{R}\simeq 4\times 10^{12} \GeV$, leading to a resonance of the baryon asymmetry, as can be seen from the peak in $Y_{B}$ for this solution. This resonance is shown in fig.\ref{GrapheRes111}, where we plot the baryon asymmetry as a function of $v_{R}$ and the corresponding right-handed neutrino masses.
\\
Nevertheless, this degeneracy is accidental and is not a generic feature of this model. Moreover, relying on resonances to obtain a sufficiently large baryon asymmetry, while $M_{i}\simeq 10^{9}\GeV$ may appear as extremely fine-tuned, and so we no longer consider this possibility.\\
A more robust and welcome consequence of having included the mass correction stems from the success of the $(+,-,+)$ solution, for which the observed baryon asymmetry  is reached for $10^{13}\GeV\lesssim v_{R}\lesssim 10^{14}\GeV$.\\
Consider for example the solution corresponding to the first set of $U_{m}$ (in red) of fig.\ref{GrapheUMYBALL},   which has the same high-energy phases $\phi_{u}^{2}=\pi/4$ of that in fig.\ref{GrapheYBALL}. For $v_{R}\simeq 10^{13-14}\GeV$, we see that the baryon asymmetry is enhanced by two orders of magnitude when the mass correction is included.\\
This comes from the very different behaviours of the $CP$ asymmetries and washout factors, as can be seen by comparing figs.\ref{Grapheek} and \ref{GrapheUM121EK}. In the former case the maximum $CP$ asymmetry is  reached for $v_{R}\simeq2\times 10^{13}\GeV$, and is $\eps_{1\tau}\simeq 10^{-7}$ with a washout $\kappa_{1\tau}\simeq 20$. In the latter, when the mass correction is included, the $CP$ asymmetry is comparatively enhanced by one order of magnitude, while the washout factor is reduced by one order of magnitude.
\\
The enhancement of the baryon asymmetry coming from the inclusion of the mass correction can be analytically explained from the influence of $\theta_{12}^{m}$ of the matrix $U_{m}$. Indeed, we roughly have that
\bea
M_{1}&\propto & v_{R}\left( f_{1}\,\vert \rm{cos}(\theta_{13}^{m}) \rm{cos}(\theta_{12}^{m})\vert^{2}+f_{2}\,\vert \rm{cos}(\theta_{13}^{m}) \rm{sin}(\theta_{12}^{m})\vert^{2}+... \right)\, ,\nonumber \\
M_{2}&\propto & v_{R}\left( f_{1}\,\vert \rm{cos}(\theta_{23}^{m}) \rm{sin}(\theta_{12}^{m})\vert^{2}+f_{2}\,\vert \rm{cos}(\theta_{23}^{m}) \rm{cos}(\theta_{12}^{m})\vert^{2}+... \right)\,  , 
\eea
where the dot stands for terms of the same order for $M_{1}$ and $M_{2}$. Hence, all other parameters being fixed, varying $\theta_{12}^{m}$ modifies the relative ($f_{1}-f_{2}$) weights, and so potentially increases $M_{1}$, which was previously slightly below the type I bound (cf. fig.\ref{GrapheMsol}).\\
In the following, we will only consider the favourable case which is depicted in red in fig.\ref{GrapheUMYBALL}.
\subsection{Dependence on the reheating temperature}
\begin{figure}[h!]
\begin{center}
\includegraphics[scale=0.4]{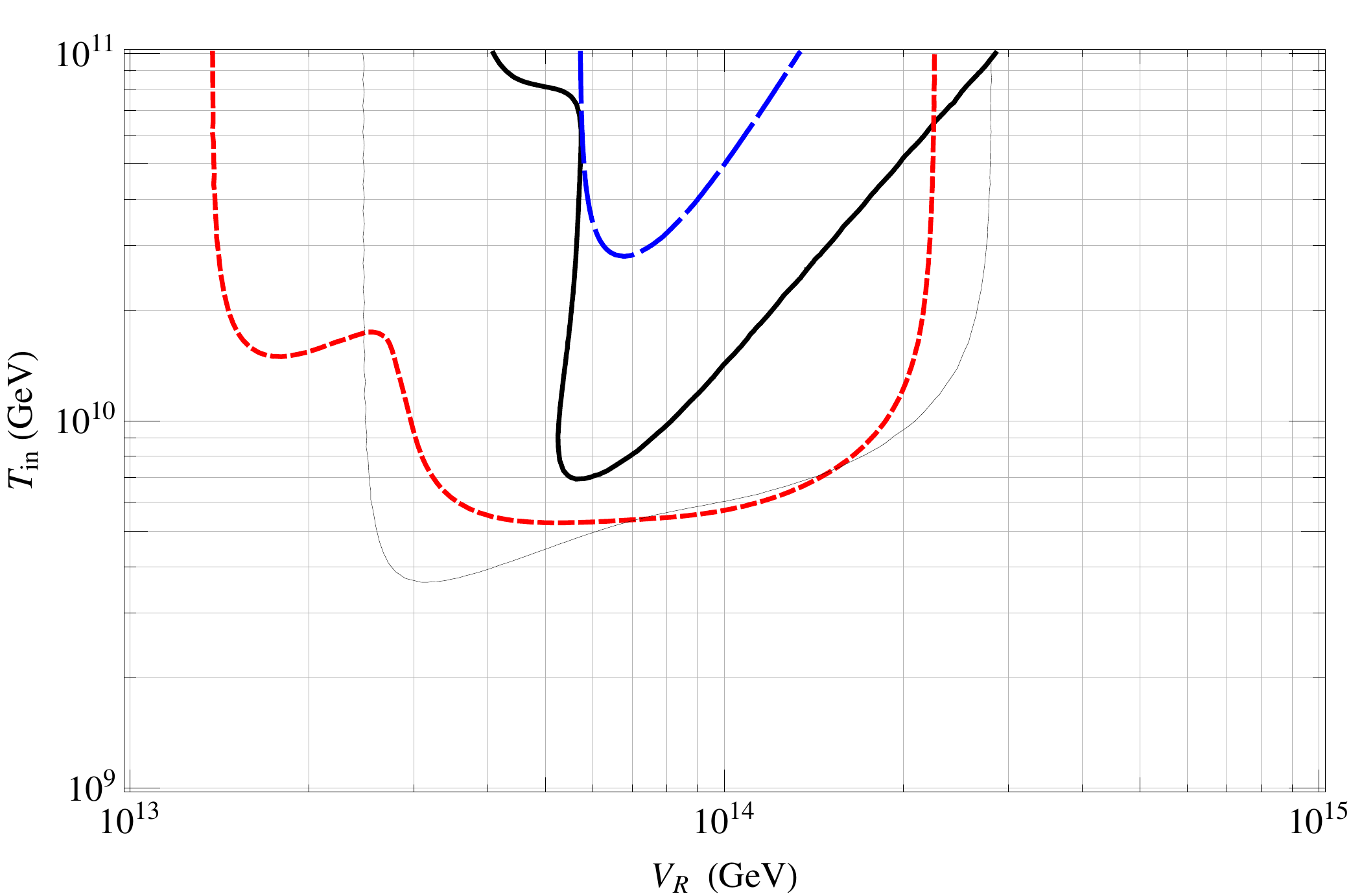}
\caption{Contour lines of the ratio of the computed baryon asymmetry over the observed one as function of  $v_R$ and $T_{in}$, for the different characteristic solutions: solution $(+ + +)$ is depicted in thick black line, solution $(+ + -)$ is in thick long-dashed blue, whereas solution $(+ - +)$ is in thick dashed red. For these three solutions, the contours represent a unit ratio, and inside the contour we have $Y_{B}^{num}/Y_{B}^{obs}\gtrsim 1$. We have depicted in a thin black line the solution $(- - -)$ for which the observed bound is not reached, and so for this solution we plotted the ratio $Y_{B}^{num}/Y_{B}^{obs}=0.1$. }
\label{TRH}
\end{center}
\end{figure}

Let us now discuss the important question in thermal leptogenesis of the lower bound on $M_{i}$, or equivalently, the lowest possible reheating temperature.
In the type I seesaw, a bound is derived on $M_{1}$, namely $M_{1}\gtrsim 2\times 10^{9}\GeV$, which follows from the constraint of having a large enough $CP$ asymmetry. This bound implies a lower bound on $T_{RH}$, which must be $\gtrsim 2\times 10^{9}\GeV$. In the type II seesaw, assuming strong hierarchy among right-handed neutrinos, the individual $CP$ asymmetries are bounded by~\cite{boundII}:
\bea
\eps_{1,\al}\lesssim \frac{3}{8\pi}\frac{M_{1}\,m_{\nu}^{\rm{max}}}{v_{u}^{2}} \, .
\eea
According to the mass pattern for the RHn we have encountered, this relation is not likely to  apply, since either the neutrino that dominantly contributes to the asymmetry production is $N_{2}$, with $M_{1}\ll M_{2}\ll M_{3}$, or else the main contribution comes from $N_{1}$, but with a mild hierarchy between $N_{1}$ and $N_{2}$.\\
Nevertheless, we partially answer this question with fig.\ref{TRH}, where, for the four solutions represented  in the $T_{RH}-v_{R}$ parameter-space, we display the limits above which the computed baryon asymmetry at least equals the observed one (except for the $(-,-,-)$ case, cf. figure caption). In order to derive this plot, we fix all other parameters, with $m_{1}=10^{-3}\eV$, and choose the first set for $U_{m}$, with $\Phi_{2}^{u}=\pi/4$, the other phases as well as $\theta_{13}$ being set to zero. 
\\
The first thing to notice is the failure of $(-,-,-)$ to produce enough baryon asymmetry.\\
Solution $(+,+,-)$ works for $T_{RH}\gtrsim 5\times 10^{10}\GeV$; for lower temperatures, $N_{2}$ becomes heavier than $T_{RH}$ and its production is Boltzmann suppressed.\\
Particularly interesting are the solutions $(+,\pm,+)$, for which the reheating temperature can be brought down to more reasonable values: the lower bound derived in these cases is $T_{RH}\gtrsim 5\times10^{9}\GeV$.\\
However, we again want to stress that we do not scan over the entire parameter space, but restrict ourselves to the case where only one high energy $CP$ violating phase is non-zero. As a consequence, the lower bound on $T_{RH}$ should  be seen only as a rough approximation of the true lower  bound.
\section{Dependence on the low energy parameters}
As  stated many times before, the conclusion of this study crucially depends on the different input parameters. The high energy parameter-space, which is 10 dimensional, exceeds our computation ability, and furthermore is not implied in low-energy experiments. On the other side, $m_{\nu}$, $\theta_{13}$ and $\delta)$ are very important low-energy parameters which are involved, for example, in a possible violation of $CP$ in neutrino oscillations. In this section we numerically investigate their influence.
\subsection{Dependence on the light-neutrino mass}
\begin{figure}[h!]
\begin{center}
\includegraphics[scale=0.5]{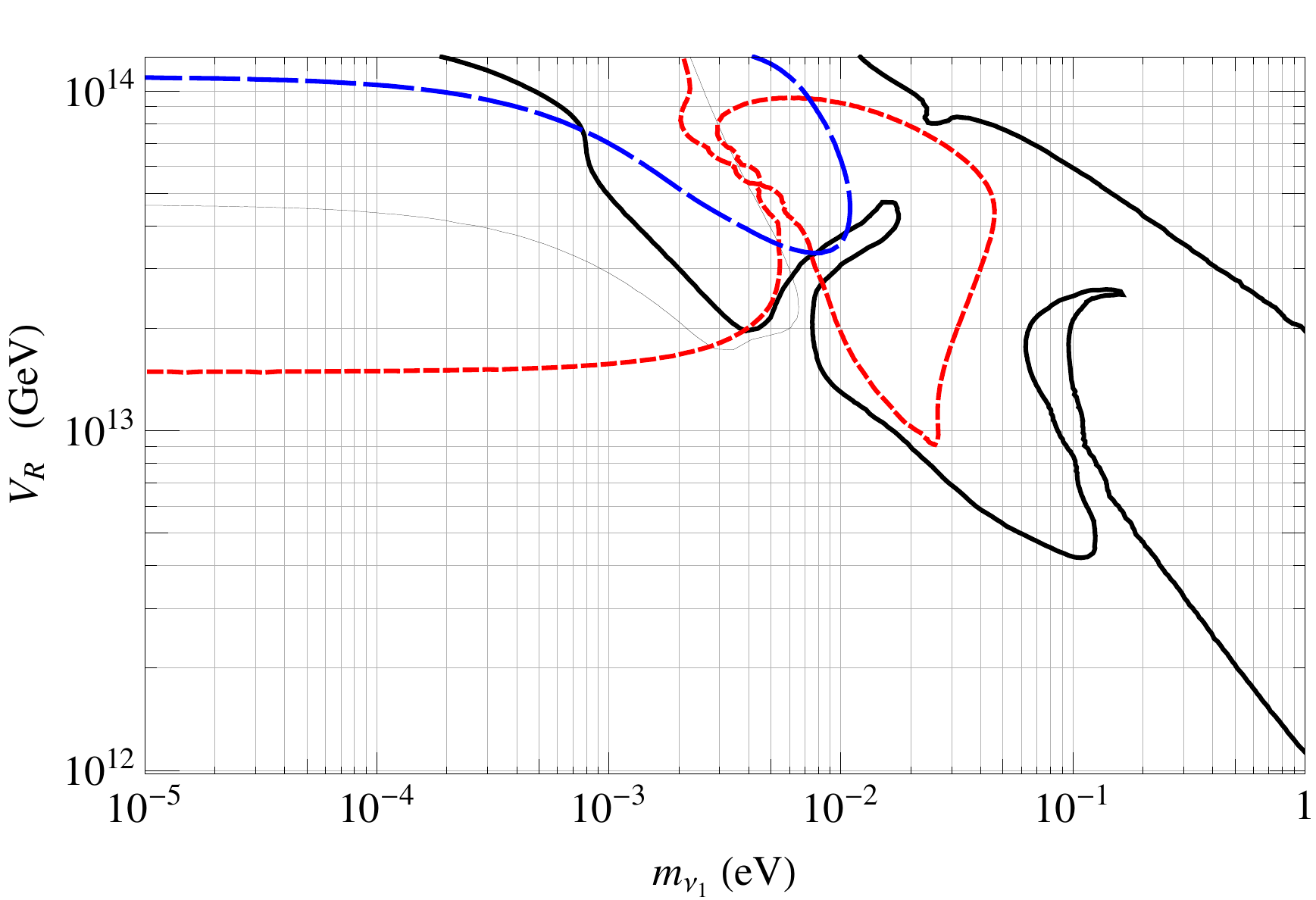} 
\caption{Contour lines of the ratio of the computed baryon asymmetry over the observed one, as functions of $v_R$ and light neutrino mass scale $m_{1}$. The ratio $Y_{B}^{num}/Y_{B}^{obs}=1$ is fulfilled for the three solutions: $(+ + +)$  depicted in thick black line, $(+ + -)$ in thick long-dashed blue, and  $(+ - +)$ in thick dashed red. Inside the contours we have $Y_{B}^{num}/Y_{B}^{obs}\gtrsim 1$. In the case of solution $(- - -)$, for which the observed bound cannot be reached, we have depicted in thin black line the ratio $Y_{B}^{num}/Y_{B}^{obs}=0.1$. }
\label{VRm1}
\end{center}
\end{figure}

We first examine the influence of the light neutrino mass scale, assuming a normal mass ordering. As can be seen from fig.\ref{GrapheFpert}, where we plot the $f_{i}$s for $m_{1}=0.1\eV$ and $10^{-5}\eV$, increasing $m_{1}$ raises the right-handed neutrino masses corresponding to a $"+"$ solution, while conversely decreasing its value lowers masses which correspond to a $"-"$ solution. This behaviour can be seen in the contour plot of fig.\ref{VRm1}, where we show for our characteristic solutions the region in the $(m_{\nu_{1}}=m_{1},v_{R})$ plane where enough baryon asymmetry is created. We choose $T_{RH}=10^{11}\GeV$, and the second set for $U_{m}-\Phi_{2}^{u}$.
\\
We notice the strong influence of $m_{1}$ on the different solutions.\\
Furthermore, even if solution $(-,-,-)$ fails to produce enough baryon asymmetry, its behaviour is similar to the one of $(+,+,-)$, as expected. These two solutions rely on the decays of $N_{2}$, but also strongly on the  washouts from inverse decays producing $N_{1}$, as we saw in the previous section. The fact is that when $m_{1}$ increases, $\kappa_{1,\al}$s also increase, hence the asymmetry produced by $N_{2}$ leptogenesis will be erased by subsequent $N_{1}$ processes. For these solutions, we have an upper-bound on $m_{1}$, namely $m_{1}\lesssim 2\times 10^{-2}\eV$.\\
In the $(+,-,+)$ case, when  $m_{1}$ increases, $M_{1}$, which corresponds to the $"-"$ solution decreases, dropping below $10^{9}\GeV$. This could be somewhat compensated with the increase of $M_{2}$, but  since $M_{2}\gtrsim 10^{11}$, its contribution is suppressed. For this solution, a similar bound on $m_{1}$ is derived, namely $m_{1}\lesssim 5\times 10^{-2}\eV$, with $v_{R}\gtrsim 10^{13}\GeV$.\\
In the $(+,+,+)$ case, increasing $m_{1}$ increases $M_{1}$, and consequently the efficiency of leptogenesis, until $M_{1}\gtrsim (2-3)\times T_{RH}$ and $Y_{B}$ drops. To avoid this decrease in $Y_{B}$, lower values of $M_{1}$ are required, and so lower values of $v_{R}$.
\subsection{Dependence on low-energy mixing angles and phases.}
\begin{figure}[h!]
\begin{center}
\includegraphics[scale=0.45]{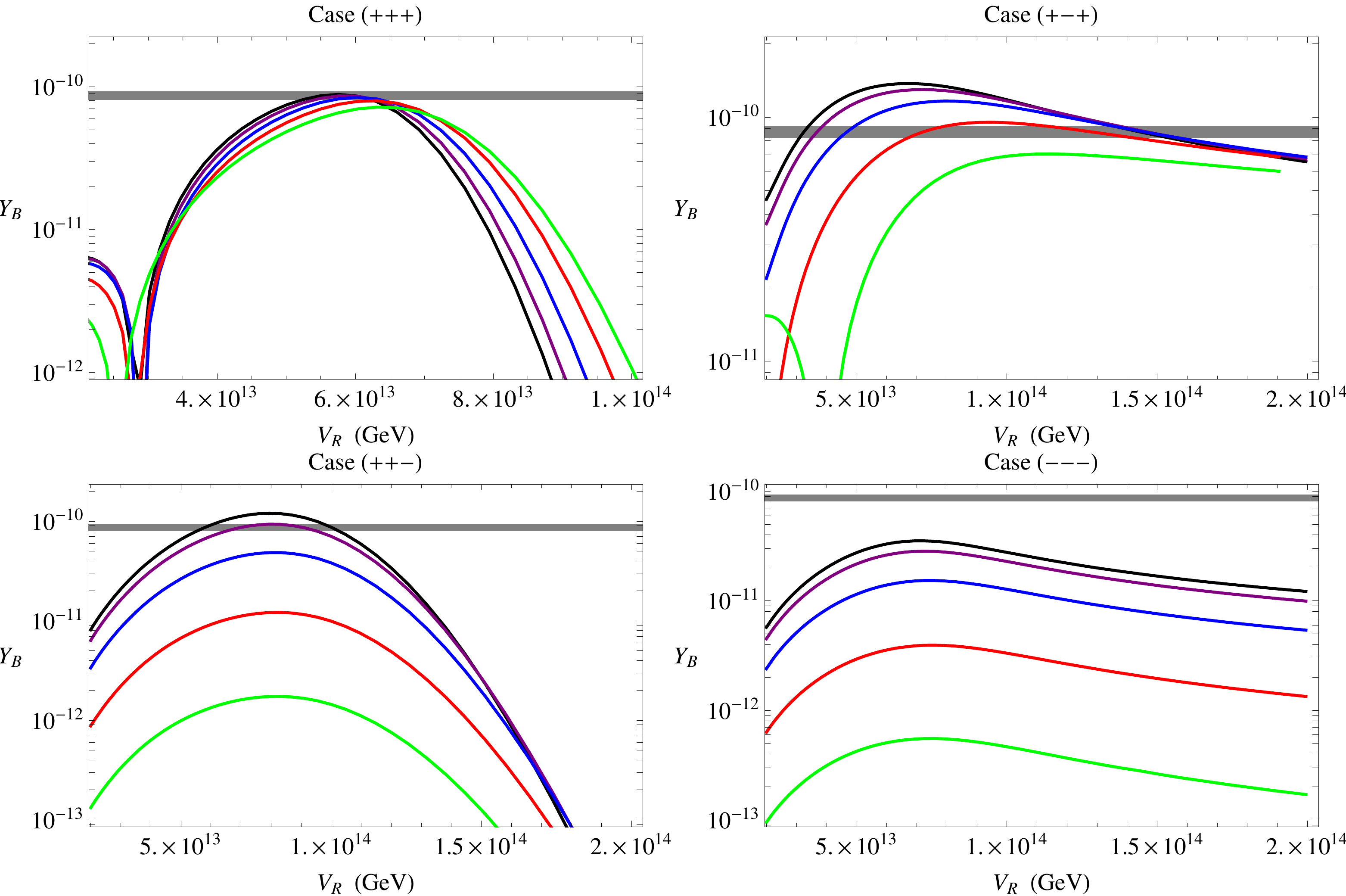} 
\caption{Baryon asymmetry as a function of $v_{R}$, for the four typical solutions. In these plots are displayed $Y_{B}$ for different values of $\theta_{13}$, taking $\delta =0$. We represent in black, purple, blue, red and green the values $\theta_{13}=0^{\circ},2^{\circ},5^{\circ},9^{\circ},13^{\circ}$, respectively. For the top panel, we choose $T_{RH}=7\times 10^{9}\GeV$, while for the bottom panel $T_{RH}=5\times 10^{10}\GeV$.}
\label{GrapheDeltaVr}
\end{center}
\end{figure}

\begin{figure}[h!]
\begin{center}
\includegraphics[scale=0.45]{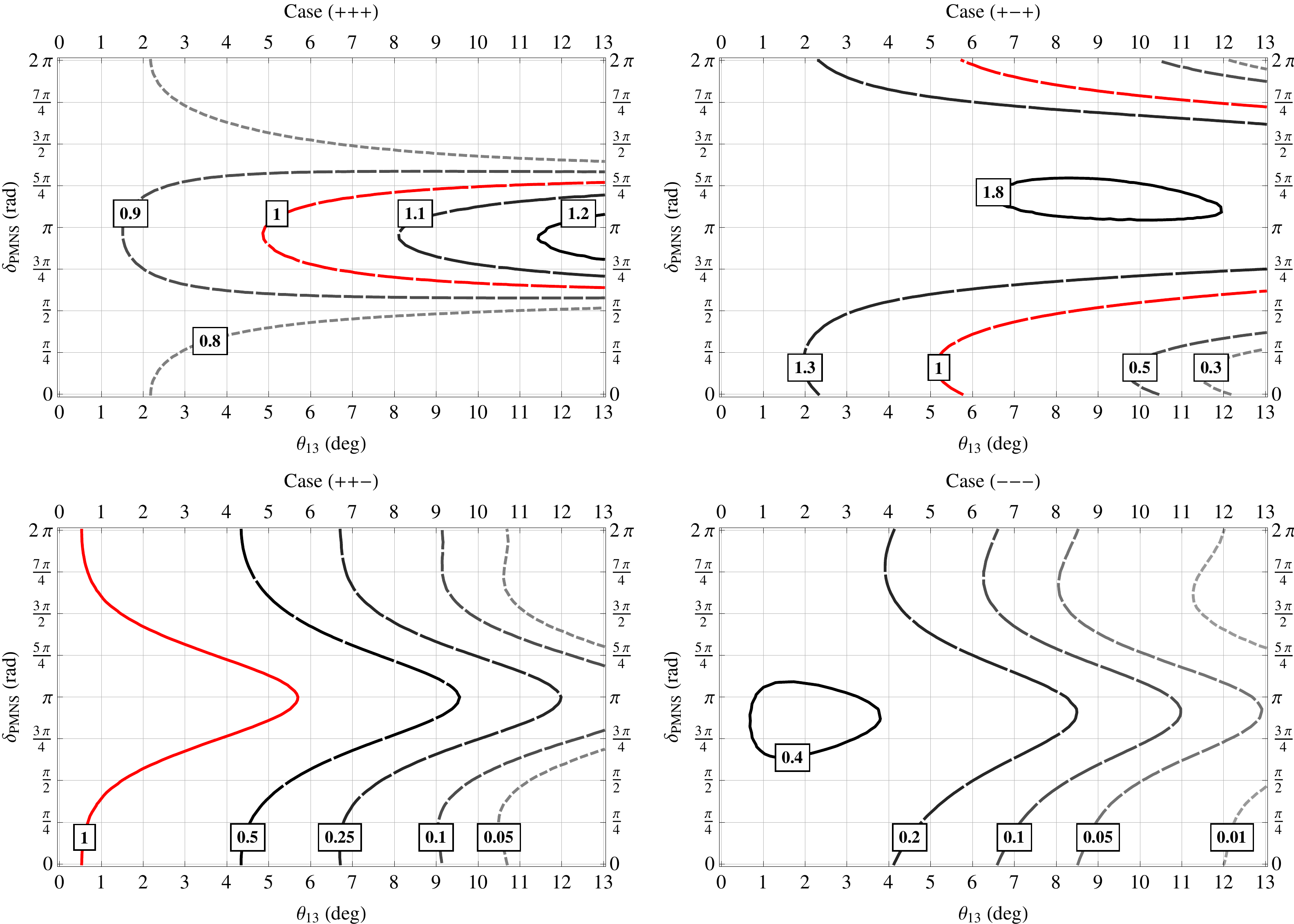}
\caption{Contour lines of the ratio of the computed baryon asymmetry over observed one, as a function of the low-energy parameters $\delta$ and $\theta_{13}$. We depicted in red lines the unit ratio, if reached.}
\label{GrapheThetaDelta}
\end{center}
\end{figure}

Let us now discuss the influence of the low-energy CP violating phase $\delta$ and mixing angle $\theta_{13}$. These two quantities are crucial for low energy neutrino physics, since $\theta_{13}=0^{\circ}$ and $\delta=0,\pi$ would imply that $CP$ is conserved in neutrino oscillations~\cite{CPosci}.\\
In order to derive constraints, we choose $T_{RH}=7\times10^{9}\GeV$ for $(+,\pm,+)$, and $T_{RH}=5\times 10^{10}\GeV$ for the remaining solutions, according to fig.\ref{TRH}. We further take $m_{1}=10^{-3}\eV$, since this value suits all solutions.\\
We first consider in fig.\ref{GrapheDeltaVr} the influence of $\theta_{13}$ on $Y_{B}$, when $v_{R}$ varies, with a fixed $CP$ phase $\delta=0$.
\\
We see that the influence of $\theta_{13}$ on $Y_{B}$ strongly depends on the type of solutions.\\
For all the solutions, we notice that for  $\delta=0$, $\theta_{13}=0$ gives the best results.
For the $(\pm,\pm,-)$ solutions, we observe a huge dependence on $\theta_{13}$: $Y_{B}^{max}$ decreases by two orders of magnitude when $\theta_{13}$ goes from $0^{\circ}$ to its upper-bound, $13^{\circ}$. We can consequently say that $\delta=0$ favours $\theta_{13}=0^{\circ}$. What about non-vanishing values of $\delta$?\\
To study the influence of $\delta$, we fix $v_{R}=6\times10^{13}\GeV$ for $(+,\pm,+)$ and $v_{R}=5\times 10^{13}\GeV$ for $(\pm,\pm,-)$, keeping the values of $T_{RH}$ quoted above; we then plot in fig.\ref{GrapheThetaDelta} $Y_{B}$ in the $\theta_{13}-\delta$ plane.
\\
We see that changing these parameters does not improve the situation for $(-,-,-)$, which still fails to produce enough baryon asymmetry. However, we notice that its behaviour is very similar to the $(+,+,-)$ case, for which the observed bound can be reached: in both cases, the baryon asymmetry is larger for small values of $\theta_{13}$, with a factor $\sim 100$ between $\theta_{13}=0^{\circ}$ and $\theta_{13}=13^{\circ}$. Moreover, we clearly see a preference of these solutions for $\delta\simeq \pi$, for which higher values of $\theta_{13}$ are possible. Actually $\delta=\pi$ is the most favourable case for $(+,+,-)$, while $(-,-,-)$ seems to favour $\delta\simeq \pi-\pi/8$.\\
This preference for $\delta\simeq \pi$ is also observed for the $(+,\pm,+)$ solutions, but in these cases non-zero values of $\theta_{13}$ maximise $Y_{B}$.\\
We can conclude that, for the choice of parameters made, our model points towards non-zero values of $\delta$, in fact values close to $\pi$. Hence it implies small $CP$ violation in oscillation experiments~\cite{CPosci}.\\
However, we stress once again that this conclusion is not general, being based on a partial scan of the 17-dimensionnal parameter space.
\section{Summary}
In this chapter we have studied leptogenesis in the type II seesaw. This model has the interesting feature of containing two sources of lepton number violation, originating from the couplings with the scalar $SU(2)_{L}$ triplet and with the right-handed neutrinos.\\
Moreover, our model is embedded in a SUSY-GUT framework: it has the advantage of constraining the neutrino Yukawa couplings, the ignorance of which is one of the weakness of the model we studied in the previous chapter. Furthermore, the fact that we assume a Left-Right symmetry enables us to invert the seesaw relation, and so to characterise right-handed neutrinos in terms of light neutrino and up-type quark properties. For the RHn spectrum, 8 solutions are found, which can be grouped in four pairs according to their different properties. We focus on one solution per pair, which exhibits either the dominance of one (pure) type of seesaw, for $(+,+,+)$ and $(-,-,-)$, or else an interesting mixing between type I and pure type II, for $(+,-,+)$ and $(+,+,-)$. We thus discriminate among this eightfold degeneracy with the requirement that leptogenesis is viable.\\
The leptogenesis criterion has already been used by the authors of~\cite{HLS}, but as they already stressed, the study was far from being definitive, mainly for two reasons. Firstly, the treatment of leptogenesis was not correct, since it was based on the single flavour picture. Secondly, the fact that only a pair of $\bf{10}$ Higgs representations had been included to give mass to light particles, imposed the relation $m_{e}=m_{d}$ at the GUT scale, which clearly does not hold. Therefore in this chapter, which is based on~\cite{AHJML}, we improved the study regarding these two points.\\
Thus, we have included the effects of lepton flavours, inclusion which is not only required~\cite{issues}, but instrumental~\cite{matters}. This is especially true in the case we considered here, where usually the total washout $K_{1}$ is strong but the individual ones are not. We further included heavy neutrino flavours, at least partially, since the heaviest RHn is usually much heavier than the other two RHns. This inclusion follows the fact that in the $(\pm,\pm,-)$ cases, while $N_{1}$ is too light to be of use for leptogenesis, $N_{2}$ is of the good order of magnitude.\\
We found that indeed, adding both light and heavy flavours, improves the situation. The most spectacular example is the $(+,+,-)$ solution, where the asymmetry is produced due to $N_{2}$ decays, and the inclusion of lepton flavours prevents this asymmetry from being completely washed-out by $N_{1}$ processes, as it would be the case in the unflavoured picture.\\
Moreover, by adding in an effective manner a non-renormalisable term involving a $\bf{45}$ Higgs representation, which is thought to be present in typical $SO(10)$-based GUT, the correction of the GUT relation $m_{e}=m_{d}$ constitutes a clear improvement of the model towards realistic fermion masses. This correction implies a redefinition of the lepton right handed neutrino Yukawa couplings $\la$, which turns out to modify the right-handed spectra, individual washouts and $CP$ asymmetries, the relevant quantities for leptogenesis. Owing to some freedom in this redefinition, we found that correcting fermion masses greatly improves the results. A typical example is the success of the $(+,-,+)$ case.\\
Requiring that leptogenesis successfully predicts the observed amount of baryon asymmetry, we thus found that 3 of our 4 characteristic solutions do work. However, these results might be pointless, since these might be a conflict with the reheating temperature. Examining this problem, we found that the solutions $(+,+,+)$ and $(+,+,-)$ work in a large part of the parameter space with $T_{RH}\lesssim 10^{10}\GeV$, whereas  $(+,+,-)$ would require $T_{RH}\gtrsim 5\times 10^{10}\GeV$. Under the constraint that $T_{RH}\lesssim 2\times 10^{10}\GeV$ we thus found that two of our four characteristic solutions satisfy the leptogenesis criterion. Hence, we can say that four of the eight solutions are compatible with leptogenesis.
Nevertheless, our model suffers from its lack of predictivity, since the 10
dimensional high-energy parameter-space remains unconstrained. Therefore, it
would be interesting to link the constraints coming from leptogenesis to
other constraints coming, for example, from lepton-flavour violation. It could
be also interesting to build a complete $SO(10)$ GUT model which provides the
LR symmetric seesaw, together with the required proton stability,
doublet-triplet splitting and realistic fermion masses and mixings. 

\newpage
%%%%%%%%%%%%%%%%%%%%%%%%%%%%%%%%%%%%%%%%%%%%%%%%%%%%%%%%%%%%%%%%%%%%%%%%%%%%%%%%%%%%%%%%%%%%%%%%%%%%%%%%%%%%%%%%%%%%%%%%%%%%%%%%%%%%%%%%%%%%%%%%%%%%%%%%%%%%%%%%%%
\renewcommand{\thechapter}{}
\renewcommand{\chaptername}{}
\chapter[Conclusions]{Conclusions}
%\addcontentsline{toc}{chapter}{Conclusions}
%\thispagestyle{empty}

%It is time to conclude.
%Douter de tout ou tout croire sont deux solutions \'egalement commodes, qui l'une et l'autre nous dispensent de r\'efl\'echir. \footnote{Henri Poincar\'e,in La Science et lhypoth\`ese.}
We have seen that under the assumption that neutrinos get their masses through the seesaw mechanism, leptogenesis qualitatively occurs; in chapter 3, we discussed the requirements under which the single flavour picture can quantitatively work. These conditions are:
\begin{itemize}
\item The matrix $R$ must be complex, and furthermore $CP$ violation has to be large enough in order to have $\eps_{CP}\gtrsim 5\times 10^{-7}$.
\item The right-handed neutrino masses are bounded: $2\times 10^{9}\GeV\lesssim M_{1}\lesssim 10^{12}\GeV$. The lower bound comes from the requirement that $\eps_{CP}$ is large enough, while the upper-bound represents the temperature below which electroweak sphalerons come in-equilibrium.
\item Assuming hierarchical heavy neutrinos, the light neutrino masses are upper-consrained: $m_{i}\lesssim 0.15\eV$.
\end{itemize}
This is for the single flavour picture.\\
However, we showed \cite{issues} that lepton flavours have to be included in leptogenesis, since for $M_{1}\lesssim 10^{12}\GeV$, the interactions involving the charged lepton Yukawa couplings are in-equilibrium. Since the different lepton flavours have different Yukawa couplings, the lepton doublet interacting with the RHns has to be decomposed into the different flavours. Consequently, one has to consider lepton flavour asymmetries.\\
In chapter 4, we show how to include the evolution of flavoured asymmetries in the Boltzmann equations, and what are the constraints under which the flavours are relevant. We showed in this chapter that, for instance, when  the tau Yukawas are in-equilibrium, then the quantum correlations between the tau flavour and the other flavours are exponentially damped, which is not the case for the correlations between the muon and the electron flavours. Regardless of whether leptogenesis is efficient or not, we identified different temperature regimes:
\begin{itemize}
\item If $10^{9}\GeV\lesssim M_{1}\lesssim 10^{12}$, only the interactions involving the tau Yukawa couplings are in-equilibrium, and therefore 2 flavours are distinguishable: the tau flavour and the coherent sum of $e+\mu$ flavours.
\item If $M_{1}\lesssim 10^{9}$, then the muon-Yukawa interactions are also in-equilibrium, and the lepton asymmetry is projected onto a three-flavour space, $(\tau,\mu,e)$.
\end{itemize}
The electron-Yukawa interactions are in-equilibrium for $M_{1}\lesssim 10^{6}\GeV$, that is for masses far below what is required by leptogenesis.\\
We subsequently applied these flavoured recipes to the type I seesaw in the SM model, and to the type II seesaw in a GUT model, respectively in chapter 5 and 6.\\
The chapter 5 is based on \cite{issues,matters,StudyOf}, and provides the main results obtained during my thesis, which can be compared with the constraints of the single-flavour picture above.
\begin{itemize}
\item In the flavoured context, the $R$-matrix can be real, with non-zero lepton flavoured $CP$ asymmetries, since the latter now depend on the $CP$ phases of the lepton mixing matrix $U_{PMNS}$. However, the lower bound on the $CP$ asymmetries is still $\eps\gtrsim 5\times 10^{-7}$. A noticeable modification, which may provide one of the most important effects of lepton flavours, is the modification of the upper-bound on the $CP$ asymmetry in the regime of degenerate light neutrinos. Indeed, while in the single flavour picture $\eps_{CP}\propto m^{-1}$ and is thus suppressed, when flavours are included the upper-bound rather reads $\eps_{\al}\simeq m$ and is even enhanced in this limit.
\item Regarding the lower bound on right-handed neutrinos, lepton flavours do not significantly modify the picture, with $M_{1}\gtrsim 2.5\times 10^{9}\GeV$.
\item On the contrary, since individual $CP$ asymmetries are no-longer suppressed in the regime of degenerate light neutrinos, the allowed parameter-space is enlarged, and light neutrino masses are no longer upper-constrained by leptogenesis when right-handed neutrinos are hierarchical.
\end{itemize}
The main effect of lepton flavours lies in the degenerate regime, which corresponds to a regime of strong washout. This is simply because the couplings of different flavours with the right-handed neutrino are not equal, and so the flavours interact at different rates.\\
Another aspect of leptogenesis, which perhaps constitutes one of its weakness, is the lack of predictivity regarding the different $CP$ violating phases and especially for the high-energy sector. In the single flavour picture and in the flavoured one as well, only optimal regions in the parameter-space can be inferred.\\
In order to remedy to this problem, we studied in chapter 6 supersymmetric leptogenesis in a Left-Right symmetric type II seesaw model. In this model, the GUT embedding yields the interesting result that the Dirac type mass of light neutrinos is no longer free, but related to the masses of the other fermions. This is a clear gain compared to a non-GUT scenario.\\
In this scenario, the high-energy sector was reconstructed from an inversion of the seesaw formula, but nevertheless suffers from an "eightfold" ambiguity in the solutions. We therefore investigated if leptogenesis enabled us to discriminate among these solutions. Furthermore, given the spectra we encountered, for half of the solutions it was clear that only the second lightest right-handed neutrino was relevant for leptogenesis. We therefore studied leptogenesis, including both light and heavy flavour effects. In this case, we found that 2 of our 8 solutions survived, namely the solution which correspond to a pure type II seesaw in the high $B-L$ breaking scale limit.\\
We further refined our model, correcting the GUT relation among charged lepton and down type quark masses, through the inclusion of non-renormalisable operators. We found that this step towards more realistic fermion masses strongly affects leptogenesis, increasing to 6 the number of allowed solutions. We further found a nice illustration of the importance of both light and heavy flavours for the case where the lepton asymmetry is generated by the decays of $N_{2}$, and the inclusion of lepton flavours prevents this asymmetry from being washed-out by $N_{1}$ processes.\\
However, since we were working in a supersymmetric extension, we had to have in mind reheating temperature constraints. When investigating whether or not leptogenesis could work with $T_{RH}\lesssim 10^{10}\GeV$, we found that 4 solutions were compatible, in a large part of the parameter space. We then consider whether the requirement of successful leptogenesis enabled us to make some predictions for the seesaw parameters. Unfortunatly, the "cure seems worse than the disease", since our GUT model contains more free parameters than the non-GUT one.\\
This lack of predictivity is a common plague to high-energy seesaw models, and thus to the thermal scenario of leptogenesis. One then one asks: is there a remedy?
There are at least several ways out: the first one would be to link the different observables which involve seesaw couplings, another could be to study leptogenesis in a complete and consistent GUT scenario, where the number of free parameters is dramatically reduced or even over-determined. Finally, lowering the seesaw scale could provide additionnal observable effects at the LHC or other future experiments and is therefore an interesting solution. Nevertheless, lowering the seesaw scale enters in conflict with the scenario of thermal leptogenesis that we discussed in this thesis. Hence the need to low-energy leptogenesis models.
%Finally, lowering the leptogenesis or the seesaw scale is clearly a way if one hopes for probing one day leptogenesis or the seesaw.
\newpage
%%%%%%%%%%%%%%%%%%%%%%%%%%%%%%%%%%%%%%%%%%%%%%%%%%%%%%%%%%%%%%%%%%%%%%%%%%%%%%%%%%%%%%%%%%%%%%%%%%%%%%%%%%%%%%%%%%%%%%%%%%%%%%%%%%%%%%%%%%%%%%%%%%%%%%%%%%%%%%%%%%
\renewcommand{\theequation}{A-\arabic{equation}}
\renewcommand{\thefigure}{A-\arabic{figure}}
\renewcommand{\thesection}{A.\arabic{section}}
\renewcommand{\thesubsection}{A.\arabic{section}.\arabic{subsection}}
  % redefine the command that creates the equation no.
 \setcounter{equation}{0} 
\setcounter{section}{0}
\setcounter{subsection}{0}
\setcounter{figure}{0}
 % reset counter 
\appendix
\chapter{A few words on particle physics}

\section{The Standard Model}
The Standard Model of particles relies on the local invariance under $SU(3)_{c}\times SU(2)_{L}\times U(1)_{Y}$ symmetry, the matter sector being composed of 3 generations of $15$ fundamental fermion fields which are given in the following table:
\bea
\begin{array}{|c|c|c|c|c|c|}
\hline \rm{field} &  G_{32} & Q_{T^{3}} & Q_{Y} & B & L \\
\hline  u_{L}  &  (\bf{3},\bf{2})  &  1/2  &  1/3 & 1/3 & 0\\
\hline  d_{L}  &  (\bf{3},\bf{2})  & - 1/2  &  1/3 & 1/3 & 0 \\
\hline  u_{R}  &  (\bf{3},\bf{1})  & 0 &  4/3 & 1/3 & 0\\
\hline  d_{R}  &  (\bf{3},\bf{1})  & 0 & - 2/3 & 1/3 & 0\\
\hline \nu_{L}  &  (\bf{1},\bf{2})  &  1/2  & - 1 & 0 & 1\\
\hline  e_{L}  &  (\bf{1},\bf{2})  & - 1/2  & - 1 & 0 & 1\\
\hline  e_{R}  &  (\bf{1},\bf{1})  & 0 & - 2 & 0 & 1\\
\hline
\end{array} 
\eea
The SM successfully describes the strong and electroweak interactions thanks to the exchange of vector bosons, which are in the adjoint representation of the different groups: the $8$ gluons for $SU(3)$, the $3$ $W$ bosons of $SU(2)$ and the vector $B_{\mu}$ of $U(1)_{Y}$.\\
Fermion mass terms emerge from the coupling of fermion bilinears with the scalar sector, which consists, in the SM, in a scalar field, doublet under $SU(2)$
\bea
\phi=\left(\begin{array}{c} \phi^{+} \\ \phi^{0}\end{array}\right) \, .
\eea
The neutral component of $\phi$ is singlet under $SU(3)_{c}\times U(1)_{em}$, and so it can take a vev $\left\langle\phi^{0}\right\rangle=v/\sqrt{2}\simeq 246 \GeV$, leaving the low-energy group invariant and in the meanwhile fermions aquire masses, thanks to the Yukawa couplings:
\bea
\mathcal{L}_{m}=\la_{e}\,\frac{v}{\sqrt{2}}\,\ol{e_{L}}\,e_{R}+\la_{d}\,\frac{v}{\sqrt{2}}\,\ol{d_{L}}\,d_{R}+\la_{d}\,\frac{v}{\sqrt{2}}\,\ol{u_{L}}\,u_{R}+h.c.
\eea
The SM does not contain a right-handed neutrino, hence $m_{\nu}=0$. Furthermore, all interaction terms are renormalisable. This renormabilisity, together with the field content given above, imply that besides the invariance under the local symmetry $G_{321}$, 4 global $U(1)$s are conserved: the baryon number B, and the lepton flavour number $L_{e,\mu,\tau}$, and consequently the total lepton number $L$. As a consequence of this, the neutrino remains massless to all orders of perturbation.
\section{Sphalerons}
\begin{figure}[h!]
\begin{center}
\includegraphics[scale=0.6]{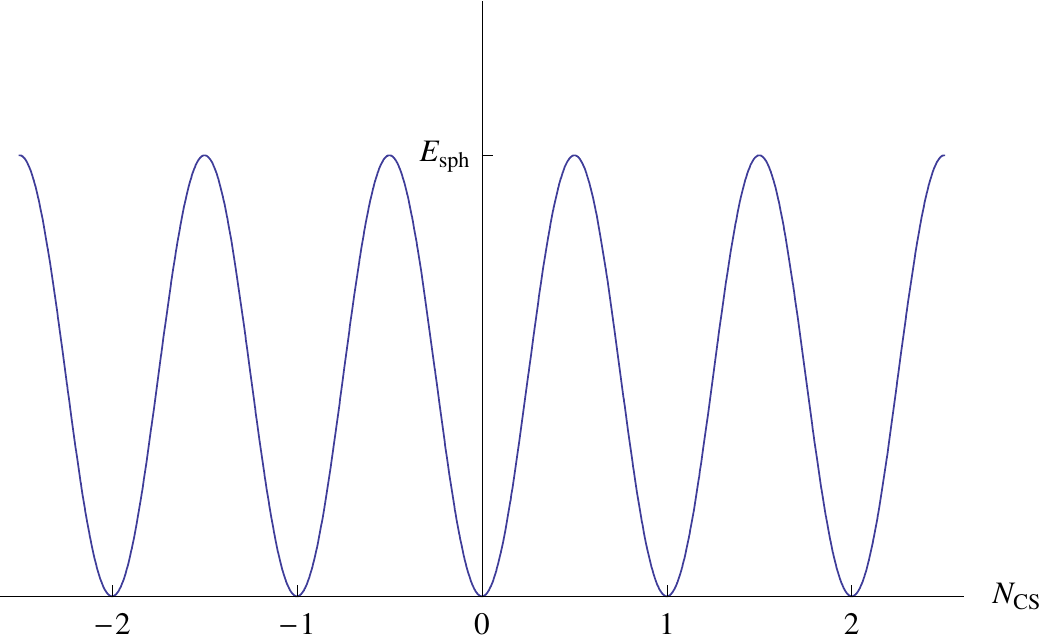} 
\caption{\footnotesize{Energy of gauge field configuration versus Chern-Simons number}}
\label{GrapheSpa}
\end{center}
\end{figure}

Baryon and Lepton numbers are accidental symmetries of the Standard Model. This means that they do not reflect a higher symmetry but are satisfied given the particle content and the renormalisable couplings of the model. B and L are conserved at tree level and to all order in perturbation theory.\\
However, 'T Hooft showed \cite{thooft} that non-perturbative effects, called instantons, can lead to the violation of $B+L$, while conserving the orthogonal $B-L$. This property is associated with the topological structure of any $SU(N)$ gauge group. The ground state of the theory is not unique, but degenerate vacua exist which are topologically inequivalent. Going from one vacuum to another vacuum can be done by tunnelling through  field configurations called instantons, but typically the probability will be highly suppressed. Another possibility arise from the existence of static but unstable field configurations that help the transition to occur: these are the sphalerons.\\
In concrete terms, the baryonic and leptonic number currents, 
\bea
J_{\mu}^{L}&=&\sum_{i} \bar{\ell_{L\,i}} \gamma_{\mu} \ell_{L\,i}-\bar{e_{L\,i}}^{c}\gamma e_{L\,i}^{c} \ , \nonumber \\
J_{\mu}^{B}&=&\frac{1}{3}\sum_{i} \bar{Q_{L\,i}} \gamma_{\mu} Q_{L\,i}-\bar{u_{L\,i}}^{c}\gamma u_{L\,i}^{c} -\bar{d_{L\,i}}^{c}\gamma d_{L\,i}^{c}\ ,
\eea
have a non-zero divergence from the ABJ triangle anomaly\cite{ABJ}:
\bea
\partial^{\mu}J_{\mu}^{B}=\partial^{\mu}J_{\mu}^{L}=\frac{N_{g}}{32 \pi^2}\left(g^2 W^{\mu \nu} \tilde{W_{\mu\nu}}-g'^2 B^{\mu \nu} \tilde{B_{\mu\nu}}\right) \ ,
\eea
where $W_{\mu\nu}$ and $B_{\mu\nu}$ are $SU(2)$ and $U(1)$ field-strengths respectively, and $N_{g}$ the number of fermions generations. Therefore, we see that $B-L$ is conserved, as $\partial{\mu}(J_{\mu}^{L}-J_{\mu}^{B})=0$.
The orthogonal combinaison, $B+L$ is violated:
\bea
\partial{\mu}(J_{\mu}^{L}+J_{\mu}^{B})=2 N_{g} \partial_{\mu}K^{\mu} \ .
\eea
The interesting point is that $K_{\mu}$ is related to the topological structure of the vacuum, by:
\bea
N_{CS}(t_i)-N_{CS}(t_0)=\int_{t_{0}}^{t_{i}} dt \int d^{3}x \partial_{\mu}K^{\mu} \equiv n \ ,
\eea
where n is an integer, and $N_{CS}$ are Chern-Simons number. Therefore we see that if during a time $\Delta t$ a transition between two vacua with distinct topological charges occurs, this will induce a baryon+lepton number violation, as 
\bea
\Delta (B+L)=N_{g} \Delta N_{CS} \ .
\eea
At zero temperature however, the transition rate between two vacua is exponentially suppressed, with 
\bea
\Gamma\simeq e^{-8 \pi^2 /g^2}\simeq e^{-173}\, ,
\eea
and effectively no transition occurs. The picture changes however at high temperature, $T\gtrsim T_{EW}$, as showed by Kuzmin, Rubakov and Schapovnikov\cite{KRS}: instead of tunnelling through the barrier, the available thermal energy allows to step over it, thanks to the Higgs and gauge boson field configurations that lie on the top of the energy barrier separating two topologically inequivalent vacua. The situation is schematically depicted in fig.\ref{GrapheSpa}.
\\
The sphaleron transition amplitude is roughly $\mathcal{A}\simeq e^{-E_{sph}/T}$, with $E_{sph}$ the height of the energy barrier
\bea
E_{sph}\simeq \frac{8\pi\,v}{g}\, .
\eea
Hence, for $T\gg E_{sph}$, the transition amplitude gets unsuppressed and rapidly occurs.\\
In the Standard Model, we have the strong sphalerons, which are related to the $SU(3)$ gauge structure, and whose density rate is given by\cite{StrongSpha}:
\bea
\Gamma_{QCD}/V\simeq 250\,\al_{s}^{5}\,T^{4}\, .
\eea
As $V\sim T^3$, the $QCD$-sphalerons enter in-equilibrium when $\Gamma_{QCD}(T)\lesssim H(T)$, which happens at $T\lesssim 10^{13}\GeV$, and then induce an effective interaction
\bea
\mathcal{ O}_{QCD}=\Pi_{i=1,2,3} (q_{L\,i}q_{L\,i}\,u_{R\,i}^{c}\,d_{R\,i}^{c}) \ ,
\eea
which relate left and right-handed quarks.\\
The SM also contains electroweak instantons, which are related to the $SU(2)$ gauge structure, with a density rate\cite{EWSpha}:
\bea
\Gamma_{EW}/V \simeq 25\,\al_{W}^{5}\, T^{4}\, .
\eea
If in-equilibrium, these sphalerons induce an effective interaction
\bea
\mathcal{O}_{EW}=\Pi_{i=1,2,3} (q_{L\,i}q_{L\,i}q_{L\,i}\ell_{L\,i}) \ ,
\eea
which consist in a $B\LRa L$ exchange. The $SU(2)$ sphalerons are faster than the Hubble expansion rate for $T\lesssim 10^{12}$ GeV. Hence, for $T_{EW} \lesssim T \lesssim 10^{12}$ GeV, any $B+L$ number will be driven to zero by the fast $B+L$ violating sphalerons. On the other hand, $B-L$ asymmetries are not affected by sphalerons transition. The early GUT baryogenesis model were based on $SU(5)$, which conserves $B-L$, hence the possible $B$ number creation was wiped out as soon as electroweak sphalerons enter in-equilibrium and this models cannot work. On the contrary of leptogenesis models, were the asymmetry is produced in the $L$ direction, which is then reprocessed into the $-B$ direction.

\newpage
\renewcommand{\theequation}{B-\arabic{equation}}
\renewcommand{\thesection}{B.\arabic{section}}
\renewcommand{\thesubsection}{B.\arabic{section}.\arabic{subsection}}
  % redefine the command that creates the equation no.

\chapter{Thermodynamics}
%The evolution of the Universe is governed by its particle content: the stress-energy tensor depends on the energy density and on the pressure or the fluids. Along the evolution of the Universe, the species are liable to interaction, hence their number density may be affected. These three quantities $ \rho_{i}, P_{i}, n_{i} $ are derived from the distribution function $f_{i}(T)$. Its full knowledge enable us to completely caracterize the Universe.\\
The evolution of the Universe is described by the Einstein equation which relate the geometry of the Universe to its content:
\bea
\mathcal{R}_{\mu \nu}-\frac{1}{2}g_{\mu \nu} \mathcal{R}=8 \pi G_{N} T_{\mu \nu}+\Lambda g_{\mu \nu} \, .
\eea
In this equation $\mathcal{R}$ is the Ricci scalar, $\mathcal{R}_{\mu \nu}$ the Ricci tensor, $T_{\mu\nu}$ the stress-energy tensor, $g_{\mu\nu}$  the space-time metric and $\Lambda$ is the cosmological constant. $G_{N}$ is the Newton constant,
\bea
G_{N}&=&\frac{1}{Mpl^{2}} \ , \\
Mpl&\simeq &  1.221\times 10^{19} \GeV \ .
\eea 
Assuming that the Universe content is a perfect fluid, we can write the stress-energy tensor as:
\bea
T_{\mu\nu}=-p g_{\mu\nu}+(p+\rho)u_{\mu}u_{\nu} \ ,
\eea
where $p$ is the pressure and $\rho$ the energy density of the perfect fluid. The velocity vector of the fluid, $u_{\mu}$ is given in the rest frame of the plasma by $u=(1,0,0,0)$. \\
Assuming further homogeneity and isotropy of the Universe, it can be described through a Robertson-Walker metric,
\bea
ds^{2}=dt^{2}-a(t)^{2} d\overline{r}^{2} \ ,
\eea
so that the (0-0) component of the Einstein equation can be written: 
\bea
\label{friedmann}
\left( \frac{\dot{a}}{a} \right)^{2}+\frac{k}{a^{2}}=\frac{8 \pi G_{N}}{3} \rho+\frac{\Lambda}{3} \ ,
\eea
where $k=0,+1,-1$ is the curvature of the Universe (resp. flat, open and closed geometry), and $a(t)$ is the scale factor of the expanding Universe. The Hubbler expansion rate $H$ is defined as: 
\bea
H\equiv \left( \frac{\dot{a}}{a} \right)^{2} \ ,
\eea
and tells us how fast the Universe is expanding.\\
%From the Einstein equation is also derived the energy momentum conservation: 
%\bea
%\dot{\rho}+3 H (\rho + p)=0 \ .
%\eea
%Then, given a simple equation of state $p=w \rho$, with $w$ independant of time, the energy density scales as $\rho(t)\sim a(t)^{-3(1+w)}$. Then we obtain standard solutions:\\
%\begin{itemize}
%\item relativist matter-radiation: $w=1/3$, $\rho \sim a^{-4}$, $a \sim t^{1/2}$.\\
%\item non-relativist matter: $w=0$, $\rho \sim a^{-3}$, $a \sim t^{2/3}$. \\
%\item cosmological constant: $w=-1$, $\rho \sim cte$, $a \sim e^{H t}$\\with $H~=~\sqrt{\frac{8 \pi G_{N} \rho_{0}}{3}}$.\\
%Along the evolution of the Universe, the species are liable to interaction, hence their number density may be affected.
The species which compose the Universe are liable to interaction. In an expanding Universe, if this rate is faster than the Hubble expansion rate, then theses particles will be maintained causally connected with the thermal bath: the species are in-equilibrium and in that case they all have the same temperature. However, Uinverse is cooling. Then it is likely that at some temperature $T_f$, interactions will not be able to keep species in equilibrium as they freeze-out: $\Gamma(T)\lesssim H(T)$ for $T\lesssim T_f $.\\
When out-of-equilibrium, the species decouple from the thermal bath and evolves independently of it.
% Such thermal relics are oftently encountered and their relic abundance constraint cosmological models (cf., \textit{e.g.}, dark matter, gravitino, moduli ...).
The criterion $\Gamma(T_f ) \simeq H(T_f )$ gives a rough picture of the decoupling, but a more accurate description require to study microscopic evolution of the particle number densities, by solving the Boltzmann equations. This appendix briefly introduce equilibrium and out-of-equilibrium thermodynamics, which is needed when studying leptogenesis. We refer the reader to refs.\cite{KolbTurner},\cite{KolbWolfram} for more detailled presentations of section 1 and 3.
\section{Equilibrium thermodynamics}
Species in-equilibrium can be described by ideal Fermi-Dirac (FD) or Bose-Einstein (BE) fluid, whose distribution function is:
\bea
f^{eq}_{i}(\textbf{p},\mu_{i},T)=\frac{g_{i}}{\exp((E_{i}-\mu_{i})/T) \pm 1} \ ,
\eea
where $g_{i}$ denotes the number of degree of freedom of the species $i$, $E_{i}~=~\sqrt{\textbf{p}^{2}+m_{i}^{2}}$ and $\mu_{i}$ is the chemical potential of the species. The $+$ ($-$) sign refers to FD (BE) statistic.
The classical approximation of Maxwell-Boltzmann (MB) statistic, which will extensively use in the network of Boltzmann equations, is:
\bea
f^{eq}_{i}=g_{i} \,\exp((E_{i}-\mu_{i})/T) \ .
\eea
Given these distribution functions, one defines the equilibrium number density $n_{i}$ and the equilibrium energy density $\rho_{i}$ are given by:
\bea
n^{eq}_{i}(T)&=&\int \frac{d^{3}\textbf{p}}{(2 \pi)^{3}} f^{eq}_{i}(\textbf{p},\mu_{i},T)=\frac{g_{i}}{2 \pi^{2}}\int_{m_{i}}^{\infty}dE \frac{\sqrt{E^2-m_{i}^2} E}{\exp((E_{i}-\mu_{i})/T) \pm 1} \\
\rho^{eq}_{i}(T)&=&\int \frac{d^{3}\textbf{p}}{(2 \pi)^{3}} f^{eq}_{i}(\textbf{p},\mu_{i},T) E(\textbf{p})=\frac{g_{i}}{2 \pi^{2}}\int_{m_{i}}^{\infty}dE \frac{\sqrt{E^2-m_{i}^2} E^{2}}{\exp((E_{i}-\mu_{i})/T) \pm 1}
%P^{eq}_{i}(T)&=&\int \frac{d^{3}\textbf{p}}{(2 \pi)^{3}} f^{eq}_{i}(\textbf{p},\mu_{i},T)\frac{p^2}{3 E}=\frac{g_{i}}{2 \pi^{2}}\int_{m_{i}}^{\infty}dE \frac{(E^2-m_{i}^2)^{3/2}}{\exp((E_{i}-\mu_{i})/T) \pm 1}
\eea
Defining $z_{i}=m_{i}/T$, the number density for a massive particle is given by:
\bea
n^{eq}_{i}(\mu, T)^{MB}=\frac{g_{i} T^{3}}{2 \pi^{2}}\,e^{-\frac{\mu_{i}}{T}}\, z_{i}^2\, K_{2}(z_{i})\, ,
\eea
for Maxwell-Boltzmann statistic, and: 
\bea
n^{eq}_{i}(\mu, T)^{BE}_{FD}=\frac{g_{i} T^{3}}{2 \pi^{2}}\sum_{k=0}^{\infty}\frac{(\pm e^{\mu_{i}/T})^k}{k+1}\,z_{i}^2 K_{2}\left((k+1)z_{i}\right) \ ,
\eea
for fermions and bosons. Here $K_{n}(x)$ is the modified Bessel function of the second kind. \\
The asymptotic expansion of Bessel functions is well known and one can therefore deduce asymptotic behaviour of the number densities.
\begin{itemize}
\item For high temperatures $T\gg m_{i}$, neglecting chemical potential:
\bea
n^{eq}_{i}(T)^{MB}&\simeq &\frac{g_{i} T^{3}}{\pi^{2}}\left(1-\frac{1}{4} z_{i}^{2}+\ldots \right) \ , \\
n^{eq}_{i}(T)^{BE}&\simeq &\frac{g_{i} T^{3}}{\pi^{2}}\left(\zeta(3)+\frac{1}{4} z_{i}^{2} \log(z_{i})+\ldots \right) \ , \\
n^{eq}_{i}(T)^{FD}&\simeq &\frac{g_{i} T^{3}}{\pi^{2}}\left(\frac{3}{4}\zeta(3)+ z_{i}^{2} \frac{\log(2)}{2}+\ldots \right) \ ,
\eea
where $\zeta(3)\simeq 1.202$.
\item The low-temperature expansion reads:
\bea
n^{eq}_{i}(T)^{MB}\simeq n^{eq}_{i}(T)^{BE}\simeq n^{eq}_{i}(T)^{FD}\simeq g_{i} \left( \frac{m_{i}T}{2 \pi} \right)^{3/2}e^{-m_{i}/T} \ .
\eea
\end{itemize}
The usual assumption is to disregard quantum statistic and handle fermions and bosons in terms of MB particle, with either 
\bea
n_{X}^{eq}=\frac{g_{X} T^{3}}{2 \pi^{2}}\, z_{X}^2\, K_{2}(z_{X})\, ,
\eea
for a massive particle, or
\bea
n_{X}^{eq}=\frac{g_{X}\,T^{3}}{\pi^{2}}\,
\eea 
for a massless one. Similarly, one obtains the asymptotic behaviour of the energy density:
\begin{itemize}
\item for the high temperature expansion $T \gg m_{i}$ one has: 
\bea
\rho^{eq}_{i}(T)^{BE}&\simeq & \frac{\pi^{2}}{30}g_{i} T^{4} \ , \\
\rho^{eq}_{i}(T)^{FD}&\simeq & \frac{7}{8}\frac{\pi^{2}}{30}g_{i} T^{4} \ .
\eea
\item For the low temperature expansion $T \ll m_{i}$, 
\bea
\rho^{eq}_{i}(T)= m_{i}\,n^{eq}_{i}(T) \propto e^{-m_{i}/T} \, .
\eea.
\end{itemize}
Therefore, at a given temperature $T$, the energy density of non-relativistic species will be exponentially suppressed, thus being safely neglected in the total energy density of the Universe 
\bea
\rho_{tot}=\frac{\pi^2}{30}g_{*}T^{4} \ ,
\eea
where $g_{*}$ counts the effective number of degrees of freedom:
\bea
g_{*}=\sum_{i=BE}g_{i} \left(\frac{T_{i}}{T}\right)^4+\frac{7}{8}\sum_{i=FD}g_{i} \left(\frac{T_{i}}{T}\right)^4 \ .
\eea
Here, $T_{i}$ is the temperature of the species $i$: $T_{i} \propto T$ with equality when the species $i$ are in-equilibrium. For the typical temperatures we  consider, $T\gg 100 \GeV$, $g_{*}=106.75$ in the Standard Model, while $g_{*}=228.75$ in the MSSM.\\
Therefore, according to eq.(\ref{friedmann}), assuming flatness and negligible cosmological constant, that is valid in the very early Universe, one deduces
\bea
H(T)=\sqrt{\frac{8 \pi \rho_{tot}}{3 M_{pl}^{2}}}\simeq 1.66 g_{*}^{1/2} \frac{T^2}{M_{pl}} \ .
\eea
In the radiation dominated epoch $a \sim t^{1/2}$ and therefore one relates time and temperature by:
\bea
\frac{t(T)}{1\,\rm{s}}\simeq 2.42\,g_{*}^{1/2}\,\left(\frac{T}{1 \MeV}\right)^{-2} \ .
\eea
Neglecting the chemical potential we have for the entropy density:
\bea
s=\frac{\rho+P}{T}=\frac{2 \pi^2}{45}\,q_{*}\,T^{3} \ ,
\eea
where
\bea
q_{*}=\sum_{i=BE}g_{i} \left(\frac{T_{i}}{T}\right)^3+\frac{7}{8}\sum_{i=FD}g_{i} \left(\frac{T_{i}}{T}\right)^3 \ .
\eea
If all relativistic species are in-equilibrium, then $T_{i}=T$ and $q_{*}=g_{*}$. Furthermore, since $T\propto a^{-1}$ and $s\propto a^{-3}$,  as long as the number of particles remains constant, the comobile number density $Y_{i}=n_{i}/s$ is constant. On the other hand, when degrees of freedom freeze and become non-relativistic, the conservation of entropy implies that $q_{*}^{<}\,T_{<}^{3}=q_{*}^{>}\,T_{>}^{3}$ and thus:
\bea
T_{>}=T_{<}\,\left(\frac{q_{*}^{<}}{q_{*}^{>}} \right)^{1/3} \, .
\eea
That is, when dof decouple from the thermal bath, their entropy is transfered to species that are in thermal equilibrium, and their temperature therefore increase: entropy creation (re)heats the Universe.\\
Given the above definitions, we further have that 
\bea
s=\frac{q_{*}\,\pi^{4}}{45\,\zeta(3)}\,n_{\gamma}\simeq 1.8 \, q_{*}\,n_{\gamma}\, ,
\eea
so that the actual entropy and photon number density are related according to
\bea
s_{0}\simeq 1.8\,q_{* 0}\,n_{\gamma 0}\,\simeq 7.04\,n_{\gamma 0}\, .
\eea
\section{Relating $B$ and $L$ numbers}
Let us see how to relate the baryon number of the Universe to the lepton number that is produced during leptogenesis~\cite{HarveyTurner}. Since that particles and antiparticles have an opposite chemical potential $\mu_{x}=-\mu_{\ol{x}}$, the net number density of a particle $x$ can be written, for small $\mu_{X}/T$
\bea
n_{x}-n_{\ol{x}}=\frac{g_{x}\,T^{3}}{6}\times \left\lbrace \begin{array}{cc}
 \mu_{X}/T & \rm{for\quad a\quad fermion} \\
2\,\mu_{X}/T & \rm{for\quad a\quad boson} \end{array}\right. \, .
\eea
Therefore, since $B$ is stored both in $SU(2)_{L}$ doublet and singlet components, the baryon number is
\bea
B=\frac{g_{x}\,T^{2}}{6}\,\sum_{i}\,\left(\mu_{u_{L}^{i}}+\mu_{d_{L}^{i}}+\mu_{u_{R}^{i}}+\mu_{d_{R}^{i}}\right)\, ,
\eea
where the sum is made over the different asymmetries that are populated.\\
Similarly, the lepton number is
\bea
L=\sum_{i}L_{i}=\frac{g_{x}\,T^{2}}{6}\,\sum_{i}\,\left(\mu_{\nu_{L}^{i}}+\mu_{e_{L}^{i}}+\mu_{e_{R}^{i}}\right)\, .
\eea
The different species undergo reactions, which if in-equilibrium enforce algebraic relations among the chemical potentials involved, since for a $i+j\LRa k+l$ interaction, the equilibrium condition implies
\bea
\mu_{i}+\mu_{j}=\mu_{k}+\mu_{l}\, .
\eea 
Then, according to the different chemical equilibriums that hold at a given temperature, $B$ and $L_{i}$ can be related. Actually in the SM, $B$ and $L$ are only related through the effective interaction induced by the electroweak sphalerons. Since this interaction is in-equilibrium for $T\lesssim 10^{12}\GeV$,  no baryon asymmetry can be generated via leptogenesis for  $T\gtrsim 10^{12}\GeV$. Furthermore, since these interactions conserve $B-L$ but violate $B+L$, which they set to zero, one could na\"ively think that 
\bea
B=\frac{B+L}{2}+\frac{B-L}{2}\Ra \frac{B-L}{2} \, ,
\eea
when electroweak sphalerons are in-equilibrium. Actually, since the latter only involve left-handed fields, while the lepton and baryon number are stored both in $SU(2)$ doublets and singlets, the situation is more involved, and relating $B$ to the $L_{i}$ require to determine which interactions are in-equilibrium when the lepton asymmetry is created. Since leptogenesis occurs at $T\sim M_{1}$, the different temperature regime then correspond to range of $M_{1}$.\\
We can list the different interactions and the temperature at which they become in-equilibrium\cite{HarveyTurner},\cite{NardiNir2}.\\
\begin{itemize}
\item Gauge interactions are always in-equilibrium. Consequently, gluons and $B^{0}$  have a null chemical potential, and furthermore  the different quark colours have a same $\mu$. Interactions with the $W^{\pm}$ enforce
\bea
\mu_{W}&=&\mu_{\phi^{-}}+\mu_{\phi^{0}} \, ,\nonumber \\
\mu_{W}&=&-\mu_{u_{L}}+\mu_{d_{L}} \, ,\nonumber\\
\mu_{W}&=&-\mu_{\nu_{L}}+\mu_{e_{L}} \, ,
\eea
where $\mu_{W}=\mu_{W^{-}}=-\mu_{W^{+}}$, and $\phi$ denotes the Higgs doublet.\\
Moreover, since $SU(2)$ is gauged, the sum of third  weak-isospin component should be zero:
\bea
\frac{3}{2}\sum_{i}\left(\mu_{u_{L}^{i}}-\mu_{d_{L}^{i}}\right)+\frac{1}{2}\sum_{i}\left(\mu_{\nu_{L}^{i}}-\mu_{e_{L}^{i}}\right)-\frac{2\,m}{2}\left(\mu_{\phi^{0}}+\mu_{\phi^{-}}\right)-4\,\mu_{W}=0\, ,
\eea
where $m$ is the number of Higgs doublet. This relation implies that $\mu_{W}=0$ and so all $SU(2)$ doublet components have an equal chemical potential $\mu_{u_{L}}=\mu_{d_{L}}=\mu_{q_{L}}$ and $\mu_{\nu_{L}}=\mu_{e_{L}}=\mu_{\ell_{L}}$.
\item The requirement that the total hypercharge is null enforces:
\bea
\sum_{i}\left(\mu_{q_{L}^{i}}+2\,\mu_{u_{R}^{i}}-\mu_{d_{R}^{i}}\right)-\sum_{i}\left(\mu_{e_{R}^{i}}+\mu_{\ell_{L}^{i}}\right)+2\,\mu_{\phi}=0 \, .
\eea
\item For $T\lesssim 10^{13}\GeV$, QCD sphalerons are in-equilibrium:
\bea
\sum_{i}\left(\mu_{q_{L}^{i}}-\mu_{u_{R}^{i}}-\mu_{d_{R}^{i}}\right)=0\, \, .
\eea
For $T\lesssim 10^{12}\GeV$, the electroweak sphalerons are in-equilibrium and so $B$ and $L$ doublets are related according to:
\bea
\sum_{i}\left(3\,\mu_{q_{L}^{i}}+\mu_{\ell_{L}^{i}}\right)=0\, \, .
\eea
\end{itemize}
Finally, the interactions involving charged fermion Yukawas, when in-equilibrium, yield the following relations:
\bea
\mu_{u_{R}^{i}}&=&\mu_{q_{L}^{i}}+\mu_{\phi^{0}} \, ,\nonumber\\
\mu_{d_{R}^{i}}&=&\mu_{q_{L}^{i}}-\mu_{\phi^{0}} \, ,\nonumber\\
\mu_{e_{R}^{i}}&=&\mu_{\ell_{L}^{i}}-\mu_{\phi^{0}} \, , 
\eea
for up-type quarks, down-type quarks and charged leptons, respectively.\\
We are interested in relating the $\D_{\al}\equiv B/3-L_{\al}$ asymmetries to the leptonic doublet ones $\ell_{L\,\al}$, and this relation depends on the temperature regime.
\begin{itemize}
\item If $M_{1}\gtrsim 10^{12}\GeV$, none of the charged lepton Yukawas are in-equilibrium. Therefore, the $SU(2)_{L}$ singlets do not store any asymmetry, since only the left-handed part interact with the gauge bosons. Consequently the lepton number and the leptonic doublet number are equal. Since no baryon number is generated, as electroweak sphalerons are out-of-equilibrium, we have $Y_{\D \al}=Y_{\D}=-Y_{L}$.
\item If $10^{9}\lesssim M_{1} \lesssim 10^{12}\GeV$, then the tau-Yukawas are in-equilibrium, as well as electroweak sphalerons. A non-zero asymmetry can thus be stored in $\tau_{R}$. Since t, b and c quarks are in-equilibrium, the conversion matrix reads:
\bea
\left(\begin{array}{c} Y_{\D e} \\ Y_{\D \mu} \\Y_{\D \tau}\end{array}\right)=\frac{T^{3}}{6\,s}\,
\left(
\begin{array}{ccc}
 -\frac{4}{9}-2 & -\frac{4}{9} & -\frac{4}{9}\\
  -\frac{4}{9}& -\frac{4}{9}-2 & -\frac{4}{9}\\
 -\frac{2}{9} & -\frac{2}{9} &-\frac{4}{9}-3+\frac{16}{45}
\end{array}
\right).\left(\begin{array}{c} \mu_{e_{L}}/T \\ \mu_{\mu_{L}}/T \\\mu_{\tau_{L}}/T\end{array}\right) \ .
\eea
However, the $e$ and $\mu$ flavours are indistinguishable, therefore one should rather consider the asymmetry produced in $e+\mu$ direction, that is, one should sum the chemical potentials $\mu_{e_{L}}+\mu_{\mu_{L}}$, so that
\bea
\left(\begin{array}{c} Y_{\D e+\mu} \\Y_{\D \tau}\end{array}\right)=\frac{T^{3}}{6\,s}\,
\left(
\begin{array}{cc}
 -\frac{26}{9} & -\frac{8}{9} \\
 -\frac{2}{9} & -\frac{139}{45}
\end{array}
\right).\left(\begin{array}{c} \mu_{e_{L}}/T+\mu_{\mu_{L}}/T \\\mu_{\tau_{L}}/T\end{array}\right) \ .
\eea
Since $Y_{\ell_{L,\al}}=\frac{T^{3}}{6\,s} \left(2\,\mu_{\ell_{L,\al}}/T\right)$, we obtain
\bea
Y_{\ell_{\al}}=\left(
\begin{array}{cc}
 -\frac{417}{589} & \frac{120}{589} \\
 \frac{30}{589} & -\frac{390}{589}
\end{array}
\right).Y_{\D \al}\, .
\eea
\item If $M_{1}\lesssim 10^{9}\GeV$, the muon Yukawas are in-equilibrium, as well as the $s$-ones. Thus, the three lepton flavours are distinguishable, and one has
\bea
\left(\begin{array}{c} Y_{\D e} \\ Y_{\D \mu} \\Y_{\D \tau}\end{array}\right)=\frac{T^{3}}{6\,s}\,
\left(
\begin{array}{ccc}
 -\frac{4}{9}-2 & -\frac{4}{9} & -\frac{4}{9}\\
  -\frac{4}{9}-\frac{1}{6}& -\frac{4}{9}-2-\frac{11}{15}& -\frac{4}{9}-\frac{4}{15}\\
 -\frac{4}{9}-\frac{1}{6}&  -\frac{4}{9}-\frac{4}{15}&-\frac{4}{9}-2-\frac{11}{15}
\end{array}
\right).\left(\begin{array}{c} \mu_{e_{L}}/T \\ \mu_{\mu_{L}}/T \\\mu_{\tau_{L}}/T\end{array}\right) \ ,
\eea
that is, 
\bea
Y_{\ell_{\al}}=\frac{1}{179}
\left(
\begin{array}{ccc}
 -151 & 20 & 20 \\
 \frac{25}{2} & -\frac{344}{3} & \frac{14}{3} \\
 \frac{25}{2} & \frac{14}{3} & -\frac{344}{3}
\end{array}
\right).Y_{\D \al}\, .
\eea
\end{itemize}
Finally, the baryon asymmetry is related to the $B/3-L_{\al}$ ones by the relation\cite{HarveyTurner}:
\bea
Y_{B}=\frac{32+4m}{98+13m}\,\sum_{\al}\,Y_{\D\al}\, ,
\eea
which gives $Y_{B}=12/37\,\sum_{\al}\,Y_{\D\al}$ in the SM with $(m=1)$ one Higgs doublet.
\section{Out-of-equilibrium thermodynamics: Boltzmann equations}
Let us see how to characterise species which undergo non-elastic collisions.
The evolution of the distribution function is given by the Boltzmann equations (BE)
\bea
L[f]=C[f] \nonumber ,
\eea
where C is the collision term and L is the Liouville operator, 
\bea
L[f]=p^{\alpha} \frac{\partial}{\partial x^{\alpha}}-\Gamma^{\alpha}_{\beta \delta}\, p^{\beta} p^{\delta} \frac{\partial}{\partial x^{\alpha}} \nonumber ,
\eea
which is conveniently express in a homogeneous and isotopric space-time as :
\bea
L[f]=E \frac{\partial f}{\partial t}-H(t)\,p\,\frac{\partial f}{\partial E} \ .
\eea
Integrating over momentum, one recover the usual BE: 
\bea
\dot{n_{i}} + 3 H(t) n_{i} = C_{i}  \ .
\label{BE1}
\eea
All the work will then to evaluate the collision term. For a generic 2-to-2 scattering $i+j\leftrightarrow k+l$, it can be written as:
\bea
C_{i}&=&(2 \pi)^{4} \int \delta(p_{i}+p_{j}-p_{k}-p_{l}) \frac{g_{i} d^{3} \textbf p_{i} }{2 (2 \pi)^{3} E_{i}} ...\frac{g_{l} d^{3} \textbf p_{l}}{2 (2 \pi)^{3} E_{l}} \nonumber \\
&\times & \vert {\mathcal{ M }}\vert ^2 \times \left( (1\pm f_{i})(1\pm f_{j}) f_{k} f_{l}-f_{i}f_{j}(1\pm f_{k})(1\pm f_{l}) \right) \, ,
\eea
where + (-) stands for FD (BE) statistics, and $ {\mathcal M } $ is the invariant amplitude. 
We can simplify this expression by using MB statistics and neglecting Pauli blocking and stimulating emission factors (\textit{i.e.}, quantum statistic), so that the collision term reads:
\bea
C_{i}&=&(2 \pi)^{4} \int \delta(p_{i}+p_{j}-p_{k}-p_{l}) \frac{g_{i} d^{3} \textbf p_{i} }{2 (2 \pi)^{3} E_{i}} ...\frac{g_{l} d^{3} \textbf p_{l}}{2 (2 \pi)^{3} E_{l}} \nonumber \\
&\times & \vert {\mathcal{ M }}\vert ^2 \times \left(f_{k} f_{l}-f_{i}f_{j} \right) \, .
\eea
Since that for what concern us, the elastic scatterings, which conserve particle densities but not their distribution, are much faster than inelastic scatterings, we can approximate $f_{X}=f_{X}^{eq}\,n_{X}/n_{X}^{eq}$, so that the collision term can be further simplified:
\bea
C_{i}=\frac{n_{k}}{n_{k}^{eq}}\frac{n_{l}}{n_{l}^{eq}} \gamma^{kl}_{ij}-\frac{n_{i}}{n_{i}^{eq}}\frac{n_{j}}{n_{j}^{eq}} \gamma^{ij}_{kl} \ ,
\eea
where we define the reaction density $\gamma^{ij}_{kl}$:
\bea
\gamma^{ij}_{kl}=(2 \pi)^{4} \int \delta(p_{i}+p_{j}-p_{k}-p_{l}) \frac{g_{i} d^{3} \textbf p_{i} }{2 (2 \pi)^{3} E_{i}} ...\frac{g_{l} d^{3} \textbf p_{l}}{2 (2 \pi)^{3} E_{l}} 
&\times & \vert {\mathcal{ M }}\vert ^2 \times f_{i}^{eq} f_{j}^{eq}  \ .
\eea
Given this, the BE reads:
\bea
\dot{n_{i}} + 3\,H(t)\,n_{i} = \sum_{k,l,j}\left( \frac{n_{k}}{n_{k}^{eq}}\frac{n_{l}}{n_{l}^{eq}} \gamma^{kl}_{ij}-\frac{n_{i}}{n_{i}^{eq}}\frac{n_{j}}{n_{j}^{eq}} \gamma^{ij}_{kl}\right) \ .
\eea
\section{Boltzmann equations in leptogenesis}
It is customary in leptogenesis to parametrise number density evolution in function of the variable 
\bea
z\equiv M_{i}/T \ ,
\eea
instead of the time t, where $M$. The relation between those two is
\bea
\frac{dz}{dt}=-\frac{M_{N_1}}{T^2}\frac{dT}{dt}=z\,H(z) \ .
\eea
It is also customary to use the comoving number density $Y_{X}$ that are defined by
\bea
Y_{X}\equiv \frac{n_{X}}{s} \ ,
\eea
where $s$ is the entropy density. Such transformation is done because both $n_{x}$ and $s$ scale as $T^3$, and therefore $Y$ is constant during the cooling of the Universe, up to non-elastic processes.\\
With these two redefinitions, the BE eq.(\ref{BE1}) translates to
\bea
\frac{dY_{i}}{dz}&=&-\frac{1}{s z H(z)} C_{i}(z) \ , \\
C_{i}(z)&=&\sum_{X,j}\left( \frac{Y_{k}}{Y_{k}^{eq}} \frac{Y_{l}}{Y_{l}^{eq}}\gamma^{k l}_{i j}-\frac{Y_{i}}{Y_{i}^{eq}}\frac{Y_{j}}{Y_{j}^{eq}}\gamma^{i j}_{k l} \right) \ .
\eea
This is the general form of the Boltzmann equations for comoving number densities. It is written here for a two-to-two scattering, but the collision term should obviously include all relevant processes of the model under consideration. In this manuscript, we only focus on decays $X\Ra i j$ and on  2-to-2 scatterings.
The density rate for a two body decay is given by: 
\bea
\gamma^{X}_{ij}(z_{x})=n_{x}^{eq}(z_{x})\times \frac{K_{1}(z_{x})}{K_{2}(z_{x})}\,\Gamma(X\Ra i j) \ ,
\eea
where $K_{1,2}$ are the modified Bessel function of the second kind, and $\Gamma(X\Ra i j)$ is the decay rate evaluated in the rest-frame of the decaying particle. For a two body scattering, one has:
\bea
\gamma^{X Y}_{k l}(z)=\frac{1}{64\pi^{4}}\frac{s_{min}^{2}}{z_{m}}\int_{1}^{\infty}dx\,\sqrt{x}\,K_{1}(\sqrt{x}\,z_{m}) \hat{\sigma}(x) \ ,
\eea
where $s_{min}= \rm{max}\left( (m_{X}+m_{Y})^2,(m_{k}+m_{l})^2\right)$, $z_{m}=\sqrt{s_{min}}/T$ and $\widehat{\sigma}(x)$ is the reduced cross-section, the cross section summed over initial phase space:
\bea
\widehat{\sigma}(s)=\frac{2 \lambda(s,m_{X}^2,m_{Y}^2)}{s}\sigma(s)=\frac{2\left(s-(m_{X}+m_{Y})^2\right)\left(s-(m_{X}-m_{Y})^2\right)}{s} \sigma(s) \ .
\eea
\subsection*{Decays/inverse decays and top-scatterings}
Let us discuss the different terms that we consider in this thesis.\\
At first order, we included decays and inverse, that are given by:
\bea
\kappa_{1}\,D(z)&\equiv &\frac{z}{s\,H(M_{1})}\,\frac{\gamma^{N}_{\ell\phi}}{Y_{N}^{eq}(z)}=\kappa_{1}\,z\,\frac{K_{1}(z)}{K_{2}(z)}\, , \nonumber \\
\kappa_{1}\,W_{id}(z)&=&\frac{\kappa_{1}}{2}\,\frac{Y_{N}^{eq}(z)}{Y_{\ell}^{eq}(z)}\,D(z)=\frac{\kappa_{1}}{4}\,z^{3}\,K_{1}(z)\, . 
\eea 
where $\kappa_{1}$ (denoted $K_{1}$ in the thesis), is the wash-out parameter, 
\bea
\kappa_{1}=\frac{\Gamma(N\Ra\ell)}{H(M_{1})}\, .
\eea
Given the high temperature expansion
\bea
K_{2}(z\ll 1)&\simeq &\frac{2}{z}\,K_{1}(z\ll 1)\simeq  \frac{2}{z^{2}}\, , \nonumber \\
\eea
we have that $D(z \ll 1)\simeq 2\,W_{id}(z \ll 1)z^{2}/2 \ll 1$: decays and inverse decays are highly suppressed. In the opposite low-temperature regime, since
\bea
K_{1}(z\gg 1)&\simeq &\frac{e^{-z}}{\sqrt{z}}\sqrt{\frac{\pi }{2}}\left(1 +\frac{3}{8\,z}\right) \, , \nonumber \\
K_{2}(z \gg 1)&\simeq& \frac{e^{-z}}{\sqrt{z}}\sqrt{\frac{\pi }{2}}\left(1 +\frac{15}{8\,z}\right)
\eea
we obtains
\bea
D(z\gg 1)&\simeq & z-\frac{3}{2}\, , \nonumber \\
W_{id}(z) & \simeq & \frac{e^{-z}}{4}\,\sqrt{\frac{\pi
   }{2}}\,z^{5/2}\left(1+\frac{3}{8\,z}\right)\, .
\eea
Hence, while decays increase with $z$, the inverse decays get Boltzmann suppressed at low temperature. The latter reach their maximum for $z\simeq 2.4$, at $W_{id}(z)\simeq 0.29 $.\\
We also include $\D L=1$ neutrino-top scatterings, which are mediated by a Higgs field in the $s-$ and $t-$ channels, as seen in chapter 3.\\
The reduced cross-section for these scatterings, involving a right-handed neutrino $N_{i}$, is given by~\cite{lepto1saveurgenerale}:
\bea
\widehat{\sigma}_{s}(x)=\frac{3\al_{t}}{2}\,\kappa_{i}\,\frac{x^{2}-a_{i}^{2}}{x^{2}}\, ,
\eea
where $x=s/M_{1}^{2}$ is the rescaled energy in the centre-of-mass frame, $a_{i}=M_{i}^{2}/M_{1}^{2}$ and $\al_{t}=h_{t}^{2}/4\pi$.\\
The $t-$channel contribution reads~\cite{lepto1saveurgenerale}:
\bea
\widehat{\sigma}_{s}(x)=3\,\al_{t}\,\kappa_{i}\,\frac{x-a_{i}}{x}\left(\frac{x-2\,a_{i}+2\,a_{h}}{x-a_{i}+a_{h}}+\frac{a_{i}-2\,a_{h}}{x-a_{h}}\,\rm{ln}\left(\frac{x-a_{i}+a_{h}}{a_{h}}\right)\right) ,
\eea
where $a_{h}$ correspond to the rescaled Higgs mass, which is used to regularise an infrared divergence, $a_{h}=(\mu/M_{1})^{2}\simeq 10^{-16}$.\\
At high temperatures, the reaction density scales as $T^{4}$, and one has~\cite{pedestrians}:
\bea
\frac{z}{s\,H(M_{1})}\frac{\gamma_{s,t}}{Y_{N}^{eq}}\stackrel{z \ll 1}{\simeq} \frac{9}{4\pi^{2}}\frac{m_{t}^{2}}{2\,v^{2}}\simeq 0.1 \ , 
\eea
and $D(z)+S(z)\simeq S(z)\simeq 0.1$, $z\ll 1$.\\
In the opposite regime of low temperatures, $D(z)+S(z)\simeq D(z)\simeq z$.

\newpage
%%%%%%%%%%%%%%%%%%%%%%%%%%%%%%%%%%%%%%%%%%%%%%%%%%%%%%%%%%%%%%%%%%%%%%%%%%%%%%%%%%%%%%%%%%%%%%%%%%%%%%%%%%%%%%%%%%%%%%%%%%%%%%%%%%%%%%%%%%%%%%%%%%%%%%%%%%%%%%%%%%
\footnotesize\newpage
\renewcommand{\textwidth}{16cm}
\renewcommand{\evensidemargin}{0cm}
\renewcommand{\oddsidemargin}{0cm}
\renewcommand{\baselinestretch}{1}
\begin{multicols}{2}

\end{multicols}
\newpage
\end{document}